\documentclass[12pt]{article}    
\usepackage{latexsym} 
\usepackage{amssymb}  
\usepackage{amsfonts} 
\usepackage{multirow}
\usepackage{epsf}
\usepackage{epsfig}
\usepackage{rotate}

\newcount\Mac  \Mac=0 

\def\rfigure#1{\if@twocolumn\local@ft{figure}{#1}\else\side@ft{figure}{#1}\fi}
\def\endrfigure{\endrside@ft}

\def\Red{}
\def\Black{}
\def\Blue{}

\newcommand{\lascia}[1]{}
\def\puttag(#1,#2)#3{\put(#1,#2){\makebox(0,0){\rm\Blue #3\Black}}}

\def\circa#1{\,\raise.3ex\hbox{$#1$\kern-.75em\lower1ex\hbox{$\sim$}}\,}

\def\putps(#1,#2)(#3,#4)#5#6{\ifnum\Mac=1 \put(#1,#2){\special{picture #5}}
\else  \put(#3,#4){\includegraphics{#6}} \fi}
\def\One{\hbox{1\kern-.24em I}}

\makeatletter
\def\art{\@ifnextchar[{\eart}{\oart}}
\def\eart[#1]#2#3#4#5#6{{\rm #2}, {\e, #3 \bf #4} {\rm (#6) #5} ({\em #1})}
\def\hepart[#1]#2{{\rm #2, \em#1}}
\newcommand{\oart}[5]{{\rm #1}, {\em #2 \bf #3} {\rm (#5) #4}}
\newcounter{alphaequation}[equation]
\def\thealphaequation{\theequation\hbox to
0.6em{\hfil\alph{alphaequation}\hfil}}
\def\eqnsystem#1{
\def\@eqnnum{{\rm (\thealphaequation)}}
\def\@@eqncr{\let\@tempa\relax \ifcase\@eqcnt \def\@tempa{& & &} \or
  \def\@tempa{& &}\or \def\@tempa{&}\fi\@tempa
  \if@eqnsw\@eqnnum\refstepcounter{alphaequation}\fi
\global\@eqnswtrue\global\@eqcnt=0\cr}
\refstepcounter{equation} \let\@currentlabel\theequation \def\@tempb{#1}
\ifx\@tempb\empty\else\label{#1}\fi
\refstepcounter{alphaequation}
\let\@currentlabel\thealphaequation
\global\@eqnswtrue\global\@eqcnt=0 \tabskip\@centering\let\\=\@eqncr
$$\halign to \displaywidth\bgroup \@eqnsel\hskip\@centering
$\displaystyle\tabskip\z@{##}$&\global\@eqcnt\@ne
\hskip2\arraycolsep\hfil${##}$\hfil& \global\@eqcnt\tw@\hskip2\arraycolsep
$\displaystyle\tabskip\z@{##}$\hfil
\tabskip\@centering&\llap{##}\tabskip\z@\cr}

\def\endeqnsystem{\@@eqncr\egroup$$\global\@ignoretrue} \makeatother

\makeatletter
\def\hdashline#1(#2){\leavevmode\hbox to \z@{\baselineskip \z@%
\lineskip \z@%
\@dashdim=#2\unitlength%
\@dashcnt=\@dashdim \advance\@dashcnt 200
\@dashdim=#1\unitlength\divide\@dashcnt \@dashdim
\ifodd\@dashcnt\@dashdim=\z@%
\advance\@dashcnt \@ne \divide\@dashcnt \tw@ 
\else \divide\@dashdim \tw@ \divide\@dashcnt \tw@
\advance\@dashcnt \m@ne
\setbox\@dashbox=\hbox{\vrule \@height \@halfwidth \@depth \@halfwidth
\@width \@dashdim}\put(0,0){\copy\@dashbox}%
\put(#2,0){\hskip-\@dashdim\copy\@dashbox}%
\multiply\@dashdim 3 
\fi
\setbox\@dashbox=\hbox{\vrule \@height \@halfwidth \@depth \@halfwidth
\@width #1\unitlength\hskip #1\unitlength}\@tempcnta=0
\put(0,0){\hskip\@dashdim \@whilenum \@tempcnta <\@dashcnt
\do{\copy\@dashbox\advance\@tempcnta \@ne }}\@tempcnta=0
}\@makepicbox(#2,0)}

\def\vdashline#1(#2){\leavevmode\hbox to \z@{\baselineskip \z@%
\lineskip \z@%
\@dashdim=#2\unitlength%
\@dashcnt=\@dashdim \advance\@dashcnt 200
\@dashdim=#1\unitlength\divide\@dashcnt \@dashdim
\ifodd\@dashcnt \@dashdim=\z@%
\advance\@dashcnt \@ne \divide\@dashcnt \tw@
\else
\divide\@dashdim \tw@ \divide\@dashcnt \tw@
\advance\@dashcnt \m@ne
\setbox\@dashbox\hbox{\hskip -\@halfwidth
\vrule \@width \@wholewidth 
\@height \@dashdim}\put(0,0){\copy\@dashbox}%
\put(0,#2){\lower\@dashdim\copy\@dashbox}%
\multiply\@dashdim 3
\fi
\setbox\@dashbox\hbox{\vrule \@width \@wholewidth 
\@height #1\unitlength}\@tempcnta0
\put(0,0){\hskip -\@halfwidth \vbox{\@whilenum \@tempcnta < \@dashcnt
\do{\vskip #1\unitlength\copy\@dashbox\advance\@tempcnta \@ne }%
\vskip\@dashdim}}\@tempcnta0
}\@makepicbox(0,#2)}
\newdimen\sideftwd
\newbox\local@box\newbox\local@hbox

\def\lfigure#1{\if@twocolumn\local@ft{figure}{#1}\else\side@ft{figure}{#1}\fi}
\def\rfigure#1{\if@twocolumn\local@ft{figure}{#1}\else\side@ft{figure}{#1}\fi}
\def\ltable#1{\if@twocolumn\local@ft{table}{#1}\else\side@ft{table}{#1}\fi}
\def\rtable#1{\if@twocolumn\local@ft{table}{#1}\else\side@ft{table}{#1}\fi}
\def\endlfigure{\endlside@ft}
\def\endrfigure{\endrside@ft}
\def\endltable{\endlside@ft}
\def\endrtable{\endrside@ft}
\def\side@ft#1#2{\par
\sideftwd=#2
\def\@captype{#1}
\setbox\@tempboxa\vtop\bgroup\textwidth=\sideftwd
\columnwidth=\sideftwd \hsize\columnwidth
\@parboxrestore}
\def\endlside@ft{\egroup
\@tempdima=\ht\@tempboxa
\advance\@tempdima by \dp\@tempboxa
\@tempcnta=\@tempdima
\divide\@tempcnta by \baselineskip
\advance\@tempcnta by 2
\global\hangindent\sideftwd
\global\hangafter-\@tempcnta
\noindent \dp\@tempboxa=\z@ \ht\@tempboxa=\z@
\hbox to \z@{
\hbox to \z@{\hss\box\@tempboxa}\hss}%
\hskip\parindent
\global\@ignoretrue}
\def\endrside@ft{\egroup
\@tempdima=\ht\@tempboxa
\advance\@tempdima by \dp\@tempboxa
\@tempcnta=\@tempdima
\divide\@tempcnta by \baselineskip
\advance\@tempcnta by 2
\global\hangindent-\sideftwd
\global\hangafter-\@tempcnta
\noindent \dp\@tempboxa=\z@ \ht\@tempboxa=\z@
\hbox to \z@{\hskip\textwidth
\hbox to \z@{\hss\box\@tempboxa}\hss}%
\hskip\parindent
\global\@ignoretrue}
\def\local@ft#1{\def\@captype{#1}
\setbox\local@box\vbox\bgroup
\boxmaxdepth\z@\hsize0.9\columnwidth}
\def\endlocal@ft{\egroup
\[\hbox{\lower1ex\box\local@box}\]
\global\@ignoretrue}
\def\localfigure{\local@ft{figure}}
\def\localtable{\local@ft{table}}
\def\endlocalfigure{\endlocal@ft}
\def\endlocaltable{\endlocal@ft}
\makeatother

\newcommand{\lbzt}{{\lambda}_R (T)}
\newcommand{\rrz}{\rho_0}

\newcommand{\rrzkt}{\rho_0(k,T)}
\newcommand{\rrzt}{\rho_0(T)}
\newcommand{\lb}{{\lambda}}

\newcommand{\mbkt}{{m}^2(k,T)}

\newcommand{\mbzt}{{m}^2_R (T)}
\newcommand{\smbzt}{{m}_R (T)}

\newcommand{\rt}{\tilde{r}}
\newcommand{\Ut}{\tilde{U}}

\newcommand{\At}{\tilde{A}}
\newcommand{\St}{\tilde{S}}
\newcommand{\Op}{{\cal W}}
\newcommand{\phibounce}{\phi_b}

\newcommand{\pha}{\phi_1}
\newcommand{\phb}{\phi_2}

\newcommand{\tcr}{T_{cr}}

\newcommand{\phit}{\tilde{\phi}}

\newcommand{\lx}{\lambda}

\newcommand{\ks}{k_s}

\newcommand{\lt}{\tilde{\lambda}}

\newcommand{\kx}{\kappa}
\newcommand{\kt}{\tilde{\kappa}}

\newcommand{\Gammak}{\Gamma_k}

\def \lta {\mathrel{\vcenter
     {\hbox{$<$}\nointerlineskip\hbox{$\sim$}}}}
\def \gta {\mathrel{\vcenter
     {\hbox{$>$}\nointerlineskip\hbox{$\sim$}}}}
\def\VEV#1{\left\langle #1\right\rangle}

\newcommand{\Ga}{\Gamma}
\newcommand{\de}{\delta}
\newcommand{\De}{\Delta}

\newcommand{\eps}{\epsilon}

\newcommand{\la}{\lambda}

\newcommand{\si}{\sigma}
\newcommand{\Si}{\Sigma}
\renewcommand{\th}{\theta}   

\newcommand{\txt}{\textstyle}
\newcommand{\dsp}{\displaystyle}

\newcommand{\e}{ {\rm e} }
\newcommand{\beq}{\begin{equation}}
\newcommand{\eeq}{\end{equation}}
\newcommand{\ba}{\begin{array}}
\newcommand{\bea}{\begin{eqnarray}}
\newcommand{\ea}{\end{array}}
\newcommand{\eea}{\end{eqnarray}}

\newcommand{\sslash}{\!\!\!/\,}

\newcommand{\ol}{\overline}

\newcommand\comment[1]{ \hbox{[{\it Comment suppressed here.}\/]} }
\newcommand\hide[1]{}

\newcommand{\skipover}[1]{}

\newcommand{\nnn} {\nonumber \vspace{.2cm} \\ }
\newcommand{\half} {{\txt {1\over 2}}}

\renewcommand{\O}{ {\cal O} }
\newcommand{\tr}{\hbox{tr}}
\newcommand{\Tr}{\hbox{Tr}}

\newcommand{\MeV}{{\rm MeV}}

\def\ZZZ{Z\kern -0.31em Z}

\def\ijmpc#1{Int. J. Mod. Phys. {\bf C#1}}

\def\npb#1{Nucl. Phys. {\bf B#1}}

\def\plb#1{Phys. Lett. {\bf B#1}}

\def\prb#1{Phys. Rev. {\bf B#1 }}

\def\prd#1{Phys. Rev. {\bf D#1 }}
\def\prle#1{Phys. Rev. Lett. {\bf #1}}

\def\zpc#1{Z. Phys. {\bf C#1}}
\def\jpa#1{J.\ Phys.\ {\bf A#1}}

\def\CMP#1{Comm. Math. Phys.~{\bf #1}}
\def\CNPP#1{Comm. Nucl. Part. Phys.~{\bf #1}}
\def\HPA#1{Helv. Phys. Acta~{\bf #1}}
\def\IJMP#1{Int. J. Mod. Phys.~{\bf #1}}

\def\MPL#1{Mod. Phys. Lett.~{\bf #1}}
\def\NP#1{Nucl. Phys.~{\bf #1}}

\def\PL#1{Phys. Lett.~{\bf #1}}
\def\PR#1{Phys. Rev.~{\bf #1}}
\def\PRP#1{Phys. Rep.~{\bf #1}}
\def\PRL#1{Phys. Rev. Lett.~{\bf #1}}
\def\PNAS#1{Proc. Nat. Acad. Sc.~{\bf #1}}
\def\PTP#1{Progr. Theor. Phys.~{\bf #1}}
\def\RMP#1{Rev. Mod. Phys.~{\bf #1}}

\def\ZP#1{Z. Phys.~{\bf #1}}

\def\appendix{\par                              
    \setcounter{section}{0}                     
    \setcounter{subsection}{0}
    \renewcommand{\theequation}{\Alph{section}.\arabic{equation}}
    \renewcommand{\thesection}{Appendix \Alph{section}
\setcounter{equation}{0}}
}

\catcode`\@=11

\def\applabel#1{\@bsphack
  \protected@write\@auxout{}%
         {\string\newlabel{#1}{{\Alph{section}}{\thepage}}}%
  \@esphack}

\def\section{
\setcounter{equation}{0}        
\@startsection {section}{1}{\z@}{-3.5ex plus -1ex minus 
 -.2ex}{2.3ex plus .2ex}{\large\bf}}
\renewcommand{\theequation}{\arabic{section}.\arabic{equation}}

\def\subsection{\@startsection{subsection}{2}{\z@}{-3.25ex plus -1ex minus 
 -.2ex}{1.5ex plus .2ex}{\normalsize\bf}}

\def\subsubsection{\@startsection{subsubsection}{3}{\z@}{-3.25ex plus
 -1ex minus -.2ex}{1.5ex plus .2ex}{\normalsize}}

\newcommand{\hal} {\frac{1}{2}}

\begin{document}
\setlength{\baselineskip}{16pt}

\title{\bf \Large Non-Perturbative Renormalization Flow \\
in Quantum Field Theory and Statistical Physics}

\author{J{\"u}rgen Berges\footnote{Email: {\tt Berges@ctp.mit.edu}}\\
{\normalsize \it Center for Theoretical Physics, 
Massachusetts Institute of Technology}\\
{\normalsize \it Cambridge, Massachusetts 02139, U.S.A.}\\[1.5ex]
Nikolaos Tetradis\footnote{Email: {\tt Tetradis@cibs.sns.it}}\\
{\normalsize \it Scuola Normale Superiore, 56126 Pisa, Italy {\rm and}}\\
{\normalsize \it Nuclear and Particle Physics Section, University of Athens}\\
{\normalsize \it 15771 Athens, Greece}\\[1.5ex]
Christof Wetterich\footnote{Email: 
{\tt C.Wetterich@thphys.uni-heidelberg.de}}\\
{\normalsize \it Institut f{\"u}r Theoretische Physik,
Universit{\"a}t Heidelberg}\\
{\normalsize \it 69120 Heidelberg, Germany}}

\date{}
\maketitle

\thispagestyle{empty}

\begin{abstract}
We review the use of an exact renormalization group equation in quantum
field theory and statistical physics. It describes the dependence of the 
free energy on an infrared cutoff for the quantum or thermal fluctuations. 
Non-perturbative solutions follow from approximations to the general
form of the coarse-grained free energy or effective average action.
They interpolate between the microphysical laws and the complex 
macroscopic phenomena. Our approach yields a simple unified description 
for $O(N)$-symmetric scalar models in two, three or four dimensions, 
covering in particular the critical phenomena for the second-order 
phase transitions, including the Kosterlitz-Thouless transition and the
critical behavior of polymer chains. We compute the aspects of the 
critical equation of state which are universal for a large variety of 
physical systems and establish a direct connection between microphysical 
and critical quantities for a liquid-gas transition. Universal
features of first-order phase transitions are studied in the context of scalar 
matrix models. We show that the quantitative treatment of coarse graining is 
essential for a detailed estimate of the nucleation rate.  
We discuss quantum statistics in thermal equilibrium or 
thermal quantum field theory with fermions and bosons
and we describe the high temperature symmetry restoration in quantum
field theories with spontaneous symmetry breaking.
In particular we explore chiral symmetry breaking and the high temperature
or high density chiral phase transition in quantum chromodynamics using 
models with effective four-fermion interactions. 

\medskip\noindent
This work is dedicated to the 60th birthday of Franz Wegner.

\end{abstract}

\vspace*{\fill}


\newpage

\setcounter{page}{1}
\vspace*{-2.cm}
\tableofcontents

\newpage


\newpage
\thispagestyle{empty}

\section{Introduction}
\subsection{From simplicity to complexity}

A few fundamental microscopic interactions govern the complexity
of the world around us. The standard model of electroweak and strong
interactions combined with gravity is a triumph for the way of
unification and abstraction. Even though some insufficiencies
become apparent in astrophysical and cosmological observations
-- the oscillations of neutrinos, the missing dark matter and
the need for cosmological inflation and baryogenesis -- there is little
doubt that no further basic interactions are needed for an understanding
of ``everyday physics''. For most phenomena the relevant
basic interactions are even reduced to electromagnetism and gravity.
Still, for many common observations there is a long way to
go before quantitative computations and predictions from the
microscopic  laws become feasible. How to go back from the
simplicity of microphysics to the complexity of macrophysics?

We will deal here only with very simple systems of many particles,
like pure water or vapor, where the interactions between molecules
are reasonably well understood. We also concentrate on the most
simple situations like thermal equilibrium. Concerning dynamics
we only touch on properties that can be calculated in equilibrium,
whereas we omit the complicated questions of
the time evolution of statistical systems. Nevertheless, it remains
a hard task to compute quantitatively such simple things as the
phase transition from water to vapor, starting from the
well-known microscopic interactions. How can we calculate the
pressure dependence of the transition temperature from atomic
physics or the  van der Waals interactions between molecules?
How do the optical properties change as we approach the endpoint
of the critical line? What would be the rate of formation of
vapor bubbles if we heat extremely pure water in space at
a given temperature $T$ slightly above the critical temperature $T_c$?
Similar questions may be asked about the temperature dependence
of the density of superfluid $He^4$ or the magnetization in
ferromagnets. One often  takes a higher level of abstraction
and asks for the properties of simplified theoretical models, like the
two-dimensional Hubbard model, which is widely believed to describe
high $T_c$-superconductivity, or the Heisenberg model for the
description of ferromagnetism or antiferromagnetism. Despite the
considerable simplifications of these models as compared to real
physics they remain very difficult to solve.

Common to all these questions is the role of fluctuations in
statistical many-body systems. Statistical physics and
thermodynamics offer a powerful framework for the macroscopic
behavior of systems with a very large number of degrees of freedom.
The statistical treatment makes the predictions about the behavior
of stationary many-body systems independent of many irrelevant details
of the microscopic dynamics. Nevertheless, one needs to bridge the
gap between the known microscopic interactions and the thermodynamic
potentials and similar quantities which embody the effective
macroscopic laws. This is the way to complexity.

For a thermodynamic equilibrium system of many identical
microscopic degrees of freedom the origin of the problems on the way to
complexity is threefold. First, there is often no small parameter
which can be used for a systematic perturbative expansion. Second,
the correlation length can be substantially larger than the
characteristic distance between the microscopic objects. Collective
effects become important. Finally, the relevant degrees of freedom which
permit a simple formulation of the macroscopic laws may be different
from the microscopic ones. A universal theoretical method for the transition
from micro- to macrophysics should be able to cope with these
generic features. We propose here a version of non-perturbative
flow equations based on an exact renormalization group equation.
This theoretical tool acts like a microscope with variable
resolution. The way from a high-resolution picture of an effectively
small piece of surface or volume (microphysics) 
to a rough resolution for large volumes (macrophysics) 
is done stepwise, where every new step in the resolution
only uses information form the previous step \cite{Kad66-1}--\cite{Has86-1}. 
We use here a
formulation in terms of the effective average action 
\cite{Wet91-1,Wet93-2}, which
permits non-perturbative approximations for practical computations.
Our method may be considered as an analytical counterpart of the
often successful numerical simulation techniques.

Modern particle physics is confronted with precisely the same problems
on the way from the beautiful simplicity of fundamental interactions
to a ``macroscopic'' description. Basically, only the relevant length
scales are different. Whereas statistical physics may have to
translate from Angstr\"oms to micrometers, particle physics
must build a bridge from the Fermi scale ($\sim (100\ {\rm GeV})^{-1}$)
to a fermi (1 fm=$10^{-15}$ m=(197.33 MeV)$^{-1}$). For interactions
with small couplings, particle physics has mastered this transition
as far as the vacuum properties are concerned -- this includes the
dynamics of the excitations, e.g. the particles. The 
perturbative renormalization
group \cite{pertrg}
interpolates between the Fermi scale and the electron mass\footnote{We
employ the word scale for distance as well as momentum or mass scales.
In our units $\hbar=c=k_B=1$ they are simply the inverse of each other.}
or even between a grand unification scale $\sim 10^{16}$ GeV
and the Fermi scale. The dynamics of electrons, positrons and
photons in vacuum can be predicted with extremely high accuracy.
This extends to weak interactions between leptons.

The situation is very different for strong interactions. The
running gauge coupling grows large at a scale below 1 GeV and the
generic problems of statistical physics reappear. Whereas quantum
electrodynamics may be compared with a dilute gas of weakly
interacting molecules, strong interactions are analogous to dense
systems with a large correlation length. At microscopic distances,
e.g. the Fermi scale, quantum chromodynamics (QCD) can give precise
predictions in terms of only one gauge coupling and the particle
masses. At the ``macroscopic scale'' of around 1 fm
numerical simulations approach only
slowly the computation of the masses and interactions of the
relevant degrees of freedom, namely baryons and mesons, and no
analytical method has achieved this goal yet.

The smallness of microscopic couplings is no guarantee for a simple
transition to ``macrophysics''. The vacuum fluctuations may
enhance considerably the relevant effective coupling at longer
distances, as in the case of QCD. In the presence of thermal fluctuations 
a similar phenomenon
happens even for the electroweak interactions. In fact, at high
temperature the electroweak interactions have a large effective
gauge coupling and show all properties of a strongly interacting
model, very similar to QCD \cite{RW93-2,W-Sintra}. The evidence
for this behavior is striking if one looks at the recently computed
spectrum of quasiparticles for the standard model at high
temperature \cite{Philipsen}. In the language of statistical physics
the hot plasma in thermal equilibrium is dense at sufficiently
high temperature, with a correlation length (typically given by the
inverse magnetic mass of the gauge bosons) substantially larger
than the inverse temperature.

The high-temperature behaviour of strong or electroweak
interactions has attracted much interest recently. It is
relevant for the hot plasma in the very early universe. Possible
phase transitions may even have left observable ``relics'' behind 
that could serve as observational tests of cosmology before
nucleosynthesis. From a statistical physics viewpoint the particle
physics systems are extremely pure -- no dirt, dotation with
other atoms, impurities, seeds of nucleation, gravitational effects
etc.\ need to be considered. In principle, tests of particle
physics in thermodynamic equilibrium would be ideal experiments
for statistical physics\footnote{The microwave background
radiation provides so far the most precise test of the spectrum
of black-body radiation.}. Unfortunately, these ``ideal
experiments'' are difficult to perform -- it is not easy
to prepare a high-temperature plasma of particles in equilibrium
in a laboratory. Nevertheless, an impressive experimental program
is already under way, aiming at a test of high temperature and
high density QCD and possible phase transitions \cite{QM99}.
This highlights the need of a theoretical understanding of the QCD phase
diagram at high temperature and density, including such interesting
issues as color superconductivity and the possibility
of (multi-) critical points with observable effects
from long-range correlations  
\cite{KrishRev,FrankLec,PhaseDia}. Renormalization
group methods should be an important tool in this attempt.

From a theoretical point of view there is actually no difference
between thermal quantum field theory and quantum-statistical
systems. In a modern language, both are formulated as
Euclidean functional integrals with ``Euclidean time'' wrapped
around a torus with circumference $T^{-1}$. (For dynamical
questions beyond the equilibrium properties the time coordinate
has to be analytically continued to real Minkowski time.) The only special
features of the particle physics systems are the very precisely
known microscopic interactions with their high degree of symmetry in
the limit of vanishing temperature. This concerns, in particular,
the Lorentz symmetry or its Euclidean counterpart of four-dimensional
rotations and Osterwalder-Schrader positivity \cite{OS}. In this line of
thought the recent high precision numerical simulations of the
high temperature electroweak interactions \cite{Laineetal,Lainecrossover}
can be considered as a fine example of a quantitative (theoretical)
transition from microphysics to macrophysics, despite the presence
of strong effective interactions and a large correlation length. They have
confirmed that the high temperature first-order phase transition
which would be
present in the standard model with modified masses turns into
a crossover for realistic masses, as has been suggested earlier by
analytical methods \cite{RW93-2, W-Sintra, Buchphil, EW-rep}.

Beyond the identical conceptual setting of particle 
and statistical physics there is also quantitative agreement for certain
questions. The critical exponents at the endpoint of the critical
line of the electroweak phase transition (the onset of crossover)
are believed to be precisely the same as for the liquid-gas
transition near the critical pressure or for magnetic transitions
in the Ising universality class\footnote{Numerical simulations
\cite{EWUC} are consistent with this picture.}. This reveals
universality as a powerful feature for the transition to complexity.
Indeed, the transition to macrophysics does not involve only complications.
For certain questions it can also bring enormous simplifications.
Due to partial fixed points in the renormalization flow of effective
couplings between short and long distances much of the microscopic
details can be lost. If the fluctuation effects are strong enough,
the long-distance behavior loses memory of the microscopic
details of the model. This is the reason why certain features
of high temperature QCD may be testable in magnets or similar
statistical systems. For example, it is possible that the
temperature and density dependence of the chiral condensate in QCD can be
approximated in a certain range by the critical
equation of state of the $O(4)$ Heisenberg model 
\cite{PW84-1}-\cite{BJW97-1} or by the Ising model 
at a nonzero density critical endpoint \cite{BerRaj99,HJSSV}. 

Exact renormalization group equations  describe the scale dependence of
some type of ``effective action''. In this context an effective
action is a functional of fields from which the physical properties at a
given length or momentum scale can be computed. The exact
equations can be derived as formal identities from the
functional integral which defines the theory. They are cast in the
form of functional differential equations.
Different versions of exact renormalization
group equations have already a long history \cite{Kad66-1}--\cite{Has86-1}.
The investigation
of the generic features of their solutions has led to a deep
insight into the nature of renormalizability. In particle
physics, the discussion of perturbative solutions has led
to a new proof of perturbative renormalizability of the $\phi^4$-theory
\cite{Pol84-1}. Nevertheless, the application of exact renormalization group
methods to non-perturbative situations has been hindered for a long time 
by the complexity of functional differential equations. First considerable
progress for the description of critical phenomena has been achieved
using the scaling-field method \cite{GRN}. In this approach the exact
renormalization group equation is transformed into an infinite hierarchy of
nonlinear ordinary differential equations for scaling fields \cite{WegSF}.
Estimates for non-trivial critical exponents and scaling functions, e.g. for  
three-dimensional scalar $O(N)$-models, are obtained from
a truncated expansion around the trivial (Gaussian) fixed point 
and certain balance assumptions constraining what operators to include
in the approximation \cite{GRN}. Another very fruitful approach
is based on evaluating the effective action 
functional for constant fields and neglecting all non-trivial
momentum dependencies. This so-called local potential approximation,
originally considered in \cite{NCS1}, was first employed in 
\cite{PaRe,Has86-1} to compute critical exponents for three-dimensional 
scalar $O(N)$-models. Unfortunately, the formulation used in that work could 
not be used for a systematic inclusion of the neglected momentum
dependencies. Some type of expansion is needed, however, if one wants to
exploit the exactness of the functional differential equation
in practice -- otherwise any reasonable guess of a realistic
renormalization group equation does as well.

Since exact solutions to functional differential equations
seem only possible for some limiting
cases\footnote{An example is the $O(N)$-model for $N\to\infty$
discussed in subsection \ref{largeNexp}.}, it is crucial to find a formulation
which permits non-perturbative approximations. Those proceed
by a truncation of the most general form of the effective
action and therefore need a qualitative understanding of
its properties. The formulation of an exact renormalization group
equation based on the effective average action \cite{Wet91-1,Wet93-2}
has been proven successful in this respect. It is the basis
of the non-perturbative flow equations which we discuss in this
review. The effective average action is the generating functional
of one-particle irreducible correlation functions in presence of an
infrared cutoff scale $k$. Only fluctuations with momenta larger
than $k$ are included in its computation. For $k\to0$
all fluctuations are included and one therefore
obtains the usual effective action from which appropriate masses and
vertices can be read off directly. The $k$ dependence is described
by an exact renormalization group equation that closely resembles
a renormalization group improved one-loop equation \cite{Wet93-2}. 
In fact, the
transition from classical propagators and vertices to effective
propagators and vertices transforms the one-loop expression into
an exact result. This close connection to perturbation theory for
which we have intuitive understanding is an important key for devising
non-perturbative approximations. Furthermore, the one-loop expression
is manifestly infrared and ultraviolet finite and can be used
directly in arbitrary number of dimensions, 
even in the presence of massless modes.

The aim of this report is to show that this version of flow equations
can be used in practice for the transition from microphysics
to macrophysics, even in the presence of strong couplings and a large
correlation length. We derive the exact renormalization group
equation and various non-perturbative truncations in section 
\ref{nonpertfloweq}. In particular, we demonstrate in section 2.5
that already an extremely simple truncation gives a unified picture
of the phase transitions in $O(N)$-symmetric scalar theories in
arbitrary dimensions $d$, including the critical exponents for $d=3$
and the Kosterlitz-Thouless phase transition for $d=2$, $N=2$. In
section 3 we discuss the solutions to the flow equation in more detail.
We present an exact solution for the limit $N\to\infty$. We also
propose a renormalization group-improved perturbation theory as an
iterative solution. We show how the effective potential becomes
convex in the limit $k\to 0$ in case of spontaneous symmetry-breaking
where the perturbative potential is nonconvex. Section 4 discusses
the universality class of $O(N)$-symmetric scalar models in more
detail. We derive the universal critical equation of state. For the
special example of carbon dioxide we explicitly connect the microphysics
with the macrophysics. In addition to the universal part this
also yields the non-universal critical amplitudes in the
vicinity of the second-order phase transition at the endpoint of the
critical line. After a short discussion of the critical behavior
of polymer chains we turn to the Kosterlitz-Thouless phase
transition for two-dimensional models with a continuous abelian
symmetry.

First-oder phase transitions are discussed in section 4.6 and section 5.
The matrix model investigated in section 5 gives an example of a
radiatively induced first-order transition as it is also characteristic
for the abelian Higgs model relevant for low $T_c$ superconductivity.
We discuss under which conditions first-order transitions are
characterized by universal behavior. In particular, we  present a universal
critical equation of state for first order transitions which involves two
scaling variables and we discuss its range of applicability.
 In section 6 we turn to the old problem
of spontaneous nucleation in first-order transitions. We show that
a detailed understanding of coarse graining is crucial for a quantitative
computation of the nucleation rate. 
We also discuss the limits of validity of spontaneous nucleation theory,
in particular for radiatively induced first order transitions.
Our results agree well with recent
numerical simulations. 

Section 7 is devoted to quantum statistics
and quantum field theory in thermal equilibrium. The flow equation
generalizes easily to the Matsubara formalism and one sees that
dimensional reduction arises as a natural consequence. Whereas for
momenta larger than the temperature the (four-dimensional) quantum statistics
are relevant, the momenta below $T$ are governed by classical
(three-dimensional) statistics. Correspondingly, the flow changes
from a four-dimensional to a three-dimensional flow as $k$ crosses
the temperature. This section also contains the flow equation for
fermions. 
We show how the renormalization flow leads to a consistent picture of a
second order phase transition for the high temperature $\phi^4$-quantum 
field theory and introduce the notion of quantum universality.
Finally, section 8 deals with the chiral phase transition
in QCD.
We discuss both the high temperature and the high density chiral phase
transition within effective fermionic models with multi-quark
interactions. In particular, we relate the universal critical behavior
at the high temperature phase transition or crossover to particle masses
and decay constants in the vacuum for QCD with two light quarks.
After reading section 2 all sections are essentially
self-contained, with the exception of section 8 relying on results
from section 7.
Section 3 contains some more advanced topics that are not mandatory for
a first understanding of the concrete models in sections 4-8. In the
second part of the introduction we briefly review the basics of the
fluctuation problem in statistical physics and quantum field
theory. This section may be skipped by the experienced reader.

Several results in statistical physics and particle physics have been
obtained first with the method presented here. This includes
the universal critical equation of state for spontaneous breaking of
a continuous symmetry in Heisenberg models \cite{BTW96-1}, 
the universal critical
equation of state for first-order phase transitions in matrix models 
\cite{BW97-1},
the non-universal critical amplitudes with an explicit connection
of the critical behavior to microphysics $(CO_2)$ \cite{B}, a
quantitatively reliable estimate of the rate of spontaneous
nucleation \cite{first,second}, 
a classification of all possible fixed points for
(one component) scalar theories in two and three dimensions in
case of a weak momentum dependence of the interactions 
\cite{Mortwod},
the second-order phase transition in the high temperature
$\varphi^4$ quantum field theory \cite{TW93-1}, 
the phase diagram for the abelian
Higgs model for $N$ charged scalar fields \cite{BLL95-1,Tet96-1}, 
the prediction that the
electroweak interactions become strong at high temperature,
with the suggestion that the standard model may show a crossover
instead of a phase transition \cite{RW93-2,W-Sintra}; 
in strong interaction physics
the interpolation between the low temperature chiral perturbation
theory results and the high temperature critical behavior for 
two light quarks in an effective model \cite{BJW97-1}.  
All these results are in the non-perturbative domain. In addition,
the approach has been tested successfully through a comparison with
known high precision results on critical exponents and universal
critical amplitude ratios and effective couplings.

Our main conclusion can already be drawn at this place: the
method works in practice. If needed and with sufficient
effort high precision results can be obtained in many
non-perturbative situations. New phenomena become accessible 
to a reliable analytical treatment. 

We do not attempt to give an overview over all relevant results.
Rather we concentrate on a systematic development which should
enable the reader to employ the method by himself. For an extensive
review on work for scalar field theories and a comprehensive
reference list we refer the reader to 
C.\ Bagnuls and C.\ Bervillier \cite{BB}. For a review 
including the basis and origins of renormalization group ideas 
in statistical physics and condensed matter theory we refer to
M.\ E.\ Fisher \cite{FisherRev}. We have omitted
two important issues: the formulation of the flow equations for gauge
theories \cite{RW93-1,RW93-2}, \cite{Bec96-1}--\cite{FreireLitim} and for 
composite operators \cite{EW94-1}. The latter is important
in order to achieve a change of effective degrees of freedom
during the flow. 

\subsection{Fluctuations and the infrared problem}
\label{intro2}

(This introductory subsection may be skipped by experienced readers.)

\noindent
The basic object in statistical physics is the (canonical) partition
function
\beq\label{NS1}
Z={\rm Tr}\  e^{-\beta H}\eeq
with $H$ the Hamiltonian and $\beta=T^{-1}$. The trace involves
an integration over all microscopic degrees of freedom. For classical
statistics it typically stands for a (generalized) phase space
integral and the classical Hamiltonian is simply a function of the
integration variables. In most circumstances it can be decomposed
as $H=H_1+H_2$ with $H_2$ quadratic in some momentum-type variable
on which $H_1$ does not depend. The Gaussian
integration over the momentum-type
variable is then trivial and usually omitted. We will be concerned
with many-body systems where the remaining degrees of freedom
$\chi_a(\vec x)$ can be associated with points $\vec x$ in space.
For simplicity we discuss in this introduction only a single real
variable $\chi(x)$. The partition function can then be written
in the form of a ``functional integral''
\beq\label{NS2}
Z=\int D\chi\ e^{-S[\chi]},\eeq
where $S=\beta H_1$. If the space points $\vec x$ are on some
discrete lattice with finite volume, the functional
measure is simply the product of integrations at every point
\beq\label{NS3}
\int D\chi\equiv \prod_{\vec x}\int d\chi(\vec x)\eeq
This can be generalized to the limits of continuous space (when
the lattice distance goes to zero) and of infinite volume. In this
review we concentrate on continuous space with the appropriate
translation and rotation symmetries. (Our formalism is, however,
not restricted to this case.) As a typical example one may associate
$\chi(x)$ with a density field $n(x)$ by $\chi(x)=a+b\ n(x)$ and consider
a Hamiltonian containing local and gradient interactions\footnote{The
constants $a,b$ can be chosen such that there is no cubic term
$\sim\chi^3(x)$ and the gradient term has standard normalization.
A linear term will be included below as a ``source'' term.}
\beq\label{NS3a}
S=\int d^3x\left\{\frac{1}{2}\vec\nabla
\chi(x)\vec\nabla\chi(x)+\frac{m^2}{2}\chi^2
(x)+\frac{\lambda}{8}\chi^4(x)\right\}.
\eeq
The mean values or expectation values can be computed as weighted integrals,
e.g.
\beq\label{NS3b}
\langle \chi(x)\chi(y)\rangle
=Z^{-1}\int D\chi\ \chi(x)\chi(y)e^{-S[\chi]}\, .\eeq

A few generalizations are straightforward: In presence of a chemical
potential $\mu$ for some conserved quantity $N$ we will consider
the grand canonical partition function
\beq\label{NS4}
Z={\rm Tr}\ e^{-\beta(H-\mu N)}\eeq
The $\mu$-dependent part can either be included in the
definition of $S=\beta(H_1-\mu N)$ or, if linear in
$\chi$, be treated as a source (see below and subsection \ref{AverAct}). 
For quantum
statistics $H$ is an operator acting on the microphysical states.
Nevertheless, the partition function can again be written as a functional
integral, now in four dimensions. We will discuss this case in section 
\ref{QuantStat}.
Furthermore, the relevant physics may only involve degrees of freedom
on a surface, a line or a single point. Classical statistics is then given
by a $D$-dimensional functional integral and quantum statistics by a
$D+1$ dimensional functional integral, with $D=2,1,0$, respectively.
Particle physics can be derived from a four-dimensional
functional integral, the Feynman path integral in Euclidean space. Except
for the dimensionality, we therefore can treat particle physics and
classical statistical physics on the same footing. There is no
difference\footnote{In the modern view, quantum field theory
is not considered to be valid up to arbitrarily short distances.
Similarly to statistical physics, a given model should be considered
as an effective description for momenta below some ultraviolet
cutoff $\Lambda$. This cutoff typically appears in all momentum
integrals. For the standard model it indicates the onset of new physics
like grand unification or unification with gravity.}
between the formulation of particle physics -- i.e. quantum field theory
-- and the theory of many-body quantum statistical systems, besides the
symmetries particular to particle physics.

The thermodynamic potential
\beq\label{NS5}
W[J]=\ln Z[J]\eeq
is related to an extension of the partition function in the
presence of arbitrary inhomogeneous
``sources'' or ``external fields'' $J(x)$ that multiply
a term linear in $\chi$:
\beq\label{NS6}
Z[J]=\int D\chi\exp\left\{-S[\chi]+\int d^dx\, \chi(x)J(x)\right\}. \eeq
$W$ and $Z$ are functionals of $J(x)$. The (functional) derivatives
of $W$ with
respect to $J(x)$ generate the connected correlation functions,
e.g. the average density or the density-density correlation
\beq\label{NS7}
\frac{\delta W}{\delta J(x)}=\langle\chi(x)\rangle\equiv\varphi(x)\eeq
\beq\label{NS8}
\frac{\delta^2W}{\delta J(x)\delta J(y)}=\langle\chi(x)\chi(y)
\rangle-\langle\chi(x)\rangle\langle\chi(y)\rangle . \eeq
Here the mean values depend on the sources, e.g. $\varphi\equiv\varphi[J]$.
The effective action $\Gamma[\varphi]$ is another thermodynamic potential,
related to $W[J]$ by a Legendre transform
\beq\label{NS9}
\Gamma[\varphi]=-W[J]+\int d^dx\,\varphi(x)J(x),\eeq
with $J[\varphi]$ obtained by the inversion of $\varphi[J]$ from
eq. (\ref{NS7}). $\Gamma[\varphi]$ is easier to compute than
$W[J]$ and the physical observables can be extracted very simply
from it (see section \ref{AverAct}). The effective action can be written
as an implicit functional integral in the presence of a
``background field'' $\varphi$
\beq\label{NS11}
\exp\{-\Gamma[\varphi]\}=\int D\chi'\exp \left\{ -S[\varphi
+\chi']+\int d^dx\frac{\delta\Gamma}{\delta\varphi}(x)\chi'(x) \right\}.\eeq

A perturbative expansion treats the fluctuations $\chi'$ around
the background $\varphi$ in a saddle point approximation. In lowest
order (tree approximation) one has $\Gamma^{(0)}[\varphi]=
S[\varphi]$. For the one-loop order we expand
\bea\label{NS12}
S[\varphi+\chi']&=&S[\varphi]+\int d^dx\frac{\delta S}{\delta\varphi}
(x)\chi'(x)+\nonumber\\
&&\frac{1}{2}\int d^dxd^dy\ \chi'(x)\ S^{(2)}(x,y)\ \chi'(y)+...\eea
with
\beq\label{NS13}
S^{(2)}(x,y)=\frac{\delta^2S}{\delta\varphi(x)\delta\varphi(y)}.
\eeq
The linear terms cancel in this order and one finds from the Gaussian
integral
\beq\label{NS14}
\Gamma[\varphi]=S[\varphi]+\frac{1}{2}\ \Tr \ln\ S^{(2)}[\varphi]+...\eeq
For constant $\varphi$ this yields the one-loop effective potential
$U_0=\Gamma/V_d$
\bea\label{NS15}
U_0(\varphi)&=&\frac{1}{2}m^2\varphi^2+\frac{1}{8}
\lambda\varphi^4+U_0^{(1)}(\varphi)+...\nonumber\\
U_0^{(1)}(\varphi)&=&\frac{1}{2}\int \frac{d^dq}{(2\pi)^d}\ \ln
\left( q^2+m^2+\frac{3}{2}\lambda\varphi^2 \right), \eea
with $V_d=\int d^dx$. A typical Landau theory results from
the assumption that the fluctuation effects induce a change of
the ``couplings'' $m^2$ and $\lambda$ (``renormalization'') without
a modification of the quartic form of the potential.

Let us consider classical statistics $(d=3)$ and perform the
momentum integration with some ultraviolet cutoff\footnote{On a
cubic lattice, $\Lambda$ would be related to the lattice distance
$a$ by $\Lambda=\pi/a$ and one has to replace for the inverse
propagator $q^2 \to (2/a^2) \sum_{\mu} (1-{\rm cos} a q_{\mu})$.} 
$q^2\leq \Lambda^2$
\bea\label{NS16}
U_0^{(1)}(\rho)=\frac{3\lambda\Lambda}{4\pi^2}\rho-\frac{1}
{12\pi}(m^2+3\lambda\rho)^{3/2}\ ,~~~~~~~
\rho=\frac{1}{2}\varphi^2.\eea
Here we have neglected corrections which are suppressed by
powers of $(m^2+3\lambda\rho)/\Lambda^2$.
Defining renormalized couplings\footnote{We omit here for simplicity
the wave function renormalization.}
\beq\label{NS17}
m^2_R=\frac{\partial U_0}{\partial\rho}(0)\ ,~~~~~~~ \lambda_R=
\frac{\partial^2U_0}{\partial\rho^2}(0)\eeq
one obtains $(m^2\geq0)$
\bea\label{NS18}
m^2_R&=&m^2+\frac{3\lambda\Lambda}{4\pi^2}-\frac{3\lambda}{8\pi}m\nonumber\\
\lambda_R&=&\lambda-\frac{9\lambda^2}{16\pi m}.\eea
As expected, the corrections are large for large couplings $\lambda$.
In addition, the correction to $\lambda$ diverges as $m\to0$.
Since the correlation length $\xi$ is given by $m_R^{-1}$, we
see here the basic reasons why the transition from microphysics
to macrophysics becomes difficult for large
couplings and large correlation length.

Due to the linear ``ultraviolet divergence'' in the mass
renormalization\footnote{For finite $\Lambda$ there is of course
no divergence. We employ here the language of particle physics
which refers to the limit $\Lambda\to\infty$.}
$\sim 3\lambda\Lambda/4\pi^2$, a large correlation length $m_R^{-1}$
actually requires a negative value of $m^2$. On the other hand, we
note that the saddle point expansion is valid only for $m^2+3\lambda
\rho>0$ and breaks down at $\rho=0$ if $m^2$ becomes negative.
The situation can be improved by using in the
one-loop expression (\ref{NS16}) the renormalized parameters
$m_R$ and $\lambda_R$ instead of $m$ and $\lambda$. For the second
term in eq. (\ref{NS16}) the justification\footnote{For the correction to
$m^2\sim \lambda\Lambda$ this replacement is not justified.} for the
``renormalization group improvement'' arises from the observation
that the momentum integral is dominated by momenta $q^2\approx m^2$. Through
an iterative procedure corresponding to the inclusion of higher
loops, the physical infrared cutoff should be replaced by $m_R$. We
will see later in more detail how this renormalization improvement
arises through the formulation of flow equations.

Writing $U_0$ in terms of renormalized parameters
\beq\label{NS19}
U_0=\left( m^2_R+\frac{3\lambda_Rm_R}{8\pi} \right) \rho
+\frac{1}{2} \left( \lambda_R+\frac{9\lambda_R^2}{16\pi m_R} \right) \rho^2
-\frac{1}{12\pi} \left( m^2_R+3\lambda_R\rho \right)^{3/2}\eeq
we can compute the deviations from the Landau theory, i.e.
\beq\label{NS20}
\frac{\partial^3U_0}{\partial\rho^3}(0)=\frac{27\lambda^3_R}{32\pi m_R^3}.
\eeq
We can also formulate a criterion for the validity of the renormalization
group-improved saddle-point approximation, namely that the one-loop
contributions to $m_R$ and $\lambda_R$  should not dominate
these couplings. This yields
\beq\label{NS21}
\frac{\lambda_R}{m_R}<\frac{16\pi}{9}.\eeq

The renormalized coupling $\lambda_R$ is not independent of
$m_R$. If we take the renormalization group improvement literally,
we could solve the relation (cf. eq. (\ref{NS18}))
\beq\label{NS22}
\lambda=\lambda_R+\frac{9\lambda^2_R}{16\pi m_R}\eeq
for fixed $\lambda$ and find
\beq\label{NS23}
\frac{\lambda_R}{m_R}=\frac{8\pi}{9}\left(\sqrt{1+\frac{9\lambda}
{4\pi m_R}}-1\right).\eeq
We see that for an arbitrary positive $\lambda$ the condition
(\ref{NS21}) breaks down in the limit of infinite correlation
length $m_R\to0$. For a second-order phase transition this is exactly
what happens near the critical temperature, 
and we encounter here the infrared problem for
critical phenomena.

One concludes that fluctuation effects beyond the Landau theory
become important for
\beq\label{NS23a}
m_R\stackrel{\scriptstyle<}{\sim}\frac{3\lambda}{4\pi}. 
\eeq
If one is close (but not too close) to the phase transition,
a linear approximation $m_R=A(T-T_c)$ remains a good guide.
This provides a typical temperature interval around the critical
temperature for which the Landau theory fails, namely
\beq\label{NS23b}
\frac{|T-T_c|}{T_c}\stackrel{\scriptstyle<}{\sim}\frac{3\lambda}
{4\pi AT_c}. \eeq
The width of the interval
depends on the dimensionless quantities $A$ and $\lambda/T_c$. Inside
the interval (\ref{NS23b}) the physics is governed by the universal
critical behavior. In fact, we should not trust the relation
(\ref{NS22}) for values of $m_R$ 
for which (\ref{NS21}) 
is violated.
The correct behavior of $\lambda_R(m_R)$ will be given by the
renormalization group and leads to a fixed point for the ratio
$\lim_{m_R\to0}\left(\lambda_R/m_R\right)\to {\rm const}$.
As an example of universal behavior we observe
that the $\varphi^6$-coupling
(\ref{NS20}) is completely determined by the ratio $\lambda_R/m_R$,
independently of the value of the microphysical coupling $\lambda$.
This is generalized to a large class of microscopic potentials. We see
how universality is equivalent to the ``loss of memory''
of details of the microphysics. Similar features in four dimensions
are the basis for the impressive predictive power of particle
physics.

In this report we present a version of the renormalization group
equation where an infrared cutoff $k$ is introduced for the
momentum integral (\ref{NS15}). In a rough version we use a sharp
cutoff $k^2<q^2<\Lambda^2$ in order to define the scale-dependent
potential $U_k$ \cite{Wet91-1,Wet93-2}. 
The dependence of the ``average potential''
$U_k$ on $k$ is simply computed $(d=3)$ as
\beq\label{NS24}
\partial_kU_k(\rho)=-\frac{k^2}{4\pi^2}\ln\left(\frac{k^2
+V'+2\rho V''}{k^2}\right),\eeq
where we have introduced the classical potential
\beq\label{NS25}
V=m^2\rho+\frac{1}{2}\lambda\rho^2\eeq
and subtracted an irrelevant $\rho$-independent constant. (Primes
denote derivatives with respect to $\rho$.) The renormalization group
improvement\footnote{For a detailed justification see refs. 
\cite{Wet91-1,Wet93-2}.}
replaces $V(\rho)$ by $U_k(\rho)$ and therefore leads to the
nonlinear partial differential equation
\beq\label{NS26}
\partial_kU_k(\rho)=-\frac{k^2}{4\pi^2}\ln\left\{
1+\frac{U_k'(\rho)}{k^2}+\frac
{2\rho U_k''(\rho)}{k^2}\right\}.\eeq
This simple equation already describes correctly the qualitative
behavior of $U_0=\lim_{k\to0}U_k$. We will encounter it later
(cf. eq. (\ref{2.52})) as an approximation to the exact renormalization
group equation. The present report motivates this
equation and provides a formalism for computing 
corrections to it\footnote{Eq.\ (\ref{NS26}) can also be obtained as the sharp
cutoff limit of the Polchinski equation \cite{Pol84-1} and was discussed in
ref.\ \cite{Has86-1}. However, in this approach it is difficult to include the
wave function renormalization and to use eq. (\ref{NS26})
as a starting point of a systematic procedure.}. We also generalize
this equation to other forms of the infrared cutoff. In section \ref{SimpEx} 
we show how this type of equation, even in the
simple quartic approximation for
the potential, gives a unified picture of the phase transitions for 
$O(N)$-symmetric scalar 
theories in arbitrary\footnote{For the Kosterlitz-Thouless
phase transition for $d=2, N=2$ the wave function renormalization
must be included.} dimensions.

\section{Non-Perturbative flow equation}
\label{nonpertfloweq}

\subsection{Average action}
\label{AverAct}

We will concentrate on a flow equation
which describes the scale dependence of the effective
{average action} $\Gamma_k$ \cite{Wet93-2}. The latter
is based on the quantum field theoretical concept of the effective
action $\Gamma$, i.e.\ the generating functional of
the Euclidean one-particle irreducible ($1 PI$) correlation functions
or proper vertices (cf.\ eq.\ (\ref{NS9})). 
This functional is obtained after ``integrating
out'' the quantum fluctuations. The scattering amplitudes and cross
sections follow directly from an analytic continuation of the $1PI$
correlation functions in a standard way. Furthermore, the
field equations derived from the
{\em effective action} are exact as all quantum effects are
included.
In thermal and chemical equilibrium
$\Gamma$ includes in addition the thermal
fluctuations and depends on the temperature $T$ and chemical potential
$\mu$.
In statistical physics $\Gamma$ is related to
the free energy as a functional
of some space-dependent order parameter $\varphi(x)$. For vanishing
external fields the equilibrium state is given by the
minimum of $\Gamma$. More generally, in the presence of (spatially
varying) external fields or sources the equilibrium state obeys
\beq\label{2.1}
\frac{\delta\Gamma}{\delta\varphi(x)}=J(x),\eeq
and the precise relation to the thermodynamic potentials like the
free energy $F$ reads
\beq
\label{2.2}
F=T\Gamma_{eq}+\mu N-T\int dx \, \varphi_{eq}(x)J(x).
\eeq
Here $\varphi_{eq}(x)$ solves (\ref{2.1}), $\Gamma_{eq}=\Gamma[\varphi
_{eq}]$, and $N$ is the conserved quantity to which the chemical potential
is associated. For homogeneous $J=j/T$ 
the equilibrium
value of the order parameter $\varphi$ is often also homogeneous.
In this case the energy density $\epsilon$, entropy density $s$,
``particle density'' $n$ and pressure $p$ can be simply expressed
in terms of the effective potential $U(\varphi)=T\Gamma/V$, namely
\bea\label{2.3}
&&\epsilon=U-T\frac{\partial U}{\partial T}-\mu\frac{\partial U}
{\partial\mu}\ ,\ s=-\frac{\partial U}{\partial T}+\frac{j\varphi}{T},
\nonumber\\
&&n=-\frac{\partial U}{\partial\mu}\ ,\ p=-U=-T\Gamma/V
\eea
Here $U$ has to be evaluated for the solution of $\partial U/\partial
\varphi=j$, $n=N/V$ and $V$ is the total volume of (three-dimensional)
space. Evaluating
$U$ for arbitrary $\varphi$ yields the equation of state in presence
of homogeneous magnetic fields or other appropriate 
sources\footnote{For the special case where $\varphi(x)$
corresponds to the density of a conserved quantity and $j=\mu$
one has $F=T\Gamma_{eq}$. The thermodynamic relations appropriate
for this case are specified in sect. 4.4.
Our notation is adapted to classical statistics
where $\int dx \equiv \int d^3x$. For quantum statistics or
quantum field theory one has to use $\int dx \equiv \int d^4x$
where the ``Euclidean time'' is a torus with circumference $1/T$.
The relations (\ref{2.3}) remain valid for a homogeneous source
$J=j$.}.

More formally, the effective action $\Gamma$ follows from a Legendre
transform of the logarithm of the partition function in presence of
external sources or fields (see below). Knowledge of $\Gamma$ is
in a sense equivalent to the ``solution'' of a theory. Therefore
$\Gamma$ is the
macroscopic quantity on which we will concentrate. In particular,
the effective potential $U$ contains already a large part of
the macroscopic information relevant for homogeneous states. We
emphasize that the concept of the effective potential
is valid universally for classical
statistics and quantum statistics, or quantum field theory in
thermal equilibrium, or the vacuum\footnote{The only difference
concerns the evaluation of the partition function $Z$ or $W=\ln Z=
-(F-\mu N)/T$. For classical statistics it involves a $D$-dimensional
functional integral, whereas for quantum statistics the dimension
in the Matsubara formalism is $D+1$. The vacuum in quantum field
theory corresponds to $T\to0$, with $V/T$ the volume of Euclidean
``spacetime''.}.

The average action $\Gamma_k$ is a simple
generalization of the effective action, with the distinction that only
fluctuations with momenta $q^2 \gtrsim k^2$ are included.
This is achieved
by implementing an infrared (IR) cutoff $\sim k$ in the functional
integral that defines the effective action $\Gamma$. In the
language of statistical physics, $\Gamma_k$ is a type of
coarse-grained free
energy with a coarse graining length scale $\sim k^{-1}$. As long as
$k$ remains large enough, the possible complicated effects of
coherent long-distance fluctuations play no role and $\Gamma_k$ is close
to the microscopic action.
Lowering $k$ results in
a successive inclusion of fluctuations with momenta
$q^2 \gtrsim k^2$ and therefore permits to explore the theory on
larger and larger length scales. The average action $\Gamma_k$ can be viewed
as the effective action for averages of fields over a volume
with size $k^{-d}$ \cite{Wet91-1} 
and is similar in spirit to the action for block--spins
on the sites of a coarse lattice.

By definition, the average action equals the standard effective action
for $k=0$, i.e.\ $\Gamma_{0}=\Gamma$, as the IR cutoff is
absent in this limit
and all fluctuations are included. On the other hand,
in a model with a physical
ultraviolet (UV) cutoff $\Lambda$ we
can associate $\Gamma_\Lambda$ with the microscopic or
classical action $S$.
No fluctuations with momenta below $\Lambda$ are effectively included
if the IR cutoff equals the UV cutoff.
Thus the average action $\Gamma_k$ has the important property that
it interpolates between the classical
action $S$
and the effective action $\Gamma$ as $k$ is lowered from the ultraviolet
cutoff $\Lambda$ to zero:
\beq\label{2.4}
\Gamma_\Lambda \approx S\ ,
\,\, \lim\limits_{k \to 0} \Gamma_k = \Gamma.\eeq
The ability to follow the evolution to $k\rightarrow0$ is equivalent to the
ability to solve the theory. Most importantly, the dependence of the
average action on the scale $k$ is described by an exact
non-perturbative flow equation which is presented in the next 
subsection.\\

Let us consider the construction of $\Gamma_k$ for a
simple model with real scalar fields $\chi_a$, $a=1 \ldots N$, in $d$
Euclidean dimensions with classical action $S$.
We start with the path integral representation of the
generating functional for the connected correlation functions in
the presence of an IR cutoff. It is given by the logarithm of the
(grand) canonical partition function in the presence of inhomogeneous
external fields or sources $J_{a}$
\beq\label{2.5}
W_k[J]=\ln Z[J]=\ln \int D \chi \exp\left(-S_{}[\chi]-\De S_k[\chi]+
\int d^dx J_a(x)\chi^a(x) \right)\; .
\label{genfunc}
\eeq
In classical statistical physics $S$ is related to the Hamiltonean
$H$ by $S=H/T$, so that $e^{-S}$ is the usual Boltzmann factor.
The functional integration $\int D\chi$ stands for the sum
over all microscopic states. In turn, the field $\chi_a(x)$ can
represent a large variety of physical objects like a (mass-)
density field $(N=1)$, a local magnetisation $(N=3)$ or
a charged order parameter $(N=2)$. The only modification as
compared to the construction of the standard effective action
is the addition of an
IR cutoff term $\De S_k[\chi]$ which is
quadratic in the fields and reads
in momentum space $(\chi_a(-q)\equiv\chi_a^*(q))$
\beq\label{2.6}
\De S_k[\chi]=\hal \int
\frac{d^dq}{(2\pi)^d} R_k(q)\chi_a(-q)\chi^a(q).
\eeq
Here the IR cutoff function $R_k$ is required to vanish
for $k \to 0$ and to diverge for $k \to \infty$ (or $k\to\Lambda)$
and fixed $q^2$.
This can be achieved, for example, by the
exponential form
\beq\label{2.7}
 R_k(q) \sim \frac{q^2}{\displaystyle{e^{q^2/k^2} - 1}}
 \label{Rk(q)}.
\eeq
For fluctuations with small momenta
$q^2\ll k^2$ this cutoff behaves as $R_k(q)\sim k^2$ and allows
for a simple interpretation: Since
$\De S_k[\chi]$ is quadratic in the fields,
all Fourier modes of $\chi$ with momenta smaller than
$k$ acquire an effective mass $\sim k$. This additional mass term
acts as an effective IR cutoff for the low momentum modes.
In contrast, for $q^2\gg k^2$ the function $R_k(q)$ vanishes
so that the functional integration of the high momentum modes
is not disturbed. The term $\De S_k[\chi]$ added to the classical
action is the main ingredient for the construction of an effective
action that includes all fluctuations with momenta $q^2 \gtrsim k^2$,
whereas fluctuations with $q^2 \lesssim k^2$ are suppressed.

The expectation value of $\chi$, i.e.\ the macroscopic field $\phi$,
in the presence of $\De S_k[\chi]$ and $J$ reads
\beq
\label{2.8}
\phi^a(x) \equiv \langle\chi^a(x)\rangle =
\frac{\de W_k[J]}{\de J_a(x)}.
\eeq
We note that the relation between $\phi$ and $J$ is
$k$--dependent, $\phi=\phi_k[J]$ and therefore \mbox{$J=J_k[\phi]$}.
In terms of $W_k$ the average action is
defined via a modified Legendre transform
\beq
\label{2.9}
\Ga_k[\phi]=-W_k[J]+\int d^dx J_a(x)\phi^a(x)-\De S_k[\phi],
\eeq
where we have subtracted the term $\De S_k[\phi]$ in the rhs.
This subtraction of the IR cutoff term as a function of the
macroscopic field $\phi$ is crucial for the definition of a
reasonable coarse-grained free energy with the property
$\Gamma_\Lambda\approx S$. It guarantees
that the only difference between $\Gamma_k$ and $\Gamma$ is the
effective infrared cutoff in the fluctuations.
Furthermore, it has the consequence that $\Gamma_k$ does not
need to be convex, whereas a pure Legendre transform is always convex by
definition. The coarse-grained free energy has to
become convex \cite{RingWet90,TetWet92}
only for $k\rightarrow0$. These considerations are important for an
understanding of spontaneous symmetry breaking and, in particular,
for a discussion of nucleation in a first-order phase transition.

In order to establish the property $\Gamma_{\Lambda}\approx S$ we consider an
integral equation for $\Gamma_k$ that is equivalent to (\ref{2.9}).
In an obvious matrix notation, where
$J\chi\equiv\int d^dxJ_a(x)\chi^a(x)=
\int\frac{d^dp}{(2\pi)^d}J_a(-p)\chi_a(p)$
and $R_{k,ab}(q,q')=R_k(q)\delta_{ab}(2\pi)^d\delta(q-q')$, we represent
(\ref{2.5}) as
\beq\label{2.10}
\exp\Big(W_k[J]\Big)= \int D \chi \exp\left(-S[\chi]+ J \chi
-\frac{1}{2} \chi\, R_k \chi \right)\; .
\eeq
As usual, we can invert the Legendre transform (\ref{2.9}) to express
\beq\label{2.11}
J=\frac{\de \Gamma_k}{\de \phi}+\phi\, R_k.
\eeq
It is now straightforward to insert the definition
(\ref{2.9}) into (\ref{2.10}). After a variable substitution
$\chi'=\chi-\phi$ one obtains the functional integral representation
of $\Gamma_k$
\beq
\label{2.12}
\exp(-\Gamma_k[\phi])=\dsp{\int D \chi' \exp\left(-S[\phi+\chi']
+ \frac{\de \Gamma_k}{\de \phi}
\chi'  -\frac{1}{2}
\chi' \, R_k\, \chi'  \right)}
\; .\eeq
This expression resembles closely the background field formalism
for the effective action which is modified only by the term $\sim
R_k$.
For $k \to \infty$ the cutoff function $R_k$ diverges and
the term $\exp( -\chi'  R_k \chi' /2) $ behaves
as a delta functional $\sim \de[\chi']$, thus leading to the
property $\Gamma_k \to S$
in this limit. For a model with a sharp UV cutoff $\Lambda$ it is
easy to enforce the identity $\Gamma_\Lambda=S$ by choosing a cutoff
function $R_k$ which diverges for $k\to\Lambda$, like
$R_k\sim q^2(e^{q^2/k^2}-e^{q^2/\Lambda^2})^{-1}$.
We note, however,
that the property $\Gamma_\Lambda=S$ is not essential, as the short
distance laws may be parameterized by $\Gamma_\Lambda$ as
well as by $S$. For momentum scales much smaller than
$\Lambda$ universality implies
that the precise form of $\Gamma_\Lambda$ is irrelevant,
up to the values of a
few relevant renormalized couplings.
Furthermore, the microscopic action may be formulated on a lattice
instead of continuous space and can involve even variables different
from $\chi_a(x)$. In this case one can still compute $\Gamma_\Lambda$
in a first step by evaluating the functional integral (\ref{2.12})
approximately. Often a saddle point expansion will do, since no
long-range fluctuations are involved in the
transition  from $S$ to $\Gamma_\Lambda$. In
this report we will assume that the first step of the computation
of $\Gamma_\Lambda$ is done and consider $\Gamma_\Lambda$ as the
appropriate parametrization of the microscopic physical laws.
Our aim is the computation of the effective action $\Gamma$
from $\Gamma_\Lambda$ -- this step may be called ``transition
to complexity'' and involves fluctuations on all scales. We emphasize
that for large $\Lambda$ the average action $\Gamma_\Lambda$
can serve as a formulation of the microscopic laws also for situations
where no physical cutoff is present, or where a momentum UV cutoff
may even be in conflict with the symmetries, like the important
case of gauge symmetries.

A few properties of the effective average action are worth mentioning:
\begin{enumerate}
\item All symmetries of the model which are respected by the IR cutoff
  $\Delta S_k$ are automatically symmetries of $\Gamma_k$. In
  particular this concerns translation and rotation invariance, and
  the approach is not plagued by many of the problems encountered by a
  formulation of the block-spin action on a lattice. 
  Nevertheless, our method is not restricted to continuous space.
  For a cubic lattice with lattice distance $a$ the propagator 
  only obeys the restricted lattice translation and rotation
  symmetries, e.g.\ a next neighbor interaction leads in momentum
  space to 
\beq
S=\frac{2}{a^2} \int \frac{d^dq}{(2 \pi)^d} \sum\limits_{\mu}
\left(1-\cos a q_{\mu}\right) \chi^*(q) \chi(q) + \ldots 
\eeq  
  The momentum cutoff $|q_{\mu}|\le \Lambda$, $\Lambda = \pi / a$
  also reflects the lattice symmetry. A rotation and translation
  symmetric cutoff $R_k$ which only depends on $q^2$ obeys 
  automatically all possible lattice symmetries. The only 
  change as compared to continuous space will be the reduced symmetry
  of $\Gamma_k$.
\item In consequence, $\Gamma_k$ can be expanded in terms of
  invariants with respect to these symmetries with couplings depending
  on $k$. For the example of a scalar $O(N)$-model in continuous space
  one may use a derivative
  expansion ($\rho=\Phi^a\Phi_a/2$)
  \begin{equation}
    \label{2.13}
    \Gamma_k=\int d^d x\left\{
    U_k(\rho)+\frac{1}{2}Z_{\Phi,k}(\rho)
    \partial^\mu\Phi_a\partial_\mu\Phi^a+\ldots\right\}
  \end{equation}
  and expand further in powers of $\rho$
  \begin{eqnarray}
    \label{2.14}
    \dsp{U_k(\rho)} &=& \dsp{
    \frac{1}{2}\ol{\la}_k\left(
    \rho-\rho_0(k)\right)^2+
    \frac{1}{6}\ol{\gamma}_k\left(
    \rho-\rho_0(k)\right)^3+\ldots}\nnn
    \dsp{Z_{\Phi,k}(\rho)} &=& \dsp{
    Z_{\Phi,k}(\rho_0)+Z_{\Phi,k}^\prime(\rho_0)
    \left(\rho-\rho_0\right)+\ldots}\; 
  \end{eqnarray}
  Here $\rho_0$ denotes the ($k$--dependent) minimum of the effective average
  potential $U_k(\rho)$.  We see that $\Gamma_k$ describes infinitely many
  running couplings.
\item Up to an overall scale factor the limit $k\to 0$ of $U_k$
corresponds to the effective potential $U=T\Gamma /V$, from
which the thermodynamic quantities can be derived for homogeneous
situations according to eq.\ (\ref{2.3}).
The overall scale factor is fixed by dimensional considerations.
Whereas the dimension of $U_k$ is (mass)$^d$ the dimension of
$U$ in eq.\ (\ref{2.3}) is (mass)$^4$ (for $h\!\!\bar{}\,\,=c=k_B=1$).
For classical statistics in $d=3$ dimensions one has
$U_k=\Gamma_k/V$ and $U=T\lim_{k\to 0}U_k$. For two dimensional
systems an additional factor $\sim$ mass appears since
$U_k=\Gamma_k/V_2=L \Gamma_k/V$ implies
$U=T L^{-1} \lim_{k\to 0} U_k$. Here  $L$ is the typical thickness
of the two dimensional layers in a physical system.
In the following we will often omit these scale factors.
\item The functional $\tilde{\Gamma}_k[\Phi]=\Gamma_k[\Phi]
+\Delta S_k[\Phi]$ is the Legendre transform of $W_k$ and 
therefore convex. This implies that all eigenvalues 
of the matrix of second functional derivatives $\Gamma^{(2)}+R_k$
are positive semi-definite. In particular, one finds for
a homogeneous field $\Phi_a$ and $q^2=0$ the simple exact bounds for
all $k$ and $\rho$
\bea
\label{PosDef}
U_k^{\prime}(\rho) &\ge& -R_k(0) \nnn
U_k^{\prime}(\rho) + 2 \rho U_k^{\prime\prime}(\rho) &\ge& -R_k(0),
\eea
where primes denote derivatives with respect to $\rho$.
Even though the potential $U(\phi)$ becomes convex for $k\to 0$
it may exhibit a minimum at $\rho_0(k) >0$ for all $k>0$.
Spontaneous breaking of the $O(N)$-symmetry is characterized by
$\lim_{k\to 0} \rho_0(k) > 0$.
\item For a formulation which respects the reparametrization invariance
of physical quantities under a rescaling of the variables
$\chi_a(x)\to \alpha\chi_a(x)$ the infrared cutoff should contain
a wave function renormalization, e.g.
\beq\label{2.15}
R_k(q)=\frac{Z_kq^2}{e^{q^2/k^2}-1} \; .
\eeq
One may choose $Z_k=Z_{\phi,k}(\rho_0)$. This choice guarantees
that no intrinsic scale is introduced in the inverse average propagator
\beq\label{2.16}
Z_kq^2+R_k=Z_kP(q)=Z_kq^2p\left(\frac{q^2}{k^2}\right).\eeq
This is important in order to obtain scale-invariant flow equations
for critical phenomena.
\item There is no problem incorporating chiral fermions, since a chirally
  invariant cutoff $R_k$ can be formulated \cite{Wet90-1,CKM97-1}
  (cf.\ section \ref{FlowFerm}).
\item Gauge theories can be formulated along similar
  lines\footnote{See also \cite{gravity} for applications to gravity. } 
  \cite{RW93-1,RW93-2}, \cite{Bec96-1}--\cite{FreireLitim}
  even though $\Delta S_k$ may not be gauge
  invariant\footnote{For a manifestly gauge invariant formulation in terms
  of Wilson loops see ref. \cite{Mor98-1}.}. In this case the usual Ward
  identities receive corrections for which one can derive closed
  expressions \cite{EHW94-1}. These corrections vanish for
  $k\rightarrow0$. On the other hand they appear as ``counterterms''
  in $\Gamma_\Lambda$ and are crucial for preserving the gauge
  invariance of physical quantities. 
\item For the choice (\ref{2.16}) the
  high momentum modes are very effectively integrated out because of
  the exponential decay of $R_k$ for $q^2\gg k^2$.  Nevertheless, it is
  sometimes technically easier to use a cutoff without this fast decay
  property, e.g.~$R_k\sim k^2$ or $R_k\sim k^4/q^2$.
  In the latter cases one
  has to be careful with possible remnants of an
  incomplete integration of the
  short distance modes. An important technical
  simplification can also be achieved by a sharp
  momentum cutoff \cite{WH73-1}. This guarantees complete integration
  of the short distance modes, but poses certain problems
  with analyticity \cite{Wet91-1,Alf94-1,Mor96,BZeta,BBMZ}. 
  In contrast, a smooth 
  cutoff like (\ref{2.16}) does not introduce any
  non--analytical behavior.
  The results for physical quantities are independent
  of the choice of the cutoff scheme $R_k$. On the other hand,
  both $\Gamma_\Lambda$ and the flow with $k$ are scheme-dependent.
  The scheme dependence of the final results is a good check for
  approximations
  \cite{Wet91-1,BHLM95-1,LitimS,LPS00,SSATM}. 
\item Despite a similar spirit and many analogies, there is a
  conceptual difference to the Wilsonian effective action
  $S_\Lambda^{\rm W}$.
  The Wilsonian effective action describes a set of different actions
  (parameterized by $\Lambda$) for one and the same
  model --- the $n$--point
  functions are independent of $\Lambda$ and have to be computed from
  $S_\Lambda^{\rm W}$ by further functional integration. In contrast,
  $\Gamma_k$ can be viewed as the effective action for
  a set of  different
  ``models'' --- for any
  value of $k$ the effective average action is related to the generating
  functional of $1PI-n$-point functions for
  a model with a different action $S_k=S+\Delta S_k$. The
  $n$--point functions depend on $k$. The Wilsonian effective action does
  not generate the $1PI$ Green functions \cite{KKS92-1}.
\item Because of the incorporation of an infrared cutoff, $\Gamma_k$ is
  closely related to an effective action for averages of
  fields \cite{Wet91-1} where the average is taken over a volume $\sim
  k^{-d}$.
\end{enumerate}

\subsection{Exact flow equation}
\label{ExactFlow}

The dependence of the average action $\Gamma_k$ on the
coarse graining scale $k$ is
described by an exact non-perturbative flow equation 
\cite{Wet93-2,BAM,Mor94,El}
\begin{equation}\label{2.17}
  \frac{\partial}{\partial k}\Ga_k[\phi] =  \hal\Tr\left\{\left[
  \Ga_k^{(2)}[\phi]+R_k\right]^{-1}\frac{\partial}{\partial k}
  R_k\right\} \; .
\end{equation}
The trace involves an integration over momenta or coordinates
as well as a summation over internal
indices. In momentum space it reads
$\Tr=\sum_a \int d^dq/(2\pi)^d$, as appropriate for the unit matrix
${\bf 1}=(2 \pi)^d \delta(q-q^\prime )\delta_{ab}$. 
The exact flow equation describes
the scale dependence of $\Gamma_k$ in terms of the inverse
average propagator $\Ga_k^{(2)}$, given by
the second functional derivative of $\Ga_k$ with respect
to the field components
\begin{equation}\label{2.18}
  \left(\Gamma_k^{(2)}\right)_{a b}(q,q^\prime)=
  \frac{\delta^2\Gamma_k}
  {\delta\phi^a(-q)\delta\phi^b(q^\prime)}\; .
\end{equation}
It has a simple graphical expression as a one-loop equation 

\vspace*{0.2in}
\hspace*{5.5cm}
$\displaystyle{\frac{\partial \Gamma_k}
{\partial k}\,\,\,\, =\,\,\,\,\frac{1}{2}\,\,\,\,}$
\parbox{1.in}{
\epsfxsize=.4in
\epsffile{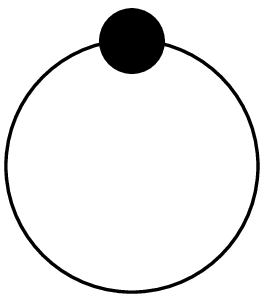}}
\vspace*{0.2in}

\noindent
with the full $k$--dependent propagator associated to the propagator
line and the dot denoting the insertion $\partial_k R_k$.

Due to the appearance
of the exact propagator $(\Gamma_k^{(2)}+R_k)^{-1}$, eq. (\ref{2.17}) is
a functional differential equation. It is remarkable that the
transition from the classical propagator in presence of the
infrared cutoff, $(S^{(2)}+R_k)^{-1}$, to
the full propagator turns the one-loop expression into an exact
identity which incorporates effects of arbitrarily high loop order
as well as genuinely non-perturbative effects\footnote{We note
that anomalies which arise from topological obstructions in
the functional measure manifest themselves already in the
microscopic action $\Gamma_\Lambda$. The long-distance
non-perturbative effects (``large-size instantons'') are,
however, completely described by the flow equation (\ref{2.17}).}
like instantons in QCD.

The exact flow equation (\ref{2.17}) can be derived in a
straightforward way \cite{Wet93-2}. Let us write
\beq\label{2.19}
\Gamma_k[\phi]=\tilde{\Gamma}_k[\phi]-\Delta S_k[\phi],
\eeq
where, according to (\ref{2.9}),
\beq\label{2.20}
\tilde{\Gamma}_k[\phi]=-W_k[J]+\int d^dx J(x)\phi(x)
\eeq
and $J=J_k(\phi)$. We consider for simplicity a one--component
field and derive first the scale dependence of $\tilde{\Gamma}$:
\beq\label{2.21}
\frac{\partial}{\partial k} \tilde{\Gamma}_k[\phi] =
-\left(\frac{\partial W_k}{\partial k}  \right) [J]
-\int d^dx \frac{\delta W_k}{\delta J(x)}
\frac{\partial J(x)}{\partial k} + \int d^dx \phi(x)
\frac{\partial J(x)}{\partial k} \; .
\eeq
With $\phi(x)=\delta W_k/\delta J(x)$ the last two terms
in (\ref{2.21}) cancel. The $k$--derivative
of $W_k$ is obtained from its defining functional integral
(\ref{2.5}). Since only $R_k$ depends on $k$ this
yields
\beq\label{2.22}
\frac{\partial}{\partial k} \tilde{\Gamma}_k[\phi]=
\langle \frac{\partial}{\partial k}\Delta S_k[\chi] \rangle =
\langle \frac{1}{2} \int d^dx d^dy \chi(x)
\frac{\partial}{\partial k}R_k(x,y) \chi(y) \rangle \; .
\eeq
where $R_k(x,y)\equiv R_k(i\partial_x) \delta(x-y)$ and
\beq\label{2.23}
\langle A[\chi] \rangle =Z^{-1}\int D\chi A[\chi]\exp(-S[\chi]-
\Delta_kS[\chi]+\int d^dxJ(x)\chi(x)).\eeq
Let
$G(x,y)=\delta^2 W_k/\delta J(x)\delta J(y)$
denote the connected $2$--point function and decompose
\beq\label{2.24}
\langle \chi(x) \chi(y) \rangle = G(x,y) + \langle \chi(x) \rangle
\langle \chi(y) \rangle \equiv G(x,y) + \phi(x) \phi(y) \; .
\eeq
Plugging this decomposition into (\ref{2.22}) the scale
dependence of $\tilde{\Gamma}_k$ can be expressed as
\bea\label{2.25}
\frac{\partial}{\partial k} \tilde{\Gamma}_k[\phi] &=&
\frac{1}{2} \int d^dx d^dy \left\{   \frac{\partial}{\partial k}
R_k(x,y)G(y,x)
+ \phi(x) \frac{\partial}{\partial k} R_k(x,y) \phi(y) \right\} \nnn
&\equiv&
\frac{1}{2} \Tr\left\{ G \frac{\partial}{\partial k} R_k\right\}
+\frac{\partial}{\partial k} \Delta S_k[\phi] \; .
\eea
The exact flow equation for the average action $\Gamma_k$
follows now through (\ref{2.19})
\beq\label{2.26}
\frac{\partial}{\partial k} \Gamma_k[\phi] =
\frac{1}{2} \Tr\left\{ G \frac{\partial}{\partial k} R_k\right\}
= \frac{1}{2} \Tr\left\{ \left[
  \Ga_k^{(2)}[\phi]+R_k\right]^{-1} \frac{\partial}{\partial k} R_k
\right\}
\eeq
For the last equation we have used that
$\tilde{\Gamma}_k^{(2)}(x,y)\equiv
\delta^2\tilde{\Gamma}_k/\delta\phi(x)\delta\phi(y)=
\delta J(x)/\delta \phi(y)$
is the inverse of $G(x,y)\equiv \delta^2 W_k/\delta J(x)\delta J(y)=
\delta \phi(x)/\delta J(y)$:
\beq\label{2.26a}
\int d^dyG(x,y)(\Gamma_k^{(2)}+R_k)(y,z)=\delta(x-z)\eeq
It is straightforward to write
the above identities in momentum space and to generalize them to
$N$ components by using the matrix notation introduced above.

Let us point out a few properties of the exact flow equation:
\begin{enumerate}
\item For a scaling form of the evolution equation and a formulation
closer to the usual $\beta$-functions one may replace the partial
$k$-derivative in (\ref{2.17}) by a partial derivative with
respect to the logarithmic variable $t=\ln (k/\Lambda)$.

\item Exact flow equations for $n$--point functions can be easily obtained
from (\ref{2.17}) by differentiation. The flow equation for the
two--point function $\Gamma_k^{(2)}$ involves the three and
four--point functions, $\Ga_k^{(3)}$ and $\Ga_k^{(4)}$, respectively.
One may write schematically
\bea\label{2.28}
\frac{\partial}{\partial t} \Gamma_k^{(2)} &=&
\frac{\partial}{\partial t} \frac{\partial^2\Gamma_k}
  {\partial\phi\partial\phi}\nnn
&=& -\frac{1}{2} \Tr \left\{ \frac{\partial R_k}{\partial t}
\frac{\partial}{\partial \phi}
\left(\left[\Ga_k^{(2)}+R_k\right]^{-1} \Ga_k^{(3)}
\left[\Ga_k^{(2)}+R_k\right]^{-1}\right)\right\}\nnn
&=& \Tr \left\{ \frac{\partial R_k}{\partial t}
\left[\Ga_k^{(2)}+R_k\right]^{-1} \Ga_k^{(3)}
\left[\Ga_k^{(2)}+R_k\right]^{-1} \Ga_k^{(3)}
\left[\Ga_k^{(2)}+R_k\right]^{-1}\right\}\nnn
&-&\frac{1}{2} \Tr \left\{ \frac{\partial R_k}{\partial t}
\left[\Ga_k^{(2)}+R_k\right]^{-1} \Ga_k^{(4)}
\left[\Ga_k^{(2)}+R_k\right]^{-1}\right\} \; .
\eea
Evaluating this equation for $\phi=0$ one sees immediately
the contributions to the flow of the two-point function
from diagrams with three- and four-point vertices. 
Below we will see in more detail that the diagramatics
is closely linked to the perturbative graphs.
In general, the flow equation for $\Ga_k^{(n)}$ involves
$\Ga_k^{(n+1)}$ and $\Ga_k^{(n+2)}$.

\item As already mentioned, the flow
equation (\ref{2.17}) closely resembles a one--loop equation.
Replacing $\Gamma_k^{(2)}$ by the second functional derivative of the
classical action, $S^{(2)}$, one obtains the corresponding one--loop
result. Indeed, the one--loop formula for
$\Gamma_k$ reads
\begin{equation}
  \label{2.27}
  \Gamma_k[\phi]=S[\phi]+
  \frac{1}{2}\Tr\ln\left(
  S^{(2)}[\phi]+R_k\right)
\end{equation}
and taking a $k$--derivative of
(\ref{2.27}) gives a one--loop flow equation very similar to
(\ref{2.17}). The ``full renormalization
group improvement'' $S^{(2)}\rightarrow\Gamma_k^{(2)}$ turns the one--loop
flow equation into an exact non-perturbative flow equation.
Replacing the propagator and vertices appearing in
$\Gamma_k^{(2)}$ by the ones derived from the classical
action, but with running $k$--dependent couplings, and expanding
the result to lowest non--trivial order in the coupling constants, one
recovers standard renormalization group improved one--loop
perturbation theory.

\item The additional cutoff function $R_k$ with a form like
the one given in eq.\ (\ref{2.15}) renders the momentum integration
implied in the trace of (\ref{2.17}) both
infrared and ultraviolet finite. In particular, for $q^2\ll k^2$
one has an additional
mass--like term $R_k \sim k^2$
in the inverse average propagator. This makes the formulation suitable
for dealing with theories which are plagued by infrared problems
in perturbation theory.
For example, the flow equation can be used in three dimensions
in the phase with spontaneous symmetry breaking despite the
existence of massless Goldstone bosons for $N>1$. We recall
that all eigenvalues of the matrix $\Gamma^{(2)}+R_k$ must be
positive semi-definite (cf.\ eq.\ (\ref{PosDef})). We note that
the derivation of the exact flow equation does not
depend on the particular choice of the cutoff function.
Ultraviolet finiteness, however, is related
to a fast decay of $\partial_t R_k$ for $q^2\gg k^2$.
If for some other
choice of $R_k$ the rhs of the flow equation would not
remain ultraviolet finite this would indicate
that the high momentum modes have
not yet been integrated out completely in the computation of
$\Gamma_k$. Unless stated otherwise we will always assume a
sufficiently fast decaying choice of $R_k$ in the following.

\item Since no infinities appear in the flow equation, one may
``forget'' its origin from a functional integral. Indeed,
for a given choice of the cutoff function $R_k$ all microscopic
physics is encoded in the microscopic effective action $\Gamma_\Lambda$.
The model is completely specified by the flow equation (\ref{2.17})
and the ``initial value'' $\Gamma_\Lambda$. In a quantum field
theoretical sense the flow equation defines a regularization
scheme. The ``ERGE''-scheme is specified by the flow equation,
the choice of $R_k$ and the ``initial condition'' $\Gamma_\Lambda$.
This is particularly important for gauge theories where other
regularizations in four dimensions and in the presence of chiral
fermions are difficult to construct. For gauge theories $\Gamma_\Lambda$
has to obey appropriately modified Ward identities. In the context
of perturbation theory a first proposal for how to regularize gauge
theories by use of flow equations can be found in \cite{Bec96-1}. We note that
in contrast to previous versions of exact renormalization group
equations there is no need in the present formulation to construct
an ultraviolet momentum cutoff -- a task known to be
very difficult in non-Abelian gauge theories.

As for all regularizations the physical quantities should be
independent of the particular regularization scheme. In our case
different choices of $R_k$ correspond
to different trajectories in the space of effective actions along
which the unique infrared limit $\Gamma_0$ is reached.
Nevertheless, once approximations are applied not
only the trajectory but also its end point may depend on the precise
definition of the function $R_k$. As mentioned above,
this dependence may be used
to study the robustness of the approximation.

\item
Extensions of the
flow equations to gauge fields \cite{RW93-1,RW93-2}, \cite{Bec96-1}--\cite{FreireLitim}
and fermions~\cite{Wet90-1,CKM97-1} are available.

\item
We emphasize that the flow equation (\ref{2.17}) is formally equivalent
to the Wilsonian exact renormalization group
equation \cite{Wil71-1,WH73-1,NC77,Wei76-1,Pol84-1,Has86-1}.
The latter describes how
the Wilsonian effective action $S_\Lambda^{\rm W}$ changes with an
ultraviolet cutoff $\Lambda$. Polchinski's continuum
version of the Wilsonian flow
equation~\cite{Pol84-1}\footnote{For a detailed presentation
see e.g.\ \cite{BaTh}.} 
can be transformed into eq.\ (\ref{2.17})
by means of a Legendre transform,  a suitable field
redefinition and the association $\Lambda=k$~\cite{BAM,ELL93,WetIJMP94}.
Although the formal relation is simple, the practical calculation
of $S^W_k$ from $\Gamma_k$ (and vice versa)
can be quite involved\footnote{If
this problem could be solved, one would be able to construct
an UV momentum cutoff which preserves gauge invariance by
starting from the Ward identities for $\Gamma_k$.}. In the presence
of massless particles the Legendre transform of $\Gamma_k$
does not remain local and $S^W_k$ is a comparatively
complicated object. We will argue below that the crucial
step for a practical use of the flow equation in a non-perturbative
context is the ability to device a reasonable approximation
scheme or truncation. It is in this context that the close
resemblence of eq. (\ref{2.17}) to a perturbative expression is
of great value.

\item
In contrast to the Wilsonian effective action no information
about the short-distance physics is effectively lost as $k$ is
lowered. Indeed, the effective average action for fields with high
momenta $q^2\gg k^2$ is already very close to the effective action.
Therefore $\Gamma_k$ generates quite accurately the vertices
with high external momenta. More precisely, this is the case
whenever the external momenta act effectively as an independent
``physical'' IR cutoff in the flow equation for the vertex. There
is then only a minor difference between $\Gamma_k^{(n)}$ and the
exact vertex $\Gamma^{(n)}$.

\item
An exact equation of the type (\ref{2.17}) can be derived
whenever $R_k$ multiplies a term quadratic in the fields,
cf. (\ref{2.6}). The feature that $R_k$ acts as a good infrared
cutoff is not essential for this. In particular, one can easily
write down an exact equation for the dependence of the effective
action on the chemical potential \cite{BJW99chem}. Another interesting
exact equation describes the effect of a variation of the
microscopic mass term for a field, as, for example, the
current quark mass in QCD. In some cases an additional UV-regularization
may be necessary since the UV-finiteness of the momentum
integral in (\ref{2.17}) may not be given.
\end{enumerate}

\subsection{Truncations}

Even though intuitively simple, the replacement of the (RG--improved)
classical propagator by the full propagator turns the solution of the flow
equation~(\ref{2.17}) into a difficult mathematical problem: The evolution
equation is a functional differential equation. Once $\Gamma_k$ is expanded
in
terms of invariants (e.g.~Eqs.(\ref{2.13}), (\ref{2.14})) this is equivalent
to a coupled system of non--linear partial differential equations for
infinitely many couplings. General methods for the solution of functional
differential equations are not developed very far. They are mainly
restricted
to iterative procedures that can be applied once some small expansion
parameter is identified.  This covers usual
perturbation theory in the case of
a small coupling, the $1/N$--expansion or expansions in the dimensionality
$4-d$ or $d-2$. It may also be extended to less familiar expansions like a
derivative expansion which is related in critical three dimensional scalar
theories to a small anomalous dimension \cite{ILF}. 
In the absence of a clearly
identified small parameter one nevertheless needs to truncate the most
general
form of $\Gamma_k$ in order to reduce the infinite system of coupled
differential equations to a (numerically) manageable size. This truncation
is
crucial. It is at this level that approximations have to be made and, as for
all non-perturbative analytical methods, they are often not easy to control.

The challenge for non-perturbative systems like
critical phenomena in statistical physics or low momentum QCD is to find
flow equations which (a) incorporate all the relevant dynamics so that
neglected effects make only small changes, and (b) remain of manageable
size.
The difficulty with the first task is a reliable estimate of the error. For
the second task the main limitation is a practical restriction for numerical
solutions of differential equations to functions depending only on a small
number of variables.  The existence of an exact functional differential flow
equation is a very useful starting point and guide for this task. At this
point the precise form of the exact flow equation is quite important.
Furthermore, it can be used for systematic expansions through enlargement of
the truncation and for an error estimate in this way.  Nevertheless, this is
not all. Usually, physical insight into a model is necessary to device a
useful non-perturbative truncation!

Several approaches to non-perturbative truncations
have been explored so far ($\rho\equiv\frac{1}{2}\Phi_a\Phi^a$):
\begin{itemize}
\item[(i)] {\it Derivative expansion}. We can classify
invariants by the number of derivatives 
\begin{equation}
\label{2.29}
  \Gamma_k[\Phi]=\int d^d x\left\{
  U_k(\rho)+\frac{1}{2}Z_{k}(\rho)
  \partial_\mu\Phi^a\partial^\mu\Phi_a+
  \frac{1}{4}Y_{k}(\rho)\partial_\mu\rho
  \partial^\mu\rho+
  \O(\partial^4)\right\} \, .
\end{equation}
The lowest level only includes the scalar potential and a standard
kinetic term. The first correction includes 
the $\rho$-dependent wave function renormalizations 
$Z_{k}(\rho)$ and $Y_{k}(\rho)$. The next level involves then
invariants with four derivatives etc.

One may wonder if a derivative expansion has any chance to account
for the relevant physics of critical phenomena, in a situation where we
know that the critical propagator is non-analytic in the
momentum\footnote{See \cite{Myerson,Golner} for 
early applications of the derivative expansion to critical phenomena.
For a recent study on convergence properties of the derivative
expansion see \cite{MTi99}.}. 
The reason why it can work is that the nonanalyticity builds
up only gradually as $k\to 0$. For the critical temperature a typical
qualitative form of the inverse average propagator is
\beq\label{2.31}
\Gamma_k^{(2)}\sim q^2(q^2+ck^2)^{-\eta/2}\eeq
with $\eta$ the anomalous dimension. Thus the behavior for $q^2\to 0$
is completely regular. In addition, the contribution
of fluctuations with small momenta $q^2\ll k^2$ to the flow
equation is suppressed by the
IR cutoff $R_k$. For $q^2\gg k^2$ the ``nonanalyticity'' of the
propagator is already manifest. The contribution of this region
to the momentum integral in (\ref{2.17}) is, however, strongly
suppressed by the derivative $\partial_k R_k$. For cutoff
functions of
the type (\ref{2.15}) only a small momentum range centered around
$q^2\approx k^2$ contributes substantially to the momentum integral
in the flow equation. This suggests the use of a hybrid derivative
expansion where the momentum dependence of $\Gamma_k-\int d^dxU_k$ is
expanded around $q^2=k^2$. Nevertheless, due to the qualitative
behavior (\ref{2.31}), also an expansion around $q^2=0$ should yield
valid results. We will see in section \ref{secsec} that the first
order in the derivative expansion (\ref{2.29}) gives a quite
accurate description of critical phenomena in three dimensional
$O(N)$ models, except for an (expected) error in the anomalous
dimension. 
\item[(ii)] {\it Expansion in powers of the fields}. As an
alternative ordering principle one may expand $\Gamma_k$
in $n$-point functions $\Gamma_k^{(n)}$ 
\begin{equation}
\label{2.30}
  \Gamma_k[\Phi]=
  \sum_{n=0}^\infty\frac{1}{n!}\int
  \left(\prod_{j=0}^n d^d x_j
  \left[\Phi(x_j)-\Phi_0\right]\right)
  \Gamma_k^{(n)}(x_1,\ldots,x_n)\; .
\end{equation}
If one chooses \cite{Wet91-1}\footnote{See also \cite{Aoki,AMSST96}
  for the importance of expanding around $\Phi=\Phi_0$ instead of
$\Phi=0$ and refs.\ \cite{MOP,HKLM,Mor94-2}.}
$\Phi_0$ as the $k$--dependent expectation value of $\Phi$, 
the series~(\ref{2.30})
starts effectively at $n=2$. 
The flow equations for the $1PI$ Green
functions $\Gamma_k^{(n)}$ are obtained by functional differentiation
of~(\ref{2.17}).  Similar equations have been discussed first 
in \cite{Wei76-1} from a somewhat different viewpoint. They can also be
interpreted
as a differential form of Schwinger--Dyson equations \cite{DS49-1}.
\item[(iii)] {\it Expansion in the canonical dimension.} We can classify
the couplings according to their canonical dimension. For this 
purpose we expand $\Gamma_k$ around some constant field $\rho_0$
\bea
\Gamma_k[\Phi] &=& 
\int d^d x \Big\{ U_k(\rho_0)+U_k^\prime (\rho_0)(\rho-\rho_0)
+\frac{1}{2} U_k^{\prime\prime}(\rho_0)(\rho-\rho_0)^2 + \ldots 
\nonumber\\
&& -\frac{1}{2}\Big(Z_k(\rho_0)+Z_k^\prime (\rho_0)(\rho-\rho_0)
+\frac{1}{2} Z_k^{\prime\prime}(\rho_0)(\rho-\rho_0)^2 + \ldots \Big)
\Phi^a \partial_\mu \partial^\mu \Phi_a \nonumber\\
&& + \frac{1}{2} \Big(\dot{Z}_k(\rho_0)+\dot{Z}_k^\prime (\rho_0)(\rho-\rho_0)
+ \ldots \Big) \Phi^a (\partial_\mu \partial^\mu)^2 \Phi_a \nonumber\\
&& - \frac{1}{4} Y_k(\rho_0) \rho \partial_\mu \partial^\mu \rho + \ldots 
\Big\} \, .
\eea
The field $\rho_0$ may depend on $k$. In particular,
for a potential $U_k$ with minimum at $\rho_0(k)>0$
the location of the minimum can be used as one of the 
couplings. In this case $\rho_0(k)$ replaces the 
coupling $U_k^\prime(\rho_0)$ since $U_k^\prime(\rho_0(k))=0$.
In three dimensions one may start by considering an approximation
that takes into account only couplings with positive canonical mass 
dimension, i.e.\ $U^\prime_k(0)$ with mass dimension $M^2$ 
and $U^{\prime\prime}_k(0)$ with dimension $M^1$
in the symmetric regime (potential minimum at $\rho=0$). 
Equivalently, in the spontaneously
broken regime (potential minimum for $\rho\not =0$)
we may take $\rho_0(k)$ and $U^{\prime\prime}_k(\rho_0)$.  
The first correction includes then the dimensionless
parameters $U^{\prime\prime\prime}_k(\rho_0)$ and
$Z_k(\rho_0)$. The second correction includes
$U_k^{(4)}(\rho_0)$, $Z_k^\prime(\rho_0)$ and
$Y_k(\rho_0)$ with mass dimension $M^{-1}$ and so on. 
Already the inclusion of the dimensionless couplings
gives a very satisfactory description of critical
phenomena in three dimensional scalar theories
(see section \ref{onpot}).
\end{itemize}

\subsection{Flow equation for the average potential}
\label{FlowAveragePot}

For a discussion of the ground state, its preserved or sponataneously
broken symmetries and the mass spectrum of excitations the
most important quantity is the average potential $U_k(\rho)$.
In the absence of external sources
the minimum $\rho_0$ of $U_{k\to0}$ determines the expectation
value of the order parameter. The symmetric phase with unbroken
$O(N)$ symmetry is realized if $\rho_0(k\to0)=0$ whereas spontaneous
symmetry breaking occurs for $\rho_0(k\to0)>0$. Except for the wave
function renormalization to be discussed later the squared particle
masses $M^2$ are given by $M^2\sim U'(\rho_0=0)$ for
the symmetric phase. Here primes denote derivatives with
respect to $\rho$. For $\rho_0\not=0$ one finds a radial mode
with $M^2\sim U'(\rho_0)+2\rho_0U''(\rho_0)$ and $N-1$
Goldstone modes with $M^2\sim U'(\rho_0)$. For vanishing
external sources the Goldstone modes are massless.

We therefore want to concentrate on the flow of $U_k(\rho)$.
The exact flow equation is obtained by evaluating eq. (\ref{2.17}) for a
constant value of $\varphi_a$, say $\varphi_a(x)=\varphi\delta_{a1},
\rho=\frac{1}{2}\varphi^2$. One finds the exact equation
\beq\label{2.32}
\partial_tU_k(\rho)=\frac{1}{2}\int\frac{d^dq}{(2\pi)^d}
\frac{\partial R_k}{\partial t}\left(\frac{N-1}{M_0}+
\frac{1}{M_1}\right)\eeq
with
\bea\label{2.33}
M_0(\rho,q^2)&=&Z_k(\rho,q^2)q^2+R_k(q)+U_{k}'(\rho)
\nonumber\\
M_1(\rho,q^2)&=&\tilde Z_k(\rho,q^2)q^2+R_k(q)+U_{k}'(\rho)
+2\rho U_k''(\rho)\eea
parametrizing the $(a,a)$ and (1,1) element of $\Gamma_k^{(2)}
+R_k\ (a\not=1)$.

As expected, this equation is not closed
since we need information about the $\rho$ and $q^2$-dependent
wave function renormalizations $Z_k$ and $\tilde Z_k$ for
the Goldstone and radial modes, respectively.
The lowest order in the derivative expansion would take
$\tilde Z_k=Z_k=Z_k(\rho_0,k^2)$ independent of $\rho$
and $q^2$, so that only the
anomalous dimension
\beq\label{2.34}
\eta=-\frac{\partial}{\partial t}\ln Z_k\eeq
is needed in addition to the partial differential equation (\ref{2.32}).
For a first discussion let us also neglect the contribution 
$\sim \partial_t Z_k$ in $\partial_t R_k$ and write
\beq\label{A.34}
\frac{\partial}{\partial t}U_k(\rho)=2v_dk^d\left[(N-1)l^d_0\left(
\frac{U_k'(\rho)}{Z_kk^2}\right)+l^d_0\left(\frac{U_k'(\rho)+2\rho
U_k''(\rho)}{Z_kk^2}\right)\right]\eeq
with
\beq\label{AA.34}
v_d^{-1}=2^{d+1}\pi^{d/2}\Gamma\left(\frac{d}{2}\right)\ ,
v_2=\frac{1}{8\pi}\ ,\
v_3=\frac{1}{8\pi^2}\ ,\ v_4=\frac{1}{32\pi^2}\eeq
Here we have introduced the dimensionless threshold function
\beq\label{B.34}
l^d_0(w)=\frac{1}{4}v^{-1}_dk^{-d}\int\frac{d^dq}{(2\pi)^d}
\frac{\partial_t(R_k(q)/Z_k)}{q^2+Z^{-1}_kR_k(q)+k^2w}
\eeq
It depends on the renormalized particle mass $w=M^2/(Z_kk^2)$ and
has the important property that it decays rapidly for $w\gg1$.
This describes the decoupling of modes with renormalized squared mass
$M^2/Z_k$ larger than $k^2$. In consequence, only modes with mass
smaller than $k$ contribute to the flow. The flow equations ensure
automatically the emergence of effective theories for the low
mass modes! The explicit form of the threshold functions (\ref{B.34})
depends on the choice of $R_k$. We will discuss several
choices in section \ref{SecThresh}. For a given explicit form
of the threshold functions eq.\ (\ref{A.34}) turns into a 
nonlinear partial differential
equation for a function $U$ depending on the two variables
$k$ and $\rho$. This can be solved numerically by appropriate
algorithms \cite{ABBFTW95-1} as is shown in later sections.    

Eq.\ (\ref{A.34}) was first derived
\cite{Wet91-1} as a renormalization group improved 
perturbative expression and its intuitive form close to perturbation theory
makes it very suitable for practical investigations. 
Here it is important to note that the use of the average action 
allows for the inclusion of propagator corrections
(wave function renormalization effects) in a direct and
systematic way.
Extensions to more complicated scalar models
or models with fermions \cite{Wet90-1} are straightforward.
In the limit of a sharp cutoff (see section \ref{SecThresh})
and for vanishing anomalous dimension eq.\ (\ref{A.34})
coincides with the Wegner-Houghton equation \cite{WH73-1} for the
potential, first discussed in \cite{NCS1} (see also
\cite{NC2,PaRe,Has86-1,CT97}). 

\begin{figure}[h]
\unitlength1.0cm
\begin{center}
\begin{picture}(13.,9.)
\put(6.4,-0.2){\footnotesize $\varphi$ }
\put(1.75,8.){$U_k(\varphi)$}

\put(-0.5,0.){
\epsfysize=13.cm
\epsfxsize=9.cm
\rotate[r]{\epsffile{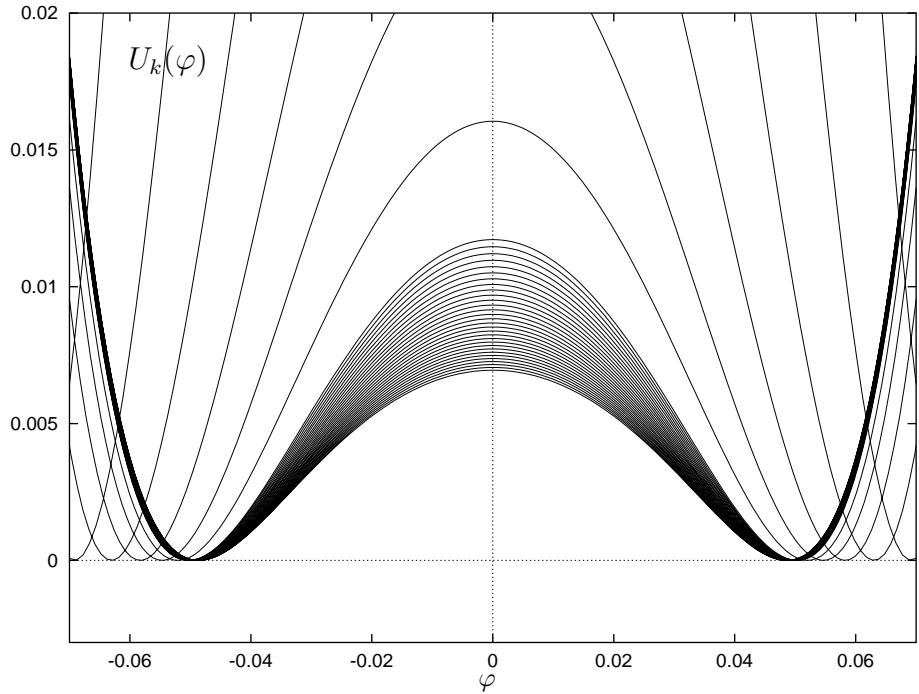}}
}
\end{picture}
\end{center}
\caption[]{\label{apptoconvexity}
\em Average potential $U_k(\varphi)$ for different scales $k=e^t$.
The shape of $U_k$ is stored
 in smaller intervals $\Delta t=-0.02$ after the minimum has settled.
 This demonstrates the approach to convexity in the ``inner region'', while
the ``outer region'' becomes $k$-independent for $k \to 0$.  }
\end{figure}

Eq.\ (\ref{A.34}) can be used as a practical
starting point for various systematic expansions.
For example, it is the lowest order in the derivative 
expansion. The next order includes $q^2$-independent
wave function renormalizations $Z(\rho), \tilde Z(\rho)$
in eq. (\ref{2.33}). For $N=1$ the first order in the
derivative expansion leads therefore to coupled partial
nonlinear differential equations for two functions $U_k(\rho)$ and
$Z_k(\rho)$ depending on two variables $k$ and $\rho$.
We have solved these differential equations numerically
and the result is plotted in fig.\ \ref{apptoconvexity}. 
The initial values of 
the integration correspond to the phase with spontaneous symmetry
breaking. More details can be found in sect.\ \ref{secsec}. 


\subsection{A simple example: the quartic potential}
\label{SimpEx}

Before we describe the solutions of more sophisticated truncations 
of the flow equation (\ref{2.32}) we consider here a very simple polynomial 
approximation to $U_k$, namely
\begin{equation}
  \label{eq:BBB002}
  U_k(\rho)=\frac{1}{2}\ol{\lambda}_k
  \left(\rho-\rho_0(k)\right)^2\; .
\end{equation}
The two parameters $\rho_0(k)$ and $\bar\lambda_k$ correspond
to the renormalizable couplings in four dimensions.
It is not surprising that the flow of $\bar\lambda_k$ will 
reproduce for $d=4$ the usual one-loop $\beta$-function. As 
compared to standard perturbation theory, one gets in
addition a flow equation for the $k$-dependent potential
minimum $\rho_0(k)$ which reflects the quadratic renormalization
of the mass term. It is striking, however, that with the inclusion
of the anomalous dimension $\eta$ the simple ansatz (\ref{eq:BBB002})
also describes \cite{Wet91-1} the physics for lower dimensions
$d=3$ or $d=2$! This includes the second-order phase transition
for $d=3$ with nontrivial critical exponents and even
the Kosterlitz-Thouless transition for $d=2$, $N=2$! In
consequence, we obtain a simple unified picture for the $\phi^4$-model
in all dimensions. This is an excellent starting point for
more elaborate approximations.

All characteristic features of the flow can already be discussed
with the approxi\-mation (\ref{eq:BBB002}). 
We use this ansatz to rewrite eq.\ (\ref{A.34}) as a coupled set
of ordinary differential equations for the minimum of the
potential $\rho_0(k)$ and the quartic coupling $\ol{\lambda}_k$.
This system is closed for a given anomalous dimension $\eta$
(\ref{2.34}).
The flow of the potential minimum can be inferred from the identity
\begin{equation}
  \label{eq:BBB003}
  0=\frac{d}{d t}U_k^\prime(\rho_0(k))=
  \partial_t U_k^\prime(\rho_0(k))+
  U_k^{\prime\prime}(\rho_0(k))\partial_t\rho_0(k)\; .
\end{equation}
Here $\partial_t U_k^\prime(\rho)$ is the partial 
$t$--derivative with $\rho$ held
fixed which is computed by differentiating eq.~(\ref{A.34}) with respect
to $\rho$. Defining the ``higher''  threshold functions by
\begin{eqnarray}
  \label{eq:BBB004}
  \dsp{l_1^d(w)} &=& \dsp{
    -\frac{\partial}{\partial w}l_0^d(w)
    }\nnn
  \dsp{l_{n+1}^d(w)} &=& \dsp{
    -\frac{1}{n}\frac{\partial}{\partial w}
    l_n^d(w)\;,\;\;\;
    n\ge1
    }
\end{eqnarray}
and constants $l_n^d=l_n^d(0)$ one obtains
\begin{equation}
  \label{eq:BBB005}
  \partial_t\rho_0=2v_d k^{d-2}Z_{k}^{-1}
  \Bigg\{3l_1^d\left(\frac{2\rho_0\ol{\lambda}}{Z_{k}k^2}\right)+
  (N-1)l_1^d\Bigg\}\; .
\end{equation}
We conclude that $\rho_0(k)$ always decreases as the infrared cutoff $k$ is
lowered. For $d=2$ and $N\ge2$ only a $Z$--factor increasing without bounds
can prevent $\rho_0(k)$ from reaching zero at some value $k>0$. We will see
that this unbounded $Z$-factor occurs
for $N=2$ in the low temperature phase. If $\rho_0(k)$ reaches
zero for $k_s>0$ the flow for $k<k_s$ can be continued with
a truncation $U_k(\rho)=m^2_k\rho
+\frac{1}{2}\bar\lambda_k\rho^2$ with $m^2_k>0$.
This situation corresponds to the symmetric or disordered
phase with $N$ massive excitations. On the other hand, the phase
with spontaneous symmetry breaking or the ordered phase is 
realized for $\lim_{k\to0}\rho_0(k)>0$.

It is convenient to introduce renormalized dimensionless couplings as
\begin{equation}
  \label{eq:BBB007}
  \kappa=Z_{k}k^{2-d}\rho_0\;,\;\;\;
  \lambda=Z_{k}^{-2}k^{d-4}\ol{\lambda}\; .
\end{equation}
For our simple truncation one has
\begin{equation}
  \label{eq:BBB009}
  u=\frac{1}{2}\lambda(\tilde{\rho}-\kappa)^2
\end{equation}
and the flow equations for $\kappa$ and $\lambda$ read
\begin{eqnarray}
  \label{eq:BBB010}
  \dsp{\partial_t\kappa} &=& \dsp{
    \beta_\kappa=(2-d-\eta)\kappa+
    2v_d\Bigg\{3l_1^d(2\lambda\kappa)+
    (N-1)l_1^d\Bigg\}
    }\\[2mm]
  \label{eq:BBB010a}
  \dsp{\partial_t\lambda} &=& \dsp{
    \beta_\lambda=(d-4+2\eta)\lambda+
    2v_d\lambda^2\Bigg\{9l_2^d(2\lambda\kappa)+
    (N-1)l_2^d\Bigg\}
    }\; .
\end{eqnarray}
In this truncation the anomalous dimension $\eta$ is given
by \cite{Wet91-1}
\begin{equation}
  \label{eq:BBB011}
  \eta=\frac{16v_d}{d}\lambda^2\kappa
  \, m_{2,2}^d(0,2\lambda\kappa)
\end{equation}
where $m_{2,2}^d$ is another threshold function 
defined in sect.\ \ref{onpot}. It has the property
\begin{equation}
  \label{eq:BBB012a}
  \eta=\frac{1}{4\pi\kappa}\;\;{\rm for}\;\;
  d=2\;,\;\;\lambda\kappa\gg1\; .
\end{equation}

The universal critical behavior of many systems of statistical mechanics is
described by the field theory for scalars with $O(N)$ symmetry. This covers
the gas--liquid and many chemical transitions
described by Ising models with a
discrete symmetry $Z_2\equiv O(1)$, superfluids with continuous abelian
symmetry $O(2)$, Heisenberg models for magnets with $N=3$, etc. In $2<d\le4$
dimensions all these models have a continuous second order phase transition.
In two dimensions one observes a second order transition for the Ising model,
a Kosterlitz--Thouless phase transition \cite{KT73-1} for $N=2$ and no
phase transition for non--abelian symmetries $N\ge3$. It is
known \cite{MW66-1} that a continuous symmetry cannot be broken in two
dimensions in the sense that the expectation value of the unrenormalized
scalar field vanishes in the limit of vanishing sources,
$\Phi_a=\langle \chi_a(x) \rangle =0$.
We want to demonstrate that the two differential
equations~(\ref{eq:BBB010}), (\ref{eq:BBB010a}) describe all 
qualitative features of phase transitions
in two or three dimensions correctly. A quantitative numerical analysis
using more sophisticated truncations will be presented in
section \ref{secsec}.

In four dimensions the anomalous dimension and the product $\lambda
\kappa$ become rapidly small quantities as $k$ decreases. We then
recognize in eq.\ (\ref{eq:BBB010a}) the usual perturbative one-loop
$\beta$-function for the quartic coupling
\begin{equation}\label{eq:2.51a}
\partial_t\lambda=\frac{N+8}{16\pi^2}\lambda^2\end{equation}
Eqs. (\ref{eq:BBB010}), (\ref{eq:BBB010a}) also exhibit the well-known
property of ``triviality'' which means that the quartic coupling
vanishes for $k\to 0$ for the massless model. 
 
For $d=3$ the equations~(\ref{eq:BBB010}) and (\ref{eq:BBB010a}) 
exhibit a fixed point
$(\kappa_*,\lambda_*)$ where $\beta_\kappa=\beta_\lambda=0$. 
This is a first example of a ``scaling solution'' for which
all couplings evolve according to an effective dimension which
is composed from their canonical dimension and anomalous dimension,
i.e.\
\beq
Z_k \sim k^{-\eta_*} \, , \quad 
\rho_0 \sim k^{d-2+\eta_*} \, , \quad
\overline{\lambda} \sim k^{4-d- 2 \eta_*} \,\, .
\eeq
From the generic form
$\beta_\lambda=-\lambda+\lambda^2(c_1+c_2(\lambda\kappa))$ one concludes 
that $\lambda$
essentially corresponds to an infrared stable coupling which is attracted
towards its fixed point value $\lambda_*$ as $k$ is lowered. On the other hand,
$\beta_\kappa=-\kappa+c_3+c_4(\lambda\kappa)$ shows that $\kappa$ is 
essentially
an infrared unstable or relevant coupling. Starting for given 
$\lambda_\Lambda$ with
$\kappa_\Lambda=\kappa_*(\lambda_\Lambda)+\delta\kappa_\Lambda$, 
$\delta\kappa_\Lambda=\kappa_T(T_c-T)$, $\kappa_T>0$,
one either ends in the symmetric phase for $\delta\kappa_\Lambda<0$, or 
spontaneous
symmetry breaking occurs for $\delta\kappa_\Lambda>0$. The fixed point 
corresponds precisely to the
critical temperature of a second order phase transition. Critical exponents
can be computed from solutions in the vicinity of the scaling solution. The
index $\nu$ characterizes the divergence of the correlation length for $T\to
T_c$, i.e., $\xi\sim m_R^{-1}\sim |T-T_c|^{-\nu}$ with
$m_R^2=\lim_{k\to0} 2\rho_0 \ol{\lambda}Z_{k}^{-1}$. 
It corresponds to the negative
eigenvalue of the ``stability matrix'' $A_{ij}=
(\partial\beta_i/\partial\lambda_j)(\kappa_*,\lambda_*)$ with 
$\lambda_i\equiv(\kappa,\lambda)$.
(This can be generalized for more than two couplings.)
The critical exponent
$\eta$ determines the long distance behavior of the two--point function for
$T=T_c$. It is given by the anomalous dimension at the fixed point,
$\eta=\eta(\kappa_*,\lambda_*)$. It is remarkable that already in a very
simple polynomial truncation the critical exponents come out with reasonable
accuracy \cite{Wet91-1,TW94-1}. In three dimensions the anomalous dimension
comes out to be small and can be neglected for
a rough treatment, further simplifying the flow equations
(\ref{eq:BBB010}), (\ref{eq:BBB010a}). As an example, for the critical
exponent $\nu$ for $N=3$ and $\eta=0$ one finds $\nu=0.74$, to be
compared with the known value $\nu=0.71$. 
(See also \cite{BHLM95-1} for a discussion
of the $N=1$ case in three dimensions and section \ref{secsec}.)

In two dimensions the term linear in $\kappa$
vanishes in $\beta_\kappa$. This
changes the fixed point structure dramatically as can be seen from
\begin{equation}
  \label{eq:BBB012}
  \lim_{\kappa\to\infty}\beta_\kappa=
  \frac{N-2}{4\pi}
\end{equation}
where $l_1^2=1$ was used. Since $\beta_\kappa$ is always positive for
$\kappa=0$ a fixed point requires that $\beta_\kappa$ becomes negative for
large $\kappa$. This is the case for the Ising
model \cite{Wet91-1,Mor98-1,KNP98-1} where $N=1$. On the other hand, for a
non--abelian symmetry with $N\ge3$ no fixed point and therefore no phase
transition occurs. The location of the minimum always reaches zero for some
value $k_s>0$. The only phase corresponds to a linear realization of $O(N)$
with $N$ degenerate masses $m_R\sim k_s$. It is interesting to note that the
limit $\kappa\to\infty$ describes the non--linear sigma model. The
non--abelian coupling $g$ of the non--linear model is related to $\kappa$ by
$g^2=1/(2\kappa)$ and eq.~(\ref{eq:BBB012}) reproduces the standard one--loop
beta function for $g$
\begin{equation}
  \label{eq:BBB013}
  \frac{\partial g^2}{\partial t}=
  -\frac{N-2}{2\pi}g^4
\end{equation}
which is characterized by asymptotic freedom \cite{Pol75-1}. The
``confinement scale'' where the coupling $g$ becomes strong can be associated
with $k_s$.  The strongly interacting physics of the non--linear model finds a
simple description in terms of the symmetric phase of the linear
$O(N)$--model \cite{Wet91-1}! This may be regarded as an example
of duality: the dual description of the non-linear $\sigma$-model
for large coupling is simply the linear $\varphi^4$-model.

Particularly interesting is the abelian continuous symmetry for $N=2$. Here
$\beta_\kappa$ vanishes for $\kappa\to\infty$ and $\kappa$ becomes a marginal
coupling. As is shown in more detail in section \ref{KT} one
actually finds \cite{GW95-1} a behavior consistent with a second order
phase transition with $\eta\simeq0.25$ near the critical trajectory. The low
temperature phase ($\kappa_\Lambda>\kappa_*$) is special since it has many
characteristics of the phase with spontaneous symmetry breaking, despite the
fact that $\rho_0(k\to 0)$ must vanish according to the Mermin--Wagner
theorem \cite{MW66-1}. There is a massless Goldstone--type boson
(infinite correlation length) and one massive mode. Furthermore, the exponent
$\eta$ depends on $\kappa_\Lambda$ or the temperature
(cf.~eq.~(\ref{eq:BBB012a})), since $\kappa$ flows only marginally. These are
the characteristic features of a Kosterlitz--Thouless phase
transition \cite{KT73-1}. The puzzle of the Goldstone boson in the low
temperature phase despite the absence of spontaneous symmetry breaking is
solved by the observation that the wave function renormalization never stops
running with $k$:
\begin{equation}
  \label{eq:BBB014}
  Z_{k}=\ol{Z}\left(\frac{k}{\Lambda}\right)^{-\eta}\; .
\end{equation}
Even though the renormalized field $\chi_R=Z_{k}^{1/2}\chi$ acquires a
non--zero expectation value $\langle \chi_R \rangle =\sqrt{2\kappa}$, 
for $k\to0$ the
unrenormalized order parameter vanishes due to the divergence of~$Z_{k}$,
\begin{equation}
  \label{eq:BBB015}
  \langle \chi(k) \rangle =\sqrt{\frac{2\kappa}{\ol{Z}}}
  \left(\frac{k}{\Lambda}\right)^{\frac{1}{4\pi\kappa}}\; .
\end{equation}
Also the inverse Goldstone boson propagator behaves as
$(q^2)^{1-1/(8\pi\kappa)}$ and circumvents Coleman's no--go
theorem \cite{Col73-1} for free massless scalar fields in two dimensions.
It is remarkable that all these features arise from the solution of a simple
one--loop type equation without ever invoking non-perturbative vortex
configurations.

\section{Solving the flow equation}

\subsection{Scaling form of the exact flow equation for the potential}
\label{ScalingFormPot}

In this section we discuss analytical approaches to the solution
of the exact flow equation for the average potential and for 
the propagator. They prove to be a useful guidance for the
numerical solutions of truncated partial differential
equations. After writing the exact equation for the potential
in a scale-invariant form and discussing explicitly the
threshold functions, we present an exact solution of the flow
equation in the limit $N\to\infty$. A renormalization group
improved perturbation theory is developed as the iterative solution
of the flow equations. This incorporates the usual gap
equation and can be used as a systematic procedure with the
gap equation as a starting point. We write down the exact flow equation
for the propagator as a basis for a systematic
computation of the anomalous dimension and related quantities.
Finally, we show how the average potential approaches for
$k\to0$ a convex form for the nontrivial case of spontaneous
symmetry breaking.  

Let us first come back to the exact flow equation (\ref{2.32})
and derive an explicitly scale-invariant form of it.
This will be a useful starting point for the discussion
of critical phenomena in later sections. It is convenient
to use a dimensionless cutoff function
\beq\label{2.35}
r_k\left(\frac{x}{k^2}\right)=\frac{R_k(x)}{Z_kx}\ ,\
x\equiv q^2\ ,\ Z_k=Z_k(\rho_0,k^2)\eeq
and write the flow equations as
\beq\label{2.36}
\frac{\partial}{\partial t}U_k(\rho)=v_d\int^\infty_0dx\ x^{\frac{d}{2}}
s_k\left(\frac{x}{k^2}\right)\left(\frac{N-1}
{M_0/Z_k}+\frac{1}{M_1/Z_k}\right)\eeq
with
\beq\label{2.38}
s_k\left(\frac{x}{k^2}\right)=\frac{\partial}{\partial t}r_k
\left(\frac{x}{k^2}\right)-\eta r_k\left(\frac{x}{k^2}\right)=
-2x\frac{\partial}{\partial x}r_k\left(\frac{x}{k^2}\right)-\eta 
r_k\left(\frac{x}{k^2}\right)\eeq
We parametrize the wave function renormalization by
\bea\label{2.39}
z_k(\rho)&=&\frac{Z_k(\rho,k^2)}{Z_k}\ ,\ z_k(\rho_0)\equiv1\ ,\
\rho \tilde y_k(\rho)=
\frac{\tilde Z_k(\rho,k^2)-Z_k(\rho,k^2)}{Z_k^2}k^{d-2},\nonumber \\
\Delta z_k(\rho,\frac{x}{k^2})&=&\frac{Z_k(\rho,x)-Z_k(\rho,k^2)}{Z_k}\ ,
\ \Delta z_k(\rho,1)=0
\nonumber\\
Z_kk^{2-d}
\rho\Delta \tilde y_k(\rho,\frac{x}{k^2})&=&
\frac{\tilde Z_k(\rho,x)-\tilde Z_k(\rho,k^2)}{Z_k}-\Delta z_k(\rho,
\frac{x}{k^2})\ , \ \Delta\tilde y_k(\rho,1)=0\eea
so that
\beq\label{2.40}
\frac{\partial}{\partial t}U_k=2v_dk^d[(N-1)l^d_0\left(
\frac{U_k'}{Z_kk^2};\eta,z_k\right)+l^d_0\left(
\frac{U_k'+2\rho U_k''}{Z_kk^2}; \eta,z_k+
Z_k\rho \tilde y_kk^{2-d}\right)]+\
\Delta\zeta_kk^d\eeq
where $l_0^d(w;\eta, z)$ is a
generalized dimensionless threshold function $(y=x/k^2)$
\beq\label{2.41}
l^d_0(w;\eta,z)=\frac{1}{2}\int^\infty_0dy y^{\frac{d}{2}}s_k(y)
[(z+r_k(y))y+w]^{-1}\eeq
The correction
$\Delta \zeta_k$ contributes only in second order in a derivative expansion.
Finally, we may remove the explicit dependence on $Z_k$ and $k$
by using scaling variables
\beq\label{2.43}
u_k=U_kk^{-d},\quad \tilde \rho=Z_kk^{2-d}\rho\eeq
Evaluating the $t$-derivative at fixed $\tilde\rho$ and
denoting by $u'=\partial u/\partial\tilde\rho$ etc. one
obtains the scaling form of the exact evolution equation
for the average potential
\bea\label{2.44}
\partial_tu_{|_{\tilde\rho}}&=&-du+(d-2+\eta)\tilde\rho u'
+\zeta_k\nonumber\\
&&\zeta_k=2v_d\left\{(N-1)l^d_0(u';\eta,z)
+l^d_0(u'+2\tilde\rho u'',\eta,z+\tilde\rho \tilde y)\right\}+\Delta 
\zeta_k\eea
with
\bea\label{2.44a}
\Delta\zeta_k&=&-v_d\int^\infty_0dyy^{\frac{d}{2}+1}s_k(y)\Bigl\{
\frac{(N-1)\Delta z(y)}
{[(z+r_k(y))y+u'][(z+\Delta z(y)+r_k(y))y+u']}\\
&&+\frac{\Delta z(y)+\tilde\rho\Delta \tilde
y(y)}{[(z+\tilde\rho\tilde y+r_k(y))
y+u'+2\tilde\rho u''][(z+\tilde\rho\tilde y+\Delta z(y)+\tilde\rho
\Delta \tilde y(y)+r_k(y))y+u'+2\tilde\rho u'']}\Bigr\}\nonumber\eea
All explicit dependence on the scale $k$ or the wave function
renormalization $Z_k$ has disappeared.
Reparametrization invariance
under field scaling is obvious in this form. For $\rho\to\alpha^2
\rho$ one also has $Z_k\to \alpha^{-2}Z_k$ so that
 $\tilde\rho$ is invariant. This property needs the factor $Z_k$
in $R_k$. This version is therefore most
appropriate for a discussion of critical behavior.
The universal features of the critical behavior for second-order
phase transitions are related to the existence of a
scaling solution\footnote{More
precisely, $\partial_t u'=0$ is a sufficient condition for the
existence of a scaling solution. 
The universal aspects of first-order transitions are
also connected to exact or approximate scaling solutions.}.  
This scaling solution solves the differential
equation for $k$-independent functions $u(\tilde\rho),\ z(\tilde
\rho)$ etc., which results from (\ref{2.44}) by setting $\partial_tu=0$.
For a constant wave function renormalization the scaling
potential can be directly obtained by solving the second order differential 
equation $\partial_tu=0$. Of all possible solutions it 
has been shown that the physical fixed point
corresponds to the solution $u(\tilde{\rho})$ which is non-singular 
in the field \cite{Felder,FB,Mor94-2,Mor94-1}. 

The scaling form of the evolution equation is the best starting
point for attempts of an analytical solution. In fact, in the
approximation where $\zeta_k$ can be expressed as a function
of $u'$ and $\eta$ one may find the general form of the solution
by the method of characteristics. For this purpose we consider the
$\tilde\rho$-derivative of eq. (\ref{2.44})
\beq\label{BX1}
\partial_t u'=-(2-\eta)u'+(d-2+\eta)\tilde\rho u''-\psi_k u''\ ,
\ \psi_k=-\frac{\partial\zeta_k}{\partial u'}\eeq
We further assume for a moment that the dependence of $\psi_k(u',\eta)$ on
$\eta$ has (approximately) the form
\beq\label{BX2}
\psi_k(u',\eta)=(2-\eta)|u'|^{\frac{d}{2-\eta}}\hat\psi_k(u')\eeq
For $u'>0$ one finds the solution
\beq\label{BX3}
\tilde\rho\ (u')^{1-\frac{d}{2-\eta}}+G(u')=F_+(u'\exp\{2t-\int^t_0
dt'\eta(t')\})\eeq
with $G(u')$ obeying the differential equation
\beq\label{BX4}
\frac{\partial G}{\partial u'}=\frac{1}{2-\eta}(u')^{-\frac{d}{2-\eta}}
\psi_k=\hat\psi_k(u')\eeq
and $F_+(w)$ an arbitrary function. Similarly, for $u'<0$ one has
\beq\label{BX5}
\tilde\rho\ (-u')^{1-\frac{d}{2-\eta}}+H(-u')=F_-(u'\exp
\{2t-\int^t_0dt'\eta(t')\})\eeq
with
\beq\label{BX6}
\frac{\partial H}{\partial(-u')}=\frac{1}{2-\eta}(-u')^{-\frac{d}{2-\eta}}
\psi_k=\hat\psi_k(u')\eeq
and $F_-(w)$ again arbitrary. For known $\hat\psi_k$ one can
now solve the ordinary differential equations for $G$ and $H$.
The initial value of the microscopic potential $u_\Lambda'$
fixes the free functions $F_\pm$ by evaluating (\ref{BX3}) and
(\ref{BX5}) for $t=0$. We finally note that the solution for constant
$\eta$ can be obtained from the solution for $\eta=0$
by the replacements $d \to d_{\eta}=2d/(2-\eta)$,
$t\to t_{\eta}=t(2-\eta)/2$, $G\to G_\eta=2G/(2-\eta),\ H\to
H_\eta=2H/(2-\eta)$.

We will see below that this solution becomes exact in the large $N$
limit. For finite $N$ it may still be used if the functions
$\tilde\rho u'', z, \tilde\rho\tilde y$
appearing in eq. (\ref{2.44}) can be expressed in terms of $u'$ and $\eta$.
The condition (\ref{BX2}) may actually be abandoned in regions of $k$
where $\eta$ varies only slowly. If $\hat\psi_k$ depends on $\eta$,
the corrections to the generic solution (\ref{BX3}), (\ref{BX5}) are
$\sim (\partial\hat\psi_k/\partial\eta)(\partial_t\eta)$.

\subsection{Threshold functions}
\label{SecThresh}

In situations where the momentum dependence of the propagator can
be approximated by a standard form of the kinetic term and is
weak for the other 1PI-correlation functions, the ``non-perturbative''
effects beyond one loop arise to a large extent from the threshold
functions. We will therefore discuss in this subsection their most
important properties and introduce the notations
\bea\label{2.45}
&&l^d_n(w;\eta,z)=\frac{n+\delta_{n,0}}{4}v_d^{-1}k^{2n-d}\int
\frac{d^dq}{(2\pi)^d}\partial_tR_k(q)(Z_kzq^2+R_k(q)+wk^2)^{-(n+1)},
\nonumber\\
&&l^d_{n+1}(w;\eta,z)=-\frac{1}{n+\delta_{n,0}}\frac{\partial}
{\partial w}l^d_n(w;\eta,z),\nonumber\\
&&l^d_n(w;\eta)=l^d_n(w;\eta,1)\ ,\ l^d_n(w)=l^d_n(w;0,1)\ ,\ l^d
_n=l^d_n(0)\eea
The precise form of the threshold functions depends on the
choice of the cutoff function $R_k(q)$. There are, however,
a few general features which are independent of the particular scheme:
\begin{enumerate}
\item
For $n=d/2$ one has the universal property
\beq\label{2.46}
l^{2n}_n=1\eeq
This is crucial to guarantee the universality of the perturbative
$\beta$-functions for the quartic coupling in $d=4$ or for the coupling
in the nonlinear $\sigma$-model in $d=2$.

\item
If the momentum integrals are dominated by $q^2\approx k^2$
and $R_k(q)\stackrel{\scriptstyle<}{\sim} k^2$, the
threshold functions obey for large $w$
\beq\label{2.47}
l^d_n(w)\sim w^{-(n+1)}\eeq
We will see below that this property is not realized for sharp
cutoffs where $l^d_n(w)\sim w^{-n}$.

\item
The threshold functions diverge for some negative value
of $w$. This is related to the fact that the average potential
must become convex for $k\to0$.
\end{enumerate}

It is instructive to evaluate the threshold function explicitly for
a simple cutoff function of the form
\beq\label{C.34}
R_k=Z_kk^2\Theta(k^2-q^2)\eeq
where $(x=q^2)$
\beq\label{D.34}
l^d_0(w)=k^{2-d}\int^{k^2}_0dxx^{\frac{d}{2}-1}
[x+(w+1)k^2]^{-1}
+k^{4-d}\int^\infty_0dxx^{\frac{d}{2}-1}\frac{\delta(x-k^2)}
{x+k^2\Theta(k^2-x)+k^2w}
\eeq
The second term in the expression for $l^d_0$ has to be defined
and we consider eq. (\ref{C.34}) as the limit $\gamma\to\infty$ of a family
of cutoff functions
\beq\label{E.34}
R_k=\frac{2\gamma}{1+\gamma}Z_kk^2\left(\frac{q^2}{k^2}\right)^\gamma
\left[\exp\left\{\frac{2\gamma}{1+\gamma}\left(\frac{q^2}{k^2}\right)
^\gamma\right\}-1\right]^{-1}\eeq
This yields the threshold functions
\bea\label{F.34}
l^d_0(w)&=&\ln\left(1+\frac{1}{1+w}\right)+\hat l_0^d(w)\ ,\
\hat l_0^d(w)=\int^1_0dy y^{\frac{d}{2}-1}(y+1+w)^{-1},\nonumber\\
\hat l^2_0(w)&=&\ln\left(1+\frac{1}{1+w}\right)\ ,\
\hat l_0^3(w)=2-2\sqrt{1+w}\ {\rm arctg}\left(\frac{1}{\sqrt{1+w}}
\right)\nonumber\\
\hat l_0^4(w)&=&1-(1+w)\ln\left(1+\frac{1}{1+w}\right)\ ,\
\hat l_0^{d+2}(w)=\frac{2}{d}-(1+w)\hat l_0^d(w)\eea

In order to get a more detailed overview, it is useful to discuss
some other limiting cases for the choice of the cutoff. For $\gamma=1$
the family of cutoffs (\ref{E.34}) represents the exponential 
cutoff (15), for $\gamma\to 0$
one finds a masslike cutoff $R_k=Z_kk^2$ and for $\gamma\to\infty$
we recover the step-function cutoff (\ref{C.34}). One may modify the
step-function cutoff (\ref{C.34}) to
\beq
R_k=Z_k k^2\Theta(\alpha k^2-q^2)
\eeq
with 
\beq
l^d_0(w)= \ldots + \int\limits_0^\alpha dy y^{\frac{d}{2}-1}
(y+1+w)^{-1} \; .
\eeq
In the limit $\alpha \to \infty$ one approaches again a
masslike cutoff $R_k=Z_k k^2$. We observe that $l^d_0$
does not remain finite in this limit. This reflects the fact
that for a masslike cutoff the high momentum modes are not yet fully
integrated out in the computation of $\Gamma_k$.
However, for low enough dimensions suitable differences remain
finite 
\beq
l_0^3(w)-l_0^3(0)=\pi (\sqrt{1+w} - 1) \; .
\eeq
For $d < 4$ the masslike cutoff can be used except for an overall
additive constant in the potential.

Another interesting extension of the step cutoff (\ref{C.34}) is
\beq\label{2.48}
R_k=Z_kk^2\beta\Theta(k^2-q^2)\eeq
such that
\beq\label{2.49}
l^d_0(w)=\ln\frac{\beta+1+w}{1+w}+\beta\hat l_0^d
(w+\beta-1)\eeq
It is instructive to consider the limit $\beta\to\infty $ which
corresponds to a sharp momentum cutoff, where $r_k(y>1)\to0$
and $r_k(y<1)\to\infty$. Although a sharp momentum cutoff leads
to certain problems with analyticity, it is useful because of
technical simplifications. The momentum integrals are now
dominated by an extremely narrow range $q^2\approx k^2$. (This
holds except for a field-independent constant in $\partial_t
\Gamma_k$.) We can therefore evaluate the two-point function
in the rhs of $\partial_t U_k$ by its value for $q^2=k^2$. In
consequence, the correction $\Delta\zeta_k$ vanishes in this
limit.

Furthermore, the threshold functions take a very simple form and one
infers\footnote{The functions $l^d_1(w)$ coincide with the sharp
cutoff limit of a different family of threshold functions considered
in \cite{Wet91-1}.} from (\ref{F.34}), (\ref{2.49})
\beq\label{2.50}
l^d_0(w)=\ln\frac{1}{1+w}+\ln\beta+\frac{2}{d}\ , \
l^d_1(w)=\frac{1}{1+w}\eeq
Actually, the constant part in $l^d_0$ depends on the precise
way how the sharp cutoff is defined. Nevertheless, this ambiguity
does not affect the field-dependent part of the flow equation and the
sharp cutoff limit obeys universally
\beq\label{2.51}
l^d_1(w;\eta, z)=\frac{1}{z+w}\eeq
One ends in the sharp cutoff limit with a simple exact equation
(up to an irrelevant constant)
\beq\label{2.52}
\partial_tu=-du+(d-2+\eta)\tilde\rho \
u'-2v_d\{(N-1)\ln(z+u')+\ln(z+\tilde\rho\tilde y+u'+2\tilde
\rho u'')\}\eeq
In leading order in the derivative expansion $(z=1, \tilde y
=0)$ and neglecting the anomalous dimension $(\eta=0)$,
this yields for $N=1, d=3$
\beq\label{2.65a}
\partial_tu=-3u+\tilde\rho u'-\frac{1}{4\pi^2}\ln(u'+2\tilde\rho u'')\eeq
which corresponds to the Wegner-Houghton equation \cite{WH73-1} for the
potential \cite{NCS1,NC2,PaRe,Has86-1,CT97}.

\subsection{Large-$N$ expansion}
\label{largeNexp}

In the limit $N\to\infty$ the exact flow equation for the average
potential (\ref{2.44}) can be solved analytically 
\cite{TW94-1,TL96}\footnote{See also refs.\ \cite{WH73-1,FB,CT97,DAM}
for the large $N$ limit of the Wegner-Houghton \cite{WH73-1} and 
Polchinski equation \cite{Pol84-1}.}.
In this limit the quantities
$\eta, \tilde y$ and $\Delta\zeta_k$
vanish and $z=1$. Furthermore, the correction
from $\tilde\rho u''$ in the second term in eq. (\ref{2.44})
is suppressed by $1/N$. In consequence, the right-hand side of the
flow equation for the potential only depends on $u'$ (and $\tilde
\rho$)
\beq\label{F1}
\partial_tu=-du+(d-2)\tilde\rho u'+2v_dNl^d_0(u')\eeq
We can therefore use the exact solution (\ref{BX3}), (\ref{BX4}) with
\beq\label{F2}
\psi_k(u')=2 v_d N l^d_1(u')\eeq
In particular, for the sharp cutoff (\ref{2.51}) the functions G, H obey the
differential equation
\bea\label{F3}
&&\frac{\partial G(w)}{\partial w}=v_dNw^{-d/2}
(1+w)^{-1}\ ,\nonumber\\
&&\frac{\partial H(w)}{\partial w}=v_dNw^{-d/2}
(1-w)^{-1}\eea

Consider the three-dimensional models. For $d=3$ the general
solution reads
\bea\label{F4}
\frac{\tilde\rho-2v_3N}{\sqrt{u'}}-2v_3N\ {\rm arctg}
(\sqrt{u'})&=&F_+(u'e^{2t})
\ ,\quad{\rm for}\ u'>0\nonumber\\
\frac{\tilde\rho-2v_3N}{\sqrt{-u'}}+v_3N\ \ln
\frac{1+\sqrt{-u'}}{1-\sqrt{-u'}}&=&F_-(u'e^{2t})
\ ,\quad
{\rm for}\ u'<0\eea
The functions $F_\pm(w)$ are arbitrary. They are only fixed
by the initial conditions at the microscopic scale $k=\Lambda
\ (t=0)$. As an example, consider the microscopic potential
\beq\label{F5}
u_\Lambda=\frac{1}{2}\lambda_\Lambda(\tilde\rho-\kappa_\Lambda)^2\eeq
Insertion into eq. (\ref{F4}) at $t=0$ yields
\bea\label{F6}
&&F_+(w)=\frac{1}{\sqrt w}(\frac{w}{\lambda_\Lambda}+\kappa_\Lambda
-2v_3N)-2v_3N\ {\rm arctg}(\sqrt w)\nonumber\\
&&F_-(w)=-\frac{\sqrt{-w}}{\lambda_\Lambda}+\frac{\kappa_\Lambda-2v_3N}
{\sqrt{-w}}+v_3N\ln\frac{1+\sqrt{-w}}{1-\sqrt{-w}}\eea
where we note the consistency condition
\beq\label{F7}
u_\Lambda'>-1\ ,\ \lambda_\Lambda\kappa_\Lambda<1\eeq
In consequence, the exact solution for $u'(\tilde\rho,t)$ obeys for $u'>0$
\bea\label{F8}
\tilde\rho-2v_3N&=&2v_3N\sqrt{u'}\ {\rm arctg}\ \sqrt{u'}+\frac
{u'}{\lambda_\Lambda}e^t\nonumber\\
&&+(\kappa_\Lambda-2v_3N)e^{-t}-2v_3N\sqrt{u'}\ {\rm arctg}\ (\sqrt{u'}e^t)\eea
Retranslating to the original variables $\rho=\tilde\rho k,\
U'=\partial U/\partial\rho=u'k^2$ and using $v_3=1/(8\pi^2)$, this
reads
\[\rho-\rho_0(k)=\frac{N}{4\pi^2}\sqrt{U'}\left({\rm arctg}
\left(\frac{\sqrt{U'}}{k}\right)-{\rm arctg}
\left(\frac{\sqrt{U'}}{\Lambda}\right)
\right)+\frac{U'}{\lambda_\Lambda\Lambda}\]
with
\beq\label{F9}
\rho_0(k)=(\kappa_\Lambda-\frac{N}{4\pi^2})\Lambda+\frac{N}{4\pi^2}k\eeq

One sees how the average potential interpolates between the
microscopic potential
\beq\label{F10}
U_\Lambda=\frac{1}{2}\lambda_\Lambda\Lambda(\rho-\kappa_\Lambda
\Lambda)^2\eeq
and the effective potential for $k\to0$
\beq\label{F11}
U=\frac{1}{3}\left(\frac{8\pi}{N}\right)^2(\rho-\rho_0)^3\ ,\
\rho_0=\left(\kappa_\Lambda-\frac{N}{4\pi^2}\right)\Lambda\eeq
Here the last formula is an approximation valid for $U'/\Lambda^2\ll
\left(\frac{N\lambda_\Lambda}{8\pi}\right)^2$ and
$U'/\Lambda^2\ll(\pi/2)^2$  which can easily be replaced by the exact
expression in the range of large $U'$. For vanishing source the
model is in the symmetric phase for $\rho_0<0$ with masses
of the excitations given by
\beq\label{F12}
M^2=U'(0)=\left(\frac{8\pi}{N}\right)^2\left(\kappa_\Lambda-
\frac{N}{4\pi^2}\right)^2\Lambda^2\eeq
For $\rho_0>0$ spontaneous symmetry breaking occurs with order
parameter (for $J=0$)
\beq\label{F13}
\langle \varphi\rangle =\sqrt{2\rho_0}=(\kappa_\Lambda-
\frac{N}{4\pi^2})^{1/2}(2\Lambda)^{1/2}\eeq
If we associate the deviation of $\kappa_\Lambda$ from the critical
value $\kappa_{\Lambda,c}=\frac{N}{4\pi^2}$
with a deviation from the critical temperature $T_c$
\beq\label{F14}
\kappa_\Lambda=\frac{N}{4\pi^2}+\frac{A}{\Lambda}(T_c-T)\ ,\
\rho_0=A(T_c-T)\eeq
it is straightforward to extract the critical exponents in the large
$N$ approximation
\beq\label{F15}
M\sim(T-T_c)^\nu\quad,\quad\langle \varphi\rangle =(T_c-T)^\beta\ ,\
\nu=1\quad, \quad\beta=0.5\eeq
Eq. (\ref{F11}) constitutes the critical equation of state for the
dependence of the magnetization $\varphi$ on a homogeneous magnetic
field $J$
with
\beq\label{F16}
J=\frac{\partial U}{\partial\varphi}=\varphi U'=
\left(\frac{4\pi}{N}\right)
^2\varphi(\varphi^2-2\rho_0)^2\eeq
At the critical temperature one has $\rho_0=0$ and
\beq\label{F17}
J\sim\varphi^\delta\quad, \quad\delta=5\eeq
whereas the susceptibility
\beq\label{F18}
\chi=\frac{\partial\varphi}{\partial J}=\left(\frac{N}{4\pi}\right)^2
(\varphi^2-2\rho_0)^{-1}(5\varphi^2-2\rho_0)^{-1}\eeq
obeys for $\rho_0<0$ at $J=0$
\beq\label{F19}
\chi\sim(T-T_c)^{-\gamma}\quad,\quad \gamma=2\eeq
We note that the critical amplitudes (given by
the proportionality constants
in eqs. (\ref{F15}), (\ref{F17}) and (\ref{F19}))
are all given explicitly by eq. (\ref{F11}) once the
proportionality constant $A$ in eq. (\ref{F14}) is fixed. Universal
amplitude ratios are those which
do not depend on $A$. In the large $N$
approximation the explicit solution $U(\rho,T)$ contains only one
free constant instead of the usual two. This is related to the
vanishing anomalous dimension and provides for an additional
universal amplitude ratio. We finally may define the quartic
coupling
\bea\label{F20}
&&\lambda_R=U''(0)\quad{\rm for}\quad \rho_0\leq0\nonumber\\
&&\lambda_R=U''(\rho_0)\quad{\rm for}\quad \rho_0>0\eea
and observe the constant critical ratio in the symmetric
phase
\beq\label{F21}
\frac{\lambda_R}{M}=\frac{16\pi}{N}\eeq
All this agrees with diagramatic studies of the large $N$ approximation
\cite{Zin93-1} and numerical solutions of the flow equation 
\cite{TW94-1,ABBFTW95-1,BTW96-1} for
the exponential cutoff function (\ref{2.15}).

The solution for the region $u'<0$ can be found along similar
lines
\beq\label{F22}
\rho-\rho_0(k)=\frac{N}{8\pi^2}\sqrt{-U'}\left(\ln\frac{
\Lambda+\sqrt{-U'}}{\Lambda-\sqrt{-U'}}-\ln\frac{k+
\sqrt{-U'}}{k-\sqrt{-U'}}\right)+\frac{U'}{\lambda_\Lambda\Lambda}
\eeq
Up to corrections involving inverse powers of $\Lambda$ this
yields the implicit relation
\beq\label{F23}
\sqrt{-U'}=k-(k+\sqrt{-U'})\exp\left(
-\frac{8\pi^2(\rho_0(k)-\rho)}{N\sqrt{-U'}}\right)\eeq
One infers that $\sqrt{-U'}$ is always smaller than $k$ and maximal
for $\rho=0$. In the phase with spontaneous symmetry breaking where
$\rho_0(k=0)=\rho_0>0$ the behavior near the origin is given
for small $k$ by
\beq\label{F24}
U'(\rho)=-k^2\left[1-2\exp\left(-\frac{8\pi^2(\rho_0-\rho)}{Nk}\right)
\right]^2\eeq
For $k\to0$ the validity of this region extends to the whole
region $0\leq\rho<\rho_0$. The ``inner part'' of the average
potential becomes flat\footnote{The exponential approach
to the asymptotic form $U'=-k^2$ may cause problems for a
numerical solution of the flow equation. The analytical information
provided here can be useful in this respect.}, in agreement with
the general discussion of the approach to convexity above.

It is interesting to compare these results with the large
$N$ limit of the scaling solution which obeys
\beq\label{89a}
u'=\frac{1}{2}(\tilde\rho-\frac{N}{4\pi^2}\frac{1}{1+u'})u''\eeq
For $u'(\kappa)=0$ this yields for the minimum of $u$
\beq\label{89b}
\kappa=\frac{N}{4\pi^2}\eeq
Taking $\rho$-derivatives of eq. (\ref{89a}) and defining
$\lambda=u''(\kappa),\ \gamma=u'''(\kappa)$, we find
\beq\label{89c}
u''(1-\frac{\kappa u''}{(1+u')^2})=(\tilde\rho-\frac{\kappa}{
1+u'})u'''\eeq
\beq\label{89d}
\lambda=\frac{1}{\kappa}=\frac{4\pi^2}{N}\ ,\ \gamma=\frac{2}{3}
\lambda^2=\frac{32\pi^4}{3N^2}\eeq

We emphasize that we have chosen the quartic potential (\ref{F5})
only for the simplicity of the presentation. The general solution
can be evaluated equally well for other microscopic potentials
(provided $u_\Lambda'>-1$). This can be used for an explicit
demonstration of the universality of the critical behavior. It also
may be employed for an investigation of tricritical behavior which
can happen for more complicated forms of the microscopic potential.
In summary, the exact solution of the flow equation for the average
potential in the limit $N\to\infty$ provides a very detailed
quantitative description for the ``transition to
complexity''.

\subsection{Graphical representation and resummed perturbation
theory}
\label{iterative}

One is often interested in the flow equation for some particular
$n$-point function $\Gamma_k^{(n)}$. Here $\Gamma^{(n)}$ is defined
by the $n$-th functional derivative of $\Gamma_k$ evaluated for a
fixed field, as, for example, $\varphi=0$. Correspondingly, the
flow equation for $\Gamma^{(n)}$
can be obtained by taking $n$ functional derivatives
of the exact flow equation (\ref{2.17}). We give here a simple
prescription how the rhs of the flow equation can be computed
from the usual perturbative Feynman graphs. Only one-loop diagrams
are needed, but additional vertices are present and the
propagators and vertices in the graphs correspond to full
propagators and full vertices as derived by functional
differentiation of $\Gamma_k$.

This purpose is achieved by writing eq. (\ref{2.17}) formally in the form
\beq\label{Z1}
\partial_t\Gamma_k=\frac{1}{2}\ {\rm Tr}\
\tilde\partial_t\ln (\Gamma_k^{(2)}+R_k)
\eeq
where the derivative $\tilde\partial_t$ acts only on $R_k$ and
not on $\Gamma_k$, i.e. $\tilde\partial_t=(\partial R_k/\partial t)
\partial/\partial R_k$. If we forget for a moment possible problems
of regularization, we may place $\tilde\partial_t$ in front
of the trace so that
\beq\label{Z2}
\partial_t\Gamma_k=\tilde\partial_t\hat\Gamma_k^{(1)}\eeq
where $\hat\Gamma_k^{(1)}$ is the one-loop expression
with ``renormalization group improvement'', i.e. full
vertices and propagators instead of the classical ones.
Functional derivatives commute with $\partial_t$ and 
$\tilde\partial_t$. In
consequence, the right-hand side of the exact flow equation
for $\Gamma_k^{(n)}$ can be evaluated by the
following procedure: 1) Write down the one-loop Feynman
graphs for $\Gamma^{(n)}$. 2) Insert ``renormalized''
couplings instead of the classical ones. This also
introduces a momentum dependence of the vertices which
may not be present in the classical couplings. There may
also be contributions from higher vertices (e.g. six-point
vertices) not present in the classical action. (The classical
couplings vanish, the renormalized effective ones do not!) All renormalized
vertices in the graphs correspond to appropriate functional
derivatives of $\Gamma_k$. 3) Replace the propagators by the full
average propagator $(\Gamma^{(2)}+R_k)^{-1}$ (evaluated at a
fixed field). 4) Apply the formal differentiation
$\tilde\partial_t$. More precisely, the derivative $\tilde\partial_t$
should act on the integrand of the one-loop momentum integral. This
makes the expression finite so that regularization is of no
worry. The result is the exact flow equation for $\Gamma_k^{(n)}$.
An example is provided by eq. (\ref{2.28}).

Standard perturbation theory can easily be recovered from an iterative
solution of the flow equation (\ref{Z1}). Starting from the leading
or ``classical'' contribution $\Gamma_{k(0)}\equiv\Gamma_\Lambda$
one may insert this instead of $\Gamma_k$ in the rhs of (\ref{Z1}).
Performing the $t$-integration generates the one-loop
contribution
\beq\label{G5}
\Gamma_k-\Gamma_\Lambda=\frac{1}{2} {\rm Tr}\left\{
\ln (\Gamma_\Lambda^{(2)}+R_k)-\ln(\Gamma_\Lambda^{(2)}+
R_\Lambda)\right\},\eeq
where we remind that $R_k\to0$ for $k\to 0$.
We observe that the momentum integration in the rhs of (\ref{G5})
is regularized in the ultraviolet through subtraction
of $\ln(\Gamma_\Lambda^{(2)}+R_\Lambda)$. This is a type
of implicit Pauli-Villars regularization with the heavy mass
term replaced by a momentum-dependent piece $R_\Lambda$ in the
inverse propagator. With suitable chirally invariant $R_\Lambda$
\cite{Wet90-1} this can be used for a regularization of models with chiral
fermions. Also gauge theories can, in
principle, be regularized in this way, but
care is needed since $\Gamma_\Lambda$ has to obey identities
reflecting the gauge invariance 
\cite{RW93-1,RW93-2}, \cite{Bec96-1}--\cite{FreireLitim}. 
Going further, the
two-loop contribution is obtained by inserting the one-loop expression
for $\Gamma_k^{(2)}$ as obtained from eq. (\ref{G5})
into the rhs of (\ref{Z1}). It is easy to see that this
generates two-loop integrals. Only the
classical inverse propagator $\Gamma_\Lambda^{(2)}$ and its
functional derivatives appear in the nested expressions. They are
independent of $k$ and the integration of the approximated flow equation
is straightforward.

It is often useful to replace the perturbative
iteration sketched above by a new one which involves  full
propagators and vertices instead of the classical ones.
This will amount to a systematic resummed perturbation
theory \cite{ResPT}. We start
again with the lowest order term
\beq\label{G6}
\Gamma_{k(0)}[\varphi]=\Gamma_\Lambda[\varphi]\eeq
where $\Lambda$ is now some conveniently chosen scale (not
necessarily the ultraviolet cutoff). In the next step we write equation
(\ref{Z1}) in the form
\beq\label{G7}
\partial_t\Gamma_k=\frac{1}{2}\ {\rm Tr}\ \partial_t\ln(\Gamma_k^{(2)}
+R_k)
-\frac{1}{2}\ {\rm Tr}\ \left\{\partial_t\Gamma_k^{(2)}(\Gamma_k
^{(2)}+R_k)^{-1}\right\}\eeq
Here $\partial_t\Gamma_k^{(2)}$ can be inferred by taking
the second functional derivative of eq. (\ref{Z1})  with respect
to the fields $\varphi$. Equation (\ref{G7})
can be taken as the starting point of a systematic loop expansion
by counting any $t$-derivative acting only on $\Gamma_k$
or its functional derivatives as an
additional order in the number of loops. From (\ref{Z1}) it is
obvious that any such derivative involves indeed a new momentum
loop. It will become clear below that in case of weak
interactions it also
involves a higher power in the coupling constants. The
contribution from the first step
of the iteration $\Gamma_{k(1)}'$ can now be defined by
\beq\label{G8}
\Gamma_k=\Gamma_{k(0)}+\Gamma_{k(1)}'\eeq
with
\beq\label{G9}
\Gamma_{k(1)}'=\frac{1}{2}\ Tr\ \left\{\ln(\Gamma_k^{(2)}+R_k)-
\ln(\Gamma_k^{(2)}+ R_\Lambda)\right\}\eeq
In contrast to eq. (\ref{G5}) the rhs involves now the full
(field-dependent) inverse propagator $\Gamma_k^{(2)}$, and, by
performing suitable functional derivatives, the full proper vertices.
Putting $k=0$ the resummed one-loop expression (\ref{G9})
resembles a Schwinger-Dyson \cite{DS49-1}
or gap equation, but in contrast to
those only full vertices appear! For example, for $k=0$ the
first iteration to the inverse propagator is obtained
by taking the second
functional derivative of eq. (\ref{G9})
\bea\label{G10}
\left({\Gamma'}^{(2)}_{(1)}\right)_{ab}(q,q')&=&\frac{1}{2}\ {\rm Tr}\
\left\{\left(\Gamma^{(2)}\right)^{-1}
\frac{\delta^2\Gamma^{(2)}}{\delta\varphi_a(q)
\delta\bar\varphi_b(q')}\right\}\nonumber\\
&&-\frac{1}{2}\ {\rm Tr}\ \left\{\left(\Gamma^{(2)}\right)^{-1}
\frac{\delta\Gamma^{(2)}}{\delta\varphi_a(q)}\left(\Gamma^{(2)}\right)
^{-1}\frac{\delta\Gamma^{(2)}}{\delta\bar\varphi_b(q')}\right\}
\nonumber\\
&&- {\rm regulator\ terms}\eea
and involves the proper three- and four-point vertices. Adding the
lowest order piece $\Gamma_{(0)}^{(2)}$ and approximating the vertices
by their lowest order expressions, eq. (\ref{G10})
reduces to the standard
gap equation for the propagator in a regularized form. Assuming that
this is solved (for example numerically by an iterative procedure)
we see that the resummed one-loop expression (\ref{G10}) involves
already arbitrarily high powers in the coupling constant, and,
in particular, contains part of the perturbative two-loop contribution.
The remaining part of the perturbative two-loop contribution
appears in the resummed two-loop contribution. Along these lines
a systematic resummed perturbation theory (SRPT) can be
developed \cite{ResPT}.

Systematic resummed perturbation theory is particularly
convenient for a computation of ultraviolet finite $n$-point
functions (as the $\varphi^6$ coupling) or differences of
$n$-point functions at different momenta. In this case the
momentum integrals in the loop expansion are dominated
by momenta for which renormalized vertices are 
appropriate\footnote{A direct use of the ``gap equation''
(\ref{G10}) for the mass term is no improvement as compared
to the standard Schwinger-Dyson equation since high momentum modes
play an important role in the mass renormalization.}.
For $\Lambda\to\infty$ all dependence on the effective ultraviolet
cutoff $\Lambda$ is absorbed in the renormalized couplings. In
this limit also the regulator terms vanish. In this context one can
combine SRPT with approximate solutions of the flow equation.
In fact, the ``closure'' of the flow equation by SRPT instead
of truncation constitutes an interesting alternative:
Eq. (\ref{Z1}) is a functional differential equation which cannot
be reduced to a closed system for a finite number of couplings
(for finite $N$).
For example, the beta function for the four-point vertex (the fourth
functional derivative of the rhs of eq. (\ref{Z1})) involves
not only two-, three- and four-point functions, but also up to
six-point functions. Approximate solutions to the flow equation
often proceed by truncation. For example, contributions involving
the five- and six-point function could be neglected. As an
alternative, these higher $n$-point functions can be evaluated
by SRPT. We observe that the
momentum integrals relevant for the higher $n$-point functions
(also for differences of lower $n$-point functions at different
momenta) are usually dominated by the low momentum modes with
$q^2\approx k^2$. This motivates the use of SRPT rather
than standard perturbation theory for this purpose.
A successful test of these ideas is provided by a computation
of the $\beta$-function for the quartic scalar coupling for
$d=4$ via an evaluation of the momentum dependence of
the vertices appearing in the flow equation by SRPT \cite{pw}.
This resulted for small coupling in the universal two-loop
$\beta$-function, without that two-loop momentum integrals
had ever to be computed.

\subsection{Exact flow of the propagator}

We now turn to the exact flow equation for the propagator. This can
be derived either by the graphical rules of the last subsection
or by the functional derivatives sketched in eq. (\ref{2.28}). We
concentrate here on the inverse propagator in the Goldstone direction
in a constant background field (see eq. (\ref{2.34})
\beq\label{98}
G^{-1}(\rho,q^2)=Z_k(\rho,q^2)q^2+U_k'(\rho)=M_0(\rho,q^2)-
R_k(q)\eeq
According to eq. (\ref{2.28}) its flow involves the three- and
four-point functions. We notice that the three-point function $\Gamma^{(3)}
_k$ vanishes for $\rho=0$.

We parametrize the most general form of the
inverse propagator
 in an arbitrary
constant background $\varphi_a$ by
\bea\label{99}
\Gamma_{(2)k}&=&\frac{1}{2}\int\frac{d^dq}{(2\pi)^d}\{(U_k'(\rho)+
Z_k(\rho,q^2)q^2)\ \varphi_a(q)\varphi_a(-q)\nonumber\\
&&+\frac{1}{2}\varphi_a\varphi_b(2U_k''(\rho)+Y_k(\rho,q^2)q^2)\ \varphi_a
(q)\varphi_b(-q)\},\eea
and note that eq. (\ref{99})  specifies all 1PI n-point functions with
at most two nonvanishing momenta.
Similarly the effective interactions which
involve up to four fields with nonvanishing momentum are
\bea\label{99a}
\Gamma_{(3)k}&=&\frac{1}{2}\int\frac{d^dq_1}{(2\pi)^d}
\frac{d^dq_2}{(2\pi)^d}\{\varphi_a\lambda_k^{(1)}
(\rho;q_1,q_2)\ \varphi_a(q_1)\varphi_b(q_2)\varphi_b(-q_1-q_2)
\nonumber\\
&&+\frac{1}{3}\varphi_a\varphi_b\varphi_c\gamma_k^{(1)}(\rho;q_1,q_2)\
\varphi_a(q_1)
\varphi_b(q_2)\varphi_c(-q_1-q_2)\},\nonumber\\
\Gamma_{(4)k}&=&\frac{1}{8}\int\frac{d^dq_1}{(2\pi)^d}
\frac{d^dq_2}{(2\pi)^d}\frac{d^dq_3}{(2\pi)^d}
\{\lambda^{(2)}_k(\rho;q_1,q_2,q_3)\ \varphi_a(q_1)\varphi_a(q_2)
\varphi_b(q_3)\varphi_b(q_4)\nonumber\\
&&+2\varphi_a\varphi_b\gamma_k^{(2)}(\rho;q_1,q_2,q_3)\
\varphi_a(q_1)\varphi_b(q_2)
\varphi_c(q_3)\varphi_c(q_4)\nonumber\\
&&+\frac{1}{3}\varphi_a\varphi_b\varphi_c\varphi_d\tau_k
(\rho;q_1, q_2,q_3)\ \varphi_a
(q_1)\varphi_b(q_2)\varphi_c(q_3)\varphi_d(q_4)\},\eea
with $q_4=-(q_1+q_2+q_3), \rho=\frac{1}{2}\varphi_a\varphi_a$.
The couplings $\lambda_k^{(i)},\gamma_k^{(i)}$ and $\tau_k$
are appropriately symmetrized in the momenta (including
symmetrization in the momentum of the last field, ie. $q_4$).
The effective vertices
are connected to $U_k, Z_k$ and $Y_k$ by continuity
\bea\label{100}
\lambda_k^{(1)}(\rho;q_1,q_2)&=&U_k''(\rho)+(q_2(q_1
+q_2))Z_k'(\rho,(q_2(q_1+q_2)))+
\frac{1}{2}q^2_1Y_k(\rho,q_1^2)+\Delta \lambda_k^{(1)}
(\rho;q_1,q_2)\nonumber\\
\lambda^{(2)}_k(\rho;q_1,q_2, q_3)&=&U_k''(\rho)-(q_1q_2)Z_k'
(\rho,-(q_1q_2))-(q_3
q_4)Z_k'(\rho,-(q_3q_4))\nonumber\\
&&+\frac{1}{2}(q_1+q_2)^2Y_k(\rho,(q_1+q_2)^2)+\Delta\lambda_k^{(2)}
(\rho;q_1,q_2,q_3)\nonumber\\
\gamma_k^{(1)}(\rho;q_1,q_2)&=&U_k'''(\rho)-\frac{1}{2}
\{(q_1q_2)Y_k'(\rho,-(q_1q_2))\nonumber\\
&&+
(q_1\to -(q_1+q_2))+(q_2\to -(q_1+q_2))\}+\Delta\gamma_k^{(1)}
(\rho;q_1,q_2)\nonumber\\
\gamma_k^{(2)}(\rho;q_1,q_2,q_3)&=&U_k'''
(\rho)-(q_3q_4)Z_k''(\rho,-(q_3q_4))
\nonumber\\
&&+\frac{1}{2}\{(q_1(q_1+q_2))Y_k'(\rho,(q_1(q_1+q_2))
+(q_1\to q_2)\}\\
&&-\frac{1}{2}(q_1q_2)Y_k'(\rho,-(q_1q_2))+\Delta\gamma_k^{(2)}
(\rho;q_1,q_2,q_3)\nonumber\\
\tau_k(\rho;q_1,q_2,q_3)&=&U^{(4)}(\rho)-\frac{1}{2}
\{(q_1,q_2)Y_k''(\rho,-(q_1q_2))+ 5\ {\rm permutations}\}
+\Delta \tau_k(\rho;q_1,q_2,q_3)\nonumber\eea
In fact, we require that eq. (\ref{99a}) coincides with eq.
(\ref{99}) if only two of the momenta are nonvanishing, without that the
corrections $\Delta\gamma,\Delta\lambda,\Delta\tau$ are involved. In
particular, $\Delta\lambda_k^{(1)}$ and $\Delta\gamma_k^{(1)}$ vanish
for $q_1=0, q_2=0$ or $(q_1+q_2)=0$, and,  similarly,
$\Delta\lambda_k^{(2)},\Delta
\gamma_k^{(2)}$ and $\Delta\tau_k$ vanish if two of the four momenta
$q_1,q_2,q_3$ or $q_4$ are zero.

In terms of these couplings the exact flow equation
for $G^{-1}(\rho,q^2)$
reads
\bea\label{101}
\partial_tG^{-1}(\rho,q^2)&=&\frac{1}{2}\int\frac{d^dp}{(2\pi)^d}
\partial_tR_k(p)\nonumber\\
&&[4\rho\{M^{-2}_1(\rho,p^2)M^{-1}_0(\rho,(p+q)
^2)(\lambda^{(1)}_k(\rho;p,q))^2\nonumber\\
&&+M^{-2}_0(\rho,p^2)M^{-1}_1(\rho,(p+q)^2)(\lambda_k^{(1)}
(\rho;-q-p,q))^2\}\nonumber\\
&&-M_0^{-2}(\rho,p^2)\{(N-1)\lambda_k^{(2)}(\rho;q,-q,p)+2
\lambda_k^{(2)}(\rho;q,p,-q)\}\nonumber\\
&&-M^{-2}_1(\rho,p^2)\{\lambda_k^{(2)}(\rho;q,-q,p)+2\rho\gamma_k^{(2)}
(\rho;p,-p,q)\}]\nonumber\\
&=&-\frac{1}{2}\int\frac{d^dp}{(2\pi)^d}
\tilde\partial_t\left\{\right. 4\rho
(\lambda_k^{(1)}(\rho;-q-p,q))^2M^{-1}_0(\rho,p^2)
M^{-1}_1((q+p)^2)\nonumber\\
&&-[(N-1)\lambda_k^{(2)}(\rho;q,-q,p)+2\lambda_k^{(2)}
(\rho;q,p,-q)]M^{-1}_0(\rho,p^2)\nonumber\\
&&-[\lambda_k^{(2)}(\rho;q,-q,p)+2\rho\gamma_k^{(2)}
(\rho;p,-p,q)]M_1^{-1}(\rho,p^2)\left.\right\}\eea
Subtraction of the mass term in eq. (\ref{98}) yields
\beq\label{102}
\partial_tZ_k(\rho,q^2)=\frac{1}{q^2}(\partial_tG^{-1}(\rho,q^2)-
\partial_tU_k'(\rho))\equiv-\xi_k(\rho,\frac{q^2}{k^2})Z_k\eeq
and the exact expression for the anomalous dimension is
\bea\label{103}
\eta&=&-\frac{d}{dt}\ln Z_k(\rho_0(k),k^2)\nonumber\\
&=&\xi_k(\rho_0,1)-\frac{2k^2}{Z_k}\frac{\partial}{\partial q^2}
Z_k(\rho_0,q^2)|_{q^2=k^2}-\ \frac{\partial\rho_0}
{\partial t}\frac{Z_k'(\rho_0,k^2)}{Z_k(\rho_0,k^2)}\eea
Alternative definitions of $\eta$ related to the flow
of $Z_k(\rho_0,q^2\to0)$ will be discussed in later sections.
Expressed in terms of the scaling variables the anomalous
dimension can be equivalently extracted from the condition
(\ref{2.39}), i.e. $dz_k(\kappa)/dt=0$ or
\beq\label{104}
\partial_tz(\kappa)=-z_k'(\kappa)\ \partial_t\kappa\eeq
with $\kappa=Z_kk^{2-d}\rho_0$ the minimum of $u(\tilde \rho)$.
The flow of $z_k(\tilde \rho)$ obeys
\beq\label{105}
\partial_tz(\tilde\rho)=\eta z(\tilde\rho)+(d-2+\eta)\tilde\rho z'(\tilde
\rho)-\xi_k(\tilde\rho,1)+2\frac{\partial}{\partial y}\Delta z(\tilde
\rho,y)|_{y=1}\eeq
and we note that for the scaling solution the rhs of eq. (\ref{104})
vanishes since $\partial_t\kappa=0$.

An exact expression for $\xi_k$ is computed for a sharp
cutoff in appendix \ref{adsc}. Here we evaluate $\xi_k$ in first order in the
derivative expansion. In this order the momentum dependence
of $Z',Z'', Y$ and $Y'$ is neglected and we can omit
the terms $\Delta\lambda_k^{(i)},\Delta\gamma_k^{(i)}$
and $\Delta \tau_k$. Inserting eq. (\ref{100}) in 
(\ref{101}) yields
\bea\label{A30A}
&&\xi_k(\rho,\frac{q^2}{k^2})=\frac{1}{2}\int\frac{d^dp}{(2\pi)^d}
Z^{-1}\tilde\partial_t\left\{\right. \rho(2U''+p^2Y)^2\nonumber\\
&&M_0^{-1}(p^2)[M_1^{-1}((p+q)^2)-M_1^{-1}(p^2)]/q^2\nonumber\\
&&+\rho[(4U''+2p^2Y)((1+2\frac{(pq)}{q^2})Y-2\frac{(pq)}
{q^2}Z')\nonumber\\
&&+q^2((1+2\frac{(pq)}{q^2})Y-2\frac{(pq)}{q^2}Z')^2]
M_0^{-1}(p^2)M_1^{-1}((p+q)^2)\nonumber\\
&&\left.-[(N-1)Z'+Y]M_0^{-1}(p^2)-(Z'+2\rho Z'')M_1^{-1}(p^2)\right\}
\eea
In this approximation we can also neglect the term 
$\partial \Delta z/\partial y$ in eq. (\ref{105}) so that
for the scaling solution $(\partial_tz(\tilde\rho)=0)$
the anomalous dimension is given by
\beq\label{A30B}
\eta_{y_\eta}=\xi_k(\kappa,y_{\eta})-(d-2+\eta_{y_\eta})\kappa
z'(\kappa)\eeq
Here we show the freedom in the definition of $\eta$ by the subscript
$y_{\eta}$ which indicates the value of $q^2/k^2$ for which
$Z_k$ is defined.

The optimal choice is presumably $y_{\eta}=1$ (see above)
which corresponds to the hybrid derivative expansion. An algebraic
simplification occurs, however, for $y_{\eta}=0$, corresponding
to the ``direct'' derivative expansion. For a smooth cutoff
the propagators
$M^{-1}_0((p+q)^2)$ and $M^{-1}_1((p+q)^2)$ can be expanded
for $q^2\to 0$
\bea\label{A30C}
&&M^{-1}_0((p+q)^2)=M_0^{-1}(p^2)-(q^2+2pq)(Z+\dot R_k(p))
M^{-2}_0(p^2)\nonumber\\
&&+(q^2+2(pq))^2[(Z+\dot R_k(p))^2M^{-3}_0(p^2)-\frac{1}{2}
\ddot R_k(p)M^{-2}_0(p^2)]+...\eea
with $\dot R_k=\partial R_k/\partial p^2$ and $Z$ replaced
by $Z+\rho Y$ in a similar expression for $M^{-1}_1((p+q)^2)$. Since
integrals over odd powers of $p_\mu$ vanish, one concludes
that $\eta_0$ is well defined for a smooth cutoff. On the other
hand, one observes \cite{Wet91-1} in the sharp cutoff limit a
divergence in the derivative expansion $\lim_{y_\eta\to 0}
\eta_{y_\eta}\sim(y_\eta)^{-1/2}$. 

We emphasize that the dependence of $\eta$ on the choice
of $y_\eta$ is a pure artifact of the truncation. Going 
beyond the derivative expansion, the scaling solution is characterized
by the same anomalous dimension $\eta$ independent of $y_\eta$.
Nevertheless, for practical calculations some type of a 
derivative expansion is often crucial. For the direct derivative
expansion it has been argued \cite{MTi99} that only smooth
and rapidly falling cutoffs (like the exponential cutoff (\ref{PosDef}))
and only the flow equations for $\Gamma_k$ (as opposed to the
one for the Wilsonian effective action) have satisfactory
convergence properties. The constraints on the precise
formulation seem, however, much less restrictive
for the hybrid derivative expansion. This is strongly 
suggested by the computation of $\eta$ for a sharp cutoff in
appendix \ref{adsc}. A detailed investigation is presented in \cite{ABW}.

The result for $\xi_k(\rho,0)$ in first order in the derivative
expansion can be found in \cite{Wet91-1,TW93-1}. Here we present explicitly
only the lowest order in the derivative expansion for which all terms
$\sim Z',Z''$ and $Y$ are neglected. One finds
\bea\label{A30D}
&&\eta_0=\xi_k(\kappa,0)=\frac{16v_d}{d}\lambda^2\kappa\  m^d_{2,2}(0,2
\lambda\kappa)\nonumber\\
&&m^d_{n_1,n_2}(w_1,w_2)=-\frac{1}{2} k^{2(n_1+n_2-1)-d}
Z_k^{n_1+n_2-2}
\int^\infty_0dxx^{\frac{d}{2}}\tilde\partial_t\nonumber\\
&&\left\{\dot P^2(x)(P(x)+Z_kk^2w_1)^{-n_1}(P(x)+Z_kk^2w_2)^{-n_2}
\right\}\eea
where $P(x)=Z_kx+R_k(x)$ and $\dot P=\partial P/\partial x$.
In particular, the limit $w_2\gg1$ is simple for all cutoff functions
for which $\dot R_k(x)$ does not exceed $Z_k$ by a large factor
in some region of $x$, namely
\beq\label{A30E}
\lim_{w_2\to\infty}m^d_{n_1,n_2}(w_1,w_2)
=w_2^{-n_2}m^d_{n_{1,0}}(w_1)\eeq
For $d=2$ and neglecting terms $\sim\eta$ one has the identity
\bea\label{A30F}
m^2_{2,0}(0)&=&-\frac{1}{2}\int^\infty_0dx\ x\tilde\partial_t\{\dot P^2(x)/P^2(x)\}\nonumber\\
&=&\int^\infty_0dx\frac{d}{dx}\left\{\frac{\dot P^2(x)x^2}
{P^2(x)}\right\}=
1\eea
One concludes for all cutoffs in this class
\beq\label{A30G}
\lim_{\lambda\kappa\to\infty}\eta_0=2v_2/\kappa=\frac{1}{4\pi\kappa}
\eeq
This independence of the precise choice of the cutoff is directly
related to the universality of the one-loop $\beta$-function
for the two-dimensional nonlinear $\sigma$-models (cf. eq. (\ref{eq:BBB013})).

\subsection{Approach to the convex potential for spontaneous symmetry
breaking}
\label{convexpot}

For $k\to 0$ the effective potential $U$ is a convex 
function of $\varphi$ \cite{RaiW}, i.e.\ $U_0^{\prime}(\rho) \ge 0$. 
This is not only a formal property due to the
Maxwell construction of thermodynamic potentials. The convexity reflects
directly the effect of fluctuations. Indeed, as long as not all
fluctuations are included the average potential $U_k$ for $k>0$ needs
not to be convex -- and it is actually not convex in case of spontaneous
symmetry breaking. One therefore has to understand quantitatively
how the fluctuations lead to an approach to
convexity for $k\to 0$. The ``flattening'' of the nonconvex
``inner region'' of the potential is crucial to the computation
of the nucleation rate in case of first-order phase transitions.
This rate receives an exponential suppression factor from the
free energy of the saddle point solution which interpolates
between two local minima of $U_k$ and corresponds to tunneling through
a (nonconvex!) potential barrier. We discuss this issue in detail in
section \ref{CoarseGrain}. The discussion of this subsection is
relevant for average potentials $U_k(\varphi)$ with several local
minima for all $k > 0$, and not for the case where only a single minimum
survives for small $k$ (e.g.\ the symmetric phase of $O(N)$ models). 

Different pictures can describe how fluctuations lead to a flattening
of the potential. In a first approach \cite{RingWet90} the average
potential
has been computed in a saddle point approximation. In the inner
(nonconvex) region of the potential the relevant saddle point does
not correspond to a constant field but rather to a spin wave (for $N>1$)
or a kink (for $N=1$). Due to the existence of these nontrivial extrema
the average potential shows in case of spontaneous symmetry breaking
a generic behavior
\beq\label{AA28}
U_k(\rho)=V(k)-pk^2\rho\eeq
for small $\rho$ and $k$. (Here $p$ is a positive constant
and $V(k)$ is independent of $\rho$.) Similar nontrivial
saddle points have been discussed \cite{AlexBranchPolon98} in the context
of the exact Wegner-Houghton \cite{WH73-1} equation. This ``classical
renormalization'' confirms the generic behavior (\ref{AA28}).

Since the flow equation (2.35) for the average potential is exact,
there is, in principle, no need to investigate special non-perturbative
saddle point solutions. The effects of all non-perturbative
solutions like spin waves, kinks or instantons in appropriate
models are fully included in the exact equation. For an appropriate
truncation the solution of eq. (2.35) should therefore directly exhibit
the approach to convexity. A first discussion of the flow of the
curvature of $U_k$ around the origin has indeed shown \cite{TetWet92}
the generic behavior (\ref{AA28}). We extend this discussion here to
the whole ``inner region'' of the potential. In this region
the poles of the threshold functions for negative arguments dominate
for small $k$. This leads indeed to a flattening of the average potential
and to convexity for $k\to0$, with the universal behavior (\ref{AA28}).

The approach to convexity will be dominated by a pole of the threshold
functions for negative arguments. This will allow us to find a solution of
the flow equation which is valid for a restricted range of the field
variable $\rho$. The way  how convexity is approached depends only on a
few characteristics of the infrared cutoff $R_k$ and is otherwise model
independent. 
We therefore need the
behavior of the threshold functions for negative $w$. 
In order to become independent of truncations we generalize
$l^d_0$ for a momentum-dependent wave function 
renormalization\footnote{Here we have not indicated 
in our notation the dependence of
$\bar l^d_0$ on $\eta$ and the function $z(y,\tilde \rho)=
z_k(\tilde\rho)+\Delta z_k(\tilde\rho,y)$. The exact flow equation
for the potential is obtained then from eq. (2.63) by the replacement
$l^d_0\to\bar l_0^d, \Delta\zeta_k=0$.}
\bea\label{A28}
\bar l^d_0(w)&=&\frac{1}{2}\int^\infty_0dyy^{\frac{d}{2}}
s_k(y)(p(\tilde\rho,y)+w)^{-1}\nonumber\\
p(\tilde\rho,y)&=&(z(\tilde\rho,y)+r_k(y))y\eea
We consider here a class of infrared cutoffs with the property 
 that for a given $\tilde\rho$ the function
$p(\tilde\rho,y)$ has a minimum at $y_0>0$, with $p(\tilde\rho,y_0)=p_0
(\tilde\rho)$. For a  range of $(\tilde \rho,w)$
for which
\beq\label{B28}
\epsilon=p_0+w>0\eeq
and for  small values of $\epsilon$  the integral
(\ref{A28}) is dominated by the region $y\approx y_0$. In the
vicinity of the minimum of $p(y)$   we can expand
\beq\label{C28}
p(y)=p_0+a_2(y-y_0)^2+...\eeq
and approximate   the threshold function\footnote{See ref.\
\cite{TetWet92} for a detailed discussion. 
The quantities $p_0$ and $a_2$ depend on $\tilde\rho$.
If $w(\tilde\rho)$ is monotonic in the appropriate range
of $\tilde\rho$ we can consider them as functions of $w$ and expand
$p_0(w)=p_0(-p_0)+\epsilon p_0'+...$, 
$a_2(w)=a_2(-p_0)+\epsilon a_2'+...$
Up to higher orders in $\epsilon$ we can neglect the
$\tilde\rho$-dependence of $p_0$ and $a_2$ and use constants
$p_0\equiv p_0(-p_0), a_2\equiv a_2(-p_0)$.}
(for $s_k(y_0)>0$)
\bea\label{D28}
\bar l^d_0(w)&=&\frac{1}{2}\int^\infty_0dyy_0^{d/2}s_k(y_0)[a_2(y-y_0)^2
+\epsilon]^{-1}\nonumber\\
&=&\frac{\pi}{2}y_0^{d/2}s_k (y_0)(a_2\epsilon)^{-1/2}\eea
One concludes that $\bar l^d_0$ has a singularity\footnote{Near the
singularity the correction is $\sim \ln\epsilon$ or smaller.}
 $\sim\epsilon^{-1/2}$. This is the generic behavior\footnote{It is, however,
not realized for a sharp cutoff for which the threshold function diverges only
logarithmically.}
 for smooth
threshold functions with sufficiently large $R_k(0)$. 

For $N>1$ the contribution of the radial mode can be at most as strong
as the one from the Goldstone modes and   
the exact evolution equation for the potential can  be
approximated in the vicinity of the singularity at $u'=-p_0$
by
\beq\label{G28}
\partial_tu+du-(d-2+\eta)\tilde\rho u'=
2\tilde c(p_0+u')^{-1/2}
\eeq
In the range of its validity we can solve eq. (\ref{G28}) by the method
of characteristics, using eqs. (2.56), (2.57)
with
\beq\label{H28}
\psi_k(u')=\tilde c(p_0+u')^{-\frac{3}{2}}\eeq
 For $\eta=0$ the most singular
terms yield
\beq\label{K28}
(p_0+u')^{-\frac{1}{2}}+\frac{1}{\tilde c}p_0^{d/2}\
\tilde\rho\ (-u')^{1-\frac{d}{2}}=f(u'e^{2t})
\eeq
Here $f$ is fixed by the ``initial value'' $u_0'(\tilde\rho)$
at $t$ and reads 
\beq\label{L28}
f(u'e^{2t})=(p_0+u'e^{2(t-t_0)})^{-\frac{1}{2}}
+\frac{1}{\tilde c}p_0^{\frac{d}{2}}\hat\rho(u'e^{2(t-t_0)})
(-u')^{1-\frac{d}{2}}e^{(2-d)(t-t_0)}\eeq
where the function $\hat\rho$ is obtained by inversion of $u'_0(\tilde
\rho)$,
\beq\label{M28}
\hat\rho(u_0'(\tilde\rho))=\tilde\rho\eeq
For simplicity we consider here a linear approximation $u_0'(\tilde\rho)
=\lambda_0(\tilde\rho-\kappa_0)$
with $\lambda_0\kappa_0=p_0-\epsilon_0,\ 0<\epsilon_0\ll1$, or
\beq\label{N28}
\hat\rho(u'e^{2(t-t_0)})=\frac{1}{\lambda_0}u'e^{2(t-t_0)}
+\kappa_0\eeq

For $d>2$ the approximate expression for eq. (\ref{K28}) in the limit
$k\ll k_0$ reads (with $e^{2(t-t_0)}=
k^2/k^2_0)$
\bea\label{P28}
u'&=&-p_0+\left(\frac{p_0\kappa_0}{\tilde{c}}\right
)^{-2}\left(\frac{k}{k_0}\right)^{2(d-2)}
\nonumber\\
&&\left(1-\frac{\tilde\rho}{\kappa_0}\left(\frac{k}{k_0}\right)^{d-2}+
\frac{\tilde c}{\kappa_0}p_0^{-\frac{3}{2}}
\left(\frac{k}{k_0}\right)^{d-2}\right)^{-2}
\eea
We see that $u'+p_0$ remains always positive. The singularity
is approached, but never crossed. (These features hold for
generic $u_0'(\tilde\rho)>-p_0$.) In consequence, the derivative
of the average potential is always larger than $-p_0k^2$
\beq\label{Q28}
\frac{\partial U_k}{\partial \rho}>-p_0k^2\eeq
and the potential becomes indeed convex for $k\to0$. Furthermore,
this implies that the region where eq. (\ref{Q28}) is valid
extends towards the potential minimum for $k\to0$. For small $k$
a reasonable approximate form for the phase with spontaneous
symmetry breaking and $\rho<\rho_0(k)$ is 
\beq\label{R28}
U_k(\rho)=\tilde V(k)-p_0k^2\rho+C\rho^2_0(k)k^{2(d-1)}
(\rho_0(k)-\rho)^{-1}\eeq
where the constant $C$ can be extracted by comparing with eq. (\ref{P28}).
This agrees with eq. (\ref{AA28}). The validity of the approximation
(\ref{R28}) breaks down in a vicinity of $\rho_0(k)$
which shrinks to zero as $k\to0$. In this vicinity the behavior
for $\rho<\rho_0(k)$ is essentially determined by analytic
continuation from the region $\rho\geq\rho_0(k)$.

For a nonvanishing constant anomalous dimension $\eta$
we have to replace
\beq\label{R28A}
d\to d_\eta=\frac{2d}{2-\eta}\ ,\ t\to t_\eta=\frac{2-\eta}{2}
t\ , \ \tilde c\to \tilde c_\eta=\frac{2}{2-\eta}\tilde c\eeq
This is important for $d=2, N=2$ where the anomalous dimension
governs the approach to convexity. Indeed, at the Kosterlitz-Thouless
phase transition (see section 3.9) or in the low temperature phase
the anomalous dimension remains strictly positive for all
values\footnote{This is in contrast to $d=3$ or $d=2,
N=1$ where the anomalous dimension vanishes for $k\to 0$, except
for the critical hypersurface of the phase transition.}
of $k$. The above discussion (\ref{P28}) of the approach to
convexity remains valid, with $\left(\frac{k}{k_0}\right)^{d-2}$
replaced by $\left(\frac{k}{k_0}\right)^\eta$. We
 finally turn to the case $N=1$ which we only discuss for
$\eta=0$. Near the pole at $\epsilon=u'+2\tilde\rho u''+p_0
\to 0$ the evolution equation for the potential is now approximated
by
\beq\label{T28}
\partial_tu=-du+(d-2)\tilde\rho u'+2\tilde c(p_0+
u'+2\tilde\rho u'')^{-1/2}\eeq
For $d>2$ and small $k$ an iterative solution can be found
for $|2\tilde\rho u''|\ll|p_0+u'|$.
In lowest order it is given by (\ref{P28})
\beq\label{U28}
p_0+u'=\left(\frac{p_0\kappa_0}{\tilde c}\right)^{-2}\left(\frac{k}
{k_0}\right)^{2(d-2)}\left[1-\frac{\tilde\rho}{\kappa_0}\left(\frac{k}{k_0}
\right)^{d-2}\right]^{-2}\eeq
and one finds that
\beq\label{V28}
\frac{2\tilde\rho u''}{p_0+u'}=\frac{4\tilde\rho}{\kappa_0}\left(
\frac{k}{k_0}\right)^{d-2}\left[1-
\frac{\tilde\rho}{\kappa_0}\left(\frac{k}
{k_0}\right)^{d-2}\right]^{-1}\eeq
becomes indeed negligible for $k\to 0$. For this solution one obtains
the same approach to convexity as for $N>1$.

It is instructive to compare the ``power law approach''
(\ref{R28}) towards the asymptotic
behavior for the ``inner region'' $U_k'=-p_0k^2$
with the ``exponential approach'' (\ref{F24}). It is obvious
that the first is much easier to handle for numerical solutions. Indeed,
already a tiny error in the numerical computation of (\ref{F24}) may
lead to vanishing or negative values of $p_0k^2+U_k'$ for which the
threshold functions are ill defined. This is often a source of
difficulties for numerical solutions of the partial differential
equations. For numerical investigations of the approach to convexity
it is advantageous to use threshold functions of the type (2.61) or
corresponding smooth versions like (\ref{2.15}) or (\ref{E.34}) multiplied
by $\beta>1$, say $\beta=4$. Then a simple truncation of the
momentum dependence of the propagator like $\Gamma^{(2)}\sim
Z_kq^2$ + const obeys the conditions for a power-law approach
(\ref{R28}), namely a minimum of $p(y)$ for $y_0>0$ (\ref{A28})
and $s_k(y_0)>0$ (\ref{G28}). On the other hand, the above simple
truncation fails to describe the appropriate approach to convexity
for the cutoff (\ref{2.15}). This is due to the fact that the minimum
of $p(y)$ occurs for $y_0=0$ in this case. One learns that
for this cutoff
the simple truncation of the propagator is insufficient for the inner
region of the potential. A correct reproduction of the exact
bound $U_k'\geq-Z_kk^2$ (\ref{PosDef}) needs an extension of the truncation for
the momentum dependence. For practical investigations of problems where
the precise approach to convexity is not relevant one may use
directly the knowledge of the exact result (\ref{PosDef}). Neglected
effects of an insufficient truncation for the ``inner
region'' may then be mimicked by
a modification ``by hand''
of the threshold function in the immediate vicinity
of the pole, for example by imposing the form (\ref{D28}).

Let us summarize the
most important result of this subsection.
The solution of the exact flow equation for the average
potential leads in case of spontaneous symmetry breaking to a universal
form $U_k'(\rho)\approx -p_0k^2$ for small $\rho$ and $k$. As
$k\to 0$ the region of validity of this behavior extends towards
the minimum of the potential. Eq. (\ref{AA28}) becomes valid
in a range $0\leq\rho<\rho_0(k)-\Delta(k)$
with $\Delta(k)>0$ and $\lim_{k\to0}\Delta (k)=0$. The potential
becomes therefore convex, in agreement with general properties
and the exact bound (\ref{PosDef}).

\section{$O(N)$-symmetric scalar models}
\label{secsec}
\subsection{Introduction}

In this section we study the $N$-component scalar
model with $O(N)$-symmetry in three and two dimensions. 
The case of four dimensional
quantum field theories will be considered in section \ref{Fermions}.
The $O(N)$ model serves as a prototype for investigations concerning 
the restoration of a spontaneously broken symmetry at 
high temperature. For $N=4$ the model describes the scalar
sector of the electroweak standard model in the limit of
vanishing gauge and Yukawa couplings. It is
also used as an effective model for the chiral
phase transition in QCD in the limit of two quark flavors 
\cite{PW84-1,Wilczek92,RaWi93-1,Raj95-1}, which will be discussed in 
section \ref{Fermions}. In condensed matter physics $N=3$ corresponds to the
Heisenberg model used to describe the ferromagnetic
phase transition. There are other applications like the
helium superfluid transition ($N=2$), liquid-vapor transition
($N=1$) or statistical properties of long polymer chains
($N=0$). 

For three dimensions, we will concentrate on the computation of 
the equation of state near the critical 
temperature of the second order phase transition. The equation 
of state for a magnetic system is specified by the
free energy density as a function
of arbitrary magnetization $\phi$ and temperature $T$.
All thermodynamic quantities can be derived from the function
$U(\phi,T)$, which equals the
free energy density for vanishing
source (cf. eq. (\ref{2.2})).
For example, the response of the system to
a homogeneous magnetic field $H$ follows from 
$\partial U/ \partial \phi = H$. This permits the computation
of $\phi$ for arbitrary $H$ and $T$.
There is a close analogy to quantum field theory at non-vanishing temperature.
Here $U$ corresponds to the temperature dependent effective
potential as a function of a scalar field $\phi$.
For instance, in the $O(4)$ symmetric model for the chiral
phase transition in two flavor QCD the meson field $\phi$
has four components. In this picture, the average light 
quark mass $\hat{m}$ is associated
with the source $H \sim \hat{m}$ 
and one is interested in the behavior during the phase 
transition (or crossover) for $H \not= 0$.
The temperature and source
dependent meson masses and zero momentum interactions 
are given by derivatives of $U$ (cf.\ section \ref{Fermions}).

The applicability of the $O(N)$-symmetric
scalar model to a wide class of very different
physical systems in the vicinity of the critical temperature
$T_{c}$ is a manifestation of universality of critical phenomena.
There exists a universal scaling form of the equation of state
in the vicinity of the second order phase transition.
The quantitative description of this scaling
form will be the main topic here
\cite{BTW96-1,ABBFTW95-1}.
The calculation of the effective potential $U(\phi,T)$ in the
vicinity of the critical
temperature of a second order phase transition is an 
old problem. One can prove 
through a general
renormalization group analysis \cite{Wil71-1, WH73-1} the Widom
scaling form \cite{Widom} of the 
equation of state\footnote{We frequently suppress
in our notation an appropriate power of a suitable ``microscopic'' 
length scale $\Lambda^{-1}$ which is used to render quantities 
dimensionless.}    
\beq
H = \phi^{\delta} \tilde{f}
\left((T-T_{c})/\phi^{1/\beta}\right) \label{wid}.
\eeq
Only the 
limiting cases $\phi \rightarrow 0$ and $\phi \rightarrow \infty$
are described by critical exponents and amplitudes.

For classical
statistics in three dimensions we present in
section \ref{onpot} a computation of the effective potential 
$U_0=\lim_{k\to 0} U_k=U/T$ from a derivative expansion
of the effective average action with a uniform wave function
renormalization factor. The approximation 
takes into account 
the most general field dependence of the potential term. 
This will allow us to compute the non--analytic behavior
of $U$ in the vicinity of the second order phase transition.
>From $U$ the universal scaling form of the equation of state
is extracted in section \ref{uceos}. 

We demonstrate in section \ref{IsingModel} that the non--universal aspects 
can be described by these methods as well. The example of
carbon dioxide is worked out in detail. Going beyond the
lowest order in a derivative expansion
the approximation used in these sections
takes into account the most general field dependence of the
wave function renormalization factor.     

Section \ref{Polymers} describes an application to the
critical swelling of long polymer chains ($N=0$).
For two dimensions, we present in section \ref{KT} a quantitative 
description of the ``Kosterlitz-Thouless'' transition ($N=2$).

\subsection{The running average potential}
\label{onpot}

In this section we compute the
effective average potential $U_k(\rho)$
directly in three dimensions \cite{BTW96-1,ABBFTW95-1}. Here 
$\rho = \frac{1}{2} \phi^a \phi_a$ and $\phi^a$ denotes the 
$N$-component real scalar field. For
$k \rightarrow 0$ one obtains the effective potential 
$U_0(\rho)\equiv U(\rho)T^{-1}$, where we omit
the factor $T^{-1}$ in the following. It is related to the (Helmholtz)
free energy density $f_H$ by $f_H/T=U_0-2\rho \partial U_0/\partial \rho
=U_0-\phi H$.
In the phase with spontaneous
symmetry breaking the minimum of the potential occurs for $k=0$ at
$\rho_0 \not= 0$. In the symmetric phase
the minimum of $U_k(\rho)$ ends at $\rho_0=0$ for $k=0$. 
The two phases are separated by a scaling solution for which
$U_k/k^3$ becomes independent of $k$ once expressed in terms
of a suitably rescaled field variable and the
corresponding phase transition is of second order.

Our truncation is the lowest order in a derivative
expansion of $\Gamma_k$, 
\beq
\Gamma_k = \int d^dx \bigl\{
U_k(\rho) + \frac{1}{2} Z_k \partial^{\mu} \phi_a
\partial_{\mu} \phi^a \bigr\}.
\label{three} \eeq
We keep for the
potential term the most general $O(N)$-symmetric
form $U_k(\rho)$, whereas the wave function renormalization
is approximated by one $k$-dependent parameter.
We study the effects of a field dependent $Z_k$ for the Ising
model in section \ref{IsingModel}. 
Next order in the derivative expansion would be the
generalization to a $\rho$-dependent wave function
renormalization $Z_k(\rho)$ plus a function
$Y_k(\rho)$ accounting for a possible different
index structure of the kinetic term for $N \geq 2$.
Going further would require the consideration of
terms with four derivatives and so on.
We employ in this section the exponential infrared cutoff
(\ref{2.15}).

For a study of the behavior in the vicinity of the phase transition
it is convenient to work
with dimensionless renormalized fields
\footnote{We keep the number of dimensions $d$ arbitrary and specialize
only later to $d=3$.}
\bea
\tilde{\rho} &=& Z_k k^{2-d} \rho \; , \nonumber \\
u_k(\tilde{\rho}) &=& k^{-d} U_k(\rho).
\label{four} \eea
The scaling form of the evolution equation for the effective
potential has been derived in (\ref{ScalingFormPot}). 
With the truncation of eq. (\ref{three}) the exact
evolution equation for $u'_k \equiv \partial u_k/\partial \tilde{\rho}$
reduces then to the partial differential equation
\bea
\frac{\partial u'_k}{\partial t} =~&&(-2 + \eta) u'_k +(d-2+ \eta)
\tilde{\rho} u_k''
\nonumber \\
&&- 2 v_d (N-1) u_k'' l^d_1(u_k';\eta)
- 2 v_d (3 u_k'' + 2 \tilde{\rho} u_k''') 
l^d_1(u_k'+2 \tilde{\rho} u_k'';\eta),
\label{five}
\eea
where $t = \ln \left( k/\Lambda \right)$, $v_3 = 1/8 \pi^2$,
primes denote derivatives with respect to $\tilde\rho$
and $\Lambda$ is
the ultraviolet cutoff of the theory.
The ``threshold'' functions $l^d_n(w;\eta) \equiv l^d_n(w;\eta,z=1)$
are discussed in section \ref{SecThresh}.
We evaluate these
functions numerically for the cutoff (\ref{2.15}). Finally, 
the anomalous dimension is 
defined here by the $q^2$-derivative of the inverse propagator
at $q^2=0$. According to sect. 3.5 it is given
in our truncation by\footnote{We neglect here for
simplicity the implicit, linear $\eta$-dependence of the
function $m^d_{2,2}$. We have numerically verified this approximation
to have only a minor effect on the value of $\eta$.} 
\cite{TW93-1,TW94-1,ABBFTW95-1}
\beq
\eta(k) =  \frac{16 v_d}{d} \kappa \lambda^{2} m^d_{2,2}(2 \lambda \kappa),
\label{ten} \eeq
with $\kappa$ the location of the minimum of $u_k$
and $\lambda$ the quartic coupling
\bea
u'_k(\kappa)&=&0 \nonumber\; , \\
u''_k(\kappa)&=&\lambda\; .
\label{eleven} \eea
The function $m^d_{2,2}$ is given by 
\bea
m^d_{2,2}(w) &= & \int_0^{\infty} dy y^{\frac{d}{2}-2}
\frac{1 + r +y \frac{\partial r}{ \partial y}}{(1+r)^2
\left[(1+r)y +w \right]^2 }
\nonumber \\
&&\biggl\{
2 y \frac{\partial r}{ \partial y}
+ 2 \left(y \frac{\partial }{ \partial y} \right)^2 r
- 2 y^2 \left( 1 +r + y \frac{\partial r}{ \partial y} \right)
\frac{\partial r}{ \partial y}
\left[ \frac{1}{(1+r)y} + \frac{1}{(1+r)y + w} \right]
\biggr\}.
\nonumber \\
{}~&~
\label{twelve} \eea
We point out that the argument
$2 \lambda \kappa$ turns out generically to
be of order one for the scaling solution. Therefore,
$\kappa \sim \lambda^{-1}$ and the mass effects are important,
in contrast to perturbation theory where they are
treated as small quantities $\sim \lambda$.

\begin{figure}
\label{EosNew}
\leavevmode
\centering  
\epsfxsize=4in
\epsffile{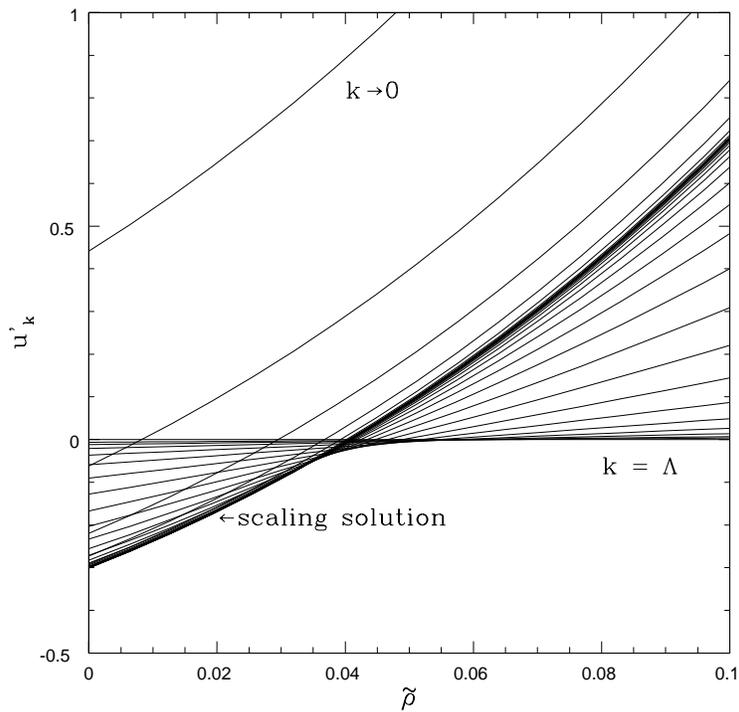}                                   
\caption{\em The evolution of $u'_k(\tilde{\rho})$ 
as $k$ is lowered from $\Lambda$ to zero for $N=1$.
The initial conditions (bare couplings)
have been chosen such that the scaling solution is approached
before the system evolves towards the symmetric phase
with $u'_k(0) > 0$. The concentration of lines near the scaling solution
(flat diagonal line) indicates that the model is close to
criticality. The scaling solution for $u(\tilde\rho)$ has a minimum for 
 $\tilde{\rho} \approx 0.04$.}
\end{figure}
At a second order phase transition 
there is no mass scale present in the theory.
In particular, one 
expects a scaling behavior of the rescaled effective 
average potential $u_k(\tilde{\rho})$. This can be studied
by following the trajectory describing the scale dependence
of $u_k(\tilde{\rho})$ as $k$ is lowered from $\Lambda$ to zero.  
Near the phase transition the   
trajectory spends most of the ``time'' $t$ in the vicinity of
the $k$-independent scaling solution of eq.\ (\ref{five}) given by
$\partial_t u'_* (\tilde{\rho}) = 0$.
Only at the end of the running the ``near-critical''
trajectories deviate from the scaling solution. For
$k \rightarrow 0$ they either end up in the symmetric phase with
$\kappa =0$ and positive constant mass term $m^2$ so that
$u'_k(0) \sim m^2/k^2$; or they lead to a non-vanishing
constant $\rho_0$ indicating spontaneous symmetry breaking with
$\kappa \rightarrow Z_0 k^{2-d} \rho_0$. The equation of state
involves the potential $U_0(\rho)$ for
temperatures away from the critical temperature.
Its computation requires the
solution for the running 
away from the critical trajectory which involves the
full partial differential equation (\ref{five}).

\begin{figure}
\label{CouplingsFig}
\leavevmode
\centering  
\epsfxsize=4in
\epsffile{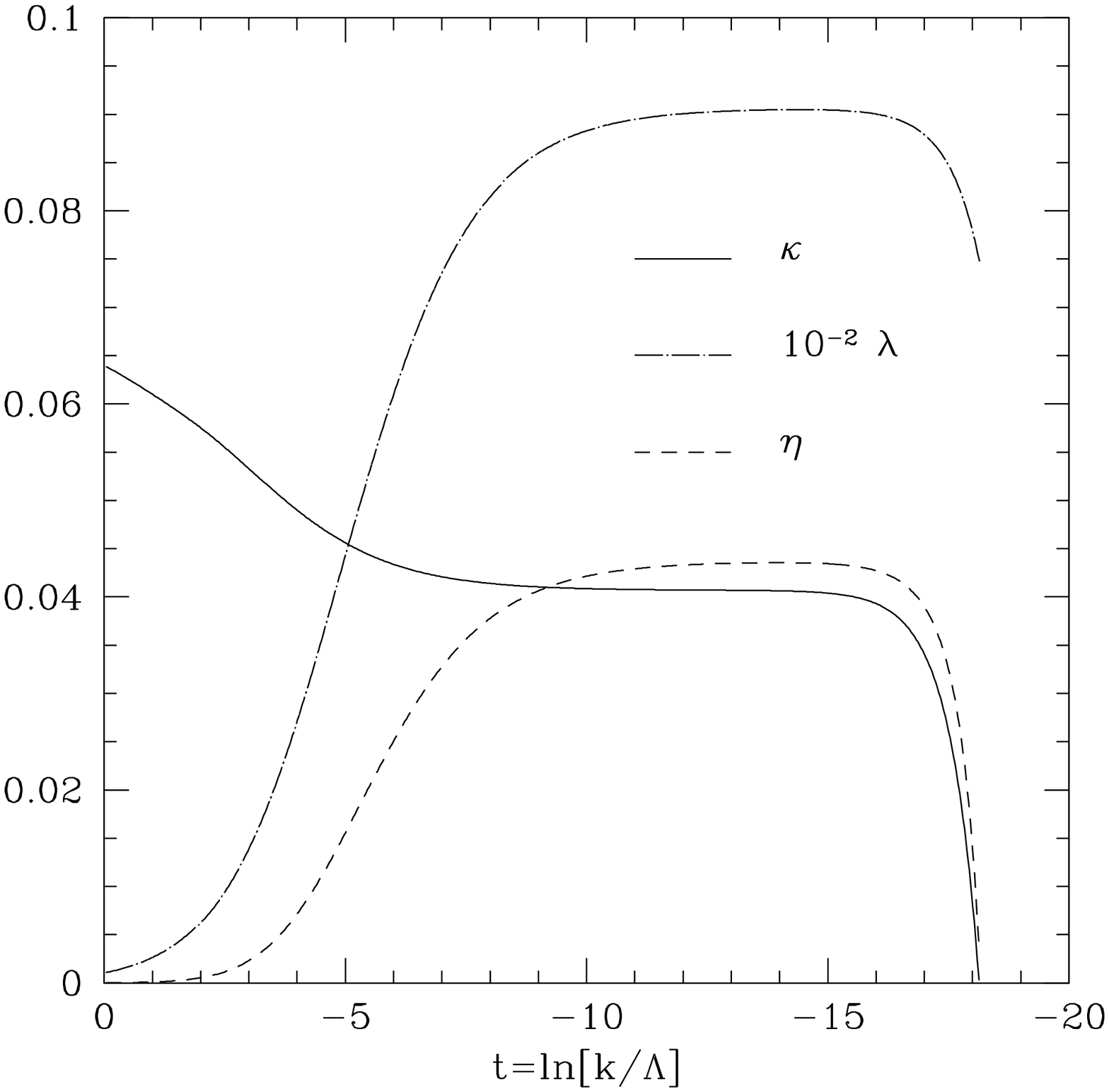}
\caption{\em The scale evolution of $\kappa$, $\lambda$ and $\eta$
for the initial conditions of Fig. 1. The plateaus correspond to the
scaling solution.}
\end{figure}
In fig.\ \ref{EosNew} we present the results of the numerical integration
of eq. (\ref{five}) for $d=3$ and $N=1$.
The function $u'_k(\tilde{\rho})$ is plotted
for various values of $t= \ln(k/\Lambda)$. The evolution starts
at $k=\Lambda$ ($t=0$)  where the average potential is equal to
the classical potential (no effective integration of modes
has been performed). We start with a quartic classical potential
parameterized as
\beq
u'_{\Lambda}(\tilde{\rho}) = \lambda_\Lambda (\tilde{\rho} - \kappa_\Lambda).
\label{eighteen} \eeq
We arbitrarily choose $\lambda_\Lambda=0.1$ and fine tune $\kappa_\Lambda$
so that a scaling solution is approached at later stages
of the evolution. There is a critical value
$\kappa_{cr} \simeq 6.396 \times 10^{-2}$
for which the evolution leads to the scaling solution.
For the results in fig.\ \ref{EosNew} a value
$\kappa_\Lambda$ slightly smaller than $\kappa_{cr}$
is used. As $k$ is lowered
(and $t$ turns negative), $u'_k(\tilde{\rho})$ deviates from its initial
linear shape. Subsequently it evolves towards a form which is
independent of $k$ and corresponds to the scaling
solution $\partial_t u'_* (\tilde{\rho}) = 0$. It spends a long ``time''
$t$ -- which can be rendered arbitrarily long through appropriate
fine tuning of $\kappa_\Lambda$ --
in the vicinity of the scaling solution. During this
``time'', the minimum of the potential $u'_k(\tilde{\rho})$
takes a fixed value $\kappa_*$,
while the minimum of $U_k(\rho)$ evolves towards zero according to
\beq
\rho_0(k) = k \kappa_* / Z_k.
\label{nineteen} \eeq
The longer $u'_k(\tilde{\rho})$ stays near the scaling solution, the
smaller the resulting value of $\rho_0(k)$ when the
system deviates from it.
As this value determines the mass scale for the
renormalized theory at $k=0$, the
scaling solution governs the behavior of the system
very close to the phase transition, where the characteristic
mass scale goes to zero.
Another important property of the ``near-critical''
trajectories, which spend a long ``time'' $t$ near
the scaling solution, is that they become insensitive
to the details of the classical theory which determine the
initial conditions for the evolution. After $u'_k(\tilde{\rho})$ has
evolved away from its scaling form $u'_*(\tilde{\rho})$, its
shape is independent of the choice of $\lambda_\Lambda$ for
the classical theory.
This property gives rise to the universal behavior
near second order phase transitions.
For the solution depicted in
fig.\ \ref{EosNew}, $u_k(\tilde{\rho})$ evolves in such a way that its minimum
runs to zero with $u'_k(0)$ subsequently increasing.
Eventually the theory settles down in the
symmetric phase with a positive constant renormalized mass term
$m^2 = k^2 u'_k(0) $ as $k \rightarrow 0$.
Another possibility is that the system ends up in the
phase with spontaneous symmetry breaking. In this case
$\kappa$ grows in such a way that
$\rho_0(k)$ approaches a constant value
for $k \rightarrow 0$.

The approach to the scaling solution and the deviation
from it can also be seen in fig.\ \ref{CouplingsFig}. 
The evolution of the
running parameters $\kappa(t)$, $\lambda(t)$ starts
with their initial classical values, leads to
fixed point values $\kappa_*$, $\lambda_*$ near the scaling solution,
and finally ends up in the symmetric phase
($\kappa$ runs to zero).
Similarly the anomalous dimension $\eta(k)$, which is given
by eq. (\ref{ten}), takes a fixed point value $\eta_*$
when the scaling solution is approached.
During this part of the evolution the wave function
renormalization is given by
\beq
Z_k \sim k^{-\eta_*}
\label{twenty} \eeq
according to eq. (\ref{2.34}). When the parts of the evolution towards
and away from the fixed point become negligible
compared to the evolution near the fixed point
-- that is, very close to the phase transition --
eq. (\ref{twenty}) becomes a very good approximation for
sufficiently low $k$.
This indicates that $\eta_*$ can be identified with the
critical exponent $\eta$.
For the solution of fig. 2 ($N=1$) we find
$\kappa_*=4.07 \times 10^{-2}$, $\lambda_*=9.04$ and
$\eta_*=4.4 \times 10^{-2}$.

\begin{table} [h]
\label{taylor}
\renewcommand{\arraystretch}{1.3}
\hspace*{\fill}
\begin{tabular}{|c|c|c|c|c|c|}  \hline

& $\kappa_*$
& $\lambda_*$
& $u^{(3)}_*$
& $\eta$
& $\nu$
\\ \hline \hline
a
& $6.57 \times 10^{-2}$
& 11.5
&
&
& 0.745
\\ \hline
b
& $8.01 \times 10^{-2}$
& 7.27
& 52.8
&
& 0.794
\\ \hline
c
& $7.86 \times 10^{-2}$
& 6.64
& 42.0
& $3.6 \times 10^{-2}$
& 0.760
\\ \hline
d
& $7.75 \times 10^{-2}$
& 6.94
& 43.5
& $3.8 \times 10^{-2}$
& 0.753
\\ \hline
e
& $7.71 \times 10^{-2}$
& 7.03
& 43.4
& $3.8 \times 10^{-2}$
& 0.752
\\ \hline
f
& $7.64 \times 10^{-2}$
& 7.07
& 44.2
& $3.8 \times 10^{-2}$
& 0.747
\\ \hline
g
& $7.765 \times 10^{-2}$
& 6.26
& 39.46
& $4.9 \times 10^{-2}$
& 0.704
\\ \hline
\end{tabular}
\hspace*{\fill}
\renewcommand{\arraystretch}{1}
\caption[y]
{\em
Truncation dependence of the scaling solution and critical exponents.
The minimum $\kappa$ of the
potential $u_k(\tilde{\rho})$ and  the quartic and six point couplings
$\lambda=u''(\kappa)$, $u_k^{(3)}(\kappa)$ are given 
for the scaling solution. We also display
the critical exponents $\eta$ and $\nu$,
in various approximations: (a)-(e) from
refs. \cite{TW93-1,TW94-1}
and (f) from the present
section \cite{ABBFTW95-1,BTW96-1}. $N=3$. \\
a) Quartic truncation where only the evolution of $\kappa$ and $\lambda$ is
considered and
higher derivatives of the potential and the anomalous
dimension are neglected (cf. section 2.5).\\
b) Sixth order truncation with 
$\kappa$, $\lambda$, $u^{(3)}_k(\kappa)$  included. \\
c) All couplings with canonical dimension $\geq 0$
are included and $\eta$ 
is approximated
by eq. (\ref{ten}). \\
d) Addition of  
$u^{(4)}_k(\kappa)$
which has negative canonical dimension. \\
e) Additional estimate of
$u^{(5)}_k(\kappa)$, $u^{(6)}_k(\kappa)$. \\
f) The partial differential equation (\ref{five})
for $u'_k(\tilde{\rho})$ is solved numerically
and $\eta$ is approximated by eq. (\ref{ten}). \\
g) First order derivative
expansion with field dependent wave function renormalizations z and y
\cite{vGW}.
}
\end{table}
As we have already mentioned the details of
the renormalized theory in the vicinity of the phase
transition are independent of the classical coupling
$\lambda_\Lambda$. Also the
initial form of the potential does not have
to be of the quartic form of eq.\ (\ref{eighteen})
as long as the symmetries are respected.
Moreover, the critical theory can
be parameterized in terms of critical exponents
\cite{WilFis72}, an example of
which is the anomalous dimension $\eta$.
These exponents are universal quantities which depend
only on the dimensionality of the system and its internal
symmetries. For our three-dimensional theory they depend only on the
value of $N$ and
can be easily extracted from our results. We concentrate here on
the exponent $\nu$, which parameterizes the behavior of the
renormalized mass in the critical region. Other exponents are
computed in the following sections 
along with the critical equation of state. 
The other exponents
are not independent quantities, but
can be determined from $\eta$ and $\nu$ through universal
scaling laws \cite{WilFis72}. We define the exponent $\nu$ through
the renormalized mass term in the symmetric phase
\beq
m^2 =  \frac{1}{Z_k} \frac{d U_k(0)}{d\rho}
= k^2 u'_k(0)~~~~~~~{\rm for}~~k \rightarrow 0.
\label{twentyone} \eeq
The behavior of $m^2$ in the critical region
depends only on the distance from the phase transition, which
can be expressed in terms of the difference of
$\kappa_\Lambda$ from the critical value $\kappa_{cr}$ for which
the renormalized theory has exactly $m^2 =0$.
The exponent $\nu$ is determined from the relation
\beq
m^2 \sim |\delta\kappa_{\Lambda}|^{2 \nu} = 
|\kappa_\Lambda - \kappa_{cr}|^{2 \nu}.
\label{twentytwo} \eeq
Assuming proportionality $\delta\kappa_\Lambda\sim T_c-T$ 
this yields the critical temperature dependence of the 
correlation length $\xi=m^{-1}$.
For a determination of $\nu$ from our results we calculate
$m^2$ for various values of $\kappa_\Lambda$ near $\kappa_{cr}$.
We subsequently plot $\ln(m^2)$ as a function
of $\ln|\delta\kappa_{\Lambda}|$. This curve becomes linear for
$\delta\kappa_{\Lambda} \rightarrow 0$ and we
obtain $\nu$ from the constant slope.

\begin{table} [h]
\renewcommand{\arraystretch}{0.8}
\hspace*{\fill}
\begin{tabular}{|c||c|c||c|c|}
\hline
$N$
&\multicolumn{2}{c||}{$\nu$}
&\multicolumn{2}{c|}{$\eta$}
\\
\hline \hline

&
&$0.5882(11)^a$
&
&$0.0284(25)^a$
\\
0
&$0.589^f$
&$0.5875(25)^{b}$
&$0.040^f$
&$0.0300(50)^{b}$
\\

&
&$0.5878(6)^c$
&
&
\\

&
&$0.5877(6)^e$
&
&
\\ \hline
&
&$0.6304(13)^a$
&
&$0.0335(25)^{a}$
\\
1
&$0.643^f$
&$0.6290(25)^{b}$
&$0.044^f$
&$0.0360(50)^{b}$
\\

&$0.6307^{g}$
&$0.6315(8)^{c}$
&$0.0467^{g}$
&
\\

&
&$0.6294(9)^e$
&
&$0.0374(14)^e$
\\ \hline

&
&$0.6703(15)^{a}$
&
&$0.0354(25)^{a}$
\\
2
&$0.697^f$
&$0.6680(35)^{b}$
&$0.042^f$
&$0.0380(50)^{b}$
\\

&$0.666^g$
&$0.675(2)^{c}$
&$0.049^g$
&
\\

&
&$0.6721(13)^e$
&
&$0.042(2)^e$
\\ \hline

&
&$0.7073(35)^{a}$
&
&$0.0355(25)^{a}$
\\
3
&$0.747^f$
&$0.7045(55)^{b}$
&$0.038^f$
&$0.0375(45)^{b}$
\\

&$0.704^g$
&$0.716(2)^{c}$
&$0.049^g$
&
\\

&
&$0.7128(14)^e$
&
&$0.041(2)^e$
\\ \hline

&
&$0.741(6)^{a}$
&
&$0.0350(45)^{a}$
\\
4
&$0.787^f$
&$0.737(8)^{b}$
&$0.034^f$
&$0.036(4)^{b}$
\\

&$0.739^g$
&$0.759(3)^{c}$
&$0.047^g$
&
\\

&
&$0.7525(10)^{e}$
&
&$0.038(1)^{e}$
\\  \hline
10
&$0.904^f$
&$0.894(4)^{c}$
&$0.019^f$
&
\\

&
&$0.877^{d}$
&
&$0.025^{d}$
\\ \hline
100
&$0.990^f$
&$0.989^{d}$
&$0.002^f$
&$0.003^{d}$
\\ \hline
\end{tabular}
\hspace*{\fill}
\renewcommand{\arraystretch}{1}
\caption[y]
{\em
Critical exponents $\nu$ and $\eta$
for various values of $N$.
For comparison we list results obtained with other methods as summarized in
\cite{GZ}, \cite{journal} and \cite{BC97-1}: \\
a) From perturbation series at fixed dimension including seven--loop 
contributions. \\
b) From the $\epsilon$-expansion at order $\epsilon^5$. \\
c) From lattice high temperature expansions \cite{BC97-1}
(see also \protect\cite{Reisz95,ZinnLaiFisher96}. \\
d) From the $1/N$-expansion at order $1/N^2$. \\
e) From lattice Monte Carlo simulations \cite{LMC0,LMC10}.\\
f) Average action in lowest order in the derivative expansion
(present section).\\
g) From first order in the derivative expansion for the average action
with field dependent wave function renormalizations (for $N=1$
see \cite{B} and section \ref{IsingModel}, and \cite{vGW} for $N>1$).
\label{critexp1}}
\end{table}
Our numerical solution of the partial differential
equation (\ref{five}) corresponds to an infinite level of
truncation in a Taylor expansion 
around the ``running'' minimum of the potential.
This infinite system may be approximately
solved by neglecting $\tilde{\rho}$-derivatives
of $u_k(\tilde{\rho})$ higher
than a given order. The apparent convergence of this procedure
can be observed from table \ref{taylor}. 
We present results obtained through the procedure
of successive truncations and through our numerical solution
of the partial differential equation for $N=3$. 
We give the values of $\kappa$, $\lambda$, $u^{(3)}_k(\kappa)$
for the scaling solution and the critical exponents
$\eta$, $\nu$. We observe how the results stabilize as more
$\tilde{\rho}$-derivatives of $u_k(\tilde{\rho})$ at 
$\tilde{\rho}=\kappa$ and the
anomalous dimension are taken into account. The last line
gives the results of our numerical solution of eq. (\ref{five}).
By comparing with the previous line we conclude
that the inclusion of all the $\tilde{\rho}$-derivatives higher than
$u_k^{(6)}(\kappa)$ and the term $\sim \eta$ in the
``threshold'' functions
generates an improvement of less than
1 \% for the results. This is smaller than the error
induced by the omission of the higher derivative terms
in the average action, which typically generates an uncertainty
of the order of the anomalous dimension. A systematic comparison 
\cite{Aoki} between the expansion around $\kappa$ presented here
and an expansion around $\tilde\rho=0$ reveals that only
the first procedure shows this convergence.

In table 2 we compare our values for the critical exponents
obtained from the numerical solution of the partial differential
equation (\ref{five}) and (\ref{ten})
with results obtained from other methods
(such as the $\epsilon$-expansion, perturbation series
at fixed dimension, lattice high temperature expansions, 
Monte Carlo simulations and the $1/N$-expansion).
As expected $\eta$ is rather poorly determined since it is
the quantity most seriously affected by the omission of the
higher derivative terms in the average action. The exponent
$\nu$ is in agreement with the known results at the
1-5 \% level, with a discrepancy roughly equal to the
value of $\eta$ for various $N$. Our results compare 
well with those obtained by similar methods using a variety of forms 
for the infrared cutoff function \cite{MT97-1,CT97,AMSST96,BHLM95-1,Alf94-1}.

In conclusion, the shape of the average potential is under
good quantitative control for every scale $k$. This
permits a quantitative understanding of the most
important properties of the system at every length
scale. We will exploit this in the following to extract the scaling
form of the equation of state.

\subsection{Universal critical equation of state\label{uceos}}

In this section we extract the Widom scaling form
of the equation of state 
from a solution \cite{BTW96-1} of eqs.\ (\ref{five}),
(\ref{ten}) for the three dimensional $O(N)$ model. 
Its asymptotic behavior yields the universal critical
exponents and amplitude ratios.
We also present fits for the scaling function for $N=3$ and $N=4$.
A detailed discussion of the universal and non-universal
aspects of the Ising model ($N=1$) is given in section
\ref{IsingModel}.

Eq.\ (\ref{wid}) establishes the scaling properties of the
equation of state. The external field $H$ is related to the derivative
of the effective potential $U'= \partial U/\partial \rho$ by
$H_a = U' \phi_a$.
The critical equation of state, relating
the temperature, the external field and the
order parameter, can then be written in the scaling
form ($\phi=\sqrt{2 \rho}$) 
\beq
\frac{U'}{\phi^{\delta-1}} =  f(x),
\qquad x=\frac{-\delta\kappa_{\Lambda}} {\phi^{1/\beta}},
\label{sixb}
\eeq
with critical exponents $\delta$ and $\beta$.
A measure of the distance from the 
phase transition is the difference
$\delta\kappa_{\Lambda} = \kappa_{\Lambda} - \kappa_{cr}$.
If $\kappa_{\Lambda}$ is interpreted as a function of temperature, 
the deviation $\delta\kappa_{\Lambda}$ 
is proportional to the deviation from the critical temperature, i.e.\
$\delta\kappa_{\Lambda} = A(T) (T_{cr}- T)$ with $A(T_{cr}) > 0$.
For $\phi \to \infty$
our numerical solution for $U'$ obeys $U' \sim \phi^{\delta-1}$ with
high accuracy. The inferred value of $\delta$ is displayed in 
table \ref{tableeos}, and we have checked the scaling relation 
$\delta=(5-\eta)/(1+\eta)$.
The value of the critical exponent $\eta$ is obtained from eq. 
(\ref{five}) for the scaling solution.
We have also verified explicitly that 
$f$ depends only
on the scaling variable $x$ for the value of $\beta$ given in table
\ref{tableeos}. 
In figs.\ 1 and 2 we plot log$(f)$
and log$(df/dx)$ 
as a function of log$|x|$ for $N=1$ and $N=3$.
Fig.\ 1 corresponds to the symmetric phase $(x > 0)$, and 
fig.\ 2 to the phase with spontaneous symmetry breaking
$(x < 0)$. 
\begin{figure}
\leavevmode
\centering  
\epsfxsize=4in
\epsffile{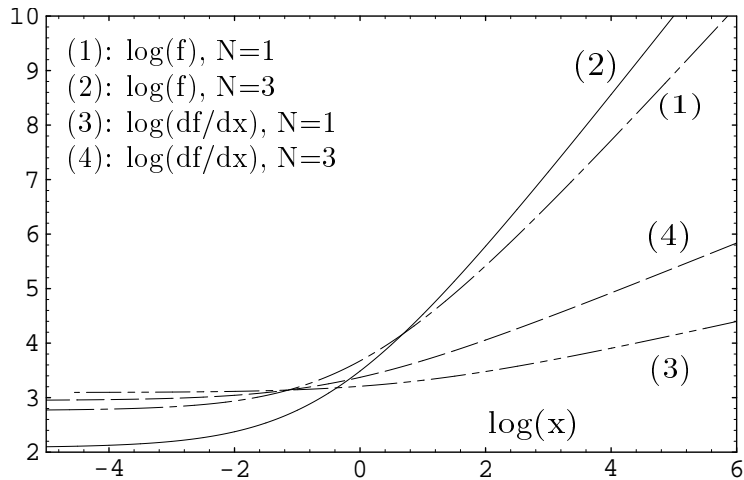}                                   
\caption{\em
Universal critical equation of state, symmetric phase: 
Logarithmic plot of $f$ and $df/dx$ for $x > 0$.}
\end{figure}
\begin{figure}
\leavevmode
\centering  
\epsfxsize=4in
\epsffile{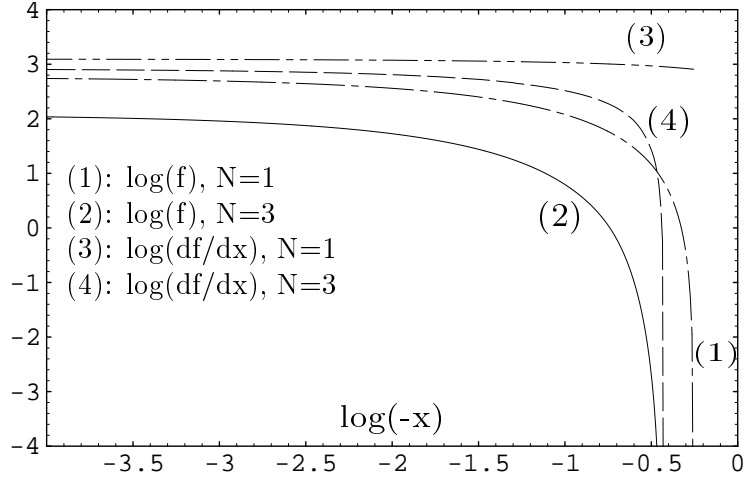}                                
\caption{\em
Universal critical equation of state, spontaneous symmetry breaking: 
Logarithmic plot of $f$ and $df/dx$ for $x < 0$.}
\end{figure}
 
One can easily extract the asymptotic behavior from the logarithmic
plots and compare with known values of critical exponents and
amplitudes.
The curves become constant, both for 
$x \to 0^+$ and $x \to 0^-$ with the same value, consistently with the
regularity of $f(x)$ at $x=0$. For the universal function one
obtains
\beq
\lim\limits_{x \to 0} f(x) = D,  \label{hd}
\eeq
and $H = D \phi^{\delta}$ on the critical isotherm.
For $x \to \infty$ one 
observes that log$(f)$ becomes a linear function of log$(x)$  
with constant slope $\gamma$.
In this limit the universal function takes the form
\beq
\lim\limits_{x \to \infty} f(x) = (C^+)^{-1} x^{\gamma} .
\eeq     
The amplitude $C^+$ and the critical exponent $\gamma$
characterize the behavior of the 'unrenormalized' squared mass
or inverse susceptibility
\beq 
\bar{m}^2=\chi^{-1}=\lim_{\phi \to 0}
\left(\frac{\partial^2 U}{\partial \phi^2}\right)=
(C^+)^{-1}|\de \kappa_{\Lambda}|^{\gamma}\phi^{\delta-1-\gamma/\beta}\, .
\eeq   
We have verified the scaling relation $\gamma/\beta=\delta-1$
that connects $\gamma$ with the exponents $\beta$ and $\delta$
appearing in the Widom scaling form (\ref{wid}). 
One observes that the zero-field magnetic susceptibility,
or equivalently the inverse unrenormalized squared mass
$\bar{m}^{-2}=\chi$, is 
non-analytic for $\delta\kappa_{\Lambda} \to 0$ in the symmetric phase:
$\chi = C^+ |\delta\kappa_{\Lambda}|^{-\gamma}$. In this phase 
we find that the correlation length
$\xi = (Z_0 \chi)^{1/2}$, which is
equal to the inverse of the renormalized mass $m_R$, 
behaves as $\xi=\xi^+|\delta\kappa_{\Lambda}|^{-\nu}$
with $\nu=\gamma/(2-\eta)$.

The spontaneously broken phase is characterized by a nonzero
value $\phi_0$ of the minimum of the effective potential 
$U$ with
$H=(\partial U/\partial \phi) (\phi_0)=0$. The appearance
of spontaneous symmetry breaking below $T_{c}$ implies
that $f(x)$ has a zero $x=-B^{-1/\beta}$ and one observes
a singularity of the logarithmic plot in fig.\ 2. In
particular, according to eq.\ (\ref{wid}) the minimum
behaves as $\phi_0 = B (\de \kappa_{\Lambda})^{\beta}$.
Below the critical temperature,
the longitudinal and transversal
susceptibilities $\chi_L$ and $\chi_T$ are different for $N > 1$
\beq
\chi_L^{-1} = \frac{\partial^2 U}{\partial \phi^2}
= \phi^{\delta-1} \bigl(\delta f(x) - \frac{x}{\beta}f'(x)\bigr),
\qquad
\chi_T^{-1} = \frac{1}{\phi}\frac{\partial U}{\partial \phi}
= \phi^{\delta-1} f(x) \label{eightb}
\eeq
(with $f'=df/dx$).
This is related to the existence of massless Goldstone modes in the 
$(N-1)$ transverse directions, which causes the transversal susceptibility
to diverge for vanishing external field.
Fluctuations of these modes 
induce the divergence of the zero-field longitudinal
susceptibility.
This can be concluded 
from the singularity 
of log$(f')$ for $N=3$
in fig.\ 2. The first $x$-derivative of the
universal function vanishes as $H \to 0$, i.e.\
$f'(x=-B^{-1/\beta})=0$ for $N > 1$.
For $N=1$ there is a non-vanishing constant value for
$f'(x=-B^{-1/\beta})$ with a finite zero-field susceptibility
$\chi = C^- (\delta\kappa_{\Lambda})^{-\gamma}$, where
$(C^-)^{-1}=B^{\delta-1-1/\beta}f'(-B^{-1/\beta})/\beta$.
For a non-vanishing physical infrared cutoff
$k$, the longitudinal susceptibility remains finite also
for $N > 1$: $\chi_L \sim (k\rho_0)^{-1/2}$.
For $N=1$ in the ordered phase, 
the correlation length behaves as
$\xi=\xi^-(\delta\kappa_{\Lambda})^{-\nu}$, and
the renormalized minimum $\rho_{0R}= Z_0 \rho_0$ of the potential $U$
scales as 
$\rho_{0R}=E (\delta\kappa_{\Lambda})^{\nu}$.

The amplitudes of singularities near the phase transition $D$, $C^{\pm}$,
$\xi^{\pm}$, $B$ and $E$ are given in table \ref{tableeos}. They are not
universal. All models in the
same universality class can be related by a multiplicative
rescaling of $\phi$ and $\de \kappa_{\Lambda}$ or $(T_{c}-T)$.
Accordingly there are only two independent amplitudes
and exponents respectively. 
Ratios of amplitudes which are invariant
under this rescaling are universal. We display the critical exponents
and the universal
combinations 
$R_{\chi}=C^+ D B^{\delta-1}$, 
$\tilde{R}_{\xi}=(\xi^+)^{\beta/\nu} D^{1/(\delta+1)} B$
and $\xi^+ E$ for $N=3,4$ in tables \ref{critex} and \ref{tableeos}.
\begin{table} [h]
\renewcommand{\arraystretch}{1.5}
\label{universalexp}
\hspace*{\fill}
\begin{tabular}{|c|c|c|c|c|c|}     \hline
$N$
& $\beta$ 
& $\gamma$ 
& $\delta$
& $\nu$
& $\eta$
\\ \hline
3
& 0.388
& 1.465
& 4.78
& 0.747
& 0.038
\\ \hline 
4
& 0.407
& 1.548
& 4.80
& 0.787
& 0.0344
\\ \hline
\end{tabular}
\hspace*{\fill}
\renewcommand{\arraystretch}{1}
\caption[]%
{\em
Universal critical exponents for $N=3,4$.}
\label{critex}
\end{table}

\begin{table} [h]
\renewcommand{\arraystretch}{1.5}
\hspace*{\fill}
\begin{tabular}{|c|c|c|c|c|c||c|c|c|}     \hline
$N$
& $C^+$ 
& $D$ 
& $B$
& $\xi^+$
& $E$
& $R_{\chi}$
& $\tilde{R}_{\xi}$
& $\xi^+ E$
\\ \hline
3
& 0.0743
& 8.02
& 1.180
& 0.263
& 0.746
& 1.11
& 0.845
& 0.196
\\ \hline
4
& 2.79
& 1.82
& 7.41
& 0.270
& 0.814
& 1.02
& 0.852
& 0.220
\\ \hline
\end{tabular}
\hspace*{\fill}
\renewcommand{\arraystretch}{1}
\caption[]%
{\em
Universal amplitude ratios $R_{\chi}$, $\tilde{R}_{\xi}$ and 
$\xi^+ E$ for $N=3,4$.
The amplitudes $C^+,D,B,\xi^+$ and $E$ are not universal.}
\label{tableeos}
\end{table}

The asymptotic behavior observed for the universal function can be
used in order to obtain a semi-analytical expression for $f(x)$.
We find that the following two-parameter
fits for $N=3,4$ reproduce the numerical values for both
$f$ and $df/dx$ with $1$--$2$\% accuracy:
\beq
f_{\rm fit}(x)=D \bigl(1+B^{1/\beta} x \bigr)^2
\bigl(1+\Theta x \bigr)^{\Delta} 
\bigl(1+c x \bigr)^{\gamma-2-\Delta} , \label{nineb}
\eeq
with $c=(C^+ D B^{2/\beta} \Theta^{\Delta} )^{-1/(\gamma-2-\Delta)}$. 
The fitting
parameters are chosen as $\Theta = 1.312$ and $\Delta = -0.595$ for $N=3$. 
For $N=4$ we find the following fit,
\begin{equation}
  \label{ffit}
  \begin{array}{rcl}
    \dsp{f_{\rm fit}(x)}&=&\dsp{1.816 \cdot 10^{-4} (1+136.1\, x)^2 \,
      (1+160.9\, \theta\,
      x)^{\Delta}}\nnn
    &&
    \dsp{(1+160.9\, (0.9446\, \theta^{\Delta})^{-1/(\gamma-2-\Delta)} 
      \, x)^{\gamma-2-\Delta}}
  \end{array}
\end{equation}
with $\theta=0.625$ $(0.656)$, $\Delta=-0.490$ $(-0.550)$ for $x
> 0$ $(x < 0)$ and $\gamma$ as given in table \ref{universalexp}.

\begin{figure}
\unitlength1.0cm
\begin{center}
\begin{picture}(17.,12.)
\put(0.3,5.5){$\displaystyle{\frac{2\overline{\sigma}_0/T_c}
{(\jmath/T_c^3 D)^{1/\delta}}}$}
\put(8.5,-0.2){$\displaystyle{\frac{(T-T_c)/T_c}
{(\jmath/T_c^3 B^{\delta} D)^{1/\beta \delta}}}$}
\put(8.,2.5){\footnotesize $\mbox{average action}$}
\put(4.19,11.2){\footnotesize $\mbox{average action}$}
\put(13.8,2.2){\footnotesize $\epsilon$}
\put(3.53,11.19){\footnotesize $\epsilon$}
\put(11.3,2.05){\footnotesize $\mbox{MC}$}
\put(3.9,10.){\footnotesize $\mbox{MC}$}
\put(13.2,1.8){\footnotesize $\mbox{mf}$}
\put(6.5,9.2){\footnotesize $\mbox{mf}$}
\put(-1.,-5.9){
\epsfig{file=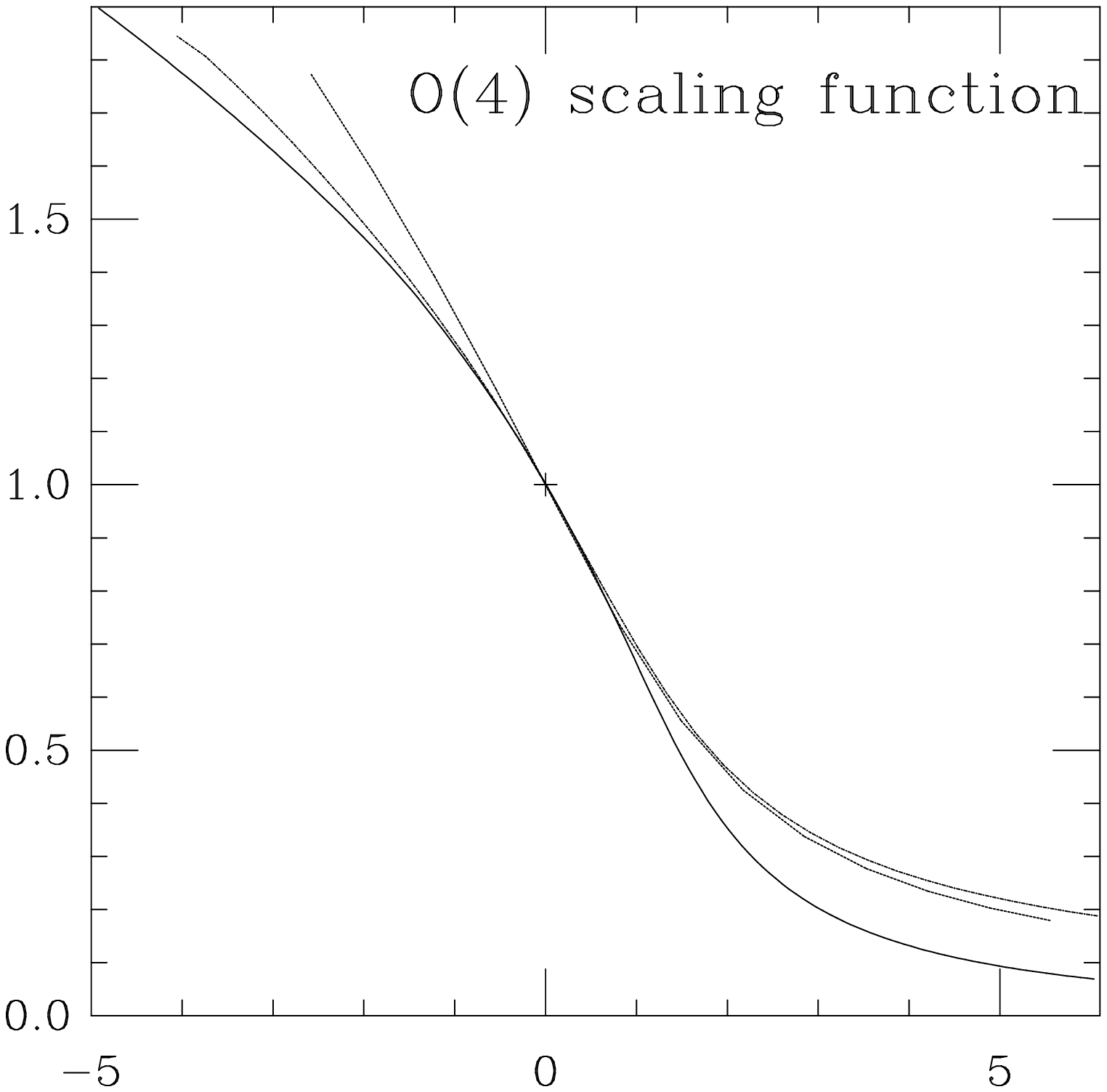,width=18.cm,height=21.cm}}
\put(1.25,0.483){
\rotate[r]{\epsfig{file=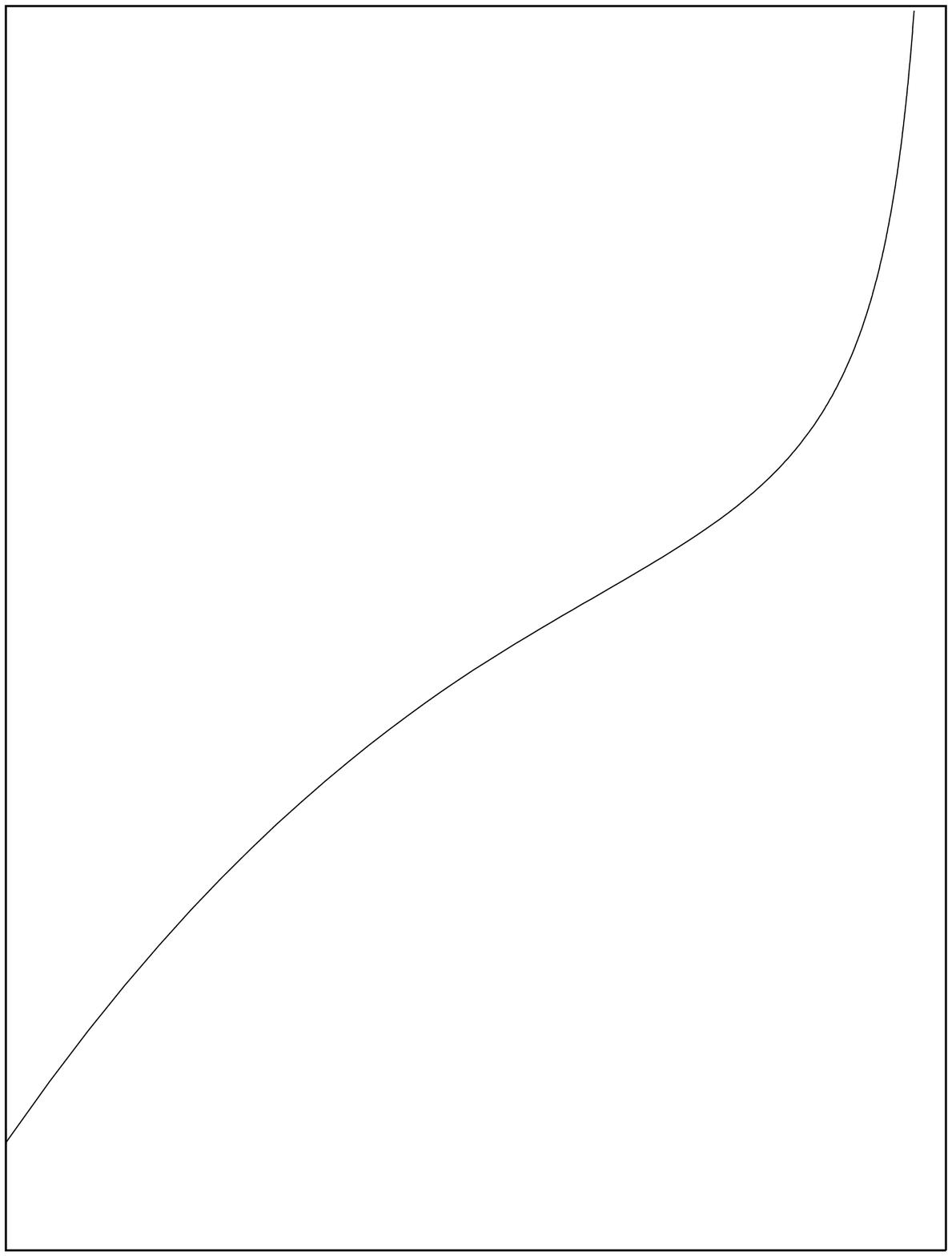,width=11.22cm,height=13.45cm}}}
\end{picture}
\end{center}
\caption[]{\em Critical equation of state for the three--dimensional 
$O(4)$ Heisenberg model. We compare our  results for the scaling
function, denoted by ``average
  action'', with results of other methods. We have labeled the
  axes  in terms of the expectation value
  $\overline{\sigma}_0$ and the source $\jmath$ 
  relevant for the chiral phase transition in QCD discussed in section
8. In this context they describe the dependence of the chiral condensate
 $\sim\bar\sigma_0$ on the quark mass $\sim\jmath$ for two flavors of quarks.
  The constants $B$ and $D$ specify the
  non--universal amplitudes of the model.  The
  curve labeled by ``MC'' represents a fit to lattice Monte Carlo
  data. The second order epsilon expansion \cite{BWW73-1} and mean
  field results are denoted by ``$\epsilon$'' and ``mf'',
  respectively.  Apart from our results the curves are taken
  from ref.~\cite{Tou}.
  \label{scalfunc}
  }
\end{figure}

The universal properties of the scaling function can be compared with
results obtained by other methods for the three--dimensional $O(4)$
Heisenberg model.  In figure \ref{scalfunc} we display our results
for $N=4$ along with those obtained from lattice Monte Carlo simulation
\cite{Tou}, second order epsilon expansion \cite{BWW73-1} and mean
field theory.

In summary, our numerical solution of eq.\ (\ref{four}) gives a very
detailed picture of the critical equation of state
of the three dimensional $O(N)$ model. The numerical 
uncertainties are estimated by 
comparison of results obtained through two independent integration
algorithms \cite{ABBFTW95-1,BW97-1}.
They are small, typically less than $0.3 \%$ for critical exponents. 
The scaling relations between the
critical exponents are fulfilled within a deviation of $2 \times 10^{-4}$.
The dominant quantitative error stems from the relatively
crude approximation of the kinetic term in (\ref{three}) and is related 
to the size of the anomalous dimension $\eta \simeq 4 \%$. We emphasize
that in contrast to most other analytical methods no scaling
hypothesis is used as an input and no resummations of series are 
needed. The scaling behavior is simply a property of the solution
of the flow equation.

In the following we improve on this approximation for the Ising model
($N=1$). We allow for a most general field dependence of the
wave function renormalization factor, and compare with the results
of this section. The Ising model is analyzed with a particular short distance
action relevant for carbon dioxide. This allows us to compare the
non-universal aspects as well as the universal aspects of the 
liquid-gas phase transition in carbon dioxide with experiment.

\subsection{Gas-liquid transition and the Ising universality class}
\label{IsingModel}

\vspace*{-1cm}
\hspace*{12cm}
\footnote{Sections
\ref{IsingModel}-\ref{cubicmodel} are based on a collaboration with
S.\ Seide \cite{B}.}
\vspace*{0.6cm}

\noindent
{\bf Field theoretical description}\\

Many phase transitions are described near the critical
 temperature by a one-component ($N=1$) scalar field theory without internal
symmetries. A typical example is the water-vapor transition where the
field $\varphi(x)$ corresponds to the average density field $n(x)$.
At normal pressure one
observes a first order transition corresponding to a jump in $\varphi$ from
high (water) to low (vapor) values as the temperature $T$ is increased.
With increasing pressure the first order transition line ends at some
critical pressure $p_{*}$ in an endpoint. For $p > p_{*}$ the phase
transition is replaced by an analytical crossover.

This behaviour is common to many systems and characterizes the universality
class of the Ising model. As another example from particle physics, the high
temperature electroweak phase transition in the early universe is described
by this universality class if the mass of the Higgs particle in the standard
model is near the endpoint value \mbox{$M_{H*}\approx 72$GeV}\cite{Rummukainen98}.
An Ising type endpoint should also exist if the high temperature
or high density chiral
phase transition in QCD
or the gas-liquid transition for
nuclear matter are of first order in some region of parameter space.

Very often the location of the endpoint - e.g. the critical $T_{*}$, $p_{*}$
and $n_{*}$ for the liquid-gas transition - is measured quite precisely.
The approach to criticality is governed by universal scaling laws with
critical exponents. Experimental information is also available about the
non-universal amplitudes appearing in this scaling behaviour. These
non-universal critical properties are specific for a given system, and the
question arises how they can be used to gain precise information about the
underlying microscopic physics.
This problem clearly involves the difficult task of an explicit connection
between the short distance physics and the collective behaviour leading to a
very large correlation length.

So far renormalization group methods {\cite{Kad66-1}--\cite{Has86-1},
\cite{Zin93-1}
have established the structure of this relation and led to a precise
determination of the universal critical properties. The non-perturbative
flow equation (\ref{2.17}) 
allows us to complete the task by mapping details of
microscopic physics to non-universal critical quantities. 
A demonstration is given for the liquid-gas
transition in carbon dioxide.

In the following, we will work with a truncation which includes 
the most general terms containing up to two derivatives,
\bea
   \Gamma_k[\varphi] & = & \dsp \int d^3 x \left\{
      U_k(\varphi(x)) + \hal Z_k(\varphi(x))\partial^{\mu}\varphi \partial_{\mu}\varphi
   \right\}.
   \label{ansatzGamma}
\eea
In contrast to the ansatz (\ref{three}) we now include the field dependence
of the wave function renormalization factor $Z_k(\varphi(x))$.
Our aim is the computation of the potential
$U_0\equiv U_{k\to 0}=U/T$
and the wave function renormalization $Z\equiv Z_{k\to 0}$ for a vanishing
infrared cutoff. For the liquid gas transition the source $J$ is
linear in the chemical potential $\mu$.
For a homogeneous situation $U_0T$ corresponds therefore 
to the free energy density.
Indeed, expressing $U_0$ as a function of the density one
finds for the liquid-gas system at a given chemical potential $\mu$
\bea
  \dsp \frac{\partial U_0}{\partial n} & = &
  \dsp \frac{\mu}{T}.
\eea
Equivalently, one may also use the more familiar form of the equation of state
in terms of the pressure $p$,
\bea
  \dsp n^2\frac{\partial}{\partial n}\left(\frac{U_0}{n}\right) & = &
  \dsp \frac{p}{T}.
\eea
(Here the additive constant in $U_0$ is fixed such that $U_0(n\!=\!0)\!=\!0$).
The wave function renormalization $Z(\varphi)$ contains the
additional information needed for a determination of the two point correlation
function at large distance for arbitrary pressure.

The computation of thermodynamic potentials, correlation length
etc.\ is done in two steps: The first is the computation of a short distance
free energy $\Gamma_{\Lambda}$. This does not involve large length scales
and can be done by a variety of expansion methods or numerical simulations.
This step is not the main emphasis here and we will use a
relatively crude approximation for the gas-liquid transition.
The second step is more difficult and will be addressed here. It involves
the relation between $\Gamma_{\Lambda}$ and $\Gamma_0$, and has to account
for possible complicated collective long distance fluctuations.

For a large infrared cutoff $k=\Lambda$ one may compute $\Gamma_{\Lambda}$
perturbatively. For example, the lowest order in a virial
expansion for the liquid-gas system yields
\bea
  \lefteqn{\dsp U_{\Lambda}(n)  =
   -n\left(1+\ln g + \frac{3}{2}\ln\frac{MT}{2\pi\Lambda^2}\right) } & & \nnn
      &  & + \dsp n \ln\left(\frac{n}{(1-b_0(\Lambda)n)\Lambda^3}\right)
      - \frac{b_1(\Lambda)}{T}n^2 + c_{\Lambda}.
  \label{virialU}
\eea
Here $\Lambda^{-1}$ should be of the order of a typical range of intermolecular
interactions, $M$ and $g$ are the mass and the number of degrees of freedom
of a molecule and $b_0$, $b_1$ parameterize the virial coefficient
$\dsp B_2(T)=b_0-b_1/T$.
\footnote{
   The Van der Waals coefficients $b_0$, $b_1$ for real gases can be
   found in the literature. These values are valid for small densities.
   They also correspond to
   $k=0$ rather than to
   $k=\Lambda$. Fluctuation effects lead to slightly different values for
   $b_i(\Lambda)$ and $b_i(k=0)$  even away from the critical line.
   We find that these differences are small for $n\ll n_*$.
   Similarly, a constant $c_{\Lambda}$ should be added to $U_{\Lambda}$
   so that $U_0(0)\!=\!0$. }
(The (mass) density $\rho$ is related to the particle density
$n$ by $\rho=Mn$.)
We emphasize that the convergence of a virial expansion is expected to improve
considerably in presence of an infrared cutoff $\Lambda$ which suppresses the
long distance fluctuations.

The field $\varphi(x)$ is related to the (space-dependent) particle density
$n(x)$ by
\bea
  \dsp \varphi(x) & = & K_{\Lambda}(n(x)-\hat{n})
\eea
with $\hat{n}$ some suitable fixed reference density. We approximate the
wave function renormalization $Z_{\Lambda}$ by a constant. It can be inferred
from the correlation length $\hat{\xi}$, evaluated
at some reference density $\hat{n}$ and temperature $\hat{T}$ away
from the critical region, through
\bea
  \dsp \hat{\xi}^{-2} & = &
  \dsp Z_{\Lambda}^{-1}\frac{\partial^2 U_0}{\partial\varphi^2}
         \left.\right|_{\hat{\varphi},\hat{T}}.
\eea
For a suitable scaling factor
\bea
  \dsp K_{\Lambda} & = &
  \dsp \left(\frac{1}{\hat{n}}+\frac{b_0(2-b_0\hat{n})}{(1-b_0\hat{n})^2}
      -\frac{2b_1}{\hat{T}}\right)^{1/2}\hat{\xi}
\eea
one has $Z_{\Lambda}=1$.

We observe that the terms linear in $n$ in eq. (\ref{virialU}) play only a
role for the relation between $n$ and $\mu$.
It is instructive to subtract from $U_{\Lambda}$ the linear piece in $\varphi$
and to expand in powers of $\varphi$:
\bea
  \dsp U_{\Lambda}(\varphi) & = &
  \dsp  \frac {m_{\Lambda}^2}{2}\varphi^2 + \frac{\gamma_{\Lambda}}{6}\varphi^3
      + \frac{\lambda_{\Lambda}}{8}\varphi^4 + \ldots
  \label{microsU}
\eea
with
\bea
  \dsp m_{\Lambda}^2 & = &
  \dsp   K_{\Lambda}^{-2}\left(\frac{1}{\hat{n}}
        +\frac{b_0(2-b_0\hat{n})}{(1-b_0\hat{n})^2}-\frac{2b_1}{T}\right) \nnn
  \dsp \gamma_{\Lambda} & = &
  \dsp  K_{\Lambda}^{-3}\left(\frac{b_0^2(3-b_0\hat{n})}{(1-b_0\hat{n})^3}
       - \frac{1}{\hat{n}^2}\right)         \nnn
  \dsp \lambda_{\Lambda} & = &
  \dsp   \frac{2}{3}K_{\Lambda}^{-4}\left(
        \frac{b_0^3(4-b_0\hat{n})}{(1-b_0\hat{n})^4}+ \frac{1}{\hat{n}^3}
             \right).
\eea
For a convenient choice $\dsp \hat{n}=\frac{1}{3b_0}$,
$\dsp \hat{T}=\frac{8}{11}\frac{b_1}{b_0}$
one has $\gamma_{\Lambda}=0$ and
\bea
  \dsp
   K_{\Lambda}=2b_0^{1/2}\hat{\xi},\;\;
   m_{\Lambda}^2=\left(\frac{27}{16}-\frac{b_1}{2b_0T}\right)
   \hat{\xi}^{-2}, \;\;
   \lambda_{\Lambda}=\frac{243}{128}b_0\hat{\xi}^{-4}.
\eea

\medskip
\noindent
{\bf Carbon dioxide}\\

In order to be specific we will discuss the equation of state for 
carbon dioxide near the endpoint of the critical line. Typical values
of the parameters are
$m_{\Lambda}^2/\Lambda^2=-0.31$,
$\lambda_{\Lambda}/\Lambda=6.63$ for $\Lambda^{-1}=5\cdot 10^{-10}\:m$,
$\hat{\xi}=0.6\:\Lambda^{-1}$.
In the limit (\ref{microsU}) one obtains a $\varphi^4$-model. Our explicit
calculations for carbon dioxide will be performed, however, for the
microscopic free energy (\ref{virialU}).
The linear piece in the potential can be absorbed in the source term so that
the equation of state reads
\footnote{
   Note that the source term is independent of $k$. The linear piece in the
   potential can therefore easily be added to $U_{k\to 0}$ once all fluctuation
   effects are included.}
\bea
  \dsp \frac{\partial U_0}{\partial\varphi}=j &, &
  \dsp j=K_{\Lambda}^{-1}\left(\frac{\mu}{T}+1+\ln g +\frac{3}{2}\ln
      \frac{MT}{2\pi\Lambda^2}\right).
\eea
We emphasize that a polynomial microscopic potential (\ref{microsU}) with
equation
of state $\partial U_0/\partial\varphi = j$
is a good approximation for a large
variety of
different systems. For the example of magnets $\varphi$ corresponds to the
magnetization and $jT$ to the external magnetic field. For
$\gamma_{\Lambda}=0$ and $\lambda_{\Lambda}\to\infty$, with finite negative
$m_{\Lambda}^2/\lambda_{\Lambda}$, this is the $Z_2$-symmetric Ising model.

For values of $\varphi$ for which the mass term
\mbox{$m^2(\varphi)=\frac{1}{Z}\frac{\partial^2 U}{\partial\varphi^2}$}
is much larger than $\Lambda^2$ the microscopic approximation to $\Gamma_k$
remains approximately valid also for $k\to 0$, i.e.
$U(\varphi)\approx U_{\Lambda}(\varphi)$.
The contribution of the long wavelength fluctuations is suppressed by
the small correlation length or large mass.
In the range where \mbox{$m^2(\varphi) \ll \Lambda^2$},
however, long distance fluctuations become important and perturbation theory
looses its validity. Beyond the computation of universal
critical exponents and amplitude ratios we want to establish an explicit
connection between the universal critical equation of state and the 
microscopic free energy $\Gamma_{\Lambda}$.

In fig. \ref{CO2plot} we plot the results for the equation of state near the
endpoint of the critical line for carbon dioxide. For the microscopic scale
we have chosen $\Lambda^{-1}=0.5\: nm$. For $\hat{\xi}=0.6\Lambda^{-1}$,
$b_0(\Lambda)=34\: cm^3 mol^{-1}$,
$b_1(\Lambda)=3.11\cdot 10^{6} \: bar\: cm^6 mol^{-2}$ one finds
the location of the endpoint at $T_*=307.4\ K, p_*=77.6\ bar,\ \rho_*
=0.442\ gcm^{-3}$. This compares well with the experimental
values
$T_*=304.15 \; K$, $p_*=73.8 \; bar$, $\rho_*=0.468 \; g cm^{-3}$.
Comparing with literature values $b_i(0)_{ld}$ for low density
this yields $b_0(\Lambda)/b_0(0)_{ld}=0.8$,
$b_1(\Lambda)/b_1(0)_{ld}=0.86$. We conclude that the microscopic
free energy can be approximated reasonably well by a van der Waals form
even for high densities near $n_*$. The coefficients of the virial
expansion are shifted compared to this low density values by 15-20
per cent.
The comparison between the ``microscopic equation of state'' (dashed lines)
and the true equation of state (solid lines) in the plot clearly demonstrates
the importance of the fluctuations in the critical region.
Away from the critical region the fluctuation effects are less significant
and could be computed perturbatively.
\begin{figure}[h]
\unitlength1.0cm
\begin{center}
\begin{picture}(13.,9.)
\put(6.5,-0.5){$\rho \:[g \: cm^{-3}]$ }
\put(1.2,7.){$p \:[bar]$}
\put(-0.5,0.){
\epsfysize=13.cm
\epsfxsize=9.cm
\rotate[r]{\epsffile{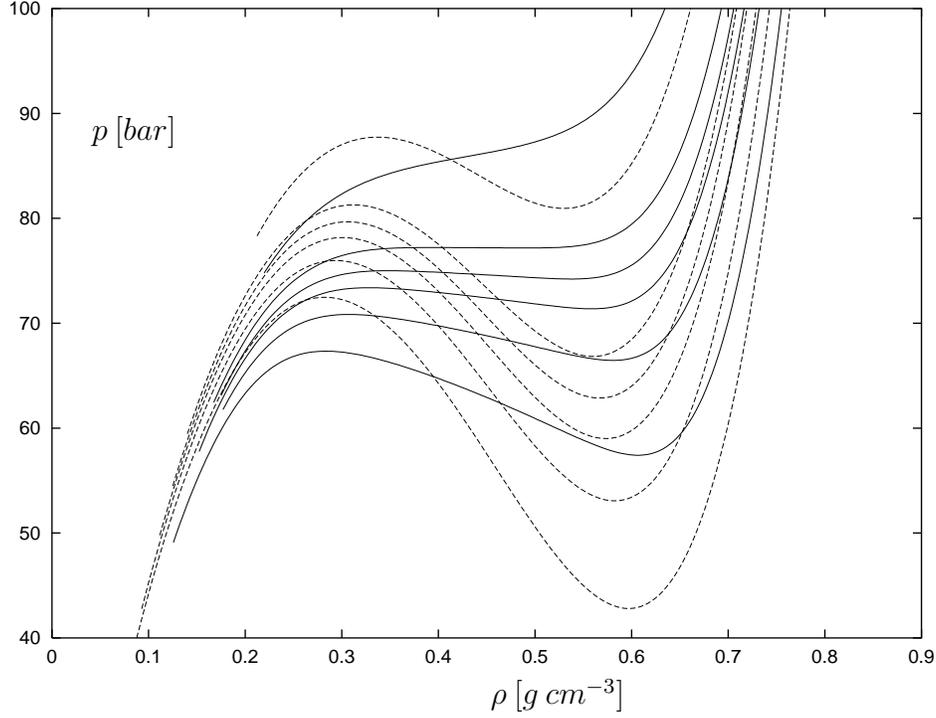}}
}
\end{picture}
\end{center}
\caption[]{
\em Liquid-vapor transition for carbon dioxide. We display
$p\!-\!\rho$-isotherms  for $T=295K$, $T=300K$,
 $T=303K$, $T=305K$, $T=307K$ and $T=315K$. The dashed lines represent the
virial expansion (at the scale
$k\!=\!\Lambda$) in 2nd order with $b_0\!=\!34 \: cm^3\: mol^{-1}$,
$b_1\!=\!3.11\cdot 10^{6} \: bar \: cm^6 \: mol^{-2}$.
The solid lines are the results at the scale $k=0$
 ($\hat{\xi}_{\Lambda}=0.6$).}
\label{CO2plot}
\end{figure}

\bigskip
\noindent
{\bf Flow equations}\\ 

Our aim is a numerical solution of the flow equation for $U_k$  
with given initial conditions at the scale $k\approx \Lambda$.
We introduce a dimensionless renormalized
field
\bea
   \dsp  \tilde{\varphi} & = & \dsp k^{\frac{2-d}{2}} Z_{0,k}^{1/2} \varphi
\eea
with wave function renormalization $Z_{0,k}=Z_k(\varphi_0(k))$ taken at 
the global potential minimum $\varphi_0(k)$. We do not choose the
field squared as a variable in order to faciliate the discussion
of first order transitions below, where terms cubic in the field 
appear\footnote{A combination of eq. (\ref{2.44})
with $\tilde y=0, \Delta\zeta_k=0$ with a suitable flow equation
for $z(\tilde\rho)$ has led to identical results.}. We also use
\bea
  \dsp u_k(\tilde{\varphi}) & = & \dsp k^{-d} U_k(\varphi)   \nnn
  \dsp \tilde{z}_k(\tilde{\varphi}) & = & Z_{0,k}^{-1} Z_k(\varphi)
\eea
and denote here by $u^{\prime}$, $\tilde{z}^{\prime}$ the derivatives with
respect to $\tilde{\varphi}$. This yields the scaling form of the flow equation
for $u_k^{\prime}$:
\bea
   \dsp \lefteqn{\frac{\partial}{\partial t} u_k^{\prime}(\tilde{\varphi})=
   \dsp -\hal (d+2-\eta_0)\cdot u_k^{\prime}(\tilde{\varphi})
       + \hal (d-2+\eta_0)\tilde{\varphi}\cdot u_k^{\prime\prime}
          (\tilde{\varphi})} & &  \nnn
   & - & \dsp  2v_d\tilde{z}_k^{\prime}(\tilde{\varphi})\cdot
       l_1^{d+2}\left(u_k^{\prime\prime}(\tilde{\varphi});\eta_0,
                      \tilde{z}_k(\tilde{\varphi}))
           - 2v_d u_k^{\prime\prime\prime}(\tilde{\varphi}\right)\cdot
       l_1^d\left(u_k^{\prime\prime}(\tilde{\varphi});
                  \eta_0,\tilde{z}_k(\tilde{\varphi})\right).
   \label{dtu}
\eea
Similarly, the evolution of $\tilde{z}_k$ is described in the truncation
(\ref{ansatzGamma}) by
\bea
   \dsp \lefteqn{\frac{\partial}{\partial t} \tilde{z}_k(\tilde{\varphi}) =
    \dsp \eta_0\cdot \tilde{z}_k(\tilde{\varphi})
        + \hal (d-2+\eta_0)\tilde{\varphi}\cdot
          \tilde{z}_k^{\prime}(\tilde{\varphi}) } & & \nnn
 & - &  \dsp \frac{4}{d} v_d \cdot
   u_k^{\prime\prime\prime}(\tilde{\varphi})^2
   \cdot m_{4,0}^d\left(u_k^{\prime\prime}(\tilde{\varphi});
                    \eta_0,\tilde{z}_k(\tilde{\varphi})\right)  \nnn
 & - &  \dsp \frac{8}{d} v_d \cdot u_k^{\prime\prime\prime}(\tilde{\varphi})
           \tilde{z}_k^{\prime}(\tilde{\varphi})
   \cdot m_{4,0}^{d+2}\left(u_k^{\prime\prime}(\tilde{\varphi});\eta_0,
                       \tilde{z}_k(\tilde{\varphi})\right) \nnn
 & - &  \dsp \frac{4}{d} v_d \cdot \tilde{z}_k^{\prime}(\tilde{\varphi})^2
   \cdot m_{4,0}^{d+4}\left(u_k^{\prime\prime}(\tilde{\varphi});
                         \eta_0,\tilde{z}_k(\tilde{\varphi})\right)
  -   \dsp 2 v_d \cdot \tilde{z}_k^{\prime\prime}(\tilde{\varphi})\cdot
    l_1^d\left(u_k^{\prime\prime}(\tilde{\varphi});
             \eta_0,\tilde{z}_k(\tilde{\varphi})\right)  \nnn
 & + & \dsp 4 v_d \cdot \tilde{z}_k^{\prime}(\tilde{\varphi})
           u_k^{\prime\prime\prime}(\tilde{\varphi})\cdot
   l_2^d\left(u_k^{\prime\prime}(\tilde{\varphi});
            \eta_0,\tilde{z}_k(\tilde{\varphi})\right)  \nnn
 &  + & \dsp \frac{2}{d}(1+2d) v_d \cdot \tilde{z}_k^{\prime}(\tilde{\varphi})^2
   \cdot
   l_2^{d+2}\left(u_k^{\prime\prime}(\tilde{\varphi});
               \eta_0,\tilde{z}_k(\tilde{\varphi})\right) \, .
 \label{dtz}
\eea
Here the mass threshold functions are
\bea
  \dsp l_{n}^d(u^{\prime\prime};\eta_0,\tilde{z}) & = &
  \dsp -\frac{1}{2}k^{2n-d}\cdot
   Z_{0,k}^n \cdot \int_{0}^{\infty}dx x^{\frac{d}{2}-1} \tilde{\partial}_t
   \left\{
      \frac{1}{(P(x)+Z_{0,k} k^2 u^{\prime\prime})^n}
   \right\}  \nnn
  \dsp m_{n,0}^d(u^{\prime\prime};\eta_0,\tilde{z}) & = &
  \dsp -\frac{1}{2}k^{2(n-1)-d}\cdot
   Z_{0,k}^{n-2} \cdot \int_{0}^{\infty}dx x^{\frac{d}{2}}\tilde{\partial}_t
   \left\{
      \frac{\dot{P}^2(x)}{(P(x)+Z_{0,k} k^2 u^{\prime\prime})^n}
   \right\}
   \label{threshfuncs}
\eea
(with $P(x)= \tilde{z}Z_{0,k} x + R_k(x)$,
 $\dot{P}\equiv\frac{dP}{dx}$ and $\tilde{\partial}_t$ acting only on $R_k$). 
The anomalous dimension
\bea
   \dsp \eta_{0,k} \equiv -\frac{d}{dt}\ln Z_{0,k} & = & \dsp
     -Z_{0,k}^{-1} \frac{\partial}{\partial t} Z_k(\varphi_0)
     - Z_{0,k}^{-1}\cdot \frac{\partial Z_k}{\partial\varphi}\left.\right|_{\varphi_0}
       \cdot \frac{d\varphi_0}{dt}
\eea
is determined by the condition $d\tilde{z}(\tilde{\varphi}_0)/dt=0$. It appears
linearly in the threshold functions due to $\tilde{\partial}_t$ acting on
$Z_{0,k}$ in $R_k$,
\beq
  R_k(x) = \frac{Z_{0,k}\, x}{\exp(x/k^2)-1}.
   \label{StfrmIRCut}
\eeq

For a computation of $\eta_0$ we need the evolution of the potential minimum
$\varphi_0(k)$, which follows from the condition
\mbox{$\frac{d}{dt}(\partial U_k/\partial\varphi(\varphi_0(k))=0$},
namely
\bea
   \dsp \lefteqn{\frac{d\tilde{\varphi}_0}{dt}  =
    \hal (2-d-\eta_{0})\tilde{\varphi_0}  } & &  \nnn
    & & \dsp +    2v_d\frac{\tilde{z}_k^{\prime}(\tilde{\varphi}_0)}
                  {u_k^{\prime\prime}(\tilde{\varphi}_0)}\cdot
         l_1^{d+2}(u_k^{\prime\prime}(\tilde{\varphi}_0);\eta_0,1)
    +   2v_d\frac{u_k^{\prime\prime\prime}(\tilde{\varphi}_0)}
              {u_k^{\prime\prime}(\tilde{\varphi}_0)}\cdot
         l_1^d(u_k^{\prime\prime}(\tilde{\varphi}_0);\eta_0,1).
   \label{dtphi0}
\eea
One infers an implicit equation for the anomalous dimension $\eta_{0,k}$,
\bea
   \dsp \lefteqn{ \eta_0 =  \frac{4}{d} v_d \cdot
   u_k^{\prime\prime\prime}(\tilde{\varphi}_0)^2
   \cdot m_{4,0}^d(u_k^{\prime\prime}(\tilde{\varphi}_0); \eta_0,1)
   +   \dsp \frac{8}{d} v_d \cdot u_k^{\prime\prime\prime}(\tilde{\varphi}_0)
   \tilde{z}_k^{\prime}(\tilde{\varphi}_0)
   \cdot m_{4,0}^{d+2}(u_k^{\prime\prime}(\tilde{\varphi}_0);\eta_0,1) } \nnn
 & & +  \dsp \frac{4}{d} v_d \cdot \tilde{z}_k^{\prime}(\tilde{\varphi}_0)^2
   \cdot m_{4,0}^{d+4}(u_k^{\prime\prime}(\tilde{\varphi}_0);\eta_0,1)
   +  \dsp 2 v_d \cdot \tilde{z}_k^{\prime\prime}(\tilde{\varphi}_0)\cdot
    l_1^d(u_k^{\prime\prime}(\tilde{\varphi}_0);\eta_0,1) \nnn
 & & - \dsp 4 v_d \cdot \tilde{z}_k^{\prime}(\tilde{\varphi}_0)
       u_k^{\prime\prime\prime}(\tilde{\varphi}_0)\cdot
   l_2^d(u_k^{\prime\prime}(\tilde{\varphi}_0);\eta_0,1) \nnn
 & &  -  \dsp \frac{2}{d}(1+2d) v_d \cdot \tilde{z}_k^{\prime}(\tilde{\varphi}_0)^2
   \cdot
   l_2^{d+2}(u_k^{\prime\prime}(\tilde{\varphi}_0);\eta_0,1)   \nnn
 & & -  \dsp 2v_d \frac{\tilde{z}_k^{\prime}(\tilde{\varphi}_0)}
       {u_k^{\prime\prime}(\tilde{\varphi}_0)}
       \cdot \left\{\tilde{z}_k^{\prime}(\tilde{\varphi}_0)
       l_1^{d+2}(u_k^{\prime\prime}(\tilde{\varphi}_0);\eta_0,1)
       + u_k^{\prime\prime\prime}(\tilde{\varphi}_0)
      l_1^d(u_k^{\prime\prime}(\tilde{\varphi}_0);\eta_0,1)
       \right\},
   \label{eta0Gl}
\eea
that can be solved by separating the threshold functions in $\eta_0$-dependent
and $\eta_0$-inde\-pen\-dent parts (c.f. eq. (\ref{threshfuncs})).
Since $\eta_0$ will turn out to be only a few
percent,
the neglect of contributions from higher derivative terms not contained
in (\ref{ansatzGamma})
induces a substantial \emph{relative} error for $\eta_0$, despite the good
convergence of the derivative expansion.
We believe that the missing higher derivative contributions to
$\eta_0$ constitute the main uncertainty in the results.

For given initial conditions $U_{\Lambda}(\varphi)$, $Z_{\Lambda}(\varphi)$ the
system of partial differential equations
(\ref{dtu}),(\ref{dtz}),(\ref{dtphi0}),(\ref{eta0Gl}) can be solved
numerically. A description of the algorithm used can be
found in \cite{BW97-1}.

\subsection{Universal and non-universal critical properties
\label{CritEqofStatIsing}}

In order to make the discussion transparent we present here first
results for polynomial initial conditions (\ref{microsU}) with
$\tilde{z}_{\Lambda}(\tilde{\varphi})=1$.
The term linear in $\varphi$ is considered as a source $j$. The special value
$\gamma_{\Lambda}=0$ realizes the $Z_2$-symmetric Ising model.
We start with the results for the universal critical behaviour for this
case.
For this particular purpose we hold $\lambda_{\Lambda}$ fixed and measure the 
deviation from the critical temperature by
\beq
   \delta m_{\Lambda}^2=m_{\Lambda}^2-m_{\Lambda,crit}^2
                       =S (T-T_c).
\eeq
For the liquid-gas system one has $S=2b_1/(K_{\Lambda}^2 T_c^2)$.

The anomalous dimension $\eta$ determines the two point function at the
critical temperature and equals $\eta_{0,k}$ for the scaling solution where
\mbox{$\partial_t u=\partial_t \tilde{z}=0$}. The results for the critical
exponents are compared with those from other methods in table
\ref{IsingExponenten}.
We observe a very good agreement for $\nu$ whereas the relative error for
$\eta$ is comparatively large as expected. Comparison with the lowest order
of the derivative expansion (f), used in section \ref{onpot}, 
shows a convincing apparent convergence of
this expansion for $\nu$.
For $\eta$ this convergence is hidden by the fact that in (\ref{ten})
a different determination of $\eta$ was used. Employing the present definition
would lead in lowest order of the derivative expansion to a value $\eta=0.11$.
As expected, the convergence of the derivative expansion is faster for the
very effective exponential cutoff than for the powerlike cutoff (g) which
would lead to unwanted properties of the momentum integrals in the next
order. 

\begin{table} [h] \centering{
\begin{tabular}
{|c|c|c|c|c|c|}
\hline
   &  $\nu$ & $\beta$ & $\gamma$ & $\eta$ \\
\hline
 (a) & $0.6304(13)$ & $0.3258(14)$ & 1.2396(13)   &
  $0.0335(25)$ \\
 (b) & $0.6290(25)$ & $0.3257(25)$ & 1.2355(50)   &
  $0.0360(50)$  \\
 (c) & $0.6315(8)$ &              & $1.2388(10)$   &
              \\
 (d) & $0.6294(9)$  &           &              &
  $0.0374(14)$   \\
\hline
 (f) & $0.643$ & $0.336$        & $1.258$ &
  $0.044$  \\
 (g) & $0.6181$ &              &             &
  $0.054$ \\
 (h) & $0.6307$  & $0.3300$  & $1.2322$ &
  $0.0467$  \\
\hline
\hline
 (i) & $0.625(6)$  & $0.316-0.327$ & $1.23-1.25$ &  \\
\hline
\end{tabular} 
\caption{\em
Critical exponents of the $(d\!=\!3)$-Ising model, calculated with various
methods.\protect\\
(a) From perturbation series at fixed dimension $d=3$ including seven--loop 
    contributions \cite{Zin93-1,GZ}. \protect \\
(b) $\epsilon$-expansion in five loop order \protect\cite{Zin93-1,GZ}. 
    \protect \\
(c) high temperature series \protect\cite{BC97-1}
    (see also \protect\cite{Reisz95,ZinnLaiFisher96}
    \protect). \protect \\
(d) Monte Carlo simulation \cite{LMC10} 
    (See also \protect\cite {Tsypin94}-
    \protect\cite{CasHas97}). \protect \\
(f)-(h): ``exact'' renormalization group equations.  \protect \\
(f) effective average action for the $O(N)$-model, $N\to 1$,
    with uniform wave function renormalization
    \protect\cite{BTW96-1} (see also
    \protect\cite{TW94-1}).   \protect \\
(g) scaling solution of equations analogous to (\protect\ref{dtu}),
    (\protect\ref{dtz}) with
    powerlike cutoff \protect \cite{Mor94-1}. \protect \\
(h) effective average action for one-component scalar field theory
    with field-dependent wave function renormalization (present section).
    \protect \\
(i) experimental data for the liquid-vapor system
    quoted from \protect \cite{Zin93-1}.
}
\label{IsingExponenten}
}
\end{table}

In order to establish the quantitative connection between the short distance
parameters $m_{\Lambda}^2$ and $\lambda_{\Lambda}$ and the universal
critical behaviour one needs the amplitudes $C^{\pm}$, $\xi^{\pm}$, etc.
For $\lambda_{\Lambda}/\Lambda=5$ we find \mbox{$C^{+}\!=\!1.033$},
 \mbox{$C^{-}\!=\!0.208$},
 \mbox{$\xi^{+}\!=\!0.981$}, \mbox{$\xi^{-}\!=\!0.484$}, \mbox{$B\!=\!0.608$},
 \mbox{$E\!=\!0.208$}.
Here and in the following all dimensionful quantities are quoted in units
of $\Lambda$.
The amplitude $D$ is given by
\mbox{$\partial U_0/\partial\varphi =  D\cdot \varphi^{\delta}$} on the critical
iso\-therme and we obtain \mbox{$D\!=\!10.213$}.
In table \ref{IsingUnivAmpTab} we present our results for the universal
amplitude ratios
$C^{+}/C^{-}$, $\xi^{+}/\xi^{-}$, \mbox{$R_{\chi}=C^{+}DB^{\delta-1}$},
 \mbox{$\tilde{R}_{\xi}=(\xi^{+})^{\beta/\nu}D^{1/(\delta+1)}B$}.

\begin{table} [h] \centering{
\begin{tabular}
{|c|c|c|c|c|c|c|c|}
\hline
 & $C^{+}/C^{-}$ & $\xi^{+}/\xi^{-}$ & $R_{\chi}$ & $\tilde{R}_{\xi}$ &
 $\xi^{+}E$ & $\lambda_R/m_R$ &
 $\hat{\lambda}_R/\hat{m}_R$ \\
\hline
\hline
 (a) & $4.79\pm 0.10$ &                & $1.669\pm 0.018$ & & & $7.88$ &\\
\hline
 (b) & $4.73\pm 0.16$ &                & $1.648\pm 0.036$ & & & $9.33$ &\\
\hline
 (c) & $4.77 \pm 0.02$ & $1.96 \pm 0.01$ 
&$1.662 \pm 0.005$ & & & $7.9 - 8.15$ &\\
\hline
 (d) & $4.75\pm 0.03$ & $1.95\pm 0.02$ &                  & & & $7.76$
     & $5.27$ \\
\hline
\hline
 (f) & 4.29 & 1.86 & 1.61 & 0.865 & 0.168  & $9.69$ & $5.55$ \\
\hline
 (h) & 4.966 & 2.027 & 1.647 & 0.903 & 0.204 & $8.11$ & $4.96$ \\
\hline
\hline
 (i) & $4.8-5.2$ &        & $1.69\pm 0.14$ & & & &\\
\hline

\end{tabular} }

\caption{\em Universal amplitude ratios and couplings of the
$(d\!=\!3)$-Ising model.
 \protect\\
(a) perturbation theory at fixed dimension $d\!=3\!$
    \protect\cite{GZ,GuidaZinn96}. \protect \\
(b) $\epsilon$-expansion \protect\cite{GZ,GuidaZinn96}. \protect \\
(c) high temperature series. Amplitude ratios from \protect\cite{CPRV,Zin93-1},
    $\lambda_R/m_R$ from \protect\cite{Reisz95,ZinnLaiFisher96,ButeraComi96}.
    \protect \\
(d) Monte Carlo simulations. Amplitude ratios from \protect\cite{CasHas97},
    $\hat{\lambda}_R/\hat{m}_R$ from \protect\cite{Tsypin96},
    $\lambda_R/m_R$ from \protect\cite{Tsypin94}. \protect \\
(f) effective average action for the $O(N)$-model, $N\to 1$,
    with uniform wave function renormalization
    \protect\cite{BTW96-1}. \protect \\
(h) present section with field-dependent wave function renormalization.\protect \\
(i) experimental data for the liquid-vapor system
    \protect\cite{PrivHohen91}.
\protect \\
}
\label{IsingUnivAmpTab}
\end{table}

The critical exponents and amplitudes only characterize the behaviour of
$U_0(\varphi)$ in the
limits \mbox{$\varphi \to \varphi_0$} and \mbox{$\varphi \to \infty$}. Our method
allows us to compute $U_0(\varphi)$ for arbitrary $\varphi$. As an example, the quartic
coupling $\lambda_R=\frac{1}{3}\frac{\partial^4 U_0}{\partial\varphi_R^4}(0)
=\frac{\partial^2U_0}{\partial\rho_R^2}(0),\ \hat\lambda_R=
\frac{\partial^2U_0}{\partial\rho^2_R}(\varphi_{0R}),\
\rho_R=\frac{1}{2}\varphi^2_R$,
becomes in the critical region proportional to $m_R$. Our results for
the universal couplings $\lambda_R/m_R$ in the symmetric and
$\hat{\lambda}_R/\hat{m}_R$ in the ordered phase can also be found in
table \ref{IsingUnivAmpTab}.
Here $m_R=\frac{\partial^2 U_0}{\partial \varphi_R^2}
\left.\right|_{\varphi_R=0}$ in
the symmetric and
$\hat{m}_R=\frac{\partial^2 U_0}{\partial \varphi_R^2}\left.\right|_{\varphi_{0R}}$
in the ordered phase.

We should emphasize that the shape of the potential in the low temperature
phase depends on $k$ in the ``inner'' region corresponding to $|\varphi|<\varphi_0$.
This is due to the fluctuations which are responsible for making the potential
convex in the limit $k\to 0$ \cite{RingWet90,TetWet92,AlexBranchPolon98}.
We illustrate this by plotting the potential for different values of $k$ in
fig. \ref{apptoconvexity}. Our results for the scaling function $f(x)$ are shown in fig. \ref{Z2Widom},
 together with the asymptotic behaviour (dashed lines) as dictated by the
critical exponents and amplitudes.

\begin{figure}[h]
\unitlength1.0cm
\begin{center}
\begin{picture}(13.,9.)
\put(6.,-0.5){$\dsp x\cdot f^{-\frac{1}{\beta\delta}}(x)$}
\put(-1.5,9.2){$\dsp \frac{\varphi}{j^{1/\delta}}=f^{-1/\delta}$}

\put(-0.5,0.){
\rotate[r]{\epsfig{file=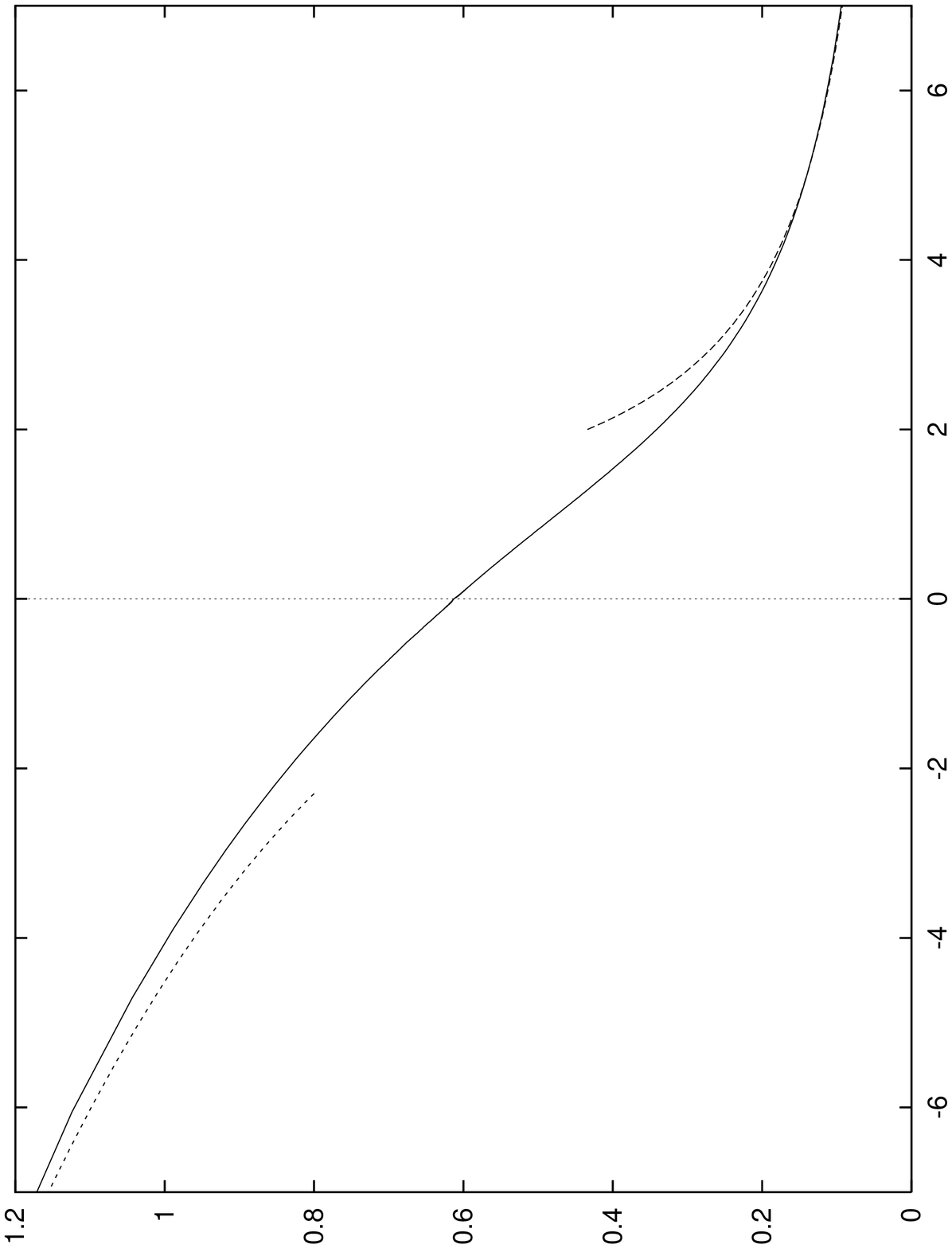,width=9.cm,height=13.cm}}
}
\end{picture}
\end{center}
\caption[]{\label{Z2Widom}
\em
Widom scaling function $f(x)$ of the $(d=3)$-Ising model. (
The present curve is generated for a quartic
short distance potential $U_{\Lambda}$ with $\lambda_{\Lambda}/\Lambda=5$).
The dashed lines indicate extrapolations of the limiting behaviour as given
by the critical exponents.}
\end{figure}

A different useful parameterization of the critical equation of state can be
given in terms of nonlinearly rescaled fields $\hat{\varphi}_R$, using a
$\varphi$-dependent wave function renormalization $Z(\varphi)=Z_{k\to 0}(\varphi)$,
\bea
   \dsp \hat{\varphi}_R & = & \dsp Z(\varphi)^{1/2}\varphi.
\label{rescaled_field}
\eea
Our numerical results can be presented in terms of a fit to the universal
function
\bea
   \dsp \hat{F}(\hat{s})=m_R^{-5/2}\frac{\partial U_0}{\partial\hat{\varphi}_R} & , &
   \dsp \hat{s}=\frac{\hat{\varphi}_R}{m_R^{1/2}}
              =\left(\frac{Z(\varphi)\varphi^2}{m_R}\right)^{1/2},  \nnn
\eea
\bea
  \dsp  \hat{F}_{Fit}(\hat{s}) & = & \left(a_0\hat{s}+a_1\hat{s}^3+a_2\hat{s}^5
        +a_3\hat{s}^7\right)\cdot f_{\alpha}(\hat{s})
        +(1-f_{\alpha}(\hat{s}))\cdot a_4\hat{s}^5.
\eea
The factors $f_{\alpha}$ and $(1-f_{\alpha})$ interpolate between a polynomial
expansion and the asymptotic behaviour for large arguments. We use
\bea
  \dsp f_{\alpha}(x) & = & \dsp \alpha^{-2} x^2\cdot
           \frac{\exp(-\frac{x^2}{\alpha^2})}{1-\exp(-\frac{x^2}{\alpha^2})}.
\eea
A similar fit can be given for
\bea
  \dsp \tilde{z}(s) = \frac{\lim_{k\to 0} Z_k(\varphi)}{Z_0} & , &
   \dsp s=\frac{\varphi_R}{m_R^{1/2}}
        =\left(\frac{Z_0\varphi^2}{m_R}\right)^{1/2}=\tilde{z}^{-1/2}\hat{s},
\eea
\bea
  \dsp  \tilde{z}_{Fit}(s) & = & \left(b_0+b_1 s^2+b_2 s^4+b_3s^6
        +b_4 s^8\right)\cdot f_{\beta}(s)
        +(1-f_{\beta}(s))\cdot b_5 |s|^{-\frac{2\eta}{1+\eta}}.
\eea
In the symmetric phase one finds (with $\eta=0.0467$) $\alpha=1.012$,  
$a_0=1.0084$, $a_1=3.1927$, $a_2=9.7076$, $a_3=0.5196$, $a_4=10.3962$
and $\beta=0.5103$, $b_0=1$, $b_1=0.3397$, $b_2=-0.8851$, \mbox{$b_3=0.8097$},
 $b_4=-0.2728$, $b_5=1.0717$, whereas the fit parameters for
the phase with spontaneous symmetry breaking are
$\alpha=0.709$, $a_0=-0.0707$, $a_1=-2.4603$, \mbox{$a_2=11.8447$},
$a_3=-1.3757$, $a_4=10.2115$ and
$\beta=0.486$, $b_0=1.2480$, $b_1=-1.4303$, $b_2=2.3865$, $b_3=-1.7726$,
$b_4=0.4904$, $b_5=0.8676$
(our fit parameters are evaluated for this phase for
$(\partial^2 U_k/\partial\varphi_R^2)(\varphi_{R,max})/k^2=-0.99$).
One observes that the coefficients $a_2$ and $a_4$ are large and of comparable
size. A simple polynomial form
$\hat{F}=\tilde{a}_0 \hat{s} + \tilde{a}_1\hat{s}^3+ \tilde{a}_2\hat{s}^5$
is not too far from the more precise result. We conclude that in terms of
the rescaled field $\hat{\varphi}_R$ (\ref{rescaled_field}) the potential is
almost a polynomial $\varphi^6$-potential.

\begin{figure}[h]
\unitlength1.0cm
\begin{picture}(6.5,6.5)
\put(3.5,-0.5){$s=\frac{\varphi_R}{m_R^{1/2}}$ }
\put(1.2,4.5){$\tilde{z}$}
\put(0.,0.){
\rotate[r]{\epsfig{file=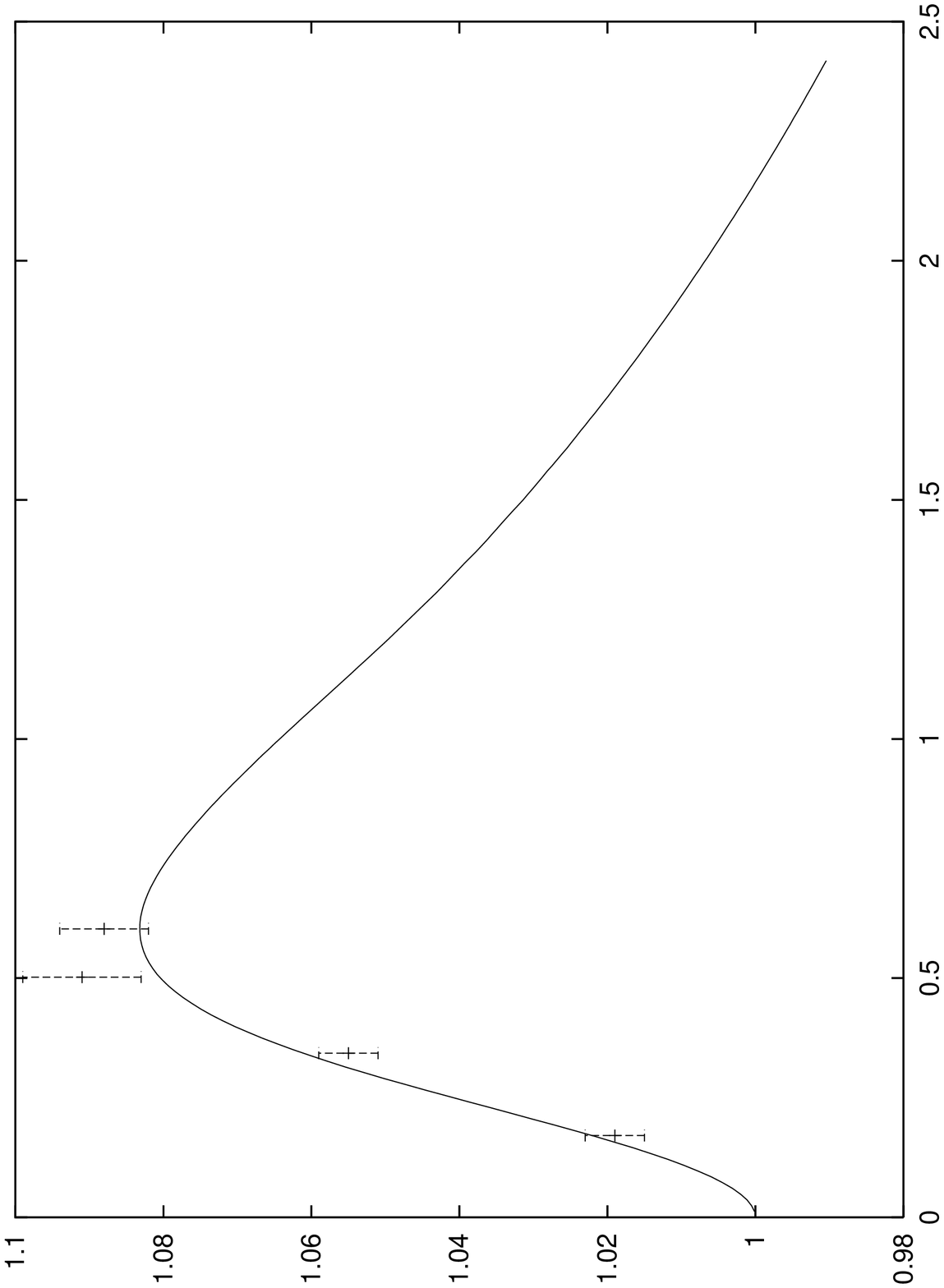,width=5.5cm,height=7.5cm}}
}
\end{picture}

\begin{picture}(6.,0.)
\put(11.5,0.){$s=\frac{\varphi_R}{m_R^{1/2}}$ }
\put(11.5,3.5){$\tilde{z}_k$}
\put(9.5,2.8){\footnotesize $c=-0.9$}
\put(10.6,5.){\footnotesize $c=-0.99$}
\put(8.5,0.5){
\rotate[r]{\epsfig{file=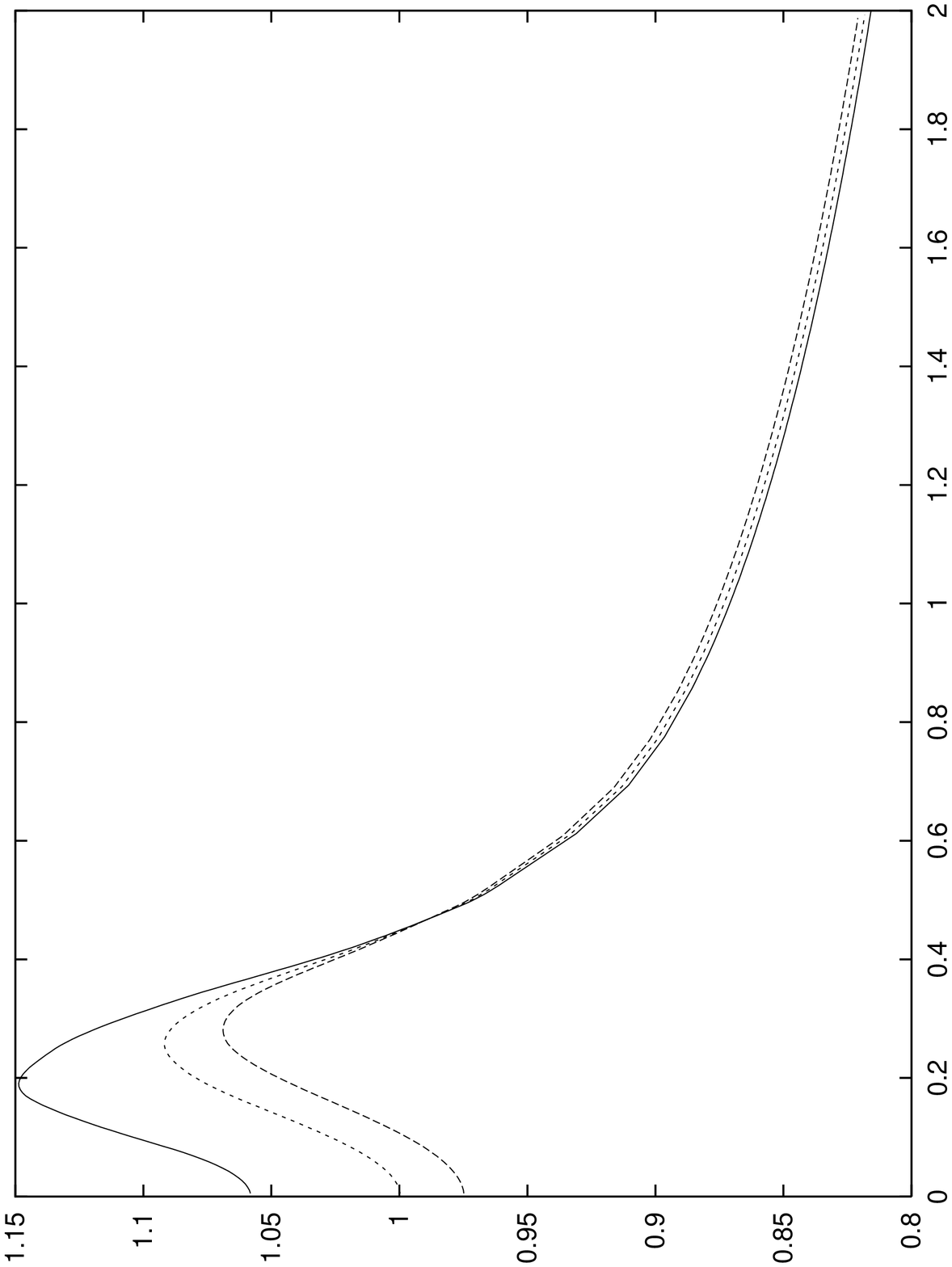,width=5.5cm,height=7.5cm}}
}
\end{picture}
\caption[]{\label{univz}
\em
Universal rescaled wave function renormalization $\tilde{z}$
in the symmetric and ordered phase of the $(d\!=\!3)$-Ising model. In the
low temperature phase the plots are for different $k$ with
$\frac{1}{k^2}\frac{\partial^2 U}{\partial \varphi_R^2}(\varphi_{R,max})\!=\!c$
($c=-0.9,-0.95,-0.99$). Here $\varphi_{R,max}$ is the location of the
local maximum of the potential in the inner (non-convex) region.
In the graph for the high temperature phase we have inserted Monte
Carlo results by M. Tsypin (private communication).
}
\end{figure}

In fig. \ref{univz} we show $\tilde{z}$ as a function of $s$ both for
the symmetric and ordered phase. Their shape is similar to the scaling
solution found in \cite{Mor94-1}.
Nevertheless, the form of $\tilde{z}$ for $k=0$ which expresses directly
information about the physical system should not be confounded with the
scaling solution which  depends on the particular infrared cutoff.
For the low temperature phase one sees
the substantial dependence of $\tilde{z}_k$ on the infrared cutoff $k$ for
small values $s<s_0$. Again this corresponds to the ``inner region'' between
the origin ($s\!=\!0$) and the minimum of the potential ($s_0\!=\!0.449$) where
the potential finally becomes convex for $k\to 0$.

Knowledge of $U_0$ and $\tilde{z}$ permits the computation of the (renormalized)
propagator for low momenta with arbitrary sources $j$. It is given by
\bea\label{4.47}
  \dsp G(q^2)=\left(\frac{\partial^2 U_0(\varphi_R)}{\partial \varphi_R^2}
             + \tilde{z}(\varphi_R)q^2\right)^{-1}
\eea
for $\tilde{z}q^2 \lesssim \partial^2 U_0/\partial \varphi_R^2$.
Here $\varphi_R$
obeys $\partial U_0/\partial\varphi_R = Z_0^{-1/2} j$.
We emphasize that the
correlation length
$\xi(\varphi_R)=\tilde{z}^{1/2}(\varphi_R)(\partial^2 U_0/\partial\varphi_R^2)^{-1/2}$
at given source $j$ requires information about $\tilde{z}$.
For the gas-liquid transition $\xi(\varphi_R)$ is directly connected to the
density dependence of the correlation length. For magnets, it expresses the
correlation length as a function of magnetization. The factor
$\tilde{z}^{1/2}$ is often omitted in other approaches. From 
\beq\label{4.47a}
\xi^{-2}=m^2_R\left\{\left(1+\frac{1}{2}\frac{\partial\ln\tilde z}{2\partial\ln s}\right)^2\frac{\partial \hat F}{\partial \hat s}+\frac{1}{2}\left
[\frac{\partial\ln\tilde z}{\partial\ln s}+\frac{1}{2}\left(\frac{\partial\ln\tilde z}
{\partial\ln s}\right)^2+\frac{\partial^2\ln\tilde z}{(\partial \ln s)^2}
\right]\frac{\hat F}{\hat s}\right\}\eeq
one can extract the behavior for $|\varphi|\to\infty$ for the
high and low temperature phases
\beq\label{4.47b}
\xi=L^\pm|\varphi|^{-\lambda}\eeq

Critical equations of state for the Ising model have been computed earlier
with several methods.
They are compared
with our result for the phase with spontaneous symmetry breaking in
\mbox{fig. \ref{compEquofStat}} and for the symmetric phase in
\mbox{fig. \ref{compEquofStatSym}}.
For this purpose we use $F(\tilde{s})=m_R^{-5/2}\partial U_0/
\partial\varphi_R$
with $\tilde{s}=\frac{\varphi_R}{\varphi_{0R}}$
in the phase with spontaneous symmetry
breaking (note $\tilde{s}\sim s$). The constant $c_F$ is chosen such that
$\frac{1}{c_F}\frac{\partial F}{\partial \tilde{s}}(\tilde{s}\!=\!1)\!=\!1$.
In the symmetric phase we take instead
$\tilde{s}\!=\!\frac{\varphi_R}{m_R^{1/2}}$ so that
$\frac{\partial F}{\partial \tilde{s}}(\tilde{s}\!=\!0)\!=\!1$.
One expects for large $\tilde{s}$ an inaccuracy of our results due to the
error in $\eta$.

\begin{figure}[h]
\unitlength1.0cm
\begin{center}
\begin{picture}(13.,9.)
\put(10.5,2.8){\footnotesize (1)}
\put(11.,5.5){\footnotesize(4)}
\put(9.9,7.){\footnotesize(3)}
\put(11.1,7.4){\footnotesize(2)}
\put(11.1,6.8){\footnotesize(5)}
\put(6.5,-0.5){$\tilde{s}$}
\put(0.7,8.){$\dsp\frac{F(\tilde{s})}{c_F}$}

\put(-0.5,0.){
\rotate[r]{\epsfig{file=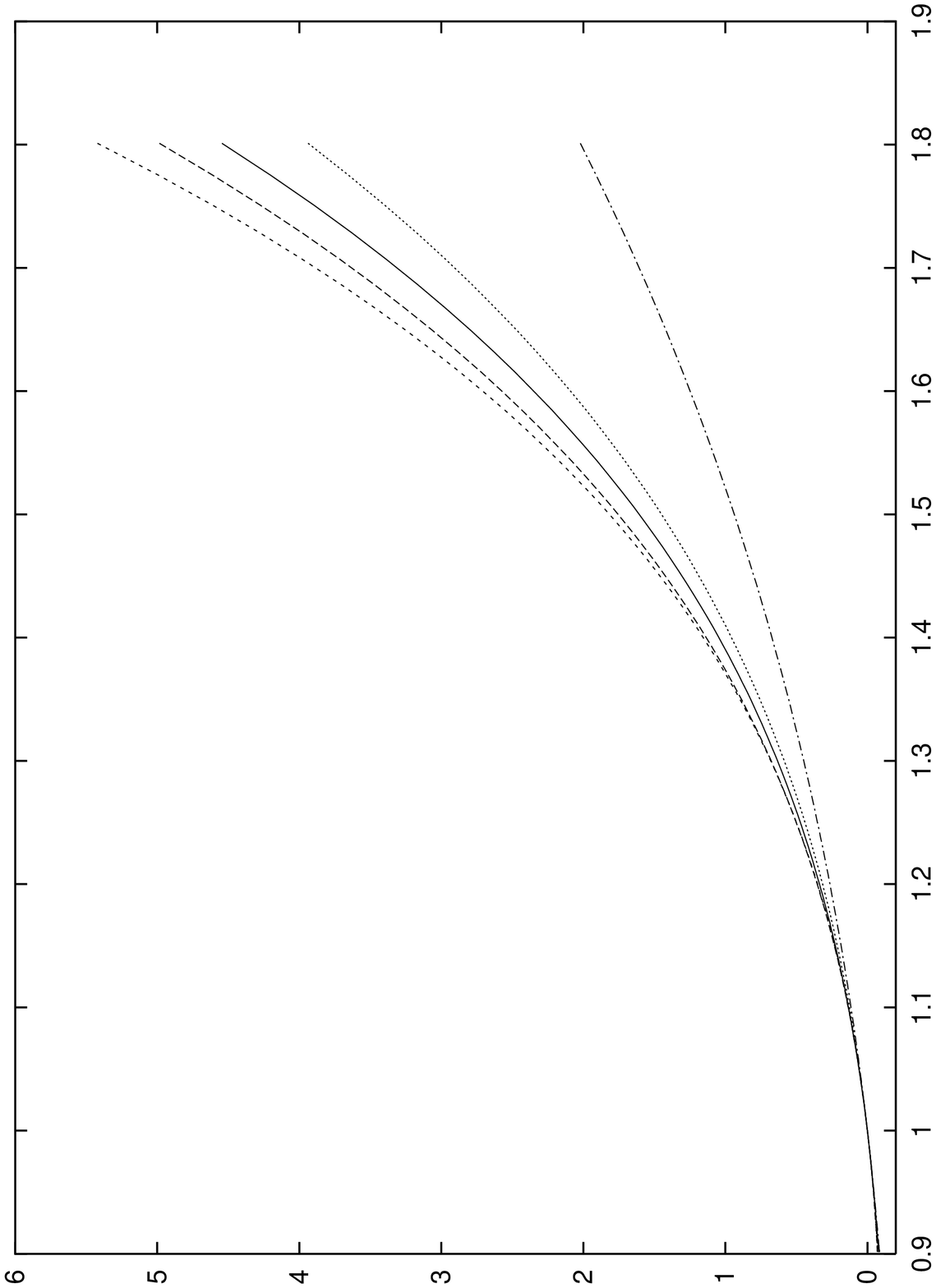,width=9.cm,height=13.cm}}
}
\end{picture}
\end{center}
\caption[]{\label{compEquofStat}
\em
The critical equation of state in the ordered phase. \protect \\
(1) mean field approximation. \protect \\
(2) effective average action with uniform wave function renormalization
    \protect\cite{BTW96-1}. \protect \\
(3) Monte Carlo simulation \cite{Tsypin96}. \protect \\
(4) resummed $\epsilon$-expansion in $O(\epsilon^3)$, five loop perturbative
    expansion and high temperature series
    \cite{GuidaZinn96}. \protect \\
(5) present section.}
\end{figure}

\begin{figure}[h]
\unitlength1.0cm
\begin{center}
\begin{picture}(13.,9.)
\put(11.4,8.2){\footnotesize (1)}
\put(10.8,7.5){\footnotesize (2)}
\put(11.3,7.4){\vector(1,-1){0.45}}
\put(10.5,7.1){\footnotesize (3)}
\put(10.9,7.0){\vector(1,-1){0.59}}
\put(10.,6.){\footnotesize(4)}
\put(10.5,6.0){\vector(1,-1){0.65}}
\put(11.15,4.5){\footnotesize (5),(6)}
\put(11.7,4.8){\vector(-1,2){0.33}}
\put(10.6,3.4){\footnotesize (7)}
\put(10.8,3.7){\vector(-1,1){0.4}}
\put(6.5,-0.5){$\tilde{s}$}
\put(1.,8.){$\dsp F(\tilde{s})$}

\put(-0.5,0.){
\rotate[r]{\epsfig{file=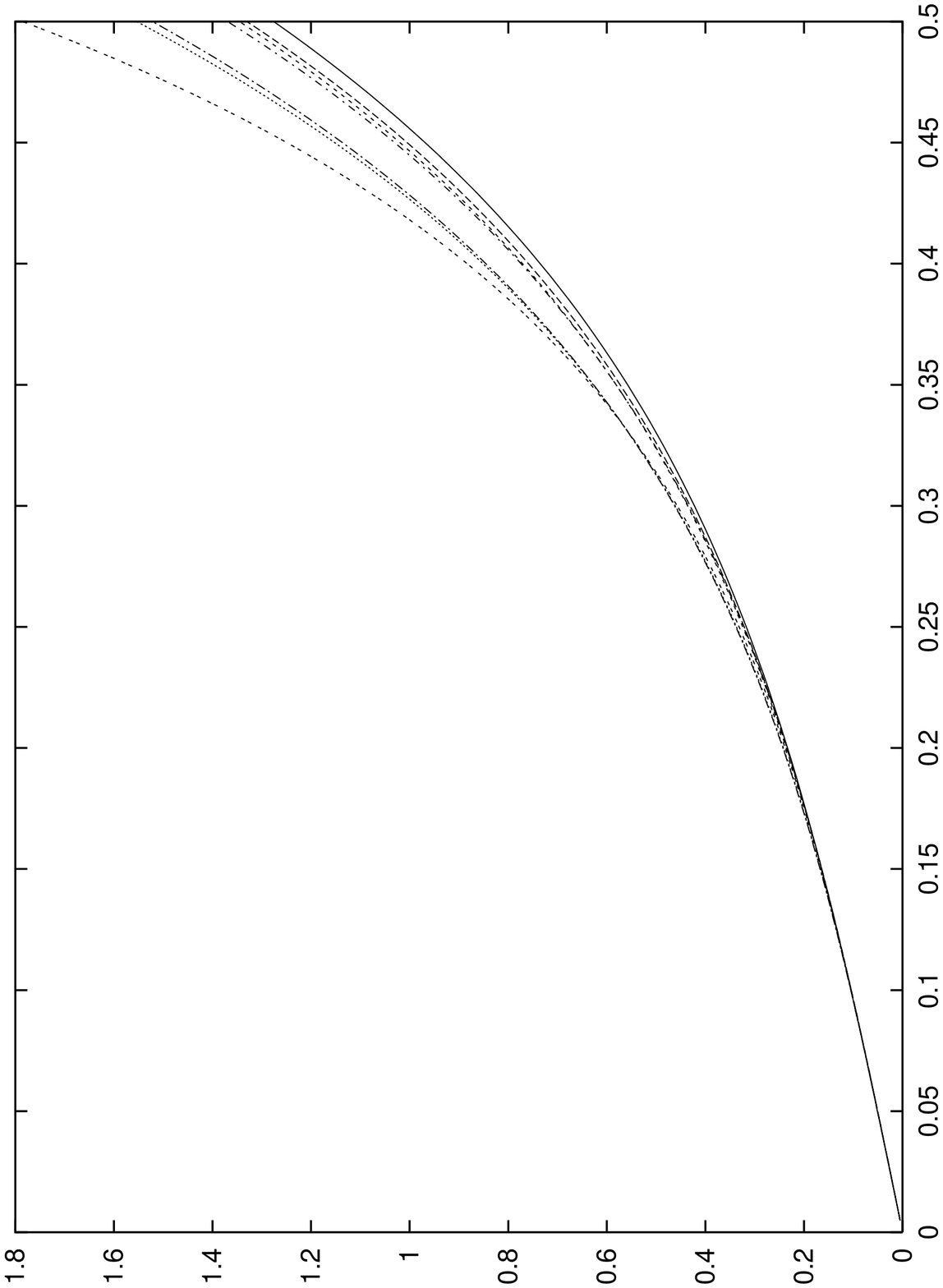,width=9.cm,height=13.cm}}
}
\end{picture}
\end{center}
\caption[]{
\em
The critical equation of state in the symmetric phase. \protect \\
(1) Monte Carlo simulation \protect\cite{KimLandau96},
(2) $\epsilon$-expansion \protect\cite{GuidaZinn96}. \protect \\
(3)  effective average action with uniform wave function renormalization
    \protect\cite{BTW96-1}. \protect \\
(4) Monte Carlo simulation \protect\cite{Tsypin94},
(5),(6) high temperature series
    \protect\cite{ButeraComi96},\protect\cite{ZinnLaiFisher96}.
    \protect \\
(7) present section.
\label{compEquofStatSym}
}
\end{figure}

\begin{table} [h] \centering
{ \begin{tabular}
{|l|c|c|c|c|c|c|c|c|}
\hline
   & $m_{\Lambda,crit}^2$ & $C^{+}$ & D \\
\hline
\hline
 $\lambda_{\Lambda}=0.1$ & $-6.4584\cdot 10^{-3}$ & $0.1655$ & $5.3317$ \\
\hline
 $\lambda_{\Lambda}=1$   & $-5.5945\cdot 10^{-2}$ & $0.485$ & 7.506 \\
\hline
 $\lambda_{\Lambda}=5$ & $-0.22134$ & $1.033$ & $10.213$ \\
\hline
 $\lambda_{\Lambda}=20$ & $-0.63875$ & $1.848$ & $16.327$ \\
\hline

\end{tabular}}

\caption{\em  The critical values
$m_{\Lambda,crit}^2$ and the non-universal amplitudes $C^{+}$, $D$ as a
function of the quartic short distance coupling $\lambda_{\Lambda}$
(all values expressed in units of $\Lambda$).
Other non-universal amplitudes can be calculated from the universal quantities
of table \ref{IsingUnivAmpTab}.
\label{nonunivtable} }
\end{table}

In summary of this section we may state that the non-perturbative flow
equations in second order in a derivative expansion lead to a critical equation
of state which is well compatible with high order expansions within other
methods. In addition, it allows to establish an explicit connection between
the parameters appearing in the microscopic free energy $\Gamma_{\Lambda}$ and
the universal long distance behaviour. For a quartic polynomial potential
this involves in addition to the non-universal amplitudes the
value of $m_{\Lambda,crit}^2$. We have listed these quantities for different
values
of $\lambda_{\Lambda}$ in table \ref{nonunivtable}. Finally, the temperature
scale is
established by $S=\partial m_{\Lambda}^2/\partial T\left.\right|_{T_c}$.

\subsection{Equation of state for first order transitions \label{cubicmodel}}

Our method is not restricted to a microscopic potential with discrete
$Z_2$-symmetry. The numerical code works for arbitrary initial potentials.
We have investigated the polynomial potential (\ref{microsU}) with
$\gamma_{\Lambda}\neq0$. The numerical solution of the flow equations
(\ref{dtu}),(\ref{dtz}) shows the expected first order
transition (in case of vanishing linear term $j$).  Quite
generally, the universal critical equation of state for first order transitions
will depend on two scaling parameters (instead of one for second order
transitions) since the jump in the order parameter or the mass introduces
a new scale. This has been demonstrated in \cite{BW97-1}
for a scalar matrix model and is discussed in more detail in
section \ref{smm}. The degree to which universality applies depends on the
properties of a given model and its parameters.
For a $\varphi^4$-model with cubic term
(\ref{microsU}) one can relate the equation of state to
the Ising model by an appropriate mapping. This allows us to
compute the universal critical equation of state for arbitrary first order
phase transitions in the Ising universality class from the critical equation
of state for the second order phase transition in the Ising model.
For other universality classes
a simple mapping to a second order equation of state
is not always possible - its existence is particular to the present model.

By a variable shift
\bea
 \dsp  \sigma & = & \dsp \varphi+\frac{\gamma_{\Lambda}}{3\lambda_{\Lambda}}
 \label{varshift}
\eea
we can bring the short distance potential (\ref{microsU}) into the form
\bea
   \dsp U_{\Lambda}(\sigma) & = &
   \dsp -J_{\gamma}\sigma + \frac{\mu_{\Lambda}^2}{2}\sigma^2
   + \frac{\lambda_{\Lambda}}{8} \sigma^4 + c_{\Lambda}
\eea
with
\bea
  \dsp J_{\gamma} & = &
  \dsp \frac{\gamma_{\Lambda}}{3\lambda_{\Lambda}}m_{\Lambda}^2
      -\frac{\gamma_{\Lambda}^3}{27\lambda_{\Lambda}^2} \nnn
  \dsp \mu_{\Lambda}^2 & = &
  \dsp  m_{\Lambda}^2 -\frac{\gamma_{\Lambda}^2}{6\lambda_{\Lambda}}.
\eea
We can now solve the flow equations in terms of $\sigma$ and reexpress the
result in terms of $\varphi$ by eq. (\ref{varshift}) at the end. The exact flow
equation (\ref{2.17}) does not involve the linear term $\sim J_{\gamma}\sigma$
in the rhs (also the constant $c_{\Lambda}$ is irrelevant).
Therefore the effective potential ($k=0$) is given by
\bea
  \dsp U_0=U^{Z_2}(\sigma)-J_{\gamma}\sigma + c_{\Lambda}& = &
  \dsp U^{Z_2}(\varphi+\frac{\gamma_{\Lambda}}{3\lambda_{\Lambda}})
      -(\varphi+\frac{\gamma_{\Lambda}}{3\lambda_{\Lambda}}) J_{\gamma}
      + c_{\Lambda},
  \label{shiftU}
\eea
where $U^{Z_2}$ is the effective potential of the Ising type model with
quartic coupling $\lambda_{\Lambda}$ and mass term $\mu_{\Lambda}^2$. The
equation of state $\partial U/\partial\varphi=j$ or, equivalently
\bea
  \dsp \frac{\partial U^{Z_2}}{\partial\varphi}\left.
           \right|_{\varphi+\frac{\gamma_{\Lambda}}{3\lambda_{\Lambda}}} & = &
  \dsp j + J_{\gamma},
  \label{shifteqofstat}
\eea
is therefore known explicitly for arbitrary $m_{\Lambda}^2$, $\gamma_{\Lambda}$
and $\lambda_{\Lambda}$ (cf. eq. (\ref{sixb}) for the
universal part). This leads immediately to the following conclusions:
\begin{itemize}
 \item[i)] First order transitions require that the combination $U_0(\varphi)-j\varphi$
           has two degenerate minima. This happens for $J_{\gamma}+j=0$ and
           $\mu_{\Lambda}^2<\mu_{\Lambda,crit}^2$ or
           \bea
           \dsp j & = & \dsp -\frac{\gamma_{\Lambda}}{3\lambda_{\Lambda}}
                \left(m_{\Lambda}^2-\frac{\gamma_{\Lambda}^2}
                {9\lambda_{\Lambda}}\right)
           \label{jcrit}
           \eea
           \bea
           \dsp m_{\Lambda}^2 & < & \dsp \mu_{\Lambda,crit}^2
                + \frac{\gamma_{\Lambda}^2}{6\lambda_{\Lambda}}.
           \eea
           Here $\mu_{\Lambda,crit}^2$ is the critical mass term of the
           Ising model.
 \item[ii)] The boundary of this region for
            \bea
            \dsp m_{\Lambda}^2 & = &
            \dsp \mu_{\Lambda,crit}^2+\frac{\gamma_{\Lambda}^2}
                {6\lambda_{\Lambda}}
            \label{2ndorderline}
            \eea
            is a line of second order phase transitions with vanishing
            renormalized mass or infinite correlation length.
\end{itemize}
For $j=0$ (e.g. magnets with polynomial potential in absence of external
fields) the equations (\ref{jcrit}), (\ref{2ndorderline}) have the solutions
\bea
  \dsp \gamma_{\Lambda;1}=0 & , &
  \dsp \gamma_{\Lambda;2,3}=\pm
       (-18\lambda_{\Lambda}\mu_{\Lambda,crit}^2)^{1/2}.
\eea
The second order phase transition for $\gamma_{\Lambda}\neq 0$ can be described
by Ising models for shifted fields $\sigma$.
For a given model, the way how a phase transition line is crossed as the
temperature is varied follows from the temperature dependence  of $j$,
 $m_{\Lambda}^2$, $\gamma_{\Lambda}$ and $\lambda_{\Lambda}$. For the
gas-liquid transition both $j$ and $m_{\Lambda}^2$ depend on $T$.

In the vicinity of the boundary of the region of first order transitions the
long range fluctuations play a dominant role and one expects universal
critical behaviour. The detailed microscopic physics is only reflected in two
non-universal amplitudes. One reflects the relation between the renormalized
and unrenormalized fields as given by $Z_0$. The other is connected to the
renormalization factor for the mass term. Expressed in terms of renormalized
fields and mass the potential $U$ looses all memory about the microphysics.

The critical equation of
state of the non-symmetric model ($\gamma_{\Lambda}\neq 0$) follows from
the Ising model (4.5). With
$\frac{\partial U^{Z_2}}{\partial\varphi}\left.\right|_{\varphi}=|\varphi|^{\delta}f(x)$,
the scaling form of the equation of state
$j=\frac{\partial U}{\partial \varphi}$ for the model with cubic coupling can be
written as
\bea
  \dsp j  & = & \dsp |\varphi+\frac{\gamma_{\Lambda}}{3\lambda_{\Lambda}}|^{\delta}
       f(x)
      - \left(\frac{\gamma_{\Lambda}}{3\lambda_{\Lambda}}\mu_{\Lambda,crit}^2
          + \frac{\gamma_{\Lambda}^3}{54\lambda_{\Lambda}^2}\right)
      - \frac{\gamma_{\Lambda}}{3\lambda_{\Lambda}}\delta\mu_{\Lambda}^2,
\eea
where $\dsp x=\frac{\delta\mu_{\Lambda}^2}
{|\varphi+\frac{\gamma_{\Lambda}}{3\lambda_{\Lambda}}|^{1/\beta}}$ and
$\delta\mu_{\Lambda}^2=m_{\Lambda}^2-\frac{\gamma_{\Lambda}^2}
{6\lambda_{\Lambda}}-\mu_{\Lambda,crit}^2$.
One may choose
\bea
  \dsp y & = & \dsp
      \frac{\gamma_{\Lambda}}{3\lambda_{\Lambda}}
      \left(\mu_{\Lambda,crit}^2+\frac{\gamma_{\Lambda}^2}{18\lambda_{\Lambda}}
        + \delta\mu_{\Lambda}^2\right)
      |\varphi+\frac{\gamma_\Lambda}{3\lambda_\Lambda}|^{-\delta}
\eea
as the second scaling variable. For small symmetry breaking cubic coupling
$\gamma_{\Lambda}$ one notes $y\sim\gamma_{\Lambda}$. The scaling form of the
equation of state for the non-symmetric model reads
\bea
  \dsp  j & = & \dsp
       |\varphi+\frac{\gamma_\Lambda}{3\lambda_\Lambda}
       |^{\delta}\left\{ f(x) - y \right\}.
\eea
This universal form of the equation of state is relevant for a large class of
microscopic free energies, far beyond the special polynomial form used for
its derivation.

It is often useful to express the universal equation
of state in terms of renormalized fields and masses. We use the variables
\bea
  \dsp \tilde{s}=\frac{\varphi_R}{\varphi_{0R}} & ; &
  \dsp v=\frac{m_R}{m_R^{Z_2}},
\eea
where $m_R=\left(\frac{\partial^2 U}{\partial\varphi_R^2}\left.\right|_{\varphi_{0R}}
\right)^{1/2}$ is the renormalized mass at the minimum
$\varphi_{0R}$ of $U(\varphi_R)$ whereas
$m_R^{Z_2}$ is the renormalized mass at the minimum of the corresponding
$Z_2$-symmetric effective potential obtained for vanishing cubic coupling
$\gamma_{\Lambda}=0$. Then the critical temperature corresponds to $v=1$.
In this parameterization the universal properties of the equation of state
for the Ising type first order transition can be compared with transitions
in other models - e.g. matrix models \cite{BW97-1} - where no simple
mapping to a second order phase transition exists (see section \ref{smm}).

A convenient universal function $G(\tilde{s},v)$ for weak first order
transitions can be defined as
\bea
  \dsp G(\tilde{s},v):=\frac{U_0(\varphi_R)}{\varphi_{0R}^6}.
\eea
We plot $G(\tilde{s},v)$ in fig. \ref{univGkub} as a function of $\tilde{s}$
for different values of $v$. For the present model all information necessary
for a universal description of first order phase transitions is already
contained in eqs. (\ref{shiftU}) or (\ref{shifteqofstat}).
The function $G(\tilde{s},v)$ can serve, however, for a comparison with other
models, for which a simple relation to 
a second order phase transition does not exist. We
discuss the function $G$ for matrix models
in sect. 5.6 (cf. fig. 25). At this place we mention that we
have actually computed the potential $U$ both by solving the flow equations
with initial values where $\gamma_{\Lambda}\neq 0$ and by a shift from the
Ising model results. We found good agreement between the two approaches.


\begin{figure}[h]
\unitlength1.0cm
\begin{center}
\begin{picture}(13.,9.)
\put(6.5,-0.5){$\tilde{s}=\frac{\varphi_R}{\varphi_{0R}}$ }
\put(1.2,7.){$G(\tilde{s},v)$}
\put(7.7,3.8){\footnotesize $v=1$}
\put(9.7,0.6){\footnotesize $v=1.195$}
\put(-0.5,0.){
\rotate[r]{\epsfig{file=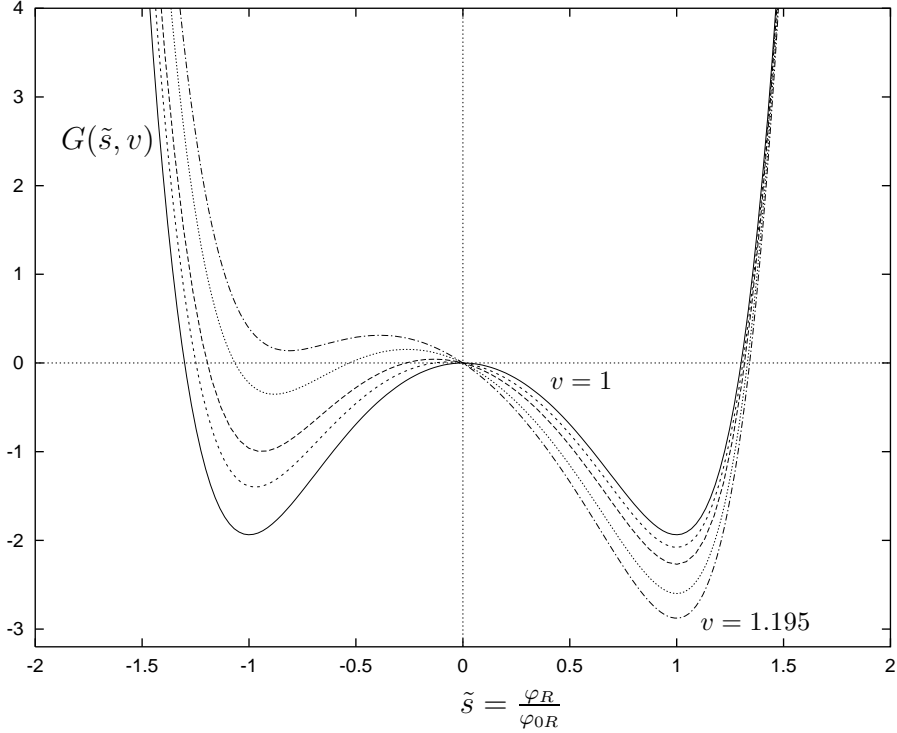,width=9.cm,height=13.cm}}
}
\end{picture}
\end{center}
\caption[]{
\em Universal critical equation of state for a first order
transition. We display 
$G(\tilde{s},v)$ for $v=1$, $v=1.037$, $v=1.072$, $v=1.137$ and 
$v=1.195$.}
\label{univGkub}
\end{figure}


In conclusion, we have employed non-perturbative flow equations in order
to compute explicitly the equation of state. We have first studied models
where the microscopic free energy can be approximated by a polynomial
approximation with terms up to quartic order. This covers second order
as well as first order transitions, both for the universal and non-universal
features. The same method can be used away from the critical hypersurface,
allowing therefore for an explicit connection between critical and
non-critical observations.

The ability of the method to deal also with a microscopic free energy which
is not of a polynomial form is demonstrated by a particular example,
namely the equation of state for carbon dioxide.
In the vicinity of the endpoint of the critical line we can
give an explicit formula for the free energy density $U(n,T)T$.
Using the fits (3.15), ((3.18), for $\hat F(\hat s)$ and
$\tilde z(s)$ one finds\footnote{Note that
$n_*$ is somewhat different form $\hat n$ and therefore
$\varphi_R$ is defined slightly different from eq. (1.7).
This variable shift (similar to (4.1) reflects the fact
that eq. (1.6) contains higher than quartic interactions
and cannot be reduced to a $\varphi^4$-potential even for
$\gamma_\Lambda=0$.}
\beq\label{5.1}
U_0(n,T)=U^{Z_2}(\hat\varphi_R(n,T),\ m_R(T))
+J(T)(n-n_*)-K(T)\eeq
\beq\label{5.2}
U^{Z_2}=\frac{1}{2}\tilde a_0m^2_R\hat\varphi_R^2+\frac{1}{4}
\tilde a_1m_R\hat\varphi^4_R+\frac{1}{6}\tilde a_2\hat\varphi^6_R\eeq
with $\tilde a_i\approx a_i$ and
\bea\label{5.3}
\hat\varphi_R(n,T)&=&\tilde z\left(\frac{\varphi_R(n,T)}
{m^{1/2}_R(T)}\right)\varphi_R(n,T)\nonumber\\
\varphi_R(n,T)&=&H_{\pm}\left|\frac{T-T_*}{T_*}\right|
^{-\eta\nu}(n-n_*)\nonumber\\
m_R(T)&=&\xi_{T\pm}\left|\frac{T-T_*}{T_*}\right|^\nu\eea
The two non-universal functions $J(T)$ and $K(T)$ enter in
the determination of the chemical potential and the critical line.

In particular, the non-universal amplitudes governing the behaviour near the
endpoint of the critical line can be extracted from the equation of state:
In the vicinity of the endpoint we find for $T=T_*$
\bea
 \dsp \rho_{>}-\rho_* = \rho_*-\rho_{<} & = &
 \dsp D_p^{-1}\left(\frac{| p-p_*|}{p_*} \right) ^{1/\delta}
\eea
with $D_p=2.8\: g^{-1} cm^3$,
where $\rho_{>}>\rho_{*}$ and $\rho_{<}<\rho_*$
refer to the density in the high and low
density region respectively. At the critical temperature $T_c < T_{*}$ and
pressure $p_c<p_*$ for a first order transition one finds for the
discontinuity in the density between the liquid $(\rho_l)$ and gas
$(\rho_g)$ phase
\bea
 \dsp \Delta \rho = \rho_l - \rho_g & = &
 \dsp B_p\left(\frac{p_*-p_c}{p_*}\right)^{\beta} =
     B_T\left(\frac{T_*-T_c}{T_*}\right)^{\beta}
\eea
with $B_p=0.85 \: g cm^{-3}$, $B_T=1.5 \: g cm^{-3}$.
This relation also defines the slope of the critical line near the endpoint.

There is no apparent limitation for the use of the flow equation for an
arbitrary microscopic free energy. This includes the case where $U_{\Lambda}$
has several distinct minima and, in particular, the interesting case of a
tricritical point. At present, the main inaccuracy arises from a
simplification of the $q^2$-dependence of the four point function which
reflects itself in an error in the anomalous dimension $\eta$.
The simplification of the momentum dependence of the effective propagator
in the flow equation plays presumably only a secondary role.
In summary, the non-perturbative flow equation appears to be a very
efficient tool for the establishment of an explicit quantitative
connection between the microphysical interactions and the
long-range properties of the free energy.

\subsection{Critical behavior of polymer chains}
\label{Polymers}

The large scale properties of isolated polymer
chains can be computed from the critical behaviour of the
$O(N)$-symmetric scalar theory using a variant of the so-called 
replica limit: the statistics of polymer
chains can be described by the $N$-component field theory in the
limit $N\to 0$ \cite{Gennes}.

In polymer theory critical behavior occurs when the size of
an isolated swollen polymer becomes infinite. 
The size of a chain can be defined by its mean square distance 
$\langle (\vec r(\lambda)-\vec r(0))^2 \rangle$.
Here $\lambda$ denotes the distance along the chain between the
end points with position $\vec r(\lambda)$ and $\vec r(0)$.
The scaling behaviour of the mean square distance is  
characterized by the exponent $\nu$, 
\beq
\langle (\vec r(\lambda)-\vec r(0))^2\rangle\sim \lambda^{2\nu} \, .
\eeq
The exponent $\nu$ for the polymer system corresponds to the 
correlation length exponent of the $N$-component field theory
in the limit $N\to 0$. There are two independent
critical exponents for the $O(N)$ model near its second order
phase transition. In polymer theory, the exponent $\gamma$ 
characterizes the asymptotic
behaviour of the number of configurations $N(\lambda)$ of the self 
avoiding chains
\beq
N(\lambda) \sim \lambda^{\gamma-1} \, .
\eeq
For a chain with independent links one has $\gamma=1$. The exponent
$\gamma$ corresponds to the critical exponent that describes the
behaviour of the magnetic susceptibility in the zero component field 
theory. 

We compute the critical exponents $\nu$ and $\gamma$, as well as
the anomalous dimension $\eta$, for the $O(N)$ model in the limit 
$N \to 0$ using the lowest order of the derivative expansion
of the effective average action (\ref{three}). The flow equations 
for the scale dependent effective potential $U_k$ and wave function
renormalization $Z_k$ (or equivalently the anomalous dimension $\eta$) 
are derived for integer $N$. The results are given in eqs.\
(\ref{five}) and (\ref{ten}). We analytically
continue the evolution equations to non-integer values of $N$.
We explicitly verify that the limit $N \to 0$ is continuously
connected to the results for integer $N$ by computing the exponents 
for numbers of components $N$ between
one and zero. The case $N=1$ is also useful in polymer theory. It 
belongs to the universality class that describes the point at the
top of the coexistence curve of a polymer 
solution \cite{Gennes}.
The first table \ref{pol1}
below shows the results for the critical exponents
$\nu,\gamma$ and $\eta$ 
for several (non-integer) values of $N$ \cite{BGW}.
\hspace{3cm}
\begin{table}[h]
\begin{center}
\begin{tabular}{|l|l|l|l|}
\hline
$N$& $\nu$ & $\gamma$ & $\eta\; [10^{-2}]$\\
\hline
0   & 0.589 & 1.155 & 4.06\\
0.1 & 0.594 & 1.165 & 4.12\\
0.2 & 0.600 & 1.175 & 4.18\\
0.3 & 0.605 & 1.185 & 4.22\\
0.4 & 0.610 & 1.195 & 4.26\\
0.5 & 0.616 & 1.205 & 4.30\\
0.6 & 0.621 & 1.216 & 4.32\\
0.7 & 0.626 & 1.226 & 4.34\\
0.8 & 0.632 & 1.237 & 4.36\\
0.9 & 0.637 & 1.247 & 4.37\\
1.0 & 0.643 & 1.258 & 4.37\\
\hline
\end{tabular}
\caption{\em Critical exponents for $0\leq N\leq 1$ 
\label{pol1}}
\end{center}
\end{table}

In the following table \ref{pol2}
we compare our results for $N=0$ with the 
epsilon expansion
\cite{GZ}, perturbation series at fixed dimension \cite{GZ}, 
lattice Monte Carlo \cite{LMC0}, high temperature series \cite{BC97-1} and
experiment \cite{Polym}.
The comparison shows 
a rather good agreement of these results.
\hspace{3cm}
\begin{table}[h]
\begin{center}
\begin{tabular}{|l|l|l|l|}
\hline
$N=0$ & $\nu$ & $\gamma$ & $\eta$\\
\hline
Average action&0.589&1.155&0.0406\\
$\epsilon$-expansion&0.5875(25)&1.1575(60) &0.0300(50)\\
$d=3$ expansion &0.5882(11) &1.1596(20)&0.0284(25)\\
lattice MC &0.5877(6)& & \\
HT series &0.5878(6)&1.1594(8)& \\
Experiment&0.586(4)& & \\
\hline
\end{tabular}
\caption{\em Critical exponents for polymer chains: Comparison
between average action, epsilon expansion
\cite{GZ}, perturbation series at fixed dimension \cite{GZ}, 
lattice Monte Carlo \cite{LMC0}, high temperature series \cite{BC97-1} and
experiment \cite{Polym}. \label{pol2}}
\end{center}
\end{table}

\subsection{Two dimensional models and the Kosterlitz-Thouless transition} 
\label{KT}

We investigate the $O(N)$-symmetric linear $\sigma$-model in two
dimensions. Apart from their physical relevance, two dimensional
systems provide a good testing ground for non-perturbative methods.
Let us consider first the $O(N)$ model in the limit $N \to 0$ motivated
in the previous section. In this 
limit the two dimensional model exhibits a second order phase transition.
The critical exponent $\nu$ describes the
critical swelling of long polymer chains \cite{Gennes}. The value of this 
exponent is known {exactly}, $\nu_{\rm exact}=0.75$ \cite{Zin93-1}.
To compare with the exact result, we study the $N=0$ model using
the lowest order derivative expansion (\ref{three}) of the effective
average action. We obtain the result $\nu=0.782$  \cite{BGW}
which already compares 
rather well to the exact result. It points out that the present 
techniques allow us to give a unified description of the $O(N)$ model 
in two and in three dimensions, as well as four dimensions which will 
be discussed in section \ref{Fermions}. 
In the following, we extend the discussion in two dimensions and show that for 
$N=2$ one obtains a good picture of the Kosterlitz-Thouless phase 
transition\footnote{For investigations in two dimensions using similar methods 
see also refs.\ \cite{Mor2d,KNP}.}.

To evaluate the equation for the potential we make a further approximation
and expand around the minimum of $u_k$ for non-zero field squared
$\tilde\rho=\kappa$ up
to the quadratic order in $\tilde\rho$:
\beq
\label{potentw}
u_k(\tilde\rho)= u_k(\kappa)+\frac{1}{2}\lambda (\tilde\rho-\kappa)^2.
\eeq
The condition $\partial u/\partial\tilde\rho\vert_{\tilde\rho=\kappa}=0$
holds independent of $t$ and gives us the evolution equation
for the location of the minimum of the potential parametrized by $\kappa$.
A similar evolution equation can be derived \cite{TW94-1} for the
symmetric regime where $\kappa=0$ and an appropriate variable is
$\partial u /\partial \tilde\rho\vert_{\tilde\rho=0}$. The flow equations
for $\kappa$ and $\lambda$ read
\bea
\label{betakappa}
\beta_\kappa &\equiv& \frac{d\kappa}{dt}  = -(d-2+\eta)\kappa +
2 v_d(N-1) l^d_1(0)
+6 v_d  l^d_1 (2\lambda\kappa)
\nonumber\\
\label{betalambda}
\beta_\lambda &\equiv& \frac{d\lambda}{dt} = (d-4+2\eta)\lambda
+2v_d(N-1)\lambda^{2} l^d_2(0)
+18 v_{d}\lambda^{2} l^d_2(2\lambda\kappa).
\eea
where the ``threshold functions'' are defined in \ref{SecThresh}.
The theory is in the
symmetric phase if $\kappa(0)=0$ -- this happens if $\kappa$ reaches zero
for some nonvanishing $k_s>0$. On the other hand, the phase with
spontaneous symmetry breaking corresponds to $\rho_0(0)>0$ where
$\rho_0(k)=k^{d-2}Z_k^{-1}\kappa(k)$. A second order phase transition is
characterized by a scaling solution corresponding to fixed points
for $\kappa$ and $\lambda$. For small deviations from the fixed point there
is
typically one
infrared unstable direction which is related to the relevant mass
parameter.
The phase transition can be studied as a function of $\kappa(\Lambda)$
with a critical value $\kappa(\Lambda)=\kappa_c$. The difference
$\kappa(\Lambda)-\kappa_c$ can be assumed to be proportional to $T_c-T$,
with $T_c$ the critical temperature. This allows to define and compute
critical exponents in a standard way. We should mention a particular
possibility for $d=2$, namely that $\kappa(0)$ remains strictly
positive whereas $\rho_0(0)$ vanishes due to $\lim_{k\to 0}Z_k\to\infty$.
This is a somewhat special form of spontaneous symmetry breaking, where
the renormalized expectation value, which determines the renormalized
mass, is different from zero whereas the expectation value of the
unrenormalized field vanishes. We will see that this scenario is indeed
realized for $d=2, N=2$. The phase with this special form of spontaneous
symmetry breaking exhibits a massive radial and a massless Goldstone
boson -- and remains nevertheless consistent with the Mermin-Wagner theorem
\cite{MW66-1} that the expectation value of the (unrenormalized) field
$\phi_a$ must vanish for $N\ge2$. The Kosterlitz-Thouless phase transition
\cite{KT73-1} describes the transition from this phase to the standard
symmetric phase of the linear $\sigma$-model, i.e. the phase where
$\kappa(0)=0$ with a spectrum of two degenerate massive modes.

In order to solve the flow equation (\ref{betalambda}) we further need
the anomalous dimension $\eta$ which is given in our truncation
by eq.\ (\ref{ten}).
We specialize to the two dimensional linear $\sigma$-model ($d=2$).
We observe that the flow equations can be solved analytically in the
limiting case
of a large mass $\omega=2\lambda\kappa$ of the radial mode.
The threshold functions vanish with powers of $\omega^{-1}$ and for
$N>1$ the leading contributions to the $\beta$-functions are those
from the Goldstone modes. Therefore this limit is called the Goldstone
regime. In this approximation the $\beta$-functions can be expanded
in powers of $\omega^{-1}$. In particular, the leading order of the
anomalous dimension can be extracted immediately from (\ref{three}):
\beq
\label{etaentw}
\eta = \frac{1}{4\pi\kappa}+{\O} (\kappa^{-2}).
\eeq
Inserting this result in (\ref{betakappa}) we have
\beq
\label{kappaent}
\beta_\kappa= \frac{(N-2)}{4\pi} + {\O}(\kappa^{-1})
\eeq
and the leading order of $\beta_\lambda$ is
\beq
\label{entbetala}
\beta_\lambda = -2\lambda+\frac{(N-1)\ln 2}{2\pi}\lambda^2  + {\O}(\kappa^{-1}).
\eeq
Eq. (\ref{entbetala}) has a fixed point solution
$\lambda_*=\frac{4\pi}{(N-1)\ln 2}
\approx 18.13/(N-1)$.

For $N>2$ there exists a simple relation between the linear and
the nonlinear $\sigma$-model:
The effective coupling between the Goldstone bosons of the nonabelian
nonlinear $\sigma$-model can be extracted directly from (\ref{three})
and reads in an appropriate normalization \cite{LWW}
\beq
g^2=\frac{1}{2\kappa}.
\eeq
The lowest order contribution to $\beta_\kappa$ (\ref{kappaent})
coincides with the one loop expression for the running of $g^2$ as
computed in the nonlinear $\sigma$-model. We emphasize in this
context the importance of the anomalous dimension $\eta$ which changes
the factor $(N-1)$ appearing in (\ref{betalambda}) into the appropriate
factor $(N-2)$ in (\ref{kappaent}). In correspondence with the
universality of the two loop $\beta$-function for $g^2$ in the nonlinear
$\sigma$-model we expect the next to leading term $\sim\kappa^{-1}$
in $\beta_\kappa$ (\ref{kappaent}) to be also proportional to
$(N-2)$. In order to verify this one has to go beyond the truncation
(\ref{three}) and systematically keep all terms contributing in the
appropriate order of $\kappa^{-1}$. (This calculation is similar
to the extraction of the two loop $\beta$-function of the linear
$\sigma$-model in four dimensions by means of an ``improved one
loop calculation'' using the flow equation (\ref{2.17}) \cite{pw}.)
We have calculated the expansion of $\beta_\kappa$ up to
the order ${\O}(\kappa^{-1})$ for the most general two derivative
action, i.e. neglecting only the momentum
dependence of $Z_k$ and $Y_k$. The result agrees with the
two loop term of the nonlinear $\sigma$-model within a few
per cent, and the discrepancy should be attributed to the neglected
momentum dependence of the wave function renormalization. The issue
of the contribution to $\beta_\kappa$ in order $\kappa^{-2}$ is less
clear:
Of course, the direct contribution of the Goldstone bosons
(combined with their contribution to $\eta$) should always vanish
for $N=2$ since no nonabelian coupling exists in this case. The
radial mode however, could generate a contribution which is not
proportional to $(N-2)$. This contribution is possibly nonanalytic
in $\kappa^{-1}$ and would correspond to a non-perturbative
contribution in the language of the nonlinear $\sigma$-model.
\begin{figure}[t]
\leavevmode
\centering
\epsfxsize=11cm
\epsffile{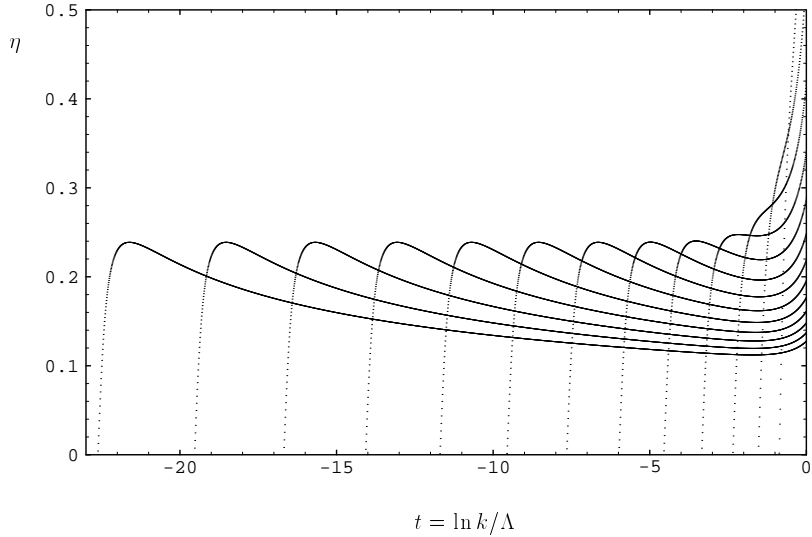}
\caption{\em
\label{n2etagraf} The scale dependent anomalous dimension $\eta(t)$
for a series of initial potentials $U_{\Lambda}$ approaching the
phase transition.}
\end{figure}

Let us now turn to the two dimensional abelian model ($d=2, N=2$)
for which we want to describe the Kosterlitz-Thouless phase transition.
In the limit of vanishing $\beta_\kappa$ for large enough $\kappa$ the
location of the minimum of $u_k(\tilde\rho)$ (\ref{potentw}) is
independent of the scale $k$. Therefore the
parameter $\kappa$, or, alternatively, the temperature difference
$T_c-T$, can be viewed as a free parameter.
If we go beyond the lowest order estimate (\ref{entbetala})
the fixed point for $\lambda$ remains, but $\lambda_*$ becomes
dependent on $\kappa$. This implies that the system has
a line of fixed points which is parametrized by $\kappa$ as
suggested by results obtained from calculations with the
nonlinear $\sigma$-model \cite{webe}.
In particular, the anomalous dimension $\eta$ depends on the
temperature $T_c-T$ (\ref{etaentw}). Even if this picture is not
fully accurate for nonvanishing $\beta_\kappa$, it is a very
good approximation for large $\kappa$: The possible running
of $\kappa$ is extremely slow, especially if $\beta_\kappa$
vanishes in order $\kappa^{-1}$.
We associate the low temperature or large $\kappa$ phase with
the phase of vortex condensation in the nonlinear $\sigma$-model.
The correlation length is always infinite due to the Goldstone boson.
Since $\eta>0$ we expect the inverse propagator of this Goldstone
degree of freedom $\sim (q^2)^{1-\eta/2}$, thus avoiding Coleman's
no go theorem \cite{Col73-1} for free massless particles in two dimensions.
On the other hand, for small values of $\lambda\kappa$ the threshold
functions can be expanded in
powers of $\lambda\kappa$. The anomalous dimension is small and
$\kappa$ is driven to zero for $k_s>0$. This corresponds to the symmetric
phase of the linear $\sigma$-model with a massive complex scalar field.
We associate this high temperature phase with the phase of vortex disorder
in the picture of the nonlinear $\sigma$-model.
The transition between the behaviour for large and small $\kappa$ is
described by the Kosterlitz-Thouless transition. In the language of the
linear $\sigma$-model it is the transition from a special type of
spontaneous symmetry breaking to symmetry restoration.
\begin{figure}[t]
\leavevmode
\centering
\epsfxsize=11cm
\epsffile{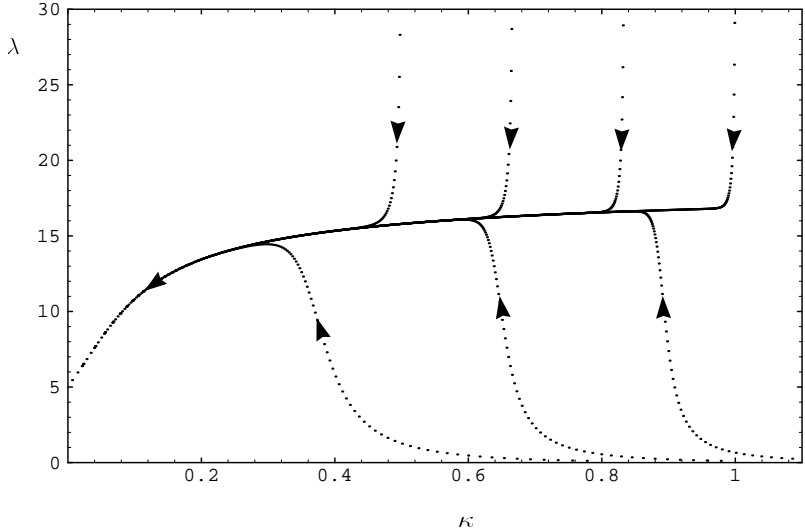}
\caption{\em \label{n2kalagraf}
Flow of the quartic coupling $\lambda$ and the minimum $\kappa$
for different $t$. One observes the (pseudo) critical line 
$\lambda(\kappa)$ as an (approximate) line of fixed points.}
\end{figure}

Finally we give a summary of the results obtained from the
numerical integration of the evolution equations (\ref{betalambda})
and (\ref{ten}) for the special
case $d=2, N=2$ \footnote{This work has also been done for $d=2, N=1$.
There we find a fixed point which corresponds to the second
order phase transition in the Ising model.}.
We use a Runge-Kutta method starting at $t=0$ with arbitrary initial
values for $\kappa$ and $\lambda$ and solve the flow equations for
large negative values of $t$.
Results are shown in figs.1-4 where we plot typical trajectories.
The distance between points corresponds to equal steps in $t$ such
that very dense points or lines indicate the very slow running in the
vicinity
of fixed points.

The understanding of the trajectories needs a
few comments:
The work of Kosterlitz and Thouless \cite{KT73-1} suggests that the
correlation length is divergent for all temperatures below
a critical temperature $T_c$ and that the critical exponent $\eta$ depends
on temperature. The consequence for our model
is that above a critical value for $\kappa$ all $\beta$-
functions should vanish for a line of fixed points parametrized by $\kappa$.
{}From the results in the Goldstone
regime and from earlier calculations \cite{webe} we conclude that
$\beta_\kappa$ should vanish faster than $\kappa^{-1}$ for large $\kappa$.
Our truncation (\ref{three}), however,
yields a function $\beta_\kappa$ which vanishes only like $\kappa^{-1}$.
The consequence is that even if the system reaches the supposed line of
fixed points the parameter $\kappa$ decreases very slowly until the
transition to the symmetric regime is reached.
The anomalous dimension first grows with decreasing $\kappa$
(\ref{etaentw})
until the critical value is reached. Then the
system runs into the symmetric regime and $\eta$ vanishes. So
we expect that $\eta$ reaches a maximum  near the phase transition.
We use this as a criterion for the critical value $\kappa_c$.
In summary, the truncation (\ref{three}) smoothens the phase
transition and this prevents a very accurate determination of the critical
value $\kappa_c$ and the corresponding anomalous dimension $\eta_c$.
{}From the numerical point of view the absence of a true phase transition
in the
truncation (\ref{three}) makes life easier: One particular trajectory can
show both the features of the low and the high temperature phase since
it crosses from one to the other.
\begin{figure}[t]
\leavevmode
\centering
\epsfxsize=11cm
\epsffile{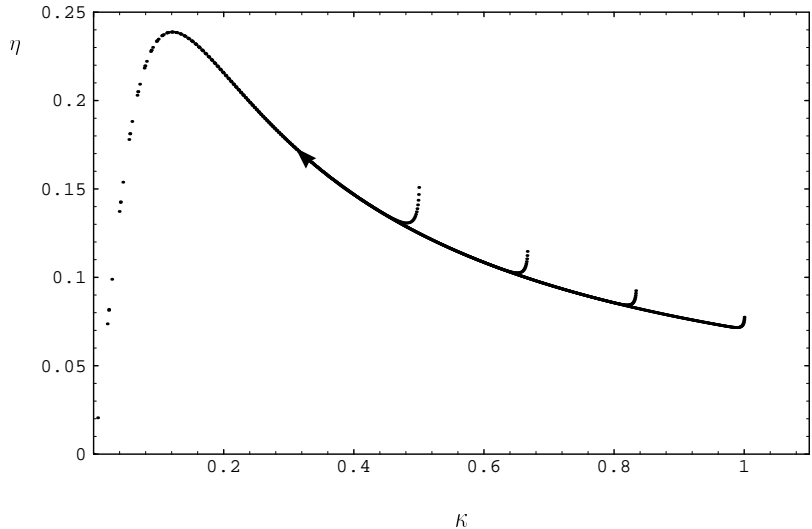}
\caption[]{\em \label{n2etakagraf} Flow of the anomalous dimension
$\eta$ and the minimum $\kappa$. One observes (approximate)
scaling solutions with the universal function $\eta(\kappa)$.}
\end{figure}

The numerical results fullfil our expectations. Fig.\ref{n2etagraf}
shows the evolution of the anomalous dimension with decreasing $t$ for
several different initial values $\kappa(\Lambda), \lambda(\Lambda)$.
The maximum is reached with $\eta_c=0.24$ which has to be compared
with the result of Kosterlitz and Thouless $\eta_c=0.25$ \cite{KT73-1}.
The approximate ``line of fixed points'' for $\kappa>\kappa_c$
$(\eta<\eta_c)$
is demonstrated by the self-similarity of the curves for large $-t$.
Trajectories with different initial conditions hit the line of fixed points
at
different $\kappa$. Subsequently they follow the line of fixed points. Except
for the value of $\kappa(k)$ all ``memory'' of the initial conditions is
lost
for $t\le-3$. Another manifestation of the line of fixed points in the
$(\kappa,\lambda)$ plane is demonstrated in fig.\ \ref{n2kalagraf}. After
some
fast ``initial running'' (dotted parts of the trajectories) all
trajectories
with large enough $\kappa(\Lambda)$ follow this line independent of the
initial
$\lambda(\Lambda)$. We emphasize that the nonlinear $\sigma$-model
corresponds to $\lambda(\Lambda)\to\infty$. Our investigation shows that
the
linear $\sigma$-model is in the same universality class, even for very
small
$\lambda(\Lambda)$. In fig.\ref{n2etakagraf} we plot $\eta(\kappa)$. Along
the line of fixed points we find perfect agreement with the analytical
estimate
(\ref{etaentw}) for large $\kappa$. For $\kappa=0.3$ the deviation from
the
lowest order result is 34\% in the present truncation. Finally we show in
fig.\ref{n2belagraf} the value of $\beta_\lambda$ for different
trajectories.
The dense parts of the ``ingoing curves'' show the fixed point behaviour at
$\lambda_*(\kappa_1)$ where $\kappa_1$ denotes the value of $\kappa$ where
the line of fixed points is hit. (For small $\kappa(\Lambda)$ there is a
substantial difference between $\kappa_1$and $\kappa(\Lambda)$ which
depends
also on $\lambda(\Lambda)$. This can be seen from the curves with
$\kappa(\Lambda)=1$.) After hitting the line of fixed points the trajectories
stay for a large $t$-interval at $\beta_\lambda$ very close to zero.
Subsequently, the ``outgoing curve'' indicates the transition to the
symmetric phase.
\begin{figure}[t]
\leavevmode
\centering
\epsfxsize=11cm
\epsffile{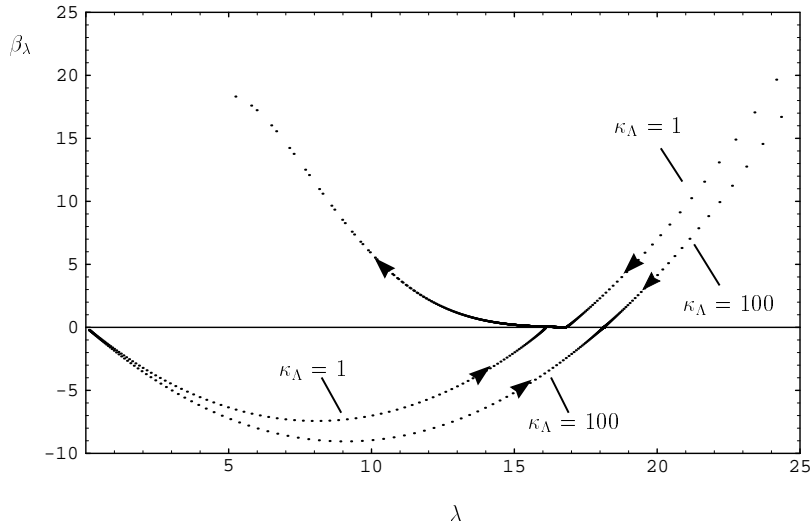}
\caption[The $\beta$-function for $\lambda$]
{\em \label{n2belagraf} The beta function for the quartic
coupling $\beta_\lambda$ for different initial $\kappa_{\Lambda}$
(see text).}
\end{figure}

In conclusion, both the analytical and the numerical investigations
demonstrate
all important characteristics of the Kosterlitz-Thouless phase transition
for the linear $\sigma$-model. This belongs to the same universality class
as the nonlinear $\sigma$-model and we have demonstrated a close
correspondence
between the linear and the nonlinear $\sigma$-model with abelian symmetry.
In
particular, the phase with vortex disorder in the nonlinear $\sigma$-model
corresponds simply to the symmetric phase of the linear $\sigma$-model. We
emphasize that we have never needed the explicit investigation of vortex
configurations. The exact non-perturbative flow equation includes
automatically
all configurations. Its ability to cope with the infrared problems of
perturbation theory is confirmed by the present work.

Despite the simple and clear qualitative picture arising from the
truncation
(\ref{three}) this letter only constitutes a first step for a
quantitative
investigation. It is not excluded that the coincidence of our critical
$\eta_c\approx 0.24$ with $1/4$ is somewhat accidental. In order to answer
this question one needs to go beyond the truncation (\ref{three}). In
view
of the relatively large value of $\eta_c$ we expect in particular that the
momentum dependence of the wave function renormalization $Z_k$ (or the
deviation of the inverse propagator from $q^2$) could play an important
role
at the phase transition. This effect should be included in a more detailed
quantitative investigation.


\section{Scalar matrix models}
\label{smm}

\subsection{Introduction \label{intro}}

Matrix models are extensively discussed in statistical
physics. Beyond the $O(N)$-symmetric
Heisenberg models (``vector models''), which we have discussed
in the previous sections, they correspond
to the simplest scalar field theories. 
There is a wide set of different applications as 
the metal insulator transition \cite{Weg}
or liquid crystals \cite{deG} or strings and
random surfaces \cite{FGZ}. 
The universal behavior of these models in the vicinity
of a second order or weak first order phase transition
is determined by the symmetries and the field content
of the corresponding field theories. We will consider
here \cite{BW97-1} models with $U(N) \times U(N)$ symmetry with
a scalar field in the $(\bar{N},N)$ representation,
described by an arbitrary complex $N \times N$ matrix
$\varphi$.\footnote{The methods presented here
have recently been applied also to the  
principle chiral model with $SO(3)\times O(3)$ symmetry \cite{PCM}
and to Heisenberg frustrated magnets with $O(N)\times O(2)$ symmetry
\cite{TDM}.} 
We do not impose nonlinear constraints for
$\varphi$ a priori  but rather use a ``classical''
potential.  This enforces nonlinear constraints
in certain limiting cases. Among those, our model
describes a nonlinear matrix model for unitary matrices
or one for singular $2 \times 2$ matrices. The universal
critical behavior does not depend on the details of the 
classical potential and there is no difference
between the linear and nonlinear models in the 
vicinity of  the limiting cases. We concentrate here
on three dimensions, relevant for statistical
physics and critical phenomena in high temperature field 
theory.

The cases $N=2$, $3$ have a relation to high
temperature strong interaction physics. At vanishing
temperature the four dimensional models can be used
for a description of the pseudoscalar and scalar mesons
for $N$ quark flavors. 
For $N=3$ the antihermitean
part of $\varphi$ describes here the pions, kaons, $\eta$
and $\eta^{\prime}$ whereas the hermitean part accounts for the 
nonet of scalar $0^{++}$ mesons.\footnote{See ref.\ \cite{JW96-1,JW97-1}
for a phenomenological analysis.} For nonzero
temperature $T$ the effects of fluctuations with momenta
$p^2 \lesssim (\pi T)^2$ are described by the 
corresponding three dimensional models. These models
account for the long distance physics and are obtained by
integrating out the short distance fluctuations.
In particular, the three dimensional models embody
the essential dynamics in the immediate vicinity
of a second order or weak first order chiral phase transition
\cite{PW84-1,Wilczek92,RaWi93-1,Raj95-1}. 
The four dimensional models at nonvanishing
temperature have also been used for investigations of
the temperature dependence of meson masses \cite{mes2,mes}.
The simple model investigated in this section is not yet 
realistic for QCD -- it neglects the effect of the axial anomaly
which reduces the chiral flavor symmetry to 
$SU(N) \times SU(N) \times U(1)$. In the present section we also neglect
the fluctuations of fermions (quarks). They play no
role for the universal aspects near the phase transition.
They are needed, however, for a realistic connection with
QCD and will be included in sect. 8.
For simplicity we will concentrate here on
$N=2$, but our methods can be generalized to $N=3$
and the inclusion of the axial anomaly.

The case $N=2$ also has a relation to the 
electroweak phase transition in models with two
Higgs doublets. Our model corresponds here
to the critical behavior in a special class
of left-right symmetric theories in the limit where 
the gauge couplings are neglected. Even though
vanishing gauge couplings are not a good 
approximation for typical realistic models
one would like to understand this limiting
case reliably.

For the present matrix models
one wants to know if the phase transition becomes second
order in certain regions of parameter space. In the context of 
flow equations this is equivalent 
to the question if the system of running couplings admits
a fixed point which is infrared stable (except one
relevant direction corresponding to $T-T_{c}$). 
We find that the phase transition for the investigated
matrix models with $N=2$ and symmetry breaking
pattern $U(2) \times U(2) \to U(2)$ is always 
(fluctuation induced) first order,
except for a boundary case with enhanced 
$O(8)$ symmetry. For a large part
of parameter space the transition is weak and one finds 
large renormalized dimensionless couplings near the critical
temperature. If the running of the couplings towards approximate
fixed points (there are no exact fixed points) is 
sufficiently fast the large distance physics looses 
memory of the details of the short distance or classical
action. In this case the physics near the phase transition
is described by an universal equation of state.
This new universal critical equation of state for first order
transitions involves two (instead of one for second order
transitions) scaling variables.

In section \ref{model}
we define the $U(2) \times U(2)$ symmetric matrix model
and we establish the connection to a matrix model
for unitary matrices and to one for
singular complex $2 \times 2$ matrices. There we also give
an interpretation of the model as the coupled system
of two $SU(2)$-doublets for the weak interaction Higgs sector.
The evolution equation
for  the average potential
$U_k$ and its scaling form is computed in section 
\ref{scale}. A detailed account on the renormalization group flow
is presented in section \ref{rg}. 
Section \ref{ps} is devoted to an overview over the phase 
structure and the coarse-grained effective potential $U_k$
for the three dimensional theory. We compute the 
universal form of the equation of state 
for weak first order phase transitions 
in section \ref{sce} and we extract critical exponents
and the corresponding index relations.

\subsection{Scalar matrix model with $U(2)\times U(2)$ symmetry
\label{model}}

We consider a $U(2) \times U(2)$ symmetric effective 
action for a scalar field
$\varphi$ which transforms in the $(2,2)$ representation
with respect to the subgroup $SU(2) \times SU(2)$. Here $\varphi$
is represented by a complex $2 \times 2$ matrix
and the transformations are
\bea
 \varphi &\rightarrow& U^{ }\varphi V^\dagger\,\, , \nnn
 \varphi^\dagger &\rightarrow& V^{ }\varphi^\dagger U^\dagger
 \label{Transformations}
\eea
where $U$ and $V$ are unitary $2 \times 2$ matrices 
corresponding to the two distinct $U(2)$ factors.

We classify the invariants for the construction of the effective 
average action by the number of derivatives. The lowest order
is given by
\beq
 \Gamma_k = \dsp{\int d^d x\left\{U_k(\varphi,\varphi^\dagger )+
 Z_k \partial_\mu \varphi^*_{ab} \partial^\mu \varphi^{ab}
  \right\}}\qquad  (a,b=1,2).
 \label{Ansatz}
\eeq
The term with no derivatives defines the scalar potential $U_k$
which is an arbitrary function of traces of powers of 
$\varphi^\dagger \varphi$. The most general $U(2) \times U(2)$ symmetric
scalar potential can be expressed as a function of only two 
independent invariants,
\bea
 \rho &=& \dsp{\tr\left(\varphi^\dagger \varphi \right) }\nonumber\\
 \tau &=& \dsp{2 \tr\left(\varphi^\dagger \varphi - \frac{1}{2} \rho \right)^2}
 = \dsp{2 \tr\left(\varphi^\dagger \varphi \right)^2 - \rho^2 }.
 \label{Invaria}
\eea
Here we have used for later convenience the traceless matrix 
$\varphi^\dagger \varphi - \frac{1}{2} \rho$ to construct the second
invariant.
Higher invariants, $\tr\left(\varphi^\dagger \varphi 
- \frac{1}{2} \rho \right)^n$ for
$n > 2$, can be expressed as functions of $\rho$ and $\tau$ 
\cite{Ju95-7}.

For the derivative part we consider a standard kinetic term with a 
scale dependent wave function renormalization constant $Z_k$. 
The first correction to the kinetic term would include field 
dependent wave function
renormalizations $Z_k(\rho,\tau)$ plus functions not
specified in eq.\ (\ref{Ansatz}) which account for a
different index structure of invariants with two
derivatives. The wave function renormalizations may be
defined at zero momentum or for $q^2=k^2$ in the hybrid derivative
expansion. 
 The next level involves 
invariants with four derivatives and so on.
We define here $Z_k$ 
at the minimum $\rho_0$, $\tau_0$ of $U_k$ and
at vanishing momenta $q^2$,
\beq
Z_k=Z_k(\rho=\rho_0,\tau=\tau_0;q^2=0). \label{zet}
\eeq
The factor $Z_k$ appearing
in the definition of the infrared cutoff $R_k$ in eq.\ (\ref{Rk(q)})
is identified with (\ref{zet}).
The $k$ dependence of this function is given by the anomalous
dimension
\beq
\eta(k)=-\frac{\mbox{d}}{\mbox{d}t}\ln Z_k .
\label{Eta} 
\eeq 

If the ansatz (\ref{Ansatz}) is inserted into the flow equation
for the effective average action (\ref{2.17}) one obtains flow equations  
for the effective average potential $U_k(\rho,\tau)$ and for the
wave function renormalization constant $Z_k$ (or equivalently the 
anomalous dimension $\eta$).
This is done in section
\ref{scale}. These flow equations 
have to be integrated starting from some short distance scale 
$\Lambda$ and one has to specify $U_{\Lambda}$ and $Z_{\Lambda}$
as initial conditions. 
The short distance potential is taken to be a quartic potential
which is parametrized by two quartic couplings $\bar{\la}_{1\Lambda}$,
$\bar{\la}_{2\Lambda}$ and a mass term. We start
in the spontaneously broken regime where the minimum
of the potential occurs at a nonvanishing field value 
and there is a negative mass term at
the origin of the potential $(\bar{\mu}_{\Lambda}^2 > 0)$, 
\beq
U_{\Lambda}(\rho,\tau)=-\bar{\mu}_{\Lambda}^2 \rho + \hal \bar{\la}_{1\Lambda}
\rho^2 +\frac{1}{4} \bar{\la}_{2\Lambda} \tau \quad
\label{uinitial}
\eeq 
and $Z_{\Lambda}=1$. 
The potential is bounded from
below provided 
$\bar{\la}_{1\Lambda} > 0$ and 
$\bar{\la}_{2\Lambda} > - 2 \bar{\la}_{1\Lambda}$.
For $\bar{\la}_{2\Lambda} > 0$ one observes 
the potential minimum for the configuration
$\varphi_{ab}=\varphi \delta_{ab}$ corresponding
to the spontaneous symmetry breaking 
down to the diagonal $U(2)$ subgroup of $U(2) \times U(2)$.
For negative $\bar{\la}_{2\Lambda}$ the potential is minimized
by the configuration $\varphi_{ab}=\varphi \delta_{a1} \delta_{ab}$
which corresponds to the symmetry breaking pattern 
$U(2) \times U(2) \longrightarrow U(1) \times U(1) \times U(1)$. 
In the special case $\bar{\la}_{2\Lambda}=0$ the theory
exhibits an enhanced $O(8)$ symmetry. This constitutes
the boundary between two phases with different symmetry
breaking patterns.

The limits of infinite couplings correspond to nonlinear
constraints in the matrix model. For 
$\bar{\la}_{1\Lambda} \to \infty$ with fixed ratio 
$\bar{\mu}_{\Lambda}^2/\bar{\la}_{1\Lambda}$ one finds the constraint
$\tr(\varphi^{\dagger}\varphi)=\bar{\mu}_{\Lambda}^2/\bar{\la}_{1\Lambda}$.
By a convenient choice of $Z_{\Lambda}$ (rescaling of $\varphi$)
this can be brought to the form $\tr(\varphi^{\dagger}\varphi)=2$.
On the other hand, the limit $\bar{\la}_{2\Lambda} \to +\infty$
enforces the constraint 
$\varphi^{\dagger}\varphi=\hal \tr(\varphi^{\dagger}\varphi)$.
Combining the limits $\bar{\la}_{1\Lambda} \to \infty$,
$\bar{\la}_{2\Lambda} \to \infty$ the constraint reads
$\varphi^{\dagger}\varphi=1$ and we deal with a matrix model for 
unitary matrices. (These considerations generalize to arbitrary
$N$.) Another interesting limit is obtained for 
$\bar{\la}_{1\Lambda}=-\hal \bar{\la}_{2\Lambda} + \Delta_{\la}$,
$\Delta_{\la} > 0$ if $\bar{\la}_{2\Lambda} \to - \infty$. In this 
case the nonlinear constraint reads
$(\tr\varphi^{\dagger}\varphi)^2=\tr(\varphi^{\dagger}\varphi)^2$ which
implies for $N=2$ that $\det \varphi =0$. This is a matrix model
for singular complex $2 \times 2$ matrices.

One can also interpret our model as the coupled 
system of two
$SU(2)$-doublets for the weak interaction Higgs
sector. This is simply done by decomposing the 
matrix $\varphi_{ab}$ into two two-component complex 
fundamental representations of one of the $SU(2)$
subgroups, $\varphi_{ab} \to \varphi_{1b},\varphi_{2b}$. The
present model corresponds to a particular
left-right symmetric model with interactions specified
by
\bea
\rho&=&\varphi_1^{\dagger}\varphi_1+\varphi_2^{\dagger}\varphi_2\\
\tau&=&\left(\varphi_1^{\dagger}\varphi_1-\varphi_2^{\dagger}\varphi_2
\right)^2+4\left(\varphi_1^{\dagger}\varphi_2\right)
\left(\varphi_2^{\dagger}\varphi_1\right) \label{th}\, .
\eea
We observe that for a typical weak interaction symmetry
breaking pattern the expectation values of $\varphi_1$ and
$\varphi_2$ should be aligned in the same direction or one
of them should vanish. In the present model this 
corresponds to the choice $\bar{\la}_{2\Lambda} < 0$.
The phase structure of a related model without the term
$\sim (\varphi_1^{\dagger}\varphi_2)(\varphi_2^{\dagger}\varphi_1)$
has been investigated previously \cite{twoscalar}
and shows second or first order 
transitions\footnote{First order phase transitions and 
coarse graining have also been discussed in a 
multi-scalar model with $Z_2$ symmetry \cite{AlMr}.
}. Combining
these results with the outcome of this work leads 
already to a detailed qualitative overview over the phase 
pattern in a more general setting with three independent
couplings for the quartic invariants 
$(\varphi_1^{\dagger}\varphi_1+\varphi_2^{\dagger}\varphi_2)^2$, 
$(\varphi_1^{\dagger}\varphi_1-\varphi_2^{\dagger}\varphi_2)^2$
and $(\varphi_1^{\dagger}\varphi_2)(\varphi_2^{\dagger}\varphi_1)$. 
We also note that the special case 
$\bar{\la}_{2\Lambda}=2\bar{\la}_{1\Lambda}$ corresponds to two
Heisenberg models interacting only by a term sensitive
to the alignment between $\varphi_1$ and $\varphi_2$, i.e.\
a quartic interaction of the form 
$(\varphi_1^{\dagger}\varphi_1)^2+(\varphi_2^{\dagger}\varphi_2)^2
+2(\varphi_1^{\dagger}\varphi_2)(\varphi_2^{\dagger}\varphi_1)$.

The model is now completely specified and it remains to extract
the flow equations for $U_k$ and $Z_k$. 

\subsection{Scale dependence of the effective average potential 
\label{scale}}
 
To obtain $U_k$ we evaluate the flow equation for   the average action 
(\ref{2.17})  for a constant field with $\Gamma_k=\Omega U_k$ where
$\Omega$ denotes the  
volume. 
With the help of  $U(2)\times
U(2)$ transformations the matrix field $\varphi$ can be turned  
into a standard diagonal form with real nonnegative eigenvalues.
Without loss of generality the evolution
equation for the effective
potential can therefore be obtained by calculating the trace 
in (\ref{2.17}) for small field fluctuations 
$\chi_{ab}$ around a constant
background configuration which is real and
diagonal, 
\beq
\varphi_{ab}=\varphi_{a} \delta_{ab} 
\,\, , \quad \varphi^*_{a}=\varphi_{a}\label{ConstConfig}.
\eeq
We separate the fluctuation field into its real and
imaginary part, $\chi_{ab}=\frac{1}{\sqrt{2}}
(\chi_{Rab}+i\chi_{Iab})$ and perform the second functional
derivatives of $\Gamma_k$ with respect to the
eight real components.
For the constant configuration
(\ref{ConstConfig}) it turns out that $\Gamma_k^{(2)}$ has a 
block diagonal form because mixed derivatives with 
respect to real and imaginary parts of the field vanish.
The remaining submatrices 
$\delta^2\Gamma_k/\delta\chi_R^{ab}\delta\chi_R^{cd}$ and 
$\delta^2\Gamma_k/\delta\chi_I^{ab}\delta\chi_I^{cd}$ can be
diagonalized in order to find
the inverse of $\Gamma_k^{(2)} + R_k$ under the trace
occuring in eq.\ (\ref{2.17}).
Here the momentum independent part
of $\Gamma_k^{(2)}$ defines the mass matrix
by the second functional derivatives
of $U_k$. 
The eight eigenvalues of the mass matrix are
\bea
(M_1^{\pm})^2 &=& U_k^{\prime}+2\left(\rho \pm
(\rho^2-\tau)^{1/2}\right)\partial_{\tau}U_k \,,\nnn
(M_2^{\pm})^2 &=& U_k^{\prime}\pm 2 \tau^{1/2} \partial_{\tau}U_k
\label{MassEigena} 
\eea
corresponding to second derivatives with respect to $\chi_I$
and
\bea
(M_3^{\pm})^2 &=& (M_1^{\pm})^2 \,,\nnn
(M_4^{\pm})^2 &=& U_k^{\prime}+\rho U_k^{\prime\prime}
+2 \rho \partial_{\tau} U_k
+4 \tau \partial_{\tau} U_k^{\prime} + 4 \rho \tau \partial_{\tau}^2 U_k 
\nnn &&
\pm \left\{ \tau \left( U_k^{\prime\prime} + 4 \partial_{\tau} U_k
+ 4 \rho \partial_{\tau} U_k^{\prime} + 4 \tau \partial_{\tau}^2 U_k \right)^2 
\right.\nnn
&& \left.
+ \left(\rho^2-\tau\right) \left( U_k^{\prime\prime} 
- 2 \partial_{\tau} U_k 
-4 \tau \partial_{\tau}^2 U_k \right)^2 \right\}^{1/2}
\label{MassEigen}
\eea
corresponding to second derivatives with respect to $\chi_R$.
Here the eigenvalues are expressed in terms of the 
invariants $\rho$ and $\tau$ using
\beq
\varphi_1^2=\hal(\rho+\tau^{1/2}),\quad
 \varphi_2^2=\hal(\rho-\tau^{1/2})
\eeq
and we adopt the convention that a prime on $U_k(\rho,\tau)$
denotes the derivative with respect to $\rho$ at fixed $\tau$ 
and $k$ and $\partial_{\tau}^n U_k \equiv \partial^nU_k/(\partial\tau)^n$.

The flow equation for the effective average potential is
simply expressed in terms of the mass eigenvalues
\bea
\lefteqn{ \dsp{\frac{\partial}{\partial t} U_k(\rho,\tau)} = \dsp{
 \hal \int \frac{d^dq}{(2\pi)^d}\frac{\partial}{\partial t} R_k(q)}}\nnn
 &&\dsp{ \left\{\frac{2}{P_k(q)+(M_1^+(\rho,\tau))^2}+
 \frac{2}{P_k(q)+(M_1^-(\rho,\tau))^2}
+\frac{1}{P_k(q)+(M_2^+(\rho,\tau))^2}\right.} \nnn
 &&+ \dsp{ \left.
 \frac{1}{P_k(q)+(M_2^-(\rho,\tau))^2}+
 \frac{1}{P_k(q)+(M_4^+(\rho,\tau))^2}+
 \frac{1}{P_k(q)+(M_4^-(\rho,\tau))^2} \right\}  }.
 \label{UkEvol}
\eea
In the rhs of the evolution equation appears the 
(massless) inverse average propagator 
\beq
 P_k(q)=Z_k q^2+R_k(q)=\frac{Z_k q^2}{1-e^{-q^2/k^2}}
 \label{Propagator}
\eeq
which incorporates the infrared cutoff function $R_k$ given
by eq.\ (\ref{2.15}).
The only approximation so far is due to the derivative 
expansion (\ref{Ansatz}) of $\Gamma_k$ which
enters into the flow equation 
(\ref{UkEvol}) through the form of $P_k$.
The mass eigenvalues (\ref{MassEigena}) and (\ref{MassEigen}) 
appearing in the above flow equation are exact since we have kept
for the potential the most general form $U_k(\rho,\tau)$.\\

{\bf Spontaneous symmetry breaking and mass spectra}

In the following we consider spontaneous symmetry 
breaking patterns
and the corresponding mass spectra for
a few special cases.
For the origin at $\varphi_{ab}=0$ all eigenvalues equal 
$U_k^{\prime}(0,0)$. If the origin is the absolute minimum of the
potential we are in the symmetric regime where all excitations have
mass squared $U_k^{\prime}(0,0)$. 

Spontaneous symmetry breaking to the diagonal $U(2)$ subgroup
of $U(2) \times U(2)$ 
can be observed for a field configuration which is proportional
to the identity matrix, i.e. $\varphi_{ab}=\varphi \delta_{ab}$.
The invariants $(\ref{Invaria})$ take on values  
$\rho=2 \varphi^2$ and $\tau=0$. The relevant
information for this symmetry breaking pattern is contained
in $U_k(\rho) \equiv U_k(\rho,\tau=0)$.
In case of spontaneous symmetry breaking there is a 
nonvanishing value for the minimum $\rho_0$ of the potential.
With $U_k^{\prime}(\rho_0)=0$ one finds the expected four 
massless Goldstone bosons with 
$(M_1^-)^2=(M_2^{\pm})^2=(M_3^-)^2=0$.
In addition there are three massive scalars in the 
adjoint representation of the unbroken diagonal
$SU(2)$ with mass squared 
$(M_1^+)^2=(M_3^+)^2=(M_4^-)^2=
4 \rho_0 \partial_{\tau} U_k$ and one singlet with mass squared
$(M_4^+)^2=2 \rho_0 U_k^{\prime\prime}$. The situation
corresponds to chiral symmetry breaking in two flavor
QCD in absence of quark masses and the chiral anomaly.
The Goldstone modes are the pseudoscalar pions and the
$\eta$ (or $\eta^{\prime}$), the scalar triplet has the
quantum numbers of $a_0$ and the singlet is the so-called
$\sigma$-field.

Another interesting case is the spontaneous symmetry breaking
down to a residual
$U(1) \times U(1) \times U(1)$ subgroup of $U(2) \times U(2)$
which can be observed for the configuration 
$\varphi_{ab} =\varphi \delta_{a1} \delta_{ab}$ ($\rho=\varphi^2$, 
$\tau=\varphi^4=\rho^2$). 
Corresponding to the number of broken generators 
one observes the five massless Goldstone bosons 
$(M_1^{\pm})^2=(M_2^+)^2=(M_3^{\pm})^2=0$ for the minimum 
of the potential at 
$U_k^{\prime}+2 \rho_0 \partial_{\tau} U_k = 0$. In addition there are two
scalars with mass squared 
$(M_2^-)^2=(M_4^-)^2=U_k^{\prime}-2 \rho_0 \partial_{\tau} U_k$
and one with 
$(M_4^+)^2=U_k^{\prime}+2 \rho_0 U_k^{\prime\prime} + 6 \rho_0
\partial_{\tau} U_k + 8 \rho_0^2 \partial_{\tau} U_k^{\prime} + 8 \rho_0^3
\partial_{\tau}^2 U_k$. 

We finally point out the special case where 
the potential is independent of the second invariant
$\tau$. In this case there is an enhanced 
$O(8)$ symmetry instead of 
$U(2) \times U(2)$. With $\partial^n_{\tau}U_k \equiv 0$
and $U_k^{\prime}(\rho_0)=0$ one observes the expected
seven massless Goldstone bosons and one massive mode with mass
squared $2 \rho_0 U_k^{\prime\prime}$.\\

{\bf Scaling form of the flow equations}

For the $O(8)$ symmetric  
model in the limit $\bar{\la}_{2\Lambda}=0$
one expects a region of
the parameter space which is characterized by renormalized
masses much smaller than the ultraviolet cutoff
or inverse microscopic length scale of the theory. 
In particular, in the absence of a mass scale one
expects a scaling behavior of the effective average
potential $U_k$. The behavior of $U_k$ at or near a second
order phase transition is most conveniently studied
using the scaling form of the evolution equation.
This form is also appropriate for an investigation that has 
to deal with weak first order phase transitions as 
encountered in the present model for 
$\bar{\la}_{2\Lambda} > 0$. The remaining
part of this subsection is devoted to the derivation of
the scaling form (\ref{DlessEvol}) of 
the flow equation (\ref{UkEvol}).

In the present form of eq.\ (\ref{UkEvol}) the rhs\
shows an explicit dependence on the scale $k$ once the
momentum integration is performed.
By a proper choice of 
variables we cast the evolution equation into a form
where the scale no longer appears explicitly. We
introduce a dimensionless potential $u_k=k^{-d}  U_k$
and express it in terms of dimensionless 
renormalized fields 
\bea
\tilde{\rho} &=& Z_k k^{2-d} \rho\,,\nnn
\tilde{\tau} &=& Z_k^2 k^{4-2 d} \tau\,.  
\label{dlessfield}
\eea
The derivatives of $u_k$ are given by
\beq
\partial_{\tilde{\tau}}^n u_k^{(m)}(\tilde{\rho},\tilde{\tau})=Z_k^{-2 n-m} 
k^{(2 n+m-1) d-4 n-2 m} \partial_{\tau}^n U_k^{(m)}
\left(\rho,\tau\right)\,.
\label{dlessder}
\eeq  
(Note that $u_k^{(m)}$ denotes $m$ derivatives
with respect to $\tilde{\rho}$ at fixed $\tilde{\tau}$ and $k$, while 
$U_k^{(m)}$ denotes $m$ derivatives
with respect to $\rho$ at fixed $\tau$ and $k$). With
\bea
\dsp{\frac{\partial}{\partial t} u_k(\tilde{\rho},
\tilde{\tau})_{|\tilde{\rho},\tilde{\tau}}} &=& 
\dsp{-d u_k(\tilde{\rho},\tilde{\tau}) +
(d-2+\eta) \tilde{\rho} u_k^{\prime}(\tilde{\rho},\tilde{\tau})
 + (2 d-4+2 \eta) \tilde{\tau}
\partial_{\tilde{\tau}} u_k(\tilde{\rho},\tilde{\tau})}\nnn 
&&+ \dsp{k^{-d} \frac{\partial}{\partial t}
U_k\left(\rho(\tilde{\rho}),\tau(\tilde{\tau})\right)_{|\rho,\tau}}
\label{dlesstrans}    
\eea
one obtains from (\ref{UkEvol})
the evolution equation for the 
dimensionless potential. Here the anomalous dimension 
$\eta$ arises from the $t$-derivative acting on $Z_k$ and 
is given by eq.\ (\ref{Eta}). Using the notation
$l^d_0(w;\eta,z=1)=l^d_0(w;\eta)$ for the threshold functions
defined in section \ref{ThreshApp} (see also section \ref{SecThresh})
one obtains
\bea
\lefteqn{\dsp{\frac{\partial}{\partial t} u_k(\tilde{\rho},\tilde{\tau})} = 
\dsp{-d u_k(\tilde{\rho},\tilde{\tau}) +
(d-2+\eta) \tilde{\rho} u_k^{\prime}(\tilde{\rho},\tilde{\tau})
 + (2 d-4+2 \eta) \tilde{\tau}
\partial_{\tilde{\tau}} u_k(\tilde{\rho},\tilde{\tau})}}\nnn
&&+ \dsp{4 v_d l^d_0\left((m_1^+(\tilde{\rho},\tilde{\tau}))^2;\eta\right)
+ 4 v_d l^d_0\left((m_1^-(\tilde{\rho},\tilde{\tau}))^2;\eta\right)
+ 2 v_d l^d_0\left((m_2^+(\tilde{\rho},\tilde{\tau}))^2;\eta\right)}\nnn
&&+ \dsp{2 v_d l^d_0\left((m_2^-(\tilde{\rho},\tilde{\tau}))^2;\eta\right)
+2 v_d l^d_0\left((m_4^+(\tilde{\rho},\tilde{\tau}))^2;\eta\right)
+2 v_d l^d_0\left((m_4^-(\tilde{\rho},\tilde{\tau}))^2;\eta\right)
}
\label{DlessEvol}    
\eea
where the dimensionless mass terms are related to 
(\ref{MassEigen}) according to 
\beq
(m_i^{\pm}(\tilde{\rho},\tilde{\tau}))^2=\frac
{\left(M_i^{\pm}(\rho(\tilde{\rho}),\tau(\tilde{\tau}))\right)^2}{Z_k k^2}.
\label{massrel}
\eeq
Eq.\ (\ref{DlessEvol}) is the scaling form of the
flow equation we are looking for. For a 
$\tilde{\tau}$-independent potential it reduces to the 
evolution equation for the $O(8)$ symmetric model
\cite{TW94-1,Mor94-1} given by eq. (\ref{2.44}) with 
 $z=1,\tilde y=0,\Delta{\zeta}_k=0$. The potential $u_k$ at
a second order phase transition is given by
a $k$-independent (scaling) solution 
$\partial u_k/\partial t=0$ \cite{TW94-1,Mor94-1}. For this solution 
all the $k$-dependent functions
in the rhs\ of eq.\
(\ref{DlessEvol}) become independent
of $k$. For a weak first order phase transition these functions
will show a weak $k$ dependence for $k$
larger than the inherent mass scale of the
system (cf.\ section \ref{rg}). There is no
particular advantage of the scaling form of
the flow equation for strong first order phase transitions.\\

Eq.\ (\ref{DlessEvol}) describes the scale dependence
of the effective average potential $u_k$
by a nonlinear partial differential equation for the
three variables $t$, $\tilde{\rho}$ and $\tilde{\tau}$. 
We concentrate in the following on spontaneous 
symmetry breaking with a residual $U(2)$ symmetry
group. As we have already pointed out in section \ref{model}
this symmetry breaking can be observed for a
configuration which is proportional to the 
identity and we have $\tilde{\tau}=0$. In this case
the eigenvalues (\ref{MassEigena}) and
(\ref{MassEigen}) of the mass matrix 
with (\ref{massrel}) are given by
\bea
(m_1^-)^2 &=& (m_2^{\pm})^2 = (m_3^-)^2 = u_k^{\prime}\,,\nnn 
(m_1^+)^2 &=& (m_3^+)^2 = (m_4^-)^2 = 
u_k^{\prime}+4 \tilde{\rho} \partial_{\tilde{\tau}}u_k\,,\nnn
(m_4^+)^2 &=& u_k^{\prime} + 2 \tilde{\rho} u_k^{\prime\prime}\,
\eea
and in the rhs\ of the partial differential equation 
(\ref{DlessEvol}) for $u_k(\tilde{\rho}) \equiv 
u_k(\tilde{\rho},\tilde{\tau}=0)$
only the functions $u_k^{\prime}(\tilde{\rho})$, $u_k^{\prime\prime}(\tilde{\rho})$
and $\partial_{\tilde{\tau}}u_k(\tilde{\rho})$ appear. We determine these
functions through the use of flow equations which
are obtained by taking the derivative in eq.\ 
(\ref{DlessEvol}) with respect to $\tilde{\rho}$ and
$\tilde{\tau}$ evaluated at $\tilde{\tau}=0$.
Since we are 
interested in the $\tilde{\rho}$-dependence of the potential at 
$\tilde{\tau}=0$ we shall use a truncated expansion in $\tilde{\tau}$ with 
\beq
\partial_{\tilde{\tau}}^n u_k(\tilde{\rho},\tilde{\tau}=0)=0 \quad \mbox{for}
\quad n \ge 2.
\label{Truncation}
\eeq
In three space dimensions the neglected  
($\tilde{\rho}$-dependent) operators have 
negative canonical mass dimension.
We make no expansion in terms of $\tilde{\rho}$
since the general $\tilde{\rho}$-dependence allows a description
of a first order phase transition where a second local minimum
of $u_k(\tilde{\rho})$ appears. 
The approximation (\ref{Truncation}) only affects  
the flow equations for $\partial_{\tilde{\tau}}u_k$. 
The form of the flow equation for
$u_k^{\prime}$ is not affected by the 
truncation. 
From $u_k^{\prime}$ we obtain the effective average
potential $u_k$ by simple integration.
We have tested the sensitivity of our results
for $u_k^{\prime}$ to a change in $\partial_{\tilde{\tau}}u_k$ by neglecting
the $\tilde{\rho}$-dependence of the $\tilde{\tau}$-derivative. We observed
no qualitative change of the results. We expect that the main
truncation error is due to the derivative expansion (\ref{Ansatz}) for
the effective average action.  

To simplify notation we introduce 
\bea
\eps(\tilde{\rho})&=&u_k^{\prime}(\tilde{\rho},\tilde{\tau}=0),\nnn
\la_1(\tilde{\rho})&=&u_k^{\prime\prime}(\tilde{\rho},\tilde{\tau}=0),\nnn
\la_2(\tilde{\rho})&=&4 \partial_{\tilde{\tau}} u_k(\tilde{\rho},\tilde{\tau}=0).
\label{notation}
\eea
Higher derivatives are denoted by primes on the 
$\tilde{\rho}$-dependent quartic
``couplings'',
i.e.\ $\la_1^{\prime}=u_k^{\prime\prime\prime}$,
$\la_2^{\prime}=\partial_{\tilde{\tau}}u_k^{\prime}$ etc. It is
convenient to introduce two-parameter functions
$l^d_{n_1,n_2}(w_1,w_2;\eta)$ \cite{Ju95-7}. 
For $n_1=n_2=1$ their relation to the functions
$l^d_n(w;\eta)$ can be expressed as
\bea
l^d_{1,1}(w_1,w_2;\eta) &=&
\dsp{\frac{1}{w_2 - w_1} \left[ l^d_1(w_1;\eta)
- l^d_1(w_2;\eta) \right]} \quad 
\mbox{for}\quad w_1 \ne w_2, \nnn
l^d_{1,1}(w,w ;\eta)&=& l^d_2(w;\eta)
\eea
and
\beq
l^d_{n_1 + 1, n_2}(w_1,w_2;\eta) =\dsp{
-\frac{1}{n_1}  \frac{\partial}{\partial w_1}
l^d_{n_1,n_2}(w_1,w_2;\eta)},\quad
l^d_{n_1,n_2}(w_1,w_2;\eta) =  
l^d_{n_2,n_1}(w_2,w_1;\eta).                         
\eeq

With the help of these functions the scale dependence of 
$\eps$ is described by
\bea
\dsp{\frac{\partial \eps}{\partial t}} &=& 
\dsp{(-2+\eta) \eps +
(d-2+\eta) \tilde{\rho} \la_1 - 6 v_d (\la_1+\la_2+\tilde{\rho} \la_2^{\prime}) 
l^d_1(\eps+\tilde{\rho} \la_2;\eta)}\nnn
&&- \dsp{2 v_d (3 \la_1+2 \tilde{\rho} \la_1^{\prime}) 
l^d_1(\eps+2\tilde{\rho} \la_1;\eta)   
-8 v_d \la_1 l^d_1(\eps ;\eta)}
\label{Epsi}
\eea
and for $\la_1$ one finds
\bea
\dsp{\frac{\partial \la_1}{\partial t}} &=& 
\dsp{(d-4+2\eta) \la_1 +
(d-2+\eta) \tilde{\rho} \la_1^{\prime} }\nnn
&&+\dsp{6 v_d \left[ (\la_1+\la_2+\tilde{\rho} \la_2^{\prime})^2
l^d_2(\eps+\tilde{\rho} \la_2;\eta)
-(\la_1^{\prime}+2\la_2^{\prime}+\tilde{\rho} \la_2^{\prime\prime})
l^d_1(\eps+\tilde{\rho} \la_2;\eta)\right]}\nnn
&&+\dsp{2 v_d \left[(3\la_1+2\tilde{\rho} \la_1^{\prime})^2
l^d_2(\eps+2\tilde{\rho} \la_1;\eta)
-(5\la_1^{\prime}+2\tilde{\rho} \la_1^{\prime\prime})
l^d_1(\eps+2\tilde{\rho} \la_1;\eta)\right] }\nnn
&&+\dsp{8 v_d \left[(\la_1)^2 l^d_2(\eps;\eta)
-\la_1^{\prime}l^d_1(\eps;\eta)\right]}.
\label{La1}                      
\eea
Similarly the scale dependence of $\la_2$ 
is given by
\bea
\dsp{\frac{\partial \la_2}{\partial t}} &=& 
\dsp{(d-4+2\eta) \la_2 +
(d-2+\eta) \tilde{\rho} \la_2^{\prime}       
- 4 v_d (\la_2)^2 l^d_{1,1}(\eps+\tilde{\rho} \la_2,\eps;\eta) }\nnn
&&+\dsp{2 v_d \left[ 3 (\la_2)^2+12\la_1\la_2+8 \tilde{\rho}
\la_2^{\prime}(\la_1+\la_2)+4\tilde{\rho}^2 (\la_2^{\prime})^2\right]
l^d_{1,1}(\eps+\tilde{\rho} \la_2,\eps+2\tilde{\rho} \la_1;\eta) }\nnn
&&- \dsp{14 v_d \la_2^{\prime} l^d_1(\eps+\tilde{\rho} \la_2;\eta)
-2 v_d (5 \la_2^{\prime}+2\tilde{\rho} \la_2^{\prime\prime})
l^d_1(\eps+2\tilde{\rho} \la_1;\eta) }\nnn
&&+\dsp{2 v_d \left[ (\la_2)^2 l^d_2(\eps;\eta)-4 \la_2^{\prime}
l^d_1(\eps;\eta)\right] } \, .
\label{La2} 
\eea
For the numerical solution
we evaluate the above flow equations
at different points $\tilde{\rho}_i$ for $i= 1,\ldots ,l$ 
and use a set of 
matching conditions that are described in ref.\ \cite{ABBFTW95-1}.
If there is a 
minimum of the potential at nonvanishing
$\kappa \equiv \tilde{\rho}_0$,
the condition $\eps(\kappa)=0$ can be used to 
obtain the  scale dependence of $\kappa(k)$:
\bea
\dsp{\frac{\mbox{d} \kappa}{\mbox{d} t}}
&=&\dsp{-[\lambda_1(\kappa)]^{-1}\frac{\partial \eps}{\partial t}}
|_{\tilde{\rho}=\kappa}\nnn
&=&\dsp{ -(d-2+\eta) \kappa 
+ 6 v_d \left(1+\frac{\lambda_2(\kappa) + \kappa \lambda_2^{\prime}(\kappa)}
{\lambda_1(\kappa)} \right) l^d_1\left(\kappa \lambda_2(\kappa);\eta \right) }\nnn
&+& \dsp{2 v_d \left( 3+\frac{2\kappa \lambda_1^{\prime}(\kappa)}{\lambda_1(\kappa)}
\right) l^d_1\left(2\kappa \lambda_1(\kappa);\eta \right) 
+8 v_d l^d_1\left(0;\eta\right) }.
\label{kappa}
\eea
To make contact with $\beta$-functions for the couplings 
at the potential minimum $\kappa$ we point out the relation
\beq
\dsp{\frac{\mbox{d} \lambda_{1,2}^{(m)}(\kappa)}{\mbox{d} t}}=
\dsp{\frac{\partial \lambda_{1,2}^{(m)} }{\partial t}|_{\tilde{\rho}=\kappa}
+ \lambda_{1,2}^{(m+1)}(\kappa) \frac{\mbox{d} \kappa}{\mbox{d} t} }.
\label{lakappa}
\eeq

It remains
to compute the anomalous dimension $\eta$ defined in (\ref{Eta})
which describes the scale
dependence of the wave function renormalization $Z_k$. 
We consider a space dependent distortion of the constant 
background field configuration (\ref{ConstConfig})
of the form
\beq
 \varphi_{ab}(x) = \varphi_a\delta_{ab} +
 \left[\delta\varphi e^{-iQx} + \delta\varphi^* e^{iQx}\right] 
\Si_{ab}.
 \label{ConfAnDi}
\eeq
Insertion of the above configuration into the parametrization 
(\ref{Ansatz}) of $\Gamma_k$ yields
\beq
 \dsp{
 Z_k }=
\dsp{
  Z_k(\rho,\tau,Q^2=0) }
 = \dsp{
 \hal\frac{1}{\Si_{ab}^*\Si_{ab}}
 \lim_{Q^2 \rightarrow 0}\frac{\partial}{\partial Q^2}
 \frac{\delta \Gamma_k}{\delta (\delta\varphi \delta\varphi^*)}|_{\delta\varphi=0}} .
 \label{Z}
\eeq 
\\
To obtain 
the flow equation of the wave function renormalization
one expands the effective average action around a configuration
of the form (\ref{ConfAnDi}) and evaluates the rhs\ of eq.\
(\ref{2.17}). This computation has been done in ref.\ 
\cite{Ju95-7} for a ``Goldstone'' configuration with   
\beq
 \Si_{ab}=\delta_{a1}\delta_{b2}-\delta_{a2}\delta_{b1}
\eeq
and $\varphi_a \delta_{ab}=\varphi\delta_{ab}$ corresponding to a 
symmetry breaking pattern
with residual $U(2)$ symmetry. The result of ref.\ \cite{Ju95-7}
can be easily generalized to arbitrary fixed field values 
of $\tilde{\rho}$
and we find
\bea
 \dsp{ \eta(k)}
&=&\dsp{
 4 \frac{v_d}{d}\tilde{\rho} \left[
 4(\lambda_1)^2 
m_{2,2}^d(\eps,\eps+
2\tilde{\rho} \lambda_1;\eta) 
  +(\lambda_2)^2 
m_{2,2}^d(\eps,\eps+
\tilde{\rho} \lambda_2;\eta) 
 \right] }.
 \label{EtaSkale}
\eea  
The definition of the threshold function
\beq 
m_{2,2}^d(w_1,w_2;\eta)=
m_{2,2}^d(w_1,w_2)-\eta\hat{m}_{2,2}^d(w_1,w_2)
\eeq
can be found in appendix \ref{ThreshApp}. 
For vanishing arguments the 
functions $m_{2,2}^d$ and $\hat{m}_{2,2}^d$ are of order
unity. They are symmetric with respect to their arguments
and in leading order $m_{2,2}^d(0,w) \sim 
\hat{m}_{2,2}^d(0,w) \sim w^{-2}$ for $w \gg 1$. 
According to eq.\ (\ref{zet})
we use $\tilde{\rho}=\kappa$ to define the uniform
wave function renormalization
\beq
Z_{k} \equiv Z_{k}(\kappa).
\label{wfr}
\eeq 
We point out that 
according to our truncation of the
effective average action with eq.\ (\ref{EtaSkale}) 
the anomalous
dimension $\eta$ is exactly zero at $\tilde{\rho} = 0$. 
This is an artefact of the truncation and
we expect the symmetric phase
to be more affected by truncation errors than the
spontaneously broken phase. We typically observe
small values for 
$\eta(k)=-\mbox{d} (\ln Z_k)/ \mbox{d} t$ (of the order of 
a few per cent). The smallness of $\eta$ is crucial
for our approximation of a uniform wave function
renormalization to give quantitatively reliable
results for the equation of state. For the universal
equation of state given in section\ \ref{sce}
one has $\eta=0.022$ as given by the corresponding
index of the $O(8)$ symmetric ``vector'' model.

\subsection{Renormalization group flow of couplings \label{rg}}

To understand the detailed picture of the phase structure, 
which is presented in section \ref{ps}, we will consider 
the flow of some characteristic quantities for the effective
average potential as the infrared
cutoff $k$ is lowered. 
The short distance potential $U_{\Lambda}$
given in eq.\ (\ref{uinitial}) is parametrized by 
quartic couplings, 
\beq
\bar{\la}_{1\Lambda},\bar{\la}_{2\Lambda}>0
\eeq 
and the location of its minimum is given by 
\beq
\rho_{0\Lambda}=\bar{\mu}_{\Lambda}^2/\bar{\la}_{1\Lambda}. 
\eeq
We integrate the
flow equation for the effective average potential $U_k$
for a variety of initial 
conditions $\rho_{0\Lambda},\bar{\la}_{1\Lambda}$ and
$\bar{\la}_{2\Lambda}$. In particular, for general 
$\bar{\la}_{1\Lambda},\bar{\la}_{2\Lambda}>0$ we are able to find a critical
value $\rho_{0\Lambda}=\rho_{0 c}$ for which the system exhibits
a first order phase transition. In this case the evolution of
$U_k$ leads at some scale $k_2 < \Lambda$ to the appearance 
of a second local
minimum at the origin of the effective average potential and both
minima become degenerate in the limit $k \to 0$. If $\rho_0(k)>0$
denotes the $k$-dependent outer minimum of the potential 
($U_k^{\prime}(\rho_0)=0$, where the prime on $U_k$ denotes the
derivative with respect to $\rho$ at fixed $k$) at
a first order phase transition one has
\beq
\lim\limits_{k \to 0}( U_k(0)-U_k(\rho_0))=0.
\label{critcon}
\eeq
A measure of the distance from the phase transition
is the difference $\delta\kappa_{\Lambda}=
(\rho_{0\Lambda}-\rho_{0 c})/\Lambda$.
If $\bar{\mu}_{\Lambda}^2$ and therefore
$\rho_{0\Lambda}$ is interpreted as a function of 
temperature, the deviation $\delta\kappa_{\Lambda}$ is 
proportional to the deviation from the critical 
temperature $T_{c}$, i.e.\ $\delta\kappa_{\Lambda}=A(T)
(T_{c}-T)$ with $A(T_{c}) > 0$.

We will always consider in this subsection
the trajectories for the critical ``temperature'', i.e.\
$\delta \kappa_{\Lambda}=0$,
and we follow the flow for different values of the
short distance parameters $\la_{1\Lambda}$
and $\la_{2\Lambda}$. The discussion for sufficiently
small $\delta \kappa_{\Lambda}$ is analogous.
In particular, we compare the
renormalization group flow of these quantities for a weak and a
strong first order phase transition. In some limiting cases
their behavior can be studied analytically.
For the discussion we will frequently consider the flow 
equations for the quartic ``couplings'' $\la_1(\tilde{\rho})$,
$\la_2(\tilde{\rho})$ eqs.\ (\ref{La1}), (\ref{La2}) and for the
minimum $\kappa$ eq.\ (\ref{kappa}).    

In fig.\ \ref{scaledep1}, \ref{scaledep2} we follow the flow
of the dimensionless renormalized minimum $\kappa$ and
the radial mass term $\tilde{m}^2=2 \kappa \la_{1}(\kappa)$ in
comparison to their dimensionful
counterparts $\rho_{0 R}=k \kappa$
and $m_R^2=k^2 \tilde{m}^2$ in units of the momentum scale $\Lambda$.
We also consider the dimensionless renormalized mass term
$\tilde{m}_2^2=\kappa \la_{2} (\kappa)$ corresponding to the curvature of the
potential in the direction of the second
invariant $\tilde{\tau}$. The height of the potential 
barrier $U_B(k)=k^3 u_k(\tilde{\rho}_B)$ 
with $u_k^{\prime}(\tilde{\rho}_B)=0$, $0 < \tilde{\rho}_B < \kappa$,
and the height
of the outer minimum $U_0(k)=k^3 u_k(\kappa)$ 
is also displayed. Fig.\ \ref{scaledep1} shows these quantities
as a function of $t=\ln(k/\Lambda)$ for $\la_{1 \Lambda}=2$, 
$\la_{2 \Lambda}=0.1$.
\begin{figure}[h]
\unitlength1.0cm
\begin{center}
\begin{picture}(17.,10.)
\put(8.5,2.8){\footnotesize $\dsp{\frac{U_B}{\Lambda^3}}\, 
2\times 10^{15}$}
\put(2.2,2.3){\footnotesize $\dsp{\frac{U_0}{\Lambda^3}}\, 
2\times 10^{15}$}
\put(0.7,6.9){\footnotesize $\dsp{\frac{
\rho_{0R}}{\Lambda}} 10^{5}$}
\put(7.2,9.1){\footnotesize $\dsp{\frac{
m_{R}^2}{\Lambda^2}} 10^{9}$}
\put(12.2,8.2){\footnotesize $\tilde{m}_2^2 $}
\put(10.1,5.8){\footnotesize $\tilde{m}^2 $}
\put(10.2,4.4){\footnotesize $\kappa$}
\put(6.4,-0.5){$t=\ln(k/\Lambda)$}
\put(-0.7,0.){
\rotate[r]{\epsfig{file=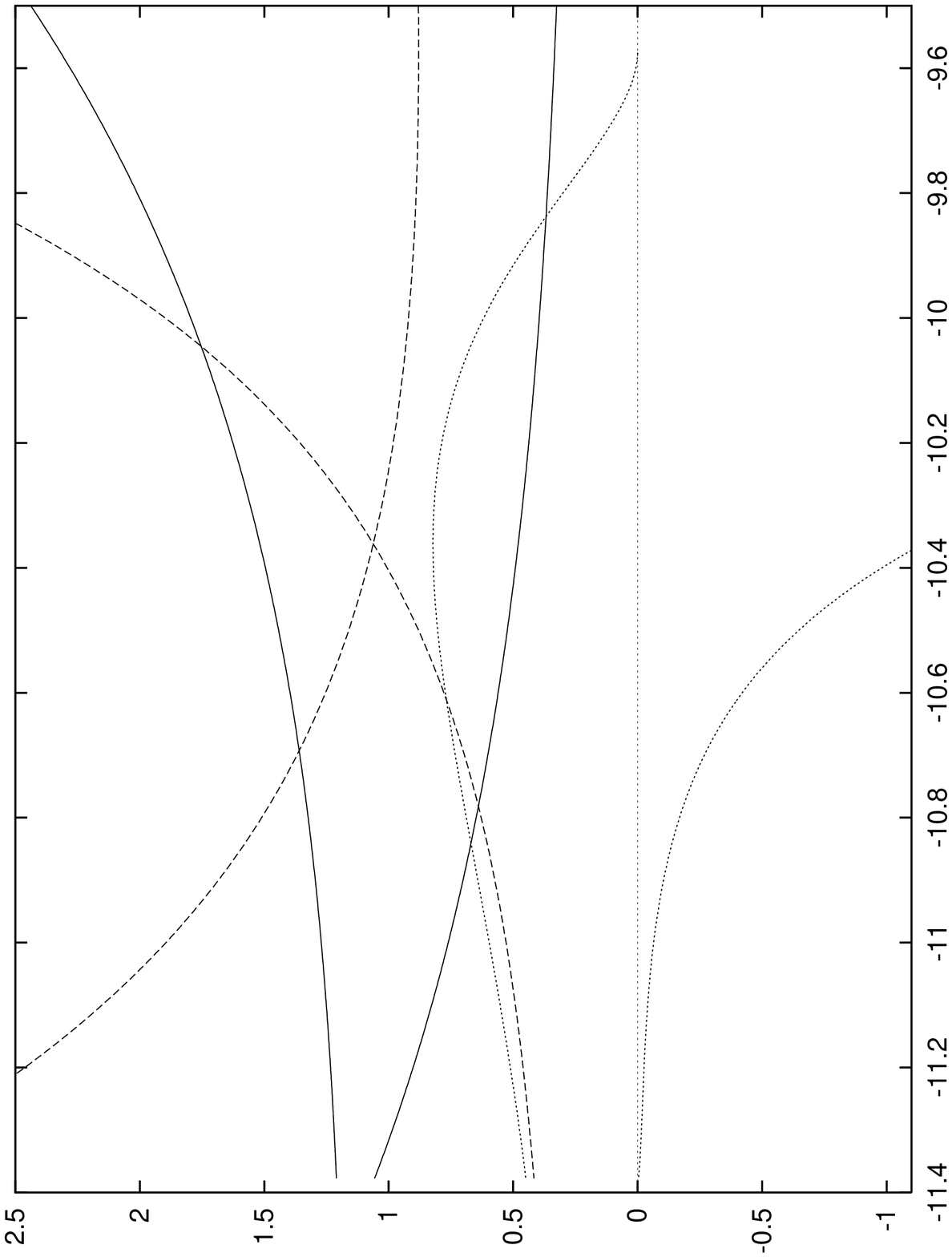,width=10.cm,height=11.7cm}}
}
\put(10.4,0.){
\rotate[r]{\epsfig{file=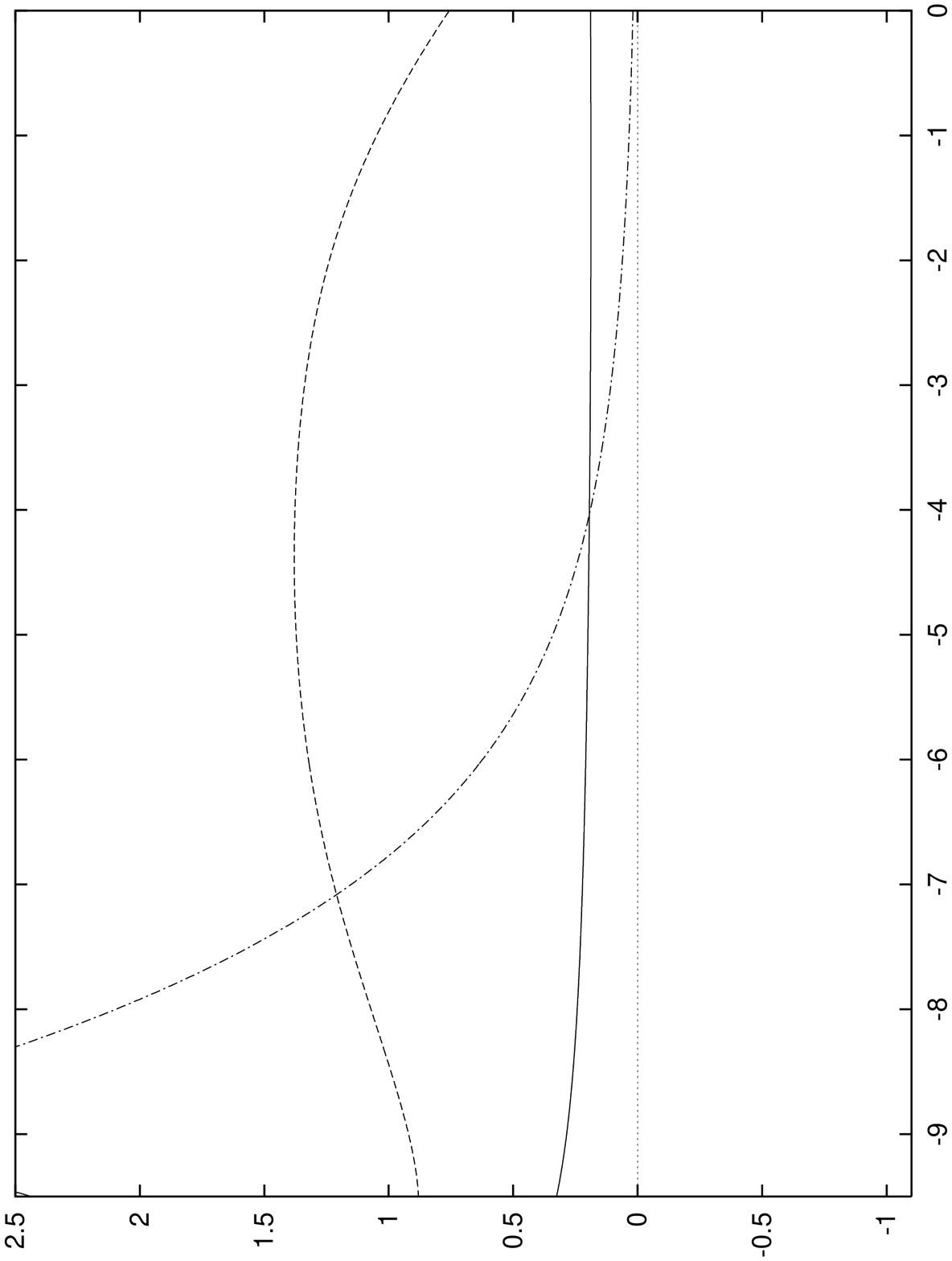,width=10.cm,height=5.5cm}}
}
\end{picture}
\end{center}
\caption[]{\em
Running couplings for a weak first order transition. We show the scale 
 dependence of 
the dimensionless renormalized 
masses $\tilde{m}^2$, $\tilde{m}_2^2$, minimum $\kappa$
and dimensionful counterparts 
$m_R^2=k^2 \tilde{m}^2$, $\rho_{0 R}=k \kappa$ 
in units of $\Lambda$. We also 
show $U_B(k)$ and
$U_0(k)$, the value of the potential at the top
of the potential barrier and at the minimum
$\rho_{0R}$, respectively. The short distance parameters are  
$\la_{1\Lambda}=2$, $\la_{2\Lambda}=0.1$ and $\delta
\kappa_{\Lambda}=0$. 
The right panel shows the approximate scaling solution.
\label{scaledep1}
}
\end{figure}
One observes that
the flow can be separated into two parts.
The first part ranging from $t=0$ to $t \simeq -6$
is characterized by $\kappa \simeq \mbox{const}$ and small
$\tilde{m}_2^2$. 
It is instructive to consider what happens in the case
$\tilde{m}_2^2 \equiv 0$. In this case $\la_{2}\equiv 0$
and the flow is governed by the Wilson-Fisher fixed
point of the O(8) symmetric theory.
At the corresponding second order phase transition
the evolution of $u_k$ leads to the scaling
solution of (\ref{DlessEvol}) which is obtained for
$\partial u_k/ \partial t=0$. As a consequence $u_k$ becomes a
$k$-independent function that takes on constant (fixed
point) values.
In particular, the minimum $\kappa$ of the potential  
takes on its fixed point value
$\kappa(k)= \kappa_{\star}$. 
The fixed point is 
not attractive in the $U(2) \times U(2)$ symmetric theory and
$\la_{2\Lambda}$ is an additional relevant parameter for the system.
For small $\la_2$ the evolution is governed by an anomalous
dimension $d\la_2/ d t=A \la_2$ with $A < 0$, leading to the
increasing $\tilde{m}_2^2$ as $k$ is lowered.

The system exhibits 
scaling behavior only for sufficiently small $\la_{2}$.
As $\tilde{m}_2^2$ increases the quartic coupling $\la_1$ and
therefore the radial mass term $\tilde{m}^2$ is driven to 
smaller values as can be observed from fig.\ \ref{scaledep1}.
For nonvanishing $\la_{2}$ the corresponding 
qualitative change in the flow equation 
(\ref{La1}) for $\la_1$ is the occurance of a term 
$\sim \la_2^2$.
It allows to drive $\la_1$ to negative values in a certain range
of $\tilde{\rho}<\kappa$ and, therefore,
to create a potential barrier inducing a first order phase
transition. We observe from the plot that at $t \lesssim -9.5$
a second minimum arises $(U_B \not = 0)$. The corresponding value
of $k=\Lambda e^t=k_2$ sets a characteristic scale for the first order
phase transition. Below this scale the dimensionless,
renormalized  quantities approximately scale 
according to their canonical dimension.
The dimensionful quantities like $\rho_{0R}$ or
$m_R^2$ show only a weak scale dependence in this range. 
In contrast to the
above example of a weak first order phase transition 
with characteristic renormalized masses much smaller
than $\Lambda$, fig.\
\ref{scaledep2} shows the flow of the corresponding quantities for
a strong first order phase transition. 
The short distance 
parameters employed are $\la_{1\Lambda}=0.1$,
$\la_{2\Lambda}=2$.
Here the range with 
$\kappa \simeq \mbox{const}$ is  absent and one observes
no approximate scaling behavior.   
\begin{figure}[h]
\unitlength1.0cm
\begin{center}
\begin{picture}(15.,10.)
\put(8.,2.15){\footnotesize $\dsp{\frac{U_B}{\Lambda^3}}\, 
2\times 10^{3}$}
\put(3.6,1.5){\footnotesize $\dsp{\frac{U_0}{\Lambda^3}}\, 
2\times 10^{3}$}
\put(1.1,3.73){\footnotesize $\dsp{\frac{
\rho_{0R}}{\Lambda}}$}
\put(6.4,8.6){\footnotesize $\dsp{\frac{
m_{R}^2}{\Lambda^2}} 10^{2}$}
\put(11.5,7.8){\footnotesize $\tilde{m}_2^2 $}
\put(13.,3.15){\footnotesize $\tilde{m}^2 $}
\put(13.,4.4){\footnotesize $\kappa$}
\put(6.4,-0.5){$t=\ln(k/\Lambda)$}
\put(-0.7,0.){
\rotate[r]{\epsfig{file=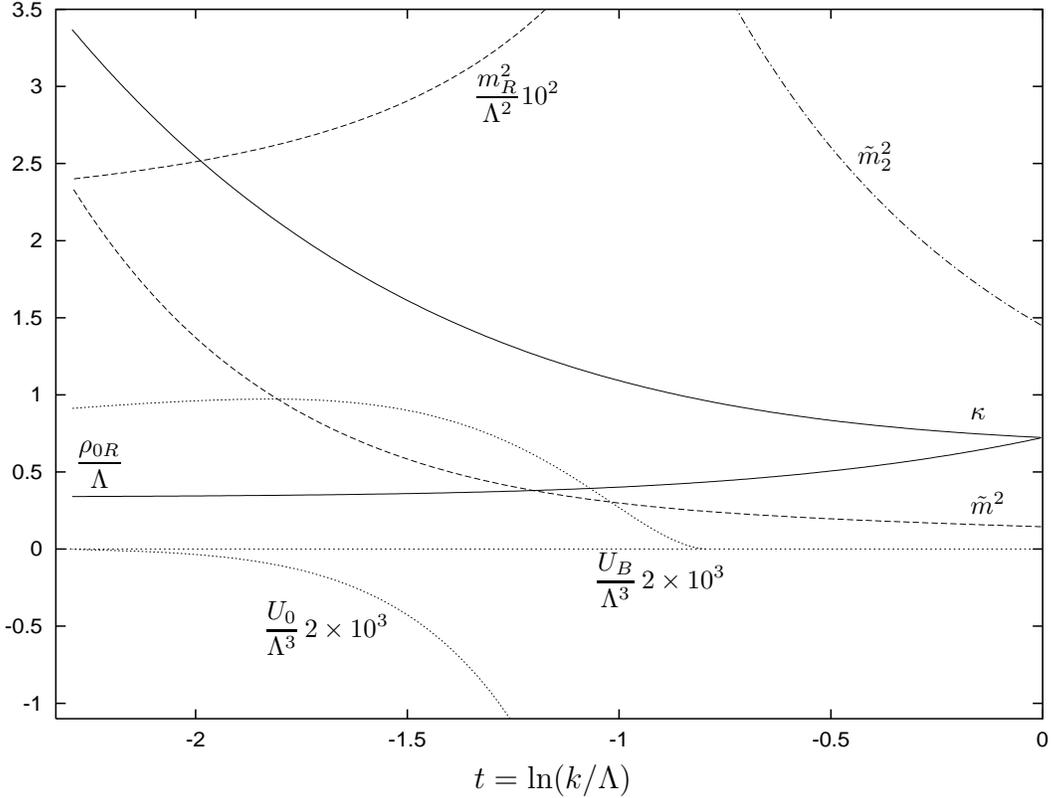,width=10.cm,height=15.cm}}
}
\end{picture}
\end{center}
\caption[]{\label{scaledep2} \em
Running couplings for a strong first order transition. We plot the same
couplings as in 
 fig.\ \ref{scaledep1}
for $\la_{1\Lambda}=0.1$ and $\la_{2\Lambda}=2$. No approximate scaling
solution is reached.
}
\end{figure}

In the discussion of the phase structure of the model in the next section we 
distinguish between the range $\la_{2\Lambda}/\la_{1\Lambda} \ll 1$
and $\la_{2\Lambda}/\la_{1\Lambda} \gg 1$ to denote the weak
and the strong first order region.
For $\la_{2\Lambda}/\la_{1\Lambda} \ll 1$ the initial   
renormalization group flow is dominated by the 
Wilson-Fisher fixed point of the $O(8)$ symmetric theory.
In this range 
the irrelevant couplings are driven close to 
the fixed point for some ``time'' $|t|=-\ln(k/\Lambda)$, loosing
their memory on the initial conditions given by the short
distance potential $u_{\Lambda}$. As a consequence we are able
to observe universal behavior as is 
demonstrated in fig.\ \ref{ratio}.
 
To discuss the case $\la_{2\Lambda}/\la_{1\Lambda} \gg 1$ we consider the
flow equations for the couplings
at the minimum $\kappa \not= 0$ of the potential
given by (\ref{kappa}) and (\ref{lakappa})
with (\ref{La1}), (\ref{La2}). 
In the limit of an infinite mass term 
$\tilde{m}_2^2=\kappa \la_{2}(\kappa) \to \infty$ the $\beta$-functions for
$\la_{1}(\kappa)$ and $\kappa$ become independent from $\la_{2}(\kappa)$ due to
the threshold functions, with 
$l^3_n(\kappa \la_{2}) \sim (\kappa \la_{2})^{-(n+1)}$ for large
$\kappa \la_{2}(\kappa)$. 
As a consequence
$\beta_{\la_1}$ and $\beta_{\kappa}$ equal the $\beta$-functions for an 
$O(5)$ symmetric model. 
We argue in the following that in this large
coupling limit fluctuations of massless Goldstone bosons
lead to an attractive fixed point for $\la_{2}(\kappa)$.
We take the flow equation (\ref{lakappa}),
(\ref{La2}) for $\la_{2} (\kappa)$ 
keeping only terms with positive canonical mass dimension
for a qualitative discussion. (This amounts to the 
approximation 
$\la_1^{(n)}(\kappa)=\la_2^{(n)}(\kappa)=0$ for $n \ge 1$.) 
To be explicit, one may consider the
case for given $\la_{1\Lambda}=2$. The critical cutoff value
for the potential minimum is $\kappa_{\Lambda}\simeq 0.2$ for
$\la_{2\Lambda}\gg 1$. For $\kappa \la_2(\kappa) \gg 1$ and taking 
$\eta \simeq 0$ the $\beta$-function for $\la_2 (\kappa)$
is to a good approximation given by $(d=3)$
\beq
\frac{\mbox{d} \la_{2} (\kappa)}{\mbox{d} t} =
-\la_{2} (\kappa)+2 v_3 (\la_{2} (\kappa))^2 l^3_2(0).
\label{fpla2}
\eeq
The second term in the rhs\ of eq.\ (\ref{fpla2})
is due to massless Goldstone modes which give the
dominant contribution in the considered range. 
The solution of (\ref{fpla2})
implies an attractive fixed point for $\la_{2} (\kappa)$ with a value
\beq
\la_{2\star} (\kappa)=\dsp{\frac{1}{2 v_3 l^3_2(0)} } \simeq 4 \pi^2.
\eeq 
Starting from $\la_{2\Lambda}$ one finds for the ``time''
$|t|$ necessary
to reach a given $\la_{2} (\kappa) > \la_{2\star} (\kappa)$ 
\beq
|t|= - \ln
\dsp{\frac{\la_{2} (\kappa)-\la_{2\star} (\kappa)}
{\la_{2} (\kappa) \left(1-\frac{\dsp{\la_{2\star} (\kappa)}}{
\dsp{\la_{2 \Lambda}}} 
\right)} }\quad .
\eeq
This converges to a 
finite value for $\la_{2 \Lambda} \to \infty$.  
The further evolution therefore 
becomes insensitive
to the initial value for $\la_{2\Lambda}$ in the large coupling
limit. The flow of $\la_{1}(\kappa)$ and $\kappa$ is not
affected by the initial running of $\la_{2}(\kappa)$
and quantities like $\Delta \rho_{0R}/\Lambda$ or 
$m_R/\Delta \rho_{0R}$ become independent of $\la_{2\Lambda}$
if the coupling is sufficiently large. 
This qualitative
discussion is confirmed by the numerical solution of the full
set of equations presented in figs.\ \ref{phase} and 
\ref{ratio} of section \ref{ps}.
For the fixed point value we obtain $\la_{2\star} (\kappa)=38.02$.
We point out that an analogous discussion for the large
coupling region of $\la_{1 \Lambda}$ cannot be made. This can be seen by
considering the mass term at the origin of the short
distance potential 
(\ref{uinitial}) given by 
$u^{\prime}_{\Lambda}(0,0)=-\kappa_{\Lambda} \la_{1\Lambda}$. Due to the
pole of $l^3_n(w,\eta)$ at $w = -1$ for $n > 1/2$
\cite{BW97-1} one obtains the constraint
\beq
\kappa_{\Lambda} \la_{1\Lambda} < 1 \label{constraint}\quad.
\eeq  
In the limit $\la_{1\Lambda} \to \infty$
the mass term $2 \kappa_{\Lambda} \la_{1\Lambda}$
at the minimum $\kappa$ of the potential
at the critical temperature therefore remains finite.

\subsection{Phase structure of the $U(2) \times U(2)$ model \label{ps}}

We study the phase structure of the
$U(2) \times U(2)$ symmetric model in three space dimensions.
We concentrate here on the spontaneous 
symmetry breaking with a residual $U(2)$ symmetry
group. We consider in the
following the effective average potential $U_k$
for a nonzero scale $k$. This allows
to observe the nonconvex part of the potential. 
As an example we show in 
fig.\ \ref{tempot} the effective average potential
$U_{k=k_f}$ for $\la_{1\Lambda}= \bar{\la}_{1\Lambda}/\Lambda=0.1$
and $\la_{2\Lambda}= \bar{\la}_{2\Lambda}/\Lambda=2$
as a function of the renormalized field 
$\varphi_R=(\rho_R/2)^{1/2}$ with $\rho_{R}=Z_{k=k_f} \rho$. 
\begin{figure}[h]
\unitlength1.0cm
\begin{center}
\begin{picture}(13.,9.)
\put(-0.6,4.4){$\dsp{\frac{U_{k_f}}{\varphi_{0R}^6}}$}
\put(6.4,-0.5){$\dsp{\frac{\varphi_R}{\varphi_{0R}}}$}
\put(-0.5,0.){
\rotate[r]{\epsfig{file=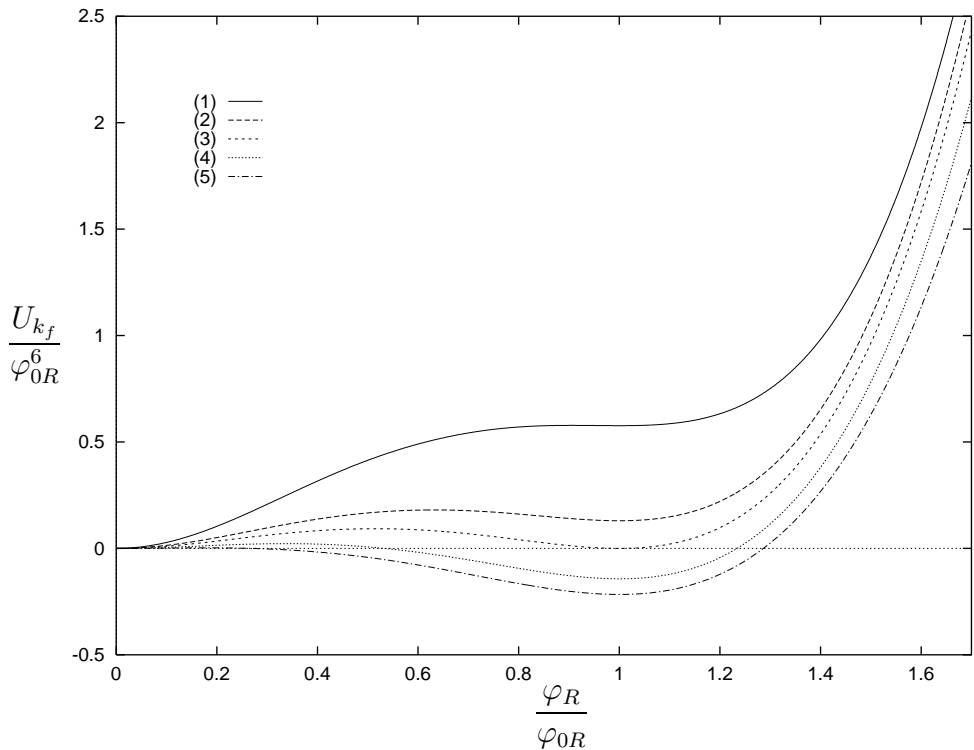,width=9.cm,height=13.cm}}
}
\end{picture}
\end{center}
\vspace{0.3cm}
\caption{\em The effective average potential
$U_{k=k_f}$ as a function of the renormalized field 
$\varphi_R$. The potential
is shown for various values of $\delta \kappa_{\Lambda} \sim T_c-T$.
The parameters for the short distance potential $U_{\Lambda}$ are
(1) $\delta \kappa_{\Lambda}=-0.03$, (2) $\delta \kappa_{\Lambda}=-0.015$,
(3) $\delta \kappa_{\Lambda}=0$, (4) $\delta \kappa_{\Lambda}=0.04$,
(5) $\delta \kappa_{\Lambda}=0.1$ and
$\la_{1\Lambda}=0.1$, $\la_{2\Lambda}=2$.
\label{tempot}
}
\end{figure}
The scale $k_f$ is some 
characteristic scale 
below which the location of the minimum $\rho_0(k)$
becomes essentially independent of $k$.  
Its precise definition is given below. We have normalized 
$U_{k_f}$ and $\varphi_R$
to powers of the renormalized minimum 
$\varphi_{0R}(k_f)= (\rho_{0R}(k_f)/2)^{1/2}$ with
$\rho_{0R}(k_f)=Z_{k_f}\rho_0(k_f)$.
The potential
is shown for various values of deviations from the critical
temperature or $\delta\kappa_{\Lambda}$. 
For the given examples $\delta \kappa_{\Lambda}=-0.03$, $-0.015$
the minimum at the origin becomes
the absolute minimum and the system is in the
symmetric (disordered) phase. Here $\varphi_{0R}$
denotes the minimum in the metastable ordered phase.
In contrast, for $\delta \kappa_{\Lambda}=0.04$, $0.1$ the 
absolute minimum is located at $\varphi_R/\varphi_{0R}=1$ which
characterizes the spontaneously broken phase.
For large enough $\delta \kappa_{\Lambda}$ the local minimum at the 
origin vanishes.
For $\delta\kappa_{\Lambda}=0$ the two distinct minima are degenerate in 
height\footnote{
We note that the critical 
temperature is determined by condition (\ref{critcon})
in the limit $k \to 0$. Nevertheless for the employed
nonvanishing scale $k=k_f$ the minima of $U_k$
become almost degenerate at the critical temperature.
}.
As a consequence the order parameter makes a discontinuous jump
at the phase transition which characterizes the transition to be 
first order.
It is instructive to consider some characteristic values
of the effective average potential.
In fig.\ \ref{temp} we consider for
$\la_{1\Lambda}=0.1,\la_{2\Lambda}=2$ the value of the renormalized
minimum $\rho_{0R}(k_f)$ and the radial mass term 
as a function of $-\delta\kappa_{\Lambda}$ or temperature.
\begin{figure}[h]
\unitlength1.0cm
\begin{center}
\begin{picture}(13.,9.)
\put(4.5,3.5){\footnotesize $\dsp{\frac{m_R}{\Lambda}}$}
\put(4.5,7.){\footnotesize $\dsp{\frac{\rho_{0R}}{\Lambda}}$}
\put(6.4,-0.5){$-\delta\kappa_{\Lambda}$}
\put(-0.5,0.){
\rotate[r]{\epsfig{file=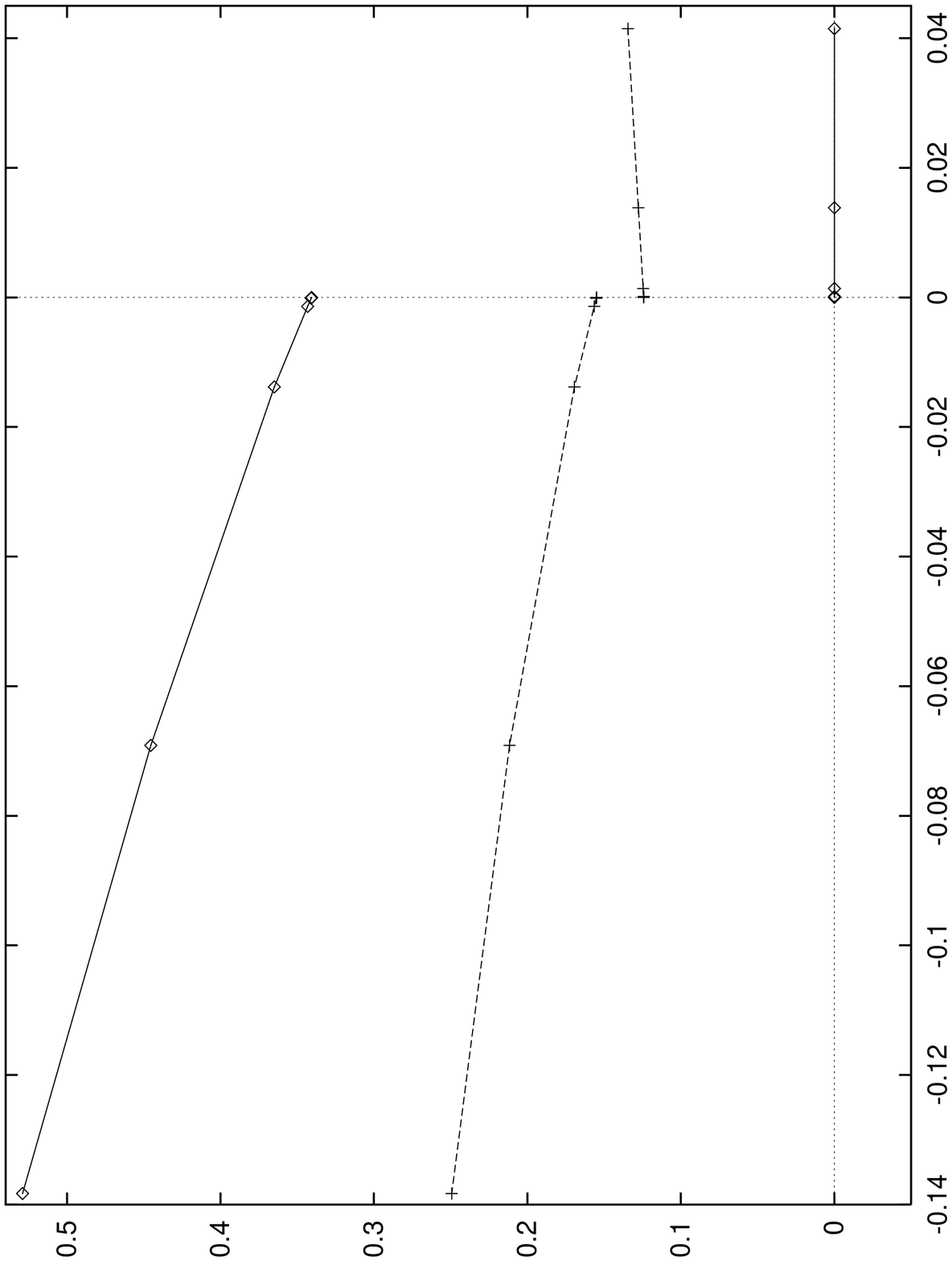,width=9.cm,height=13.cm}}
}
\end{picture}
\end{center}
\caption[]{\label{temp} 
\em The
minimum $\rho_{0R}$ 
and the mass term 
$m_R$ in units of the momentum scale $\Lambda$
as a function of $-\delta\kappa_{\Lambda}$ or temperature
($\la_{1\Lambda}=0.1$, $\la_{2\Lambda}=2$, $k=k_f$). For 
$\delta\kappa_{\Lambda}=0$ one observes the jump in the
renormalized order parameter $\Delta \rho_{0R}$
and mass $\Delta m_R$. 
}
\end{figure} 
In the spontaneously
broken phase the renormalized radial mass squared is given by
\beq
m^2_R(k_f)=2 Z_{k_f}^{-1} \rho_0 U^{\prime\prime}_{k_f}(\rho_0),
\label{radialmass}
\eeq 
whereas in the symmetric phase the renormalized mass term reads
\beq
m^2_{0R}(k_f)=Z_{k_f}^{-1} U^{\prime}_{k_f}(0).
\label{originmass}
\eeq
At the critical temperature ($\delta\kappa_{\Lambda}=0$) one observes
the discontinuity $\Delta \rho_{0R}=\rho_{0R}(k_f)$ and the jump
in the mass term $\Delta m_R=m_R(k_f)-m_{0R}(k_f)
=m_R^c-m_{0R}^c$. (Here the index ``c'' denotes
$\delta\kappa_{\Lambda}=0$).
The ratio $\Delta \rho_{0R}/\Lambda$ is a rough measure for 
the ``strength'' of the first order transition.
For $\Delta \rho_{0R}/\Lambda \ll 1$ the phase transition is
weak in the sense that typical masses are small compared to
$\Lambda$. In consequence, the long-wavelength fluctuations
play a dominant role and the system exhibits universal 
behavior, i.e.\ it becomes largely independent of the
details at the short distance scale $\Lambda^{-1}$.
We will discuss the universal behavior in more 
detail below.     

In order to characterize the strength of 
the phase transition for 
arbitrary positive values of $\la_{1\Lambda}$ and $\la_{2\Lambda}$
we consider lines of constant $\Delta \rho_{0R}/\Lambda$ in the
$\la_{1\Lambda} , \la_{2\Lambda}$ plane. In fig.\ \ref{la1la2} this is
done for the logarithms of these quantities. 
\begin{figure}[h]
\unitlength1.0cm
\begin{center}
\begin{picture}(13.,9.)
\put(-0.9,4.6){$\dsp{\ln\left(\la_{2\Lambda}\right)}$}
\put(6.4,-0.5){$\dsp{\ln\left(\la_{1\Lambda}\right)}$}
\put(9.5,6.5){\footnotesize $(1)$}
\put(9.5,5.4){\footnotesize $(2)$}
\put(9.5,3.){\footnotesize $(3)$}
\put(9.5,1.5){\footnotesize $(4)$}
\put(-0.5,0.){
\rotate[r]{\epsfig{file=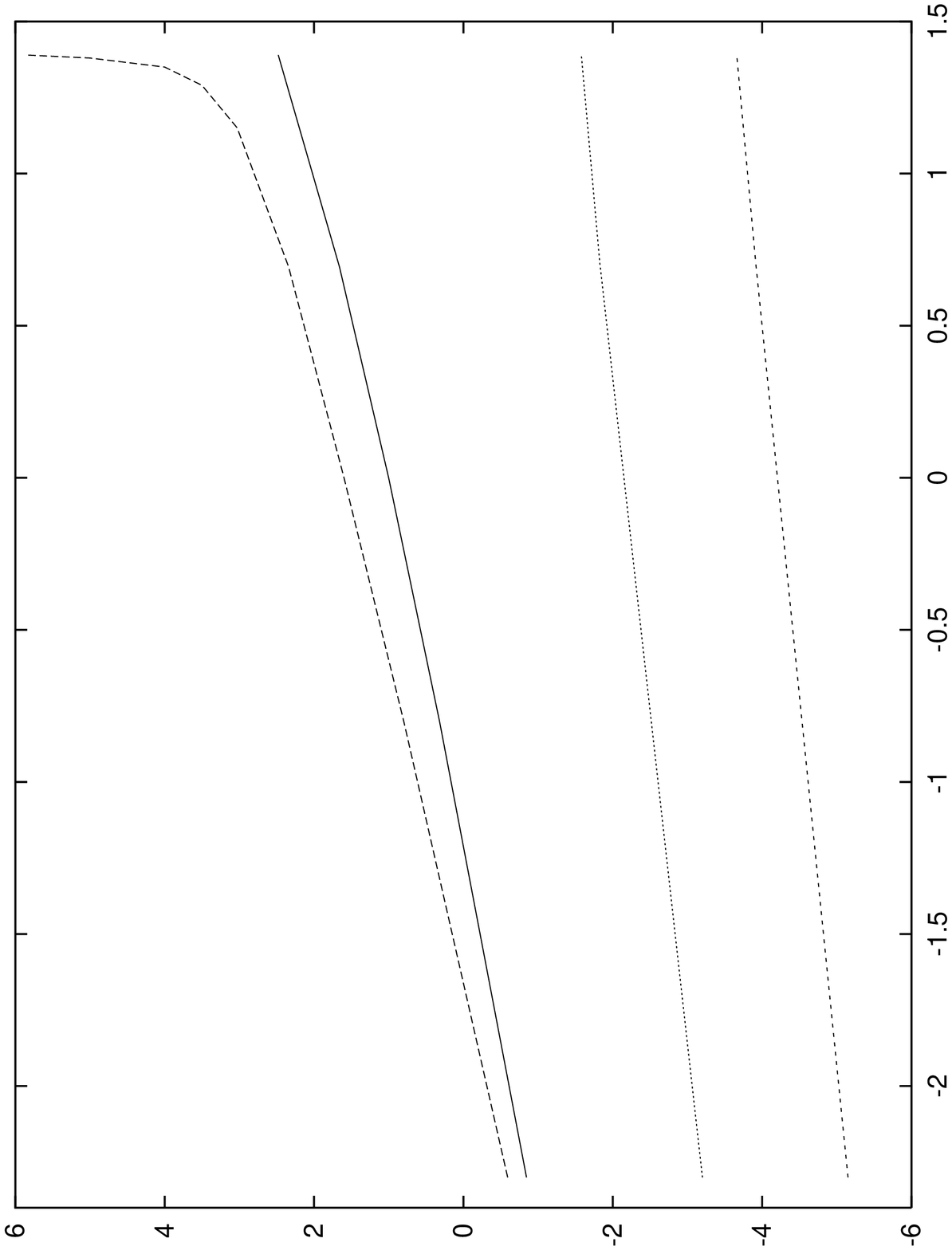,width=9.cm,height=13.cm}}
}
\end{picture}
\end{center}
\caption[]{\label{la1la2} \em
Lines of constant jump of the 
renormalized order parameter
$\Delta \rho_{0R}/\Lambda$ at the phase transition in the
$\ln(\la_{1\Lambda}) , \ln(\la_{2\Lambda})$ plane. The curves
correspond to (1) $\ln(\Delta \rho_{0R}/\Lambda)=-4.0$,
(2) $\ln(\Delta \rho_{0R}/\Lambda)=-4.4$,
(3) $\ln(\Delta \rho_{0R}/\Lambda)=-10.2$,
(4) $\ln(\Delta \rho_{0R}/\Lambda)=-14.3$.    
}
\end{figure}
For fixed
$\la_{2\Lambda}$ one observes that the discontinuity
at the phase transition weakens with increased $\la_{1\Lambda}$.
On the other hand for given $\la_{1\Lambda}$ one finds a larger jump in the
order parameter for increased $\la_{2\Lambda}$. This is true up to a 
saturation point where $\Delta \rho_{0R}/\Lambda$ becomes independent
of $\la_{2\Lambda}$. In the plot this can be observed from the
vertical part of the line of constant $\ln(\Delta \rho_{0R}/\Lambda)$.
This phenomenon occurs for arbitrary nonvanishing   
$\Delta \rho_{0R}/\Lambda$ in the strong $\la_{2\Lambda}$ coupling limit
as discussed in section \ref{rg}. 

In the following we give a detailed quantitative
description of the first order phase
transitions and a separation in weak and strong transitions.
We consider some characteristic quantities for the effective average 
potential in dependence on the short distance parameters 
$\la_{1\Lambda}$ and
$\la_{2\Lambda}$ for $\delta \kappa_{\Lambda}=0$.  
We consider the discontinuity
in the renormalized order parameter $\Delta \rho_{0R}$ 
and the inverse correlation
lengths (mass terms) 
$m_R^c$ and $m_{0R}^c$ in the ordered and the 
disordered phase respectively. 
Fig.\ \ref{phase}
shows the logarithm of  
$\Delta \rho_{0 R}$ in units of 
$\Lambda$ as a function of the logarithm of the initial 
coupling $\la_{2\Lambda}$. 
\begin{figure}[h]
\unitlength1.0cm
\begin{center}
\begin{picture}(13.,9.)
\put(-1.2,4.6){$\dsp{\ln\left(\frac{\Delta \rho_{0R}}{\Lambda}\right)}$}
\put(6.2,-0.5){$\dsp{\ln\left(\la_{2\Lambda}\right)}$}
\put(8.9,8.5){\footnotesize $(1)$}
\put(10.5,7.7){\footnotesize $(2)$}
\put(10.5,6.7){\footnotesize $(3)$}
\put(-0.2,0.){
\rotate[r]{\epsfig{file=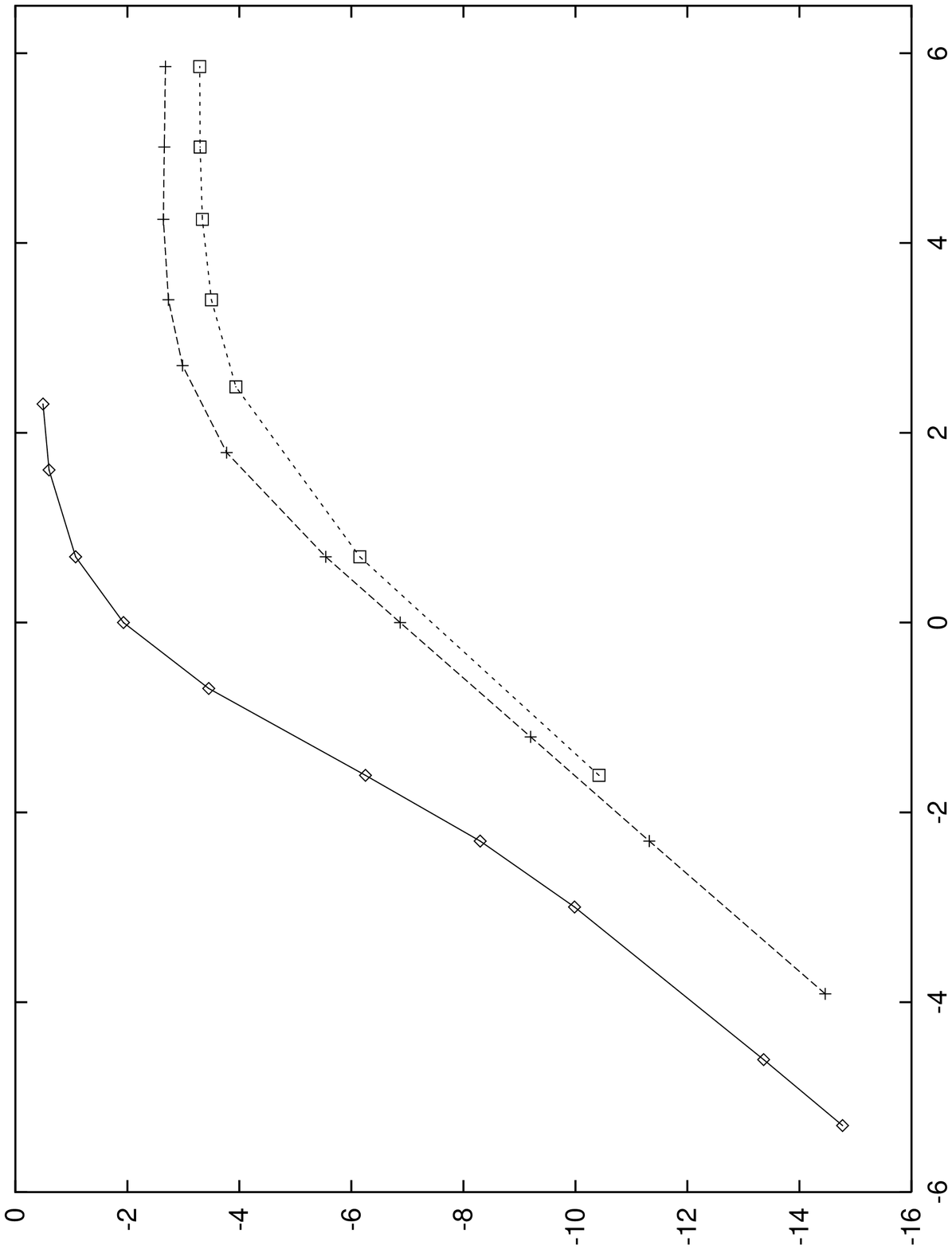,width=9.cm,height=13.cm}}
}
\end{picture}
\end{center}
\caption[]{\em 
Strength of the first order phase transition. We plot the
 logarithm of the discontinuity 
of the renormalized order parameter
$\Delta \rho_{0 R}/\Lambda$ 
as a function of $\ln(\la_{2\Lambda})$.
Data points
for fixed (1) $\la_{1\Lambda}=0.1$,
(2) $\la_{1\Lambda}=2$, (3) $\la_{1\Lambda}=4$ are
connected by straight lines. One observes a universal slope for small
 $\lambda_{2\Lambda}$ which is related to a critical exponent.
\label{phase}
}
\end{figure} 
We have connected the calculated values
obtained for various fixed $\la_{1\Lambda}=0.1$, 
$2$ and $\la_{1\Lambda}=4$
by straight lines.

For 
$\la_{2\Lambda}/\la_{1\Lambda} \lesssim 1$ the curves show constant 
positive slope. In this range $\Delta \rho_{0 R}$ follows a 
power law behavior
\beq
\Delta \rho_{0 R}=R \, (\la_{2\Lambda})^{\th}, \quad  \th=1.93 
\label{Powrho}.
\eeq
The critical exponent $\th$ is obtained from the slope
of the curve in fig.\ \ref{phase} for 
$\la_{2\Lambda}/\la_{1\Lambda}\ll 1$. The exponent is
universal and, therefore, does not depend on the specific 
value for $\la_{1\Lambda}$.
On the other hand, the amplitude $R$ grows with
decreasing $\la_{1\Lambda}$. For vanishing $\la_{2\Lambda}$
the order parameter changes continuously at the transition
point and one observes a second order phase transition
as expected for the $O(8)$ symmetric vector model. As 
$\la_{2\Lambda}/\la_{1\Lambda}$ becomes larger than one the
curves deviate substantially
from the linear behavior. The deviation
depends on the specific choice of the short distance
potential. For $\la_{2\Lambda}/\la_{1\Lambda}\gg 1$ the curves
flatten. In this range
$\Delta \rho_{0 R}$ becomes insensitive to a variation
of the quartic coupling $\la_{2\Lambda}$.

In addition to the jump in the order parameter 
we present the mass terms $m_R^c$ 
and $m_{0R}^c$ which we
normalize to $\Delta \rho_{0 R}$.
In fig.\ \ref{ratio} these ratios are
plotted versus the logarithm of the ratio of the initial
quartic couplings $\la_{2\Lambda}/\la_{1\Lambda}$. 
\begin{figure}[h]
\unitlength1.0cm
\begin{center}
\begin{picture}(13.,9.)
\put(5.5,-0.8){$\dsp{\ln\left(\frac{\la_{2 \Lambda}}
{\la_{1 \Lambda}}\right)}$}
\put(10.6,1.3){\footnotesize $(1)$}
\put(10.6,8.28){\footnotesize $(2)$}
\put(10.6,7.55){\footnotesize $(3)$}
\put(10.6,6.65){\footnotesize $(2)$}
\put(10.6,5.7){\footnotesize $(3)$}
\put(2.2,6.05){\footnotesize $\dsp{\frac{m_R^c}
{\Delta \rho_{0R}}}$}
\put(2.2,4.1){\footnotesize $\dsp{\frac{m_{0R}^c}
{\Delta \rho_{0R}}}$}
\put(-0.5,0.){
\rotate[r]{\epsfig{file=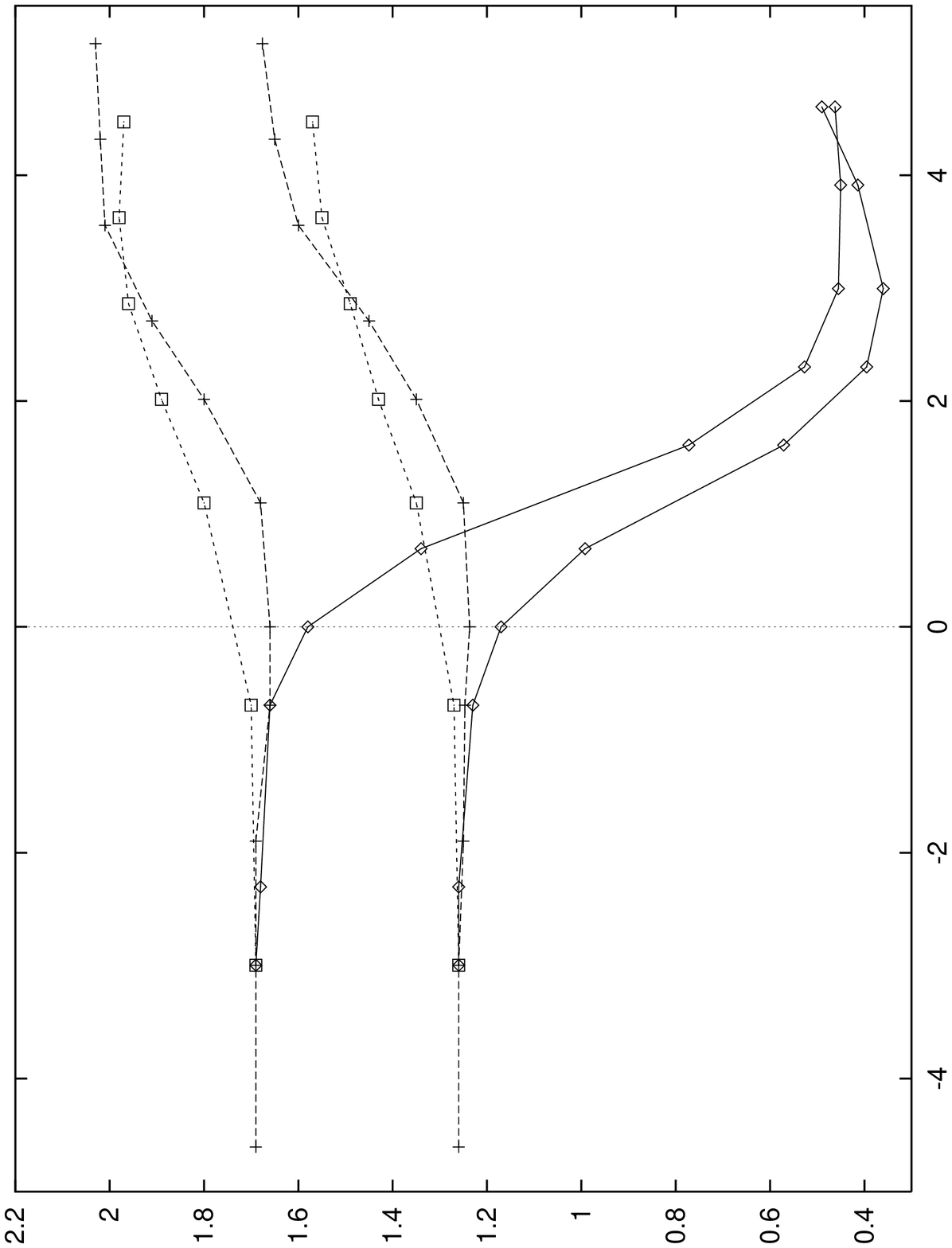,width=9.cm,height=13.cm}}
}
\end{picture}
\end{center}
\vspace{0.5cm}
\caption[]{\label{ratio} \em
The inverse correlation
lengths $m_R^c$ and $m_{0R}^c$ in the ordered and the 
disordered phase respectively. They are normalized to
$\Delta \rho_{0R}$ and given as a function
of $\ln(\la_{2\Lambda}/\la_{1\Lambda})$. Data points
for fixed (1) $\la_{1\Lambda}=0.1$,
(2) $\la_{1\Lambda}=2$, (3) $\la_{1\Lambda}=4$ are
connected by straight lines. One observes universality for small ratios
 $\lambda_{2\Lambda}/\lambda_{1\Lambda}$.
}
\end{figure}
Again values
obtained for fixed $\la_{1\Lambda}=0.1$, $2$ and $\la_{1\Lambda}=4$
are connected by straight lines. The universal range is set
by the condition $m_R^c/ \Delta \rho_{0 R} \simeq \mbox{const}$
(equivalently for $m_{0R}^c/ \Delta \rho_{0 R}$).  
The universal ratios are 
$m_R^c/ \Delta \rho_{0 R}=1.69$
and $m_{0R}^c/ \Delta \rho_{0 R}=1.26$. 
For the given curves universality holds 
approximately for
$\la_{2\Lambda}/\la_{1\Lambda} \lesssim 0.5$ 
and becomes ``exact'' in the limit 
$\la_{2\Lambda}/\la_{1\Lambda} \to 0$. In this range we obtain
\beq
m_R^c = S (\la_{2\Lambda})^{\th},\qquad 
m_{0R}^c = \tilde{S} (\la_{2\Lambda})^{\th} 
\label{Powm}.
\eeq
The 
universal features of the system are not restricted to the weak coupling 
region of $\la_{2\Lambda}$. This is demonstrated in fig.\ \ref{ratio}
for values up to  $\la_{2\Lambda} \simeq 2$. 
The ratios $m_R^c/ \Delta \rho_{0 R}$ and $m_{0R}^c/ \Delta \rho_{0 R}$
deviate from the universal
values as $\la_{2\Lambda}/\la_{1\Lambda}$ is increased. For fixed
$\la_{2\Lambda}$
a larger $\la_{1\Lambda}$ results in a weaker transition
concerning $\Delta \rho_{0 R}/\Lambda$.
The ratio $m_R^c/ \Delta \rho_{0 R}$ increases with $\la_{1\Lambda}$
for small fixed $\la_{2\Lambda}$ whereas in the asymptotic
region, $\la_{2\Lambda}/\la_{1\Lambda} \gg 1$, one observes from
fig.\ \ref{ratio} that this tendency is reversed 
and $m_R^c/ \Delta \rho_{0 R}$,
$m_{0R}^c/ \Delta \rho_{0 R}$ start to decrease at about 
$\la_{1\Lambda} \simeq 2$.  

In summary,
the above results show that though the short distance 
potential $U_{\Lambda}$ indicates a second order phase transition,
the transition becomes first order once fluctuations are taken into
account. This fluctuation induced first order phase transition
is known in four dimensions as the Coleman-Weinberg phenomenon
\cite{Col73}. 
The question of the order of the phase transition of the three dimensional 
$U(2) \times U(2)$ symmetric model has been addressed also using
the $\eps$-expansion \cite{Pisa84,PW84-1},
in lattice studies \cite{Dreher92Shen94} and in
high-temperature expansion \cite{Khl95}. All studies are 
consistent with the first order nature of the 
transition and with the absence of non-perturbative infrared
stable fixed points. Our method
gives here a clear and unambiguous answer and allows
a detailed quantitative
description of the phase transition. The universal form
of the equation of state for weak first order phase
transitions is presented in section \ref{sce}.

In the following we specify the scale $k_f$ 
for which we have given
the effective average potential $U_{k}$.
We observe that $U_k$ depends
strongly on the infrared cutoff $k$ as long as $k$ is larger
than the scale $k_2$ where 
the second minimum of the potential
appears. Below $k_2$ the two minima start to become
almost degenerate for $T$ near $T_{c}$ and the running
of $\rho_0(k)$ stops rather soon. The
nonvanishing value of $k_2$ induces
a physical infrared cutoff and represents a characteristic
scale for the first order phase transition.  
We stop the integration of the flow equation
for the effective average potential at a scale
$k_f < k_2$ which is determined in terms of the 
curvature (mass term) at the top
of the potential barrier that separates the two local minima
of $U_k$ at the origin and at $\rho_{0}(k)$.
The top of the potential barrier at
$\rho_B(k)$ is determined
by
\beq
U_k^{\prime}(\rho_B)=0 \label{bar}
\eeq
for $0 < \rho_B(k) < \rho_0(k)$ and for the 
renormalized mass term at 
$\rho_B(k)$ one obtains 
\beq
m_{B,R}^2(k)=2 Z_k^{-1} \rho_B U_k^{\prime\prime}(\rho_B) < 0. 
\label{mbr}
\eeq  
We fix our final value for the running by 
\beq
\dsp{\frac{k_f^2-|m_{B,R}^2(k_f)|}{k_f^2}}=0.01
\label{fix}
\eeq 
For this choice the coarse-grained effective potential $U_{k_f}$
essentially includes all fluctuations with momenta larger than the
mass $|m_{B,R}|$ at the top of the potential barrier. It is a
nonconvex function which is the appropriate quantity for the 
study of physical
processes such as tunneling or inflation.\\

\subsection{Universal equation of state for 
weak first order phase
transitions \label{sce}}

We presented in section \ref{ps} some characteristic
quantities for the effective average potential which
become universal at the phase transition 
for a sufficiently small 
quartic coupling $\lambda_{2\Lambda}=\bar{\lambda}_{2\Lambda}/\Lambda$
of the short distance potential $U_{\Lambda}$ (\ref{uinitial}). 
The aim of this section is to generalize this observation
and to find a universal scaling form of 
the equation of state for weak first order phase transitions.
The equation of state relates the derivative of the free energy
$U=\lim_{k\to 0}U_k$ to an external source, 
$\partial U/\partial \varphi = j$.
Here the derivative has to be evaluated in the outer
convex region of the potential. For instance, for the meson model
of strong interactions the source $j$ is proportional
to the average quark mass \cite{PW84-1,QuMa} and the 
equation of state
permits to study the quark mass dependence of properties of the
chiral phase transition. We will compute the equation
of state for a nonzero coarse graining scale $k$. It 
therefore contains information for quantities
like the ``classical'' bubble surface tension in the
context of Langer's theory of bubble formation which will be
discussed in section \ref{CoarseGrain}.
  
In three dimensions the $U(2) \times U(2)$ symmetric 
model exhibits a second order phase transition in
the limit of a vanishing quartic coupling ${\lambda}_{2\Lambda}$ due to
an enhanced $O(8)$ symmetry. In this case  
there is no scale present in the theory at the critical
temperature. 
In the vicinity of the critical temperature
(small $|\delta \kappa_{\Lambda}| \sim |T_c-T|$) and for
small enough $\lambda_{2\Lambda}$
one therefore expects a scaling behavior of
the effective average potential $U_k$ and
accordingly a universal scaling form of the equation of state. 
At the second order phase transition in the 
$O(8)$ symmetric model there are only two independent
scales corresponding to the deviation from
the critical temperature and to the external source
or $\varphi$. 
As a consequence the properly rescaled potential $U/ \rho_R^{3}$
or $U/ \rho^{(\delta +1)/2}$ (with the usual critical exponent
$\delta$) can only depend on one dimensionless ratio.
A possible set of variables
to represent the two independent scales are the renormalized
minimum of the potential $\varphi_{0R}=(\rho_{0R}/2)^{1/2}$ 
(or the renormalized mass for the symmetric phase) and
the renormalized field $\varphi_R=(\rho_{R}/2)^{1/2}$. The 
rescaled potential will then only depend on the 
scaling variable $\tilde s=\varphi_R/\varphi_{0R}$ \cite{BTW96-1}. Another
possible choice is the Widom scaling variable
$x=-\delta \kappa_{\Lambda}/\varphi^{1/\beta}$ \cite{Widom}.  
In the $U(2) \times U(2)$ symmetric 
theory $\lambda_{2\Lambda}$ is an additional relevant
parameter which renders the phase transition first
order and introduces a new scale, e.g.\ the nonvanishing
value for the jump in the renormalized order parameter
$\Delta \varphi_{0R}=(\Delta \rho_{0R}/2)^{1/2}$ at the critical 
temperature or $\delta \kappa_{\Lambda}=0$.
In the universal range we therefore observe three
independent scales and the scaling form of the equation
of state will depend on two dimensionless ratios. 

The rescaled potential
$U/\varphi_{0R}^6$ can then be written as a universal function $G$
\beq
\frac{U}{\varphi_{0R}^6}=G(\tilde s,\tilde v)
\label{scalingeos}
\eeq
which depends on the two scaling variables
\beq
\tilde s=\frac{\varphi_R}{\varphi_{0R}},\quad \tilde v= \frac{\Delta
 \varphi_{0R}}{\varphi_{0R}}\,\,.
\eeq
The relation (\ref{scalingeos}) is the scaling 
form of the equation of
state we are looking for. At a second order phase
transition the variable $\tilde v$ vanishes and 
$G(\tilde s,0)$ describes the scaling equation of state
for the model with $O(8)$ symmetry \cite{BTW96-1}.
The variable $\tilde v$ accounts for the additional scale 
present at the first order phase transition.
We note that $\tilde s=1$ corresponds to a vanishing source
and $G(1,\tilde v)$ describes the temperature dependence
of the free energy for $j=0$. In this case $\tilde v=1$ denotes the 
critical temperature $T_c$ whereas for $T < T_c$
one has $\tilde v < 1$. Accordingly $\tilde v > 1$ is obtained for
$T > T_c$ and $\varphi_{0R}$ describes here the local 
minimum corresponding to the metastable ordered phase. 
The function $G(\tilde s,1)$ accounts for the
dependence of the free energy on $j$ for $T=T_c$.

We consider the scaling form (\ref{scalingeos}) of the
equation of state for a nonzero coarse graining scale $k$
with  renormalized field given by $\varphi_R=Z_k^{1/2} \varphi$.
As we have  pointed out in section \ref{ps}  there is a
characteristic scale $k_2$ for the first order phase
transition where the second local minimum of the effective
average potential appears. For weak first order phase
transitions one finds $\rho_{0R} \sim k_2$. To observe
the scaling form of the equation of state the infrared cutoff
$k$ has to run below $k_2$ with $k \ll k_2$. For
the scale $k_f$ defined in eq.\ (\ref{fix})
we observe universal behavior to high
accuracy (cf.\ fig.\ \ref{ratio} for small 
$\lambda_{2\Lambda}/\lambda_{1\Lambda}$).
The result for the universal function 
$U_{k_f}/\varphi_{0R}^6=G_{k_f}(\tilde s,\tilde v)$ is presented in fig.\
\ref{scalfu}. 
\begin{figure}[h]
\unitlength1.0cm
\begin{center}
\begin{picture}(13.,9.)
\put(-0.6,4.4){$G_{k_f}$}
\put(6.4,-0.5){$\tilde s$}
\put(-0.5,0.){
\rotate[r]{\epsfig{file=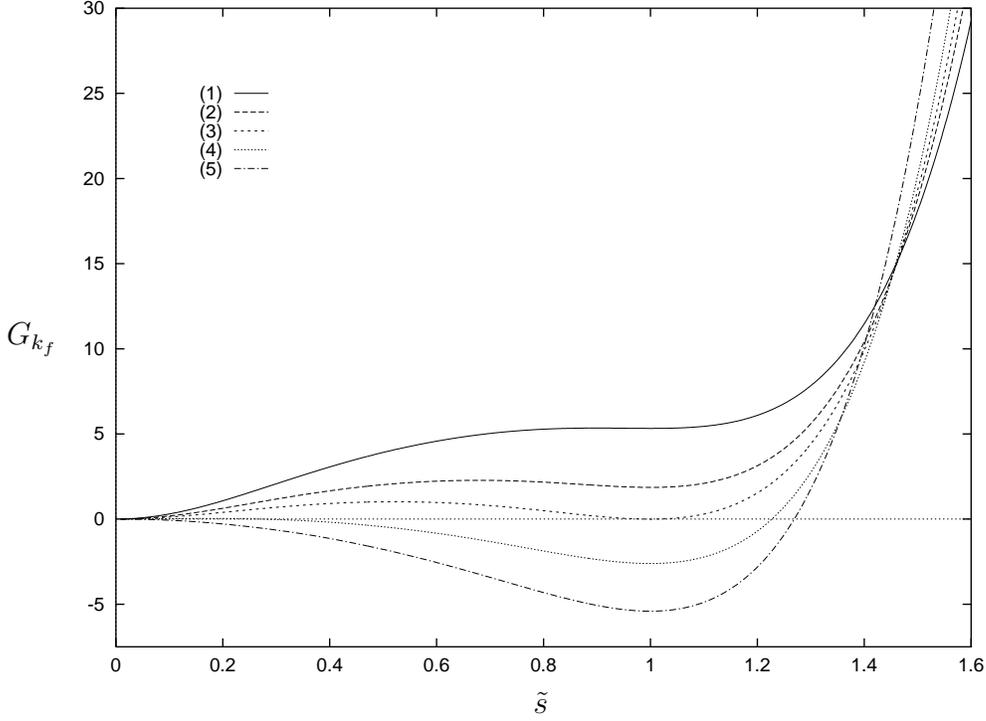,width=9.cm,height=13.cm}}
}
\end{picture}
\end{center}
\caption[]{\label{scalfu} \em
Universal shape of the coarse-grained potential $(k=k_f)$
as a function of the scaling variable $\tilde s=\varphi_R/\varphi_{0R}
=(\rho_R/\rho_{0R})^{1/2}$ for different values of
$\tilde v=\Delta \varphi_{0R}/\varphi_{0R}=(\Delta 
\rho_{0R}/\rho_{0R})^{1/2}$.
The employed values for $\tilde v$ are (1) $\tilde v=1.18$,
(2) $\tilde v=1.07$, (3) $\tilde v=1$, (4) $\tilde v=0.90$, (5) 
$\tilde v=0.74$. 
For vanishing sources one has $\tilde s=1$. In this case $\tilde v=1$
denotes the critical temperature $T_c$. Similarly
$\tilde v > 1$ corresponds to $T > T_c$ with $\varphi_{0R}$
denoting the minimum in the metastable ordered phase.
The function $G$ is symmetric for $\tilde s\to -\tilde s$ and one notes
the qualitative difference with fig. 12.
}
\end{figure}
For $\tilde v=1$ one has $\varphi_{0R}(k_f)=\Delta \varphi_{0R}(k_f)$ 
which denotes the critical temperature.
Accordingly $\tilde v > 1$ denotes temperatures above 
and $\tilde v < 1$ temperatures below the critical temperature.
One observes that $G_{k_f}(\tilde s,1)$ 
shows two almost degenerate minima. (They become exactly
degenerate in the limit $k \to 0$).
For the given examples $\tilde v=1.18$, $1.07$
the minimum at the origin becomes
the absolute minimum and the system is in the
symmetric phase. In contrast, for $\tilde v = 0.90$, $0.74$ the 
absolute minimum is located at $\tilde s=1$ which
characterizes the spontaneously broken phase.
For small enough $\tilde v$ the local minimum at the 
origin vanishes.

We have explicitly verified that the universal
function $G_{k_f}$ depends only 
on the scaling variables $\tilde s$ and $\tilde v$ by choosing various
values for $\delta \kappa_{\Lambda}$ and for the quartic couplings of
the short distance potential, $\lambda_{1\Lambda}$ and $\lambda_{2\Lambda}$.
We have seen in section \ref{ps}  that the model shows
universal behavior for a certain range of the parameter
space. For given $\lambda_{1\Lambda}$ and small enough 
$\lambda_{2\Lambda}$ one always observes universal behavior.
For $\lambda_{1\Lambda}=0.1$, $2$ and $4$ it is 
demonstrated that (approximate) universality holds
for $\lambda_{2\Lambda}/ \lambda_{1\Lambda} \lesssim 1/2$ 
(cf.\ fig.\ \ref{ratio}). 
For $\lambda_{1\Lambda}$ around $2$ one observes 
from figs.\ \ref{phase}, \ref{ratio} that
the system is to a good accuracy described by
its universal properties for even larger values 
of $\lambda_{2\Lambda}$. The corresponding phase transitions
cannot be considered as particularly weak first order. 
The universal function $G_{k_f}$
therefore accounts 
for a quite large range of the parameter space.

We emphasize that the universal form of the 
effective potential given in fig.\ \ref{scalfu} depends on
the scale $k_f$ where the integration of the flow 
equations is stopped (cf.\ eq.\ (\ref{fix})).  
A different prescription
for $k_f$ will, in general, 
lead to a different
form of the effective potential in the ``inner region''. We may interpret
this as a scheme dependence describing the effect of 
different coarse graining procedures. This is fundamentally
different from non-universal corrections since $G_{k_f}$
is independent of details of the short distance
or classical action and in this sense universal.
Also the ``outer region'' for $\tilde s \ge 1$ 
is not affected by the approach to convexity and
becomes independent of the choice of $k_f$.
A more quantitative discussion of this scheme
dependence will be presented in section \ref{CoarseGrain}.
Since  fluctuations on scales 
$k < k_f$ do not influence substantially the location of
the minima of the coarse-grained potential and the
form of $U_k(\varphi_R)$ for $\varphi_R > \varphi_{0R}$ one has
$\partial U_{k_f}/\partial \varphi=j(k_f)$ with
$j(k_f) \approx \lim_{k \to 0} j(k) = j$.\footnote{
The role of massless
Goldstone boson fluctuations for the 
universal form of the effective average potential
in the limit $k \to 0$ has been discussed previously
for the $O(8)$ symmetric model \cite{BTW96-1}.}

Let us consider the
renormalized minimum $\varphi_{0R}$ in two limits which are
denoted by  $\Delta \varphi_{0R} = \varphi_{0R}(\delta \kappa_{\Lambda}=0)$
and $\varphi^0_{0R} = \varphi_{0R}(\lambda_{2\Lambda}=0)$.  
The behavior of $\Delta \varphi_{0R}$ is described in
terms of the exponent $\theta$ according to
eq.\ (\ref{Powrho}),
\beq
\Delta \varphi_{0R} \sim (\lambda_{2\Lambda})^{\theta/2}, \qquad \theta=1.93.
\label{scath}
\eeq
The dependence of the minimum $\varphi^0_{0R}$ of the
$O(8)$ symmetric potential on the temperature is 
characterized by the critical exponent $\nu$,
\beq
\varphi^0_{0R} \sim (\delta \kappa_{\Lambda})^{\nu/2}, \qquad \nu=0.882.
\label{scanu}
\eeq
The exponent $\nu$ for the $O(8)$ symmetric model
is determined analogously to $\theta$ as described in section
\ref{phase} \footnote{For the $O(8)$ symmetric model
($\lambda_{2\Lambda}=0)$ we consider the 
minimum $\varphi^0_{0R}$ at $k=0$.}. 
We can also introduce a critical exponent $\zeta$ for the
jump of the unrenormalized order parameter
\beq
\Delta \varphi_{0} \sim (\lambda_{2\Lambda})^{\zeta}, 
\qquad \zeta=0.988 \, .
\eeq
With
\beq
\varphi^0_{0} \sim (\delta \kappa_{\Lambda})^{\beta}, 
\qquad \beta=0.451
\eeq
it is related to $\th$ and $\nu$ by the additional index
relation
\beq
\frac{\theta}{\zeta}=\frac{\nu}{\beta}=1.95 \, . 
\eeq   
We have verified this numerically. 
For the case $\delta \kappa_{\Lambda}=\lambda_{2\Lambda}=0$ one obtains
\beq
j \sim \varphi^{\delta}.
\eeq
The exponent $\delta$ is related to the 
anomalous dimension $\eta$ via the usual
index relation $\delta=(5-\eta)/(1+\eta)$. From the
scaling solution of eq.\ (\ref{EtaSkale}) we
obtain $\eta=0.0224$.

With the help of the above relations one immediately
verifies that for $\lambda_{2\Lambda}=0$ 
\beq
\tilde s \sim (-x)^{-\beta}\, , \qquad \tilde v=0
\eeq
and for $\delta \kappa_{\Lambda}=0$
\beq
\tilde s \sim y^{-\zeta} \, , \qquad \tilde v=1 \, .
\eeq
Here we have used the Widom scaling variable $x$ and the 
new scaling variable $y$ given by
\beq
x=\frac{-\delta \kappa_{\Lambda}}{\varphi^{1/\beta}}\, , 
\qquad y=\frac{\lambda_{2\Lambda}}{\varphi^{1/\zeta}}\, .
\eeq

\subsection{Summary \label{con}}

We have presented a detailed investigation
of the phase transition in three dimensional models for 
complex $2 \times 2 $ matrices. They are 
characterized by two quartic couplings
$\bar{\lambda}_{1\Lambda}$ and $\bar{\lambda}_{2\Lambda}$. 
In the limit
$\bar{\lambda}_{1\Lambda}\to \infty$, $\bar{\lambda}_{2\Lambda}\to \infty$
this also covers the model of unitary matrices.
The picture arising from this study is unambiguous:

(1) One 
observes two symmetry breaking patterns for
$\bar{\lambda}_{2\Lambda}>0$ and $\bar{\lambda}_{2\Lambda}<0$
respectively. The case $\bar{\lambda}_{2\Lambda}=0$
denotes the boundary between the two phases
with different symmetry breaking patterns. 
In this special case the theory
exhibits an enhanced $O(8)$ symmetry.  
The phase transition is always first order
for the investigated symmetry breaking
$U(2) \times U(2) \to U(2)$ ($\bar{\lambda}_{2 \Lambda}>0$). 
For $\bar{\lambda}_{2 \Lambda}=0$ 
the $O(8)$ symmetric Heisenberg model is recovered
and one finds a second order phase transition.

(2) The strength of the phase transition depends on the size
of the classical quartic couplings $\bar{\lambda}_{1\Lambda}/\Lambda$ and  
$\bar{\lambda}_{2\Lambda}/\Lambda$. They describe the short distance or
classical action at a momentum scale $\Lambda$. The strength of the
transition can be parametrized by $m_R^c/\Lambda$ with $m_R^c$ 
a characteristic inverse correlation length at the 
critical temperature.
For fixed $\bar{\lambda}_{2\Lambda}$ the strength of the
transition decreases with increasing $\bar{\lambda}_{1\Lambda}$.
This is analogous to the Coleman-Weinberg effect in 
four dimensions.

(3) For a wide range of classical couplings the critical
behavior near the phase transition is universal.
This means
that it becomes largely independent of the details of the 
classical action once everything is expressed in terms of the
relevant renormalized parameters. 
In particular, characteristic ratios like
$m_R^c/\Delta \rho_{0R}$ (critical inverse correlation length
in the ordered phase over discontinuity in the order parameter)
or $m_{0R}^c/\Delta \rho_{0R}$ (same for the disordered phase)
are not influenced by the addition of
new terms in the classical action
as far as the symmetries are respected.

(4) The range of short distance parameters
$\bar{\lambda}_{1\Lambda}$, $\bar{\lambda}_{2\Lambda}$ for 
which the phase transition exhibits universal behavior
is not only determined by the strength of the phase
transition as measured by $m_R^c/\Lambda$.
For a given $\bar{\lambda}_{1\Lambda}/\Lambda$ and 
small enough $\bar{\lambda}_{2\Lambda}/\Lambda$ one always observes
universal behavior. In the range of small 
$\bar{\lambda}_{1\Lambda}/\Lambda$ the essential criterion for universal 
behavior is given by the size of 
$\bar{\lambda}_{2\Lambda}/\bar{\lambda}_{1\Lambda}$, with approximate 
universality for $\bar{\lambda}_{2\Lambda} < \bar{\lambda}_{1\Lambda}$.
For strong couplings universality extends to larger
$\bar{\lambda}_{2\Lambda}/\bar{\lambda}_{1\Lambda}$ and occurs for much larger
$m_R^c/\Lambda$.

(5) We have investigated how various characteristic 
quantities like  
the discontinuity in the order parameter
$\Delta \rho_{0}$ or the corresponding renormalized quantity
$\Delta \rho_{0R}$ or critical correlation lengths
depend on the classical parameters. 
In particular, at the critical temperature
one finds universal 
critical exponents for not too large $\bar{\lambda}_{2\Lambda}$,
\bea
\Delta \rho_{0R} &\sim& (\bar{\lambda}_{2\Lambda})^{\theta},
\qquad \theta=1.93\, ,\nnn
\Delta \rho_{0} &\sim& (\bar{\lambda}_{2\Lambda})^{2\zeta},
\qquad \zeta=0.988 \, .
\eea 
These exponents are related by a scaling relation
to the critical 
correlation length and order parameter exponents
$\nu$ and $\beta$ of the $O(8)$ symmetric 
Heisenberg model according to 
$\theta/\zeta=\nu/\beta=1.95$ ($\nu=0.882$, 
$\beta=0.451$ in our calculation for 
$\bar{\lambda}_{2\Lambda}=0$). Small values of $\bar{\lambda}_{2\Lambda}$
can be associated with a perturbation of the 
$O(8)$ symmetric model and $\theta , \zeta$ are
related to the corresponding crossover exponents.
On the other hand, $\Delta \rho_{0R}$ ($\Delta \rho_{0}$)
becomes independent of $\bar{\lambda}_{2\Lambda}$ in the 
infinite coupling limit. 

(6) We have computed the universal equation of state
for the first order transition. It depends
on two scaling variables, e.g. $\tilde s=(\rho_R/\rho_{0R})^{1/2}$
and $\tilde v =(\Delta\rho_{0R}/\rho_{0R})^{1/2}$. 
The equation of state relates the 
derivative of the free energy
$U$ to an external source, $\partial U/\partial \varphi = j$.
>From there one can extract universal ratios
e.g.\ for the jump in the order parameter
$(\Delta \rho_{0R}/m_R^c=0.592)$ or for the ratios of critical 
correlation lengths in the disordered (symmetric) and
ordered (spontaneously broken) phase
$(m_{0R}^c/m_{R}^c=0.746)$.
It specifies critical couplings
$(\lambda_{1R}/m_R^c=0.845,\lambda_{2R}/m_R^c=15.0)$. 
The universal behavior
of the potential for large field arguments 
$\rho \gg \rho_{0}$ is
$U \sim \rho^{3/(1+\eta)}$
provided $\rho_R$ is sufficiently 
small as compared to $\Lambda$. Here the critical exponent
$\eta$ which characterizes the dependence of the potential
on the unrenormalized field $\rho$
is found to be $\eta=0.022$. For large $\rho$ the universal
equation of state equals the one for the $O(8)$ symmetric
Heisenberg model and $\eta$ specifies the anomalous
dimension or the critical exponent $\delta=(5-\eta)/(1+\eta)$.
In contrast to the Ising universality class (section 5.6) 
the first order universal equation of state cannot be
reduced to the universal equation of state for the $O(8)$ model for
general $\rho$.

Finally, we should mention that our approach can be extended
to systems with
reduced $SU(N) \times SU(N)$ symmetry. They obtain by adding to
the classical potential a term involving the invariant
$\xi=\det \varphi + \det \varphi^{\dagger}$. (Note that $\xi$
is not invariant with respect to $U(N) \times U(N)$).
This will give an even richer pattern of phase transitions
and permits a close contact to realistic meson models in
QCD where the axial anomaly is incorporated. Finally
one can extend the three dimensional treatment to a
four dimensional study of field theories at nonvanishing
temperature. How this can be used to approach
the chiral phase transition in QCD is presented in section 8.

\newpage

\section{Spontaneous nucleation and coarse graining}

\vspace*{-1cm}
\hspace*{12cm}
\footnote{This section
is based on a collaboration with A.\ Strumia \cite{first,second}.}
\label{CoarseGrain}
\vspace*{0.2cm}

\subsection{Introduction}

Let us consider a slow change with time of the ``parameters'' of the model
that describes a physical system. 
This concerns, for example, the change in temperature
in the early universe or a variation of the magnetic field in an
experiment with ferromagnets. We assume that the time scale of
the parameter change is much larger than the characteristic
equilibration time $t_{eq}$ of the system, so that the system
can follow adiabatically in (approximate) local equilibrium.
(For the example of the early universe the ratio of time scales
involves the age of the universe $H^{-1}$, i.e. the characteristic
small quantity is $Ht_{eq}$.) A second-order phase transition
can proceed under these circumstances without major non-equilibrium
effects. In this section we consider first-order phase transitions.
Due to the discontinuity in the order parameter no continuous
equilibrium evolution through the phase transition is possible.
Near the phase transition the effective average potential $U_k$ is
characterized by two separate local minima. In the course of
the evolution the minimum corresponding to the ``true vacuum''
(for late times) becomes lower than the one corresponding to
the ``false vacuum''. However, the system may not adapt 
immediately to the new equilibrium situation, and we encounter
the familiar phenomena of supercooling or hysteresis. As vapor
is cooled below the critical temperature, local droplets form
and grow until the transition is completed. The inverse evolution
proceeds by the formation of vapor bubbles in a liquid. The
transition in ferromagnets is characterized by the formation
of local Weiss domains with the magnetization corresponding to
the late time equilibrium.

The formation of ``bubbles'' of the new vacuum is similar to
a tunneling process and typically exponentially suppressed
at the early stages of the transition. The reason is the
``barrier'' between the local minima. The transition requires
at least the action of the saddle-point corresponding to
the configuration with lowest action on the barrier. One therefore
encounters a Boltzmann factor involving the action of this
``critical bubble'' that leads to exponential suppression.

A quantitative understanding of this important process is
difficult both from the experimental and theoretical side. The
theory deals mainly with pure systems, whereas in an experiment
the exponentially suppressed rate of ``spontaneous nucleation''
has to compete with processes where impurities act as seeds
for the formation of bubbles. As long as the exponential 
suppression is substantial, the theoretical treatment may 
separate the dynamics (which involves the growth of bubbles etc.)
from the computation of the exponential suppression factor. The
latter can be computed from equilibrium properties. Its quantitative
determination is by itself a hard theoretical problem for which 
we propose a solution in this section. We also discuss carefully
the range of applicability of this solution.

The problem of calculating nucleation rates 
during first-order phase transitions has
a long history. (For reviews with an extensive list of
references, see refs. \cite{review1,review2}.) 
Our present understanding of the phenomenon of nucleation
is based largely on the work of Langer~\cite{langer}.
His approach has been applied to relativistic field theory by
Coleman~\cite{coleman} and Callan~\cite{colcal} and extended by
Affleck~\cite{affleck} and Linde~\cite{linde} to finite-temperature
quantum field theory. 
The basic quantity in this approach is 
the nucleation rate $I$, which
gives the probability per unit time and volume to nucleate a certain
region of the stable phase (the true vacuum) within the metastable 
phase (the false vacuum). 
The calculation of $I$ relies on a semiclassical approximation 
around a dominant saddle-point that is identified with 
the critical bubble. 
This is a static configuration 
(usually assumed to be spherically symmetric) within the metastable phase 
whose interior consists of the stable phase. 
It has a certain radius that can be determined from the 
parameters of the underlying theory. Bubbles slightly larger 
than the critical one expand rapidly, thus converting the 
metastable phase into the stable one. 

The nucleation rate is exponentially suppressed by a suitable
effective action of the critical bubble.
Possible deformations of the critical  bubble
generate a static pre-exponential factor.
The leading contribution to it
has the form of a ratio of fluctuation determinants and corresponds to the
first-order correction to the semiclassical result. 
Apart from the static prefactor, the nucleation rate 
includes a dynamical prefactor that takes into account the expansion of
bubbles after their nucleation. In this review we concentrate only on
the static aspects of the problem and neglect the dynamical prefactor.
Its calculation requires the extension of our formalism to real time
nonequilibrium dynamics.

For a four-dimensional theory of a real scalar field 
at temperature
$T$, the nucleation rate
is given by~\cite{langer}--\cite{linde}
\beq
I=\frac{E_0}{2\pi}
\left(\frac{\Gamma_b}{2\pi}\right)^{3/2}\left|
\frac{\det'[\delta^2 \Gamma/\delta\phi^2]_{\phi=\phibounce}}
{\det[\delta^2 \Gamma/\delta\phi^2]_{\phi=0}}\right|^{-1/2}
\exp\left(-\Gamma_b\right). 
\label{rate0} \eeq
Here $\Gamma$ is the effective action (see sects. 1.2 and 2.1)
of the system for a given configuration of the
field $\phi$ that acts as the order parameter of the problem. 
The action of the critical bubble is $\Gamma_b
=\Gamma\left[\phibounce(r)\right]-\Gamma[0]$,
where $\phibounce(r)$ is the spherically-symmetric
bubble configuration and $\phi = 0$ corresponds to the false vacuum. 
The fluctuation determinants are evaluated either 
at $\phi = 0$ or around $\phi=\phibounce(r)$. 
The prime in the fluctuation determinant around
the bubble denotes that the three zero eigenvalues 
of the operator $[\delta^2 \Gamma/\delta\phi^2]_{\phi=\phibounce}$
have been removed. 
Their contribution generates the factor 
$\left(\Gamma_b/2\pi \right)^{3/2}$ and the volume factor
that is absorbed in the definition of $I$ (nucleation rate per unit volume). 
The quantity $E_0$ is the square root of
the absolute value of the unique negative eigenvalue.

In field theory, the rescaled free energy 
density of a system for homogeneous configurations 
is identified with
the temperature-dependent effective potential. This is often evaluated 
through some perturbative scheme, such as the loop expansion~\cite{Col73}. 
In this way, the profile and the action of the bubble are determined
through the potential. This approach, however,
faces three fundamental difficulties:
\begin{itemize}

\item[a)] The effective potential, being the Legendre transform of the
generating functional for the connected Green functions, 
is a convex function of the field. Consequently, it does not 
seem to be the appropriate quantity for the study of tunneling,
as no structure with more than one minima separated by a barrier
exists\footnote{It has been 
argued in ref.~\cite{wu} that the appropriate quantity for
the study of tunneling is the generating functional of the 
1PI Green functions (calculated perturbatively), 
which differs from the effective potential 
in the non-convex regions. However, as we discuss in the following,
the consistent picture must rely on the notion of coarse graining
and on the separation of the high-frequency fluctuations that may be 
responsible
for the non-convexity of the potential, from the low-frequency ones
that are relevant for tunneling. Such notions cannot be easily 
implemented in the context of perturbation theory.}.

\item[b)] The fluctuation determinants in the expression for the nucleation
rate have a form completely analogous to the one-loop correction to
the potential. The question of double-counting the effect of
fluctuations (in the potential and the prefactor)
must be properly addressed. 
This point is particularly important 
in the case of 
radiatively induced first-order phase transitions. These are 
triggered by
the appearance of a new vacuum state in the 
theory as a result of the integration of (quantum or thermal) fluctuations
~\cite{Col73}.
A radiatively induced 
first-order phase transition takes place in theories for which the 
tree-level potential has only one minimum, while a second minimum
appears at the level of radiative corrections\footnote{
In ref.~\cite{ewein} an alternative procedure was suggested for the 
treatment of radiatively-induced first-order phase transitions: The
fields whose fluctuations are responsible for the appearance 
of the new vacuum are integrated out first, so that an ``effective''
potential with two minima is generated for the remaining fields. 
Our philosophy is different: We integrate out high-frequency fluctuations
of all fields, so that we obtain an effective low-energy action which we
use for the calculation of the nucleation rate. Our procedure
involves an explicit infrared cutoff in the calculation of the 
low-energy action. This prevents the appearance of non-localities arising
from integrating out massless fields, which may be problematic for
the approach of ref.~\cite{ewein}. For example, the fields that
generate the new vacuum in radiatively-induced (fluctuation-driven)
first-order phase transitions
are usually massless or very light at the origin of
the potential.
}.

\item[(c)]
Another difficulty 
concerns the ultraviolet divergences that are inherent in
the calculation of the fluctuation determinants 
in the prefactor. An appropriate regularization scheme must be
employed in order to control them (for other approaches 
see refs.~\cite{cott}--\cite{ramos}). 
Moreover, this scheme must be consistent with the one employed
for the absorption of the divergences appearing in the 
calculation of the potential that determines the
action of the critical bubble. 
\end{itemize}

In ref.~\cite{BergTetWet97} it was argued that 
all the above issues can be resolved
through the implemention of the notion of coarse graining in the 
formalism, in agreement with Langer's philosophy.
The problem of computing the difference of the effective
action between the critical bubble and the false vacuum may be
divided into three steps:
In the first step, one only includes fluctuations with momenta
larger than a scale $k$ which is of the order of the typical gradients
of $\phi_b(r)$. For this step one can consider approximately
constant fields $\phi$ and use a derivative expansion for the resulting 
coarse-grained free energy 
$\Gamma_k[\phi]$. The second step searches for the
configuration $\phi_b(r)$ which is a saddle point of $\Gamma_k$.
The quantity $\Gamma_b$ in eq. (\ref{rate0} is identified with $\Gamma_k
[\phi_b]-\Gamma_k[0]$.
Finally, the remaining fluctuations with momenta smaller
than $k$ are evaluated in a saddle-point approximation
around $\phi_b(r)$. This yields the ratio of fluctuation
determinants with an ultraviolet cutoff $k$. Indeed, 
Langer's approach corresponds to a one-loop approximation
around the dominant saddle point
for fluctuations with momenta smaller than a coarse-graining
scale $k$. We solve here the problem of how to determine the
coarse-grained free energy  $\Gamma_k$ in a consistent way.
This is crucial for any quantitative treatment of the nucleation
rate since $\Gamma_k$ appears in an exponential.

In the following subsections we review studies of nucleation 
based on the formalism of the effective average action
$\Gamma_k$, which can be identified with the 
free energy, rescaled by the temperature, at a given coarse-graining scale 
$k$. In the simplest case, we consider a
statistical system with one space-dependent degree of freedom described
by a real scalar field $\phi(x)$.
For example, $\phi(x)$ may correspond to
the density for the gas/liquid transition,
or to a difference in concentrations for chemical phase transitions,
or to magnetization for the ferromagnetic transition.
Our discussion also applies to a quantum field theory in
thermal quasi-equilibrium.
As we will see in sect. 7, an effective three-dimensional description
applies for a thermal quantum field theory at 
scales $k$ below the temperature $T$. We assume that
$\Gamma_{k_0}$ has been computed (for example perturbatively) for $k_0=T$
\cite{TW93-1,Tet96-1} and concentrate here on the three-dimensional
(effective) theory.

We compute $\Gamma_k$ by solving the flow equation
between $k_0$ and $k$. For this purpose we approximate $\Gamma_k$ by 
a standard kinetic term and
a general potential $U_k$. This corresponds to the 
first level of the derivative expansion of eq. (\ref{2.13}), 
where we set $Z_{\Phi,k}(\rho)=1$ and neglect the higher derivative terms.
Our approximation is expected to be a good one for the models we 
consider, because the deviations of $Z_{\Phi,k}(\rho)$ from 1 and 
the size of the higher-derivative terms
are related to the anomalous dimension of the field, and this 
is small ($\eta \simeq 0.04$).
The long-range 
collective fluctuations
are not yet important at a short-distance scale\footnote{In a three
dimensional picture $k_0$ plays the role of the microscopic
scale $\Lambda$.} 
$k_0^{-1}=T^{-1}$.
For this reason, we assume here a polynomial potential
\beq
U_{k_0} (\phi) =
\frac{1}{2}m^2_{k_0} \phi^2
+\frac{1}{6} \gamma_{k_0} \phi^3
+\frac{1}{8} \lx_{k_0} \phi^4 \; .
\label{eq:two20} \eeq
The parameters $m_{k_0}^2, \gamma_{k_0}$ and $\lambda_{k_0}$
depend on $T$ . 
This potential has the typical form relevant for
first-order phase transitions in statistical systems
with asymmetric potentials or in four-dimensional quantum
field theories at high temperature \cite{TW93-1,Tet96-1}.
The two first-order critical lines are located at
$\gamma_{k_0}^2=9\lambda_{k_0} m_{k_0}^2$ and
$\gamma_{k_0}=0$, with endpoints at $m_{k_0}^2=-2\mu_{\rm cr}^2$,
$\gamma_{k_0}^2=-18\lambda_{k_0}\mu_{\rm cr}^2$ and
$m_{k_0}^2=\mu_{\rm cr}^2$, $\gamma_{k_0}=0$ respectively.
Here $\mu_{\rm cr}^2$ is the critical mass term of the Ising model
($\mu_{\rm cr}^2/k_0^2 \approx -0.0115$ for $\lambda_{k_0}/k_0=0.1$).
We point out that, 
for fixed $m_{k_0}^2$ and $\lambda_{k_0}$, opposite
values of $\gamma_{k_0}$ result in potentials related through 
$\phi \leftrightarrow -\phi$. 
Also a model with $m_{k_0}^2<0$ can be mapped onto the equivalent
model with $m^{\prime2}_{k_0}>0$ by the shift $\phi\to\phi+c$,
$\lambda_{k_0}c^2+\gamma_{k_0}c=-2m_{k_0}^2$, where
$m^{\prime 2}_{k_0}=-2m_{k_0}^2-\frac{1}{2}\gamma_{k_0}c$,
$\gamma'_{k_0}=\gamma_{k_0}+3\lambda_{k_0}c$.

As we have seen in section \ref{cubicmodel} 
a different shift $\phi\to\phi+\tilde{c}$ can eliminate
the cubic term 
in favor of a term linear in $\phi$.
Therefore, the potential (\ref{eq:two20}) also
describes statistical systems of the Ising universality class
in the presence of an external magnetic field. 
For a Hamiltonian
\beq
H=\int d^3x\,
\left\{
\frac{\hat{\lx}}{8}\left(\chi^2-1\right)^2-B\chi+\frac{\zeta}{2}
\,\partial_i\chi\,\partial^i\chi
\right\},
\label{hamil} \eeq
the parameters are 
$m^2_{k_0}=\hat{\lx}(3y^2-1)/2\zeta$, $\gamma_{k_0}=3\hat{\lx}T^{1/2}y/
\zeta^{3/2}$, $\lx_{k_0}=\hat{\lx}T/\zeta^2$, with $y$ given by 
$y(y^2-1)=2B/\hat{\lx}$.
For real magnets $k_0$ must be taken somewhat 
below the inverse lattice distance,
so that effective rotation and translation symmetries apply. 
Correspondingly, $\chi$ and $H$ are the effective normalized spin field and
the effective Hamiltonian at this scale. We emphasize that our
choice of potential encompasses a large class of field-theoretical
and statistical systems.  
The universal critical behavior of these systems has been
discussed extensively in section \ref{secsec}.
In a different context, our results can also
be applied to the problem of quantum tunneling in a (2+1)-dimensional
theory at zero temperature. In this case $k_0,m^2,\gamma$ and
$\lambda$ bear no relation to temperature.

We compute the form of the potential $U_k$ at scales $k\leq k_0$ by
integrating the evolution equation (\ref{2.33}) with 
$Z_k(\rho,q^2)=\tilde{Z}_k(\rho,q^2)=1$ in
eqs. (\ref{2.34}).
The form of $U_k$ changes as
the effect of fluctuations with momenta above the decreasing scale
$k$ is incorporated in
the effective couplings of the theory. 
We consider an arbitrary form of $U_k$ which, in general,
is not convex for non-zero $k$.
$U_k$ approaches the convex effective potential only in the limit $k\to 0$.
In the region relevant for a first-order phase 
transition, $U_k$ has two distinct
local minima, where one is lower than the other away from the
phase transition at $\gamma_{k_0}=0$ or $|\gamma_{k_0}|=3
\sqrt{|\lambda_{k_0}m_{k_0}^2|}$.
The nucleation rate should be computed for $k$ larger than
or around the scale $k_f$ at which  $U_k$ starts  receiving important
contributions from field configurations that interpolate between
the two minima. This happens when the negative curvature at the top
of the barrier becomes approximately equal to $-k^2$~\cite{RingWet90,TetWet92}
(see subsect. \ref{convexpot}).
Another consistency check for the above choice of $k$  is  
the typical length scale of a thick-wall critical
bubble which is $\gta  1/k$ for $k > k_f$.
The use of $U_k$ at a non-zero value of $k$ resolves the first fundamental
difficulty 
in the calculation of bubble-nucleation rates that we mentioned
earlier.

The other two difficulties are overcome as well.
In our approach the pre-exponential factor in eq. (\ref{rate0})
is well-defined and finite,
as an ultraviolet cutoff of order $k$ must implemented in the calculation of
the fluctuation determinants.
The cutoff must guarantee that fluctuations
with characteristic momenta $q^2 \gta k^2$ do not contribute to the
determinants. This is natural, as
all fluctuations with typical momenta above $k$ are
already incorporated in the form of $U_k$.
The choice of the ultraviolet cutoff must be consistent with
the infrared cutoff procedure that determines $\Gamma_k$
and, therefore, $U_k$. In the following subsection
we show how this is achieved. 
It is clear that our approach
resolves then automatically the problem of double-counting 
the effect of the fluctuations.

As a test of the validity of the approach, the result for the rate
$I$ must be independent of the coarse-graining scale $k$, because
the latter should be considered only as a technical device.
In the following we show that this is indeed the case when
the expansion around the saddle point is convergent and the
calculation of the nucleation rate reliable. Moreover, the 
residual $k$ dependence of the rate can be used as
a measure of the contribution of the next order in the
saddle-point expansion.

\subsection{Calculation of the nucleation rate}

In all our calculations of bubble-nucleation rates 
we employ a mass-like infrared cutoff $k$ for the fluctuations
that are incorporated in $\Gamma_k$. This corresponds to the
choice $R_k=k^2$ for the cutoff function defined in eq. (\ref{2.6}).
The reason for our choice is that the 
evaluation of the fluctuation determinants is technically simplified
for this type of cutoff. In three dimensions and for our approximation
of neglecting the effects of wave function renormalization, 
the threshold function $l^3_0(w)$, defined in eqs. (\ref{2.45}),
is\footnote{We have neglected an infinite 
$w$-independent contribution to the threshold function
that affects only
the absolute normalization of the potential.
As we are interested only in relative values of the
potential for various field expectation values, this
contribution is irrelevant
for our discussion.  
}  
$l^3_0(w)=\pi\sqrt{1+w}$. 
The evolution equation (\ref{2.40}) for the potential can now be written as
\beq
 \frac{\partial }{\partial k^2}\left[U_k(\phi)-U_k(0)\right]
= - \frac{1}{8 \pi}\left[\sqrt{k^2+U_k''(\phi)}
-\sqrt{k^2+U_k''(0)}\right]. 
\label{twofour} \eeq
In this entire section primes denote derivatives with respect to $\phi$,
similar to sections \ref{IsingModel}-\ref{cubicmodel}. 
In order to avoid confusion with the notation of other previous
sections, in which primes denote derivatives with respect to
$\rho=\phi^2/2$, we display explicitly the argument of the 
function with respect to which we differentiate.

In order to implement the appropriate ultraviolet cutoff $\sim k$
in the fluctuation determinant, let us look at
the first step of an iterative solution for $U_k$, 
discussed in subsection \ref{iterative}
\beq
U_k^{(1)}(\phi)-U_k^{(1)}(0)=U_{k_0}(\phi)-U_{k_0}(0)+
\frac{1}{2}\ln\left[\frac{\det[-\partial^2+k^2+U_{k}''(\phi)]}{
\det[-\partial^2+k^2_0+U_{k}''(\phi)]}
\frac{\det[-\partial^2+k^2_0+U_{k}''(0)]}{
\det[-\partial^2+k^2+U_{k}''(0)]}\right].
\label{iter} \eeq
For $k\to 0$, this solution is a regularized one-loop
approximation to the effective
potential. Due to the ratio of determinants, only
momentum modes with $k^2<q^2<k_0^2$ are effectively included
in the momentum integrals. The form of the infrared cutoff in eq.
(\ref{twofour}) suggests that we should implement the ultraviolet cutoff
for the fluctuation determinant in the nucleation rate (\ref{rate0})
as
\bea
I
&\equiv&A_k \exp({-S_k})
\nonumber \\
A_k&=& \frac{E_0}{2\pi}\left(\frac{S_k}{2\pi}\right)^{3/2}
\left|
\frac{\det'\left[-\partial^2+U''_k(\phibounce(r)) \right]}
{\det \left[ -\partial^2+k^2  + U''_k(\phibounce(r)) \right]}
~\frac{\det\left[-\partial^2+k^2 + U''_k(0) \right]}
{\det\left[-\partial^2+U''_k(0)\right]}
\right|^{-1/2}.
\label{rrate} \eea
where we switch the notation to $S_k=\Gamma_k[\phi_b]-\Gamma_k[0]$ instead
of $\Gamma_b$ in order to make the $k$-dependence in the
exponential suppression factor explicitly visible. A comparison
between eqs. (\ref{iter}) and (\ref{rrate}) shows that the explicitly
$k$-dependent regulator terms drop out for the combination 
$S_k-\ln\ A_k$. Our computation of $U_k$ also includes contributions 
beyond eq. (\ref{iter}). The residual $k$-dependence of the nucleation 
rate will serve as a test for the validity of our approximations.

The critical bubble configuration $\phi_b(r)$ is an 
SO(3)-invariant solution of
the classical equations of motion which interpolates between 
the local maxima of the potential
$-U_k(\phi)$. It satisfies the equation 
\beq
{d^2\phibounce\over dr^2}+\frac{2}{r}~{d\phibounce\over dr}=U_k'(\phibounce), 
\label{eom} \eeq
with the boundary conditions 
$\phibounce\rightarrow 0$ for $r \rightarrow \infty$ and
$d\phibounce/dr= 0$ for $r =0$.
The bubble action $S_k$ is given by 
\beq
S_k=4 \pi
\int_0^\infty 
\left[ \frac{1}{2}
\left(\frac{d\phibounce(r)}{dr}\right)^2
+U_k(\phibounce(r))-U_k(0) \right]
r^{2}\,dr
\equiv  S_k^t+S_k^v,
\label{action} \eeq 
where the kinetic and potential contributions, 
$S_k^t$ and $S_k^v$ respectively,
satisfy $S_k^v/ S_k^t=-1/3.$

The computation of the fluctuation determinants $A_k$ is more complicated. 
The differential operators that appear in eq.~(\ref{rrate}) 
have the general form
\beq
\Op_{\kappa\alpha}=-\partial^2 +m_\kappa^2+\alpha W_k(r), 
\label{op} 
\eeq
where $m_\kappa^2\equiv U''_k(0)+\kappa k^2$ and
$W_k(r)\equiv U''_k(\phibounce(r))-U''_k(0)$,
with $\kx,\alpha=0$ or 1.
It is convenient
to express the eigenfunctions $\psi$ in terms of spherical harmonics:
$\psi(r,\theta,\varphi)=Y_{\ell m}(\theta,\varphi)u(r)/r~$
~\cite{cott,baacke1}.
Here $\ell$ and $m$ are the usual angular momentum quantum numbers.
The Laplacian operator $\partial^2$ takes the form
\beq
-\partial^2~~~\to~~~ \frac{1}{r} 
\left[-\frac{d^2}{dr^2}+\frac{\ell(\ell+1)}{r^2}\right]r
\equiv -\frac{1}{r}\nabla^2_\ell r,
\label{sph} \eeq
so that
\bea
\det \Op_{\kappa\alpha}
&=&\prod_{\ell=0}^\infty (\det \Op_{\ell\kappa\alpha})^{2\ell+1}
\nonumber \\
\Op_{\ell\kappa\alpha}&=&-\nabla^2_\ell+m_\kappa^2+\alpha W_k(r).
\label{opl} \eea
We recall that 
$\det \Op_{\ell \kx \alpha}$ is defined as the product of all eigenvalues 
$\lambda$ that lead to solutions of 
${\cal W}_{\ell \kx\alpha} u(r)=\lambda u(r)$, 
with the function $u(r)$ vanishing at $r=0$ and $r \to \infty$.
The computation of such complicated determinants is made possible by 
a powerful theorem~\cite{erice,cott} that relates ratios
of determinants to solutions of ordinary differential equations.
In particular, we have  
\beq
g_{\ell\kappa}\equiv
\frac{\det \Op_{\ell \kappa 1}}{\det \Op_{\ell \kappa 0}}
=\frac{\det[-\nabla^2_\ell+m_\kappa^2+1\cdot W_k(r)]}
{\det[-\nabla^2_\ell+m_\kappa^2+0\cdot W_k(r)]}=
\frac{y_{\ell\kappa1}(r\to \infty)}{y_{\ell\kappa0}(r \to \infty)},
\label{theorem} \eeq
where $y_{\ell\kappa\alpha}(r)$ is the solution of the differential equation
\beq
\left[-\frac{d^2}{dr^2}+\frac{\ell(\ell+1)}{r^2}
+m_\kappa^2+\alpha W_k(r)\right]y_{\ell\kappa\alpha}(r)=0,
\label{diffeq} \eeq
with the behaviour
$y_{\ell\kappa\alpha}(r)\propto r^{\ell+1}$ for $r\to 0$.
Such equations can be easily solved numerically.
Special care is required for the treatment of the negative 
eigenvalue of the operator $\Op_{001}$ and the zero eigenvalues
of $\Op_{101}$. The details are given in ref. \cite{first}.

\newpage

\unitlength1.0cm
\begin{figure}[h]
\begin{center}
\begin{picture}(17,12)
\putps(-0.7,-0.4)(-2.,-0.6){f1}{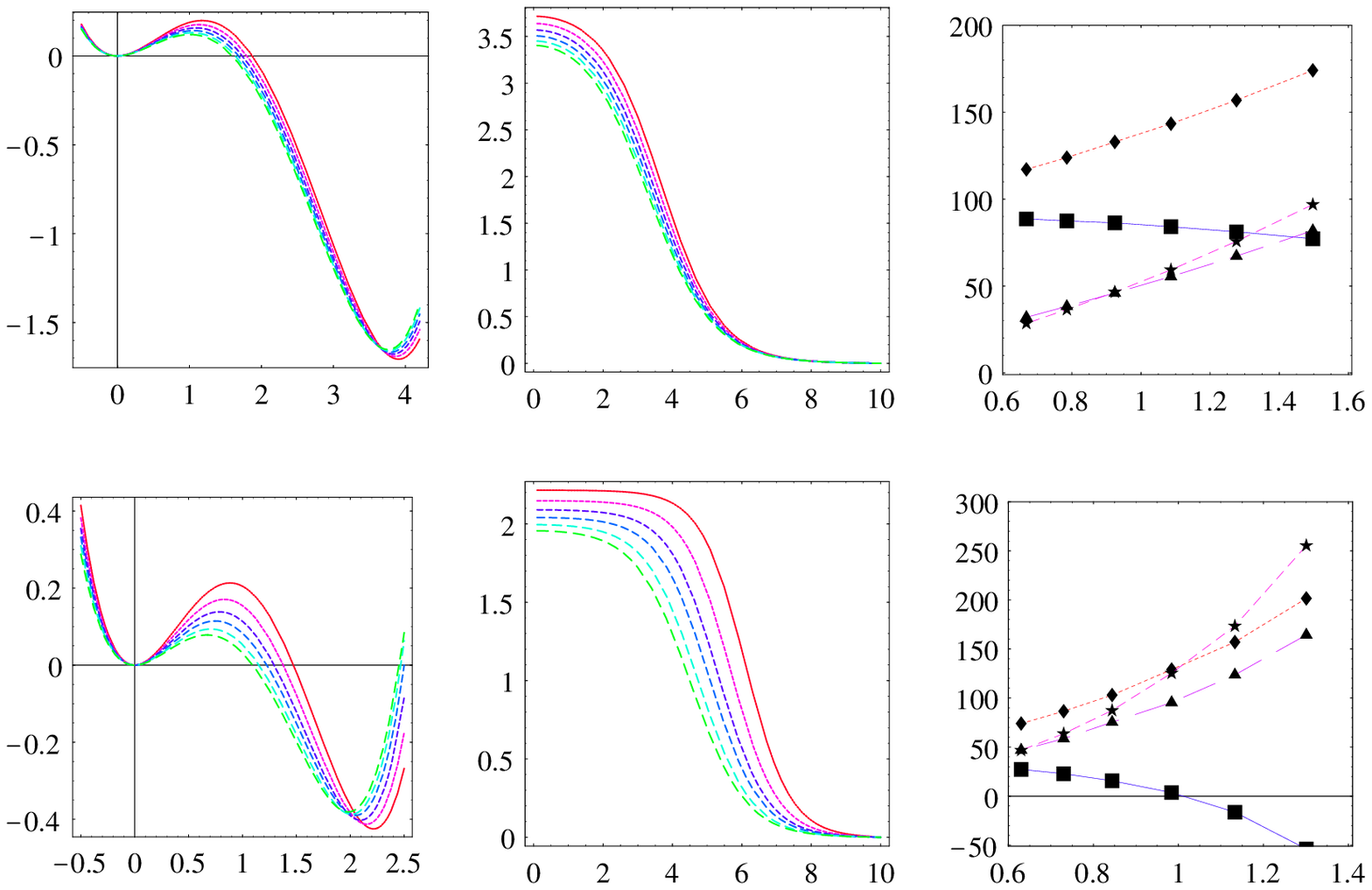}\Red
\put(2.6,12){\makebox(0,0){Coarse-grained}}
\put(2.6,11.5){\makebox(0,0){potential $U_k(\phi)$}}
\put(8.3,11.75){\makebox(0,0){Saddle-point $\phi_b(r)$}}
\put(14,12){\makebox(0,0){Saddle-point action $S_k$,}}
\put(14,11.5){\makebox(0,0){prefactor $A_k$ and nucleation rate $I$}}
\Black
\put(2.6,-0.7){\makebox(0,0){$\phi$}}
\put(8.3,-0.7){\makebox(0,0){$r$}}
\put(14,-0.7){\makebox(0,0){$k/\sqrt{U_k''(\phi_t)}$}}
\put(2.6,10.8){(a)}\put(8.3,10.8){(b)}\put(13.9,10.8){(c)}
\put(2.6,5){(d)}\put(8.2,5){(e)}\put(13.9,5){(f)}
\put(1.6,9){\small small $k\qquad$ large $k$}
\put(12,8.85){$\phantom{-}S_k$}
\put(12,8.2){$~-\ln(I/k_f^4)$}
\put(12,7.4){$~\ln (A_k/k_f^4)$}
\end{picture}
\vspace{7mm}
\caption[SP]{\em Dependence of effective potential, critical bubble and
nucleation rate on the coarse graining scale $k$.
The parameters are $\lambda_{k_0}=0.1\cdot k_0$,
$m^2_{k_0}=-0.0433\cdot k_0^2$, $\gamma_{k_0}=-0.0634~k_0^{3/2}$
(figs. a--c)
and
$m^2_{k_0}=-0.013\cdot k_0^2$, $\gamma_{k_0}=-1.61 \cdot 10^{-3}~k_0^{3/2}$
(figs. d--f).
All dimensionful quantities are given in units of $k_f$,
equal to $0.223\cdot k_0$ in the first series
and to  $0.0421\cdot k_0$ in the second series.
\label{fig3.1}
}
\end{center}\end{figure}

Sample computations are presented in fig.~\ref{fig3.1}.
The potential $U_k$ is
determined through the numerical integration
of eq.~(\ref{twofour}) between the scales $k_0$ and $k$,
using algorithms from ref.~\cite{ABBFTW95-1}.
The initial condition for the integration
is given by eq.~(\ref{eq:two20}).
Figs.~\ref{fig3.1}a--\ref{fig3.1}c correspond to
a model with
$m^2_{k_0}=-0.0433~k_0^2$,
$\gamma_{k_0}=-0.0634~k_0^{3/2}$,
$\lx_{k_0}=0.1~k_0$. 
We first show in
fig.~\ref{fig3.1}a  the evolution of the potential $U_k(\phi)$ as the scale
$k$ is lowered. 
(We always shift the metastable vacuum to $\phi=0$.)~
The solid line corresponds to $k/k_0=0.513$ while the
line with longest dashes (that has the smallest barrier height)
corresponds to $k_f/k_0=0.223$. At the scale $k_f$ the negative
curvature at the top of the barrier is slightly larger than
$-k_f^2$ and we stop the evolution.
The potential and the field have been
normalized with respect to $k_f$.
As $k$ is lowered from $k_0$ to $k_f$, the absolute minimum of the potential
settles at a non-zero value of $\phi$, while a significant barrier
separates it from the metastable minimum at $\phi=0$.
The profile of the critical bubble $\phi_b(r)$
is plotted in fig.~\ref{fig3.1}b in units of $k_f$
for the same sequence of scales.  For $k\simeq k_f$ the characteristic
length scale of the bubble profile and $1/k$ are comparable. This is expected,
because the form of the profile is determined by the barrier of the potential,
whose curvature is $\simeq -k^2$ at this point.
This is an indication that we should not proceed to coarse-graining
scales below $k_f$.
We observe a significant
variation of the value of the field $\phi$ in the interior of the bubble
for different $k$.

Our results for the nucleation rate are presented in fig.~\ref{fig3.1}c.
The horizontal axis corresponds to $k/\sqrt{U''_k(\phi_t})$,
i.e. the ratio of the scale $k$
to the square root of the positive curvature (equal to the
mass of the field) at the
absolute minimum of the potential located at $\phi_t$.
Typically, when $k$ crosses below this mass, 
the massive fluctuations of the field
start decoupling. The evolution of the convex parts of
the  potential slows down and eventually stops.
The dark diamonds give the values of the action $S_k$ 
of the critical bubble. We observe a strong
$k$ dependence of this quantity, which is expected from
the behaviour in figs.~\ref{fig3.1}a, \ref{fig3.1}b.
The stars in fig.~\ref{fig3.1}c indicate the values of
$\ln ( A_k/k^4_f )$.
Again a substantial decrease with decreasing $k$ is observed. This is expected,
because $k$ acts as the effective ultraviolet cutoff in the calculation
of the fluctuation determinants in $A_k$.
The dark squares give our results for
$-\ln(I/k^4_f )
= S_k-\ln ( A_k/k^4_f )$. It is remarkable that the
$k$ dependence of this quantity almost disappears for $k/\sqrt{U''_k(\phi_t})
\lta 1$.
The small residual dependence on $k$ can be used to estimate the
contribution of the next order in the expansion around the saddle point.
It is reassuring that this contribution is expected to be
smaller than $\ln ( A_k/k^4_f )$.

This behaviour confirms our expectation that the
nucleation rate should be independent of the scale $k$ that
we introduced as a calculational tool. It also demonstrates that
all the configurations plotted in fig.~\ref{fig3.1}b give equivalent
descriptions of the system, at least for the lower values of $k$.
This indicates that the critical bubble should not be associated only
with the saddle point of the semiclassical approximation, whose
action is scale dependent. It is the combination of
the saddle point and its possible deformations
in the thermal bath that has physical meaning.

For smaller values of $|m^2_{k_0}|$ the dependence of the nucleation
rate on $k$ becomes more pronounced. We demonstrate this in the
second series of figs.~\ref{fig3.1}d--\ref{fig3.1}f where
$\lambda_{k_0}/(-m^2_{k_0})^{1/2}$ $=0.88$ 
(instead of $0.48$ for figs.~\ref{fig3.1}a--\ref{fig3.1}c).
The value of $\lambda_{k_0}$ is the same as before, whereas
$\gamma_{k_0}=-1.61\cdot 10^{-3}k_0^{3/2}$ and $k_f/k_0=0.0421$.
The strong $k$ dependence is caused by the larger value of the dimensionless
renormalized quartic coupling for the second parameter set~\cite{BergTetWet97}.
Higher-loop contributions to $A_k$ become important and the expansion
around the saddle point does not converge any more. There are two clear
indications of the breakdown of the expansion: 
\begin{itemize}
\item[a)] The values of the 
leading and subleading contributions to the nucleation rate, 
$S_k$ and $\ln ( A_k/k^4_f )$ respectively, become comparable. 
\item[b)] The $k$
dependence of $\ln ( I/k^4_f )$ is strong and must be canceled
by the higher-order contributions.
\end{itemize}
We point out that the discontinuity in the order parameter 
at the phase transition is approximately 5 times
smaller in the second example
than in the first one. As a result, the second phase transition
can be characterized as weaker. Typically,
the breakdown of the saddle-point approximation
is associated with weak first-order phase transitions.

It is apparent from figs.~\ref{fig3.1}c and \ref{fig3.1}f
that the leading contribution to the 
pre-exponential factor increases the total nucleation rate. 
This behaviour, related to the fluctuations of the field whose expectation 
value serves as the order parameter, is observed
in multi-field models as well.
The reason can be traced to the form of the differential
operators in the prefactor of eq.~(\ref{rrate}).
This prefactor involves the ratio 
$\det'\left[-\partial^2+U''_k(\phi_b(r))\right]/
\det\left[-\partial^2+U''_k(0)\right]$ before regularization.
The function $U''_k(\phi_b(r))$ always has a minimum away from
$r=0$ where it takes negative values (corresponding to
the negative curvature at the top of the barrier), while
$U''_k(0)$ is always positive. As a result the lowest
eigenvalues of the operator $\det'\left[-\partial^2+U''
_k(\phi_b(r))\right]$
are smaller
than those of $\det\left[-\partial^2+U''_k(0)\right]$. The elimination of
the very large eigenvalues 
through regularization does not affect this
fact and the prefactor is always larger than 1. Moreover,
for weak first-order phase transitions it becomes exponentially large
because of the proliferation of low eigenvalues of the first operator.
In physical terms, this implies the 
existence of a large class of field configurations of free energy comparable
to that of the saddle-point. Despite the fact that they are not 
saddle points of the free energy 
(they are rather deformations of a saddle point)
and are, therefore, unstable, they result in an important  increase of
the nucleation rate. 
This picture is very similar to that of
``subcritical bubbles'' of ref.~\cite{gleiser}.

In figs.~\ref{fig3.1}c and \ref{fig3.1}f we also display the values of 
$\ln ( A_k/k^4_f )$ (dark triangles) predicted by the approximate expression 
\beq
\ln \frac{A_k}{k^4_f}
\approx \frac{\pi k}{2} 
\left[
- \int_0^\infty \!\!\! r^3 \left[ 
U''_k\left( \phi_b(r) \right)
-U''_k\left( 0 \right)
\right] dr
\right]^{1/2}.
\label{eq:appr}
\eeq
This expression is based on the behaviour of
the ratio of determinants~(\ref{theorem}) for large $\ell$,
for which an analytical treatment is possible \cite{first,second}.
It is apparent from figs.~\ref{fig3.1}c and \ref{fig3.1}f 
that eq.~(\ref{eq:appr}) gives a good approximation to the exact
numerical results, especially near $k_f$. It can be used for
quick checks of the validity of the expansion around the
saddle point.

\subsection{Region of validity of homogeneous nucleation theory}
\label{regionvalid}

It is useful to obtain some intuition on the behaviour of the 
nucleation rate by using the approximate expression~(\ref{eq:appr}).
We assume that the potential has a form similar
to eq.~(\ref{eq:two20}) even near $k_f$, i.e.
\beq
U_{k_f} (\phi) \approx
\frac{1}{2}m^2_{k_f} \phi^2
+\frac{1}{6} \gamma_{k_f} \phi^3
+\frac{1}{8} \lx_{k_f} \phi^4.
\label{eq:two21} \eeq
(Without loss of generality we take $m^2_{k_f}>0$.)
For systems
not very close to the endpoint of the first-order critical line,
our assumption is supported by
the numerical data, as can be verified from fig.~~\ref{fig3.1}.
The scale $k_f$ is determined by the relation
\beq
k^2_f \approx \max \left| U''_{k_f}(\phi)
\right|=\frac{\gamma_{k_f}^2}{6\lx_{k_f}}-m_{k_f}^2.
\label{abc} \eeq

Through the rescalings 
$r=\rt/m_{k_f}$, $\phi=2\phit\, m_{k_f}^2/\gamma_{k_f}$, the potential
can be written as 
$\Ut(\phit)=\phit^2/2-\phit^3/3+h\,\phit^4/18$, with
$h=9\lx_{k_f} m_{k_f}^2/\gamma_{k_f}^2$. 
For $h \approx 1$ 
the two minima of the potential have approximately
equal depth. The action of the saddle point can be expressed
as
\beq
S_{k_f} = 
\frac{4}{9} \frac{m_{k_f}}{\lx_{k_f}}\, h \St(h),
\label{action2} \eeq
where $\St(h)$ must be determined numerically through $\Ut(\phit)$.
Similarly, the pre-exponential factor can be estimated through
eq.~(\ref{eq:appr}) as
\bea
\ln \frac{ A_{k_f}}{k^4_f}  &\approx& 
\frac{\pi}{2}
\sqrt{ \frac{3}{2h}-1}\,\, \At(h),
\nonumber \\ 
\At^2(h) &=& 
\int_0^\infty \left[ \Ut'' \left(\phit_b(\rt)\right)-1 \right] \,\rt^3\, d\rt,
\label{prefactor2} \eea
with $\At(h)$ computed numerically.
Finally, the relative importance of the fluctuation
determinant is given by
\beq
R=\frac{\ln \left(A_{k_f}/k_f^4 \right)}{S_{k_f}} \approx
\frac{9\pi}{8}\frac{1}{h}\sqrt{\frac{3}{2h}-1}\,
\frac{\At(h)}{\St(h)}
\,\frac{\lx_{k_f}}{m_{k_f}}
=T(h) \,\frac{\lx_{k_f}}{m_{k_f}}.
\label{fin} \eeq
The ratio $R$ can be used as an indicator for the validity of 
the saddle point expansion, which is valid only for $R\stackrel
{\scriptstyle<}{\sim}1$.

\unitlength1.0cm
\begin{figure}[t]
\begin{center}\hspace{-5mm}
\begin{picture}(9,5)
\putps(-0.5,0)(-0.5,0){fTh}{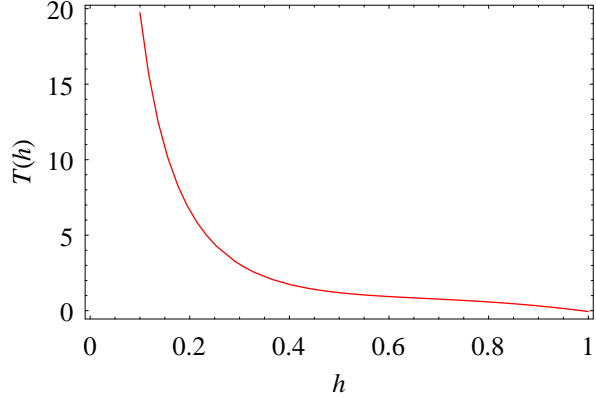}
\end{picture}
\caption[SP]{\em The parameter $T(h)$, defined in eq.~$(\ref{fin})$,
as a function of $h.$
\label{fig3.2}}
\end{center}\end{figure}

In fig.~\ref{fig3.2} we plot $T(h)$ as a function 
of $h$ in the interval (0,\,1). 
It diverges for $h\to 0$.
For $h \to 1$, our estimate of the prefactor
predicts $T(h)\to 0$. The reason is that, for our approximate
potential of eq.~(\ref{eq:two21}), the field masses  
at the two minima are equal in this limit. As a result, the
integrand in eq.~(\ref{eq:appr}) vanishes, apart from  the surface
of the bubble. The small surface contribution is negligible for
$h\to 1$, because the 
critical bubbles are very large in this limit. 
This behaviour is not expected to
persist for more complicated potentials. Instead, we expect a
constant value of $T(h)$ for $h\to 1$.
However, we point out that the approximate expression~(\ref{eq:appr}) has not
been tested for very large critical bubbles. The divergence of the 
saddle-point action in this limit results in low accuracy
for our numerical analysis. Typically, our results are reliable
for $\St(h)$ less than a few thousand.
Also, eq.~(\ref{eq:appr})
relies on a large-$\ell$ approximation and is not guaranteed to be valid
for large bubbles.
We have checked that 
both our numerical and approximate results are reliable for 
$h \lta 0.9$.

\begin{figure}[h]\begin{center}\begin{picture}(7,7)
\putps(-0.5,-1)(-0.5,-1){f2}{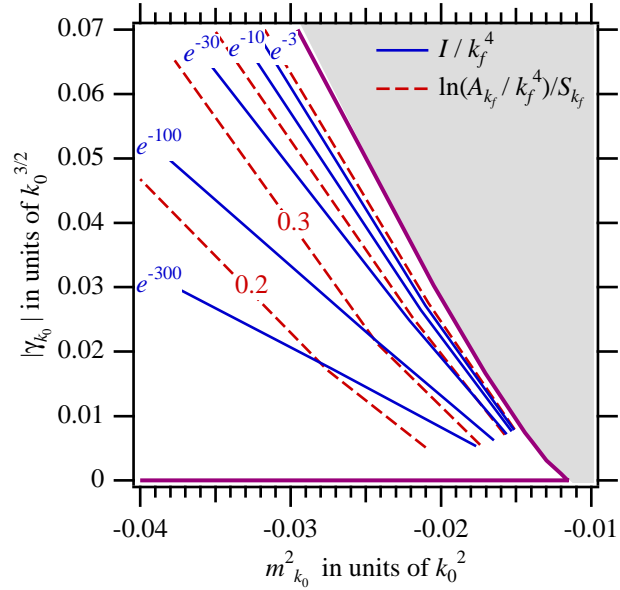}
\end{picture}\vspace{7mm}
\caption[SP]{\em Contour plots of the nucleation rate $I/k_f^4$
and of $R=\ln(A_{k_f}/k_f^4)/S_{k_f}$  in the plane
$(m_{k_0}^2,\gamma_{k_0})$, for $\lx_{k_0}/k_0=0.1$.
Regions to the right of the spinodal line
(only one minimum) are shaded.
The dashed lines correspond to $R=\{0.2,0.3,0.5,1\}$ and the solid lines
to $I/k_f^4$.\label{fig3.3}}
\end{center}\end{figure}

The estimate of eq.~(\ref{fin}) suggests two cases in which 
the expansion around the saddle point is expected to break down:
\begin{itemize}
\item[a)] For 
fixed $\lx_{k_f}/m_{k_f}$, the ratio $R$ becomes larger than 1 for
$h\to 0$. In this limit the barrier becomes negligible and the 
system is close to the spinodal line. 
\item[b)] For fixed $h$, $R$ can be large for sufficiently
large $\lx_{k_f}/m_{k_f}$. This is possible even for $h$ close
to 1, so that the system is far from the spinodal line. This 
case corresponds to weak first-order phase transitions, as 
can be verified by observing that 
the saddle-point action ~(\ref{action2}), 
the location of the true vacuum 
\beq
\frac{\phi_t}{\sqrt{m_{k_f}}}=
\frac{2}{3}\sqrt{h}\,\,
\phit_t(h)\,\sqrt{\frac{m_{k_f}}{\lx_{k_f}}},
\label{phit} \eeq
and the
difference in
free-energy density between the minima 
\beq
\frac{\Delta U}{m_{k_f}^3}=
\frac{4}{9}\,{h}\,\,
\Delta\Ut(h)\,\frac{m_{k_f}}{\lx_{k_f}}
\label{deltaU} \eeq
go to zero in the limit $m_{k_f}/\lx_{k_f}\to 0$ for fixed $h$. This
is in agreement with the discussion of fig.~\ref{fig3.1} 
in the previous subsection.
\end{itemize}

\begin{figure}[h]\begin{center}\begin{picture}(7,7)
\putps(-0.5,-1)(-0.5,-1){f2V}{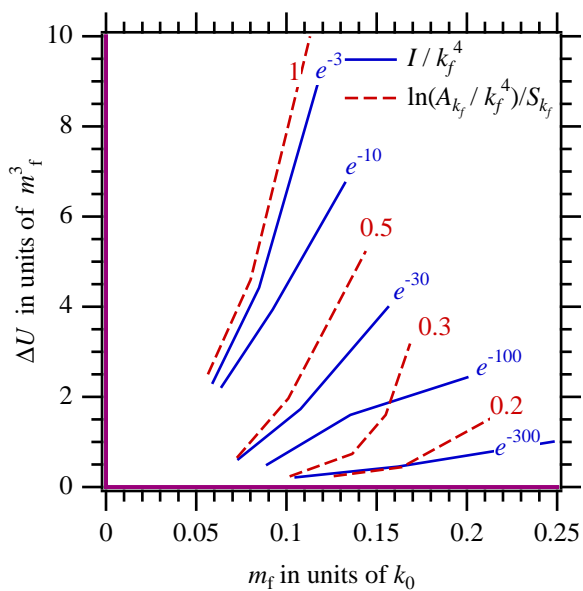}
\end{picture}\vspace{7mm}
\caption[SP]{\em 
Nucleation rate as function of thermodynamic variables. We show the 
 contour plots of fig. \ref{fig3.3}
in the plane $(m_{f},\Delta U)$, where $m_f^{-1}$ is 
the correlation length in the metastable phase
 and $\Delta U$ is
the difference in free-energy density between the metastable and stable phase.
The spinodal line corresponds to the vertical axis.
\label{fig3.4}}
\end{center}\end{figure}

The breakdown of homogeneous nucleation theory in both the above
cases is confirmed through the numerical computation of the nucleation 
rates \cite{second}. 
In fig.~\ref{fig3.3} we show contour plots for $I/k_f^4$ and for
$R=\ln(A_{k_f}/k^4_f)/S_{k_f}$ in the $(m_{k_0}^2,\gamma_{k_0})$
plane for fixed $\lambda_{k_0}/k_0=0.1$.
One can see the decrease of the rate as
the first-order critical line $\gamma_{k_0}=0$ is approached.
The spinodal line (end of the shaded region),
on which the metastable minimum of $U_k$ becomes unstable, is also
shown.
The nucleation rate becomes large before the 
spinodal line is reached.
For $-\ln (I/k_f^4)$ of order 1, 
the exponential suppression of the nucleation
rate disappears. Langer's approach can no longer be applied
and an alternative picture for the dynamical transition must be
developed~\cite{boy}.
In the region between the contour $I/k_f^4=e^{-3}$ and the
spinodal line, one expects a smooth transition from nucleation to
spinodal decomposition.
The spinodal and critical lines meet at the endpoint in the
lower right corner, which corresponds to a second order phase transition.
The figure exhibits an increasing rate as 
the endpoint is approached at a fixed distance from the critical line.

The ratio $R$ is a better measure
of the validity of the semiclassical approximation. For $R\approx 1$
the fluctuation determinant is as important as the ``classical''
exponential factor $e^{-S_k}$. There is no reason to assume that
higher loop contributions from the fluctuations around the critical
bubble can be neglected anymore. Near the endpoint in the lower right corner, 
Langer's semiclassical picture
breaks down, despite the presence of a discontinuity in the order
parameter. Requiring $I/k_f^4\lta e^{-3}$, $R\lta1$,
gives a limit of validity for Langer's theory.
For a fixed value of the nucleation rate (solid lines in fig.~\ref{fig3.3}),
the ratio $R$ grows as the endpoint in the lower right corner is
approached. This indicates that Langer's theory is not applicable
for weak first-order phase transitions, even if the predicted rate
is exponentially suppressed. The concept of nucleation of a region of
the stable phase within the metastable phase may still be relevant. However,
a quantitative estimate of the nucleation rate requires taking into account
fluctuations of the system that are not described properly by the 
semiclassical approximation \cite{gleiser}.

The parameter region discussed
here may be somewhat unusual since the critical line
of the phase
transition is approached  by varying $\gamma_{k_0}$ from negative or positive
values
towards zero. We have chosen it only for making the graph more
transparent.
However,
the results of fig.~\ref{fig3.3} 
can be mapped by a shift $\phi\to\phi+c$ to another
region with $m_{k_0}^2>0$, for which the
first-order phase transition can be approached by varying $m_{k_0}^2$ 
at fixed $\gamma_{k_0}$.
As opposite values of $\gamma_{k_0}$ result in potentials
related by
$\phi \leftrightarrow -\phi$, we can always choose 
$\gamma_{k_0}<0$. Then the phase transition proceeds 
from a metastable minimum at the origin to a stable minimum
along the positive $\phi$-axis (as in fig.~\ref{fig3.1}a). Potentials with
$m_{k_0}^2>0$, $\gamma_{k_0}<0$
are relevant for cosmological phase transitions, such as the
electroweak phase transition.

The microscopic parameters at the scale $k_0$ are often
not known in statistical systems. In order to facilitate
the interpretation of possible experiments where the
nucleation rate would be measured together with the correlation
length and the latent heat, we also give I as a function of renormalized
parameters. 
In fig.~\ref{fig3.4} we depict the region of validity
of homogeneous nucleation
theory in terms of parameters of the low-energy theory at the scale 
$k_f$. The contours
correspond to the same quantities as in fig.~\ref{fig3.3}. They are now
plotted as a function of the renormalized mass at the false 
vacuum $m_f=\sqrt{U''_{k_f}(0)}$ in units of $k_0$
and the difference in free-energy density between the two vacua in units
of $m^3_f$. 
Here $m_f^{-1}$ corresponds to the correlation length in the false 
vacuum and $\Delta U/m^3_{k_f}$ can be related to observable quantities like
the jump in the order parameter or the latent heat if $\lx_{k_0}/k_0$ is
kept fixed. 
Furthermore, for given $\lx_{k_f}/m_{k_f}$, we can relate
$\Delta U/m^3_{k_f}$  to $h$ in the approximation of eq.~(\ref{eq:two21}) 
using eq.~(\ref{deltaU}) and compute the observables from the explicit
form of the free energy density (\ref{eq:two21}). 
The spinodal line corresponds to the vertical axis, as for 
$m_f=0$ the origin of the potential turns into a maximum.
The critical line corresponds to the horizontal axis.
The origin is the endpoint of the critical line. 
All the potentials we have studied have an approximate form similar to
eq.~(\ref{eq:two21}) with $h\lta 0.9$.
>From our discussion in the
previous subsection and fig.~\ref{fig3.2} we expect that $R$ 
is approximately given by eq.~(\ref{eq:appr})
with $T(h)\gta 0.3$ for $h\lta 0.9$.
This indicates that $R \gta 1$
for $m_f/k_0\lta  0.05$ even far from the spinodal line.
This expectation is confirmed by fig.~\ref{fig3.4}. 
Even for theories with a significant exponential suppression for
the estimated nucleation rate 
we expect $R \sim 1$ near $m_f/k_0 \approx 0.05$.

Finally, we point out that realistic statistical systems often
have large
dimensionless couplings $\lx_{k_0}/k_0\sim 10$. 
Our results
indicate that Langer's homogeneous nucleation theory breaks down for
such systems even for small correlation lengths in the metastable
phase ($m_f/k_0 \sim 1$). 
For a large correlation length
the universal behavior of the potential
has been discussed in sect. \ref{cubicmodel}. One obtains
a large value $\lx_{k_f}/m_f\approx 5$, 
independently of the short-distance couplings \cite{B}. Therefore, 
a saddle-point approximation for
the fluctuations around the critical bubble will not give accurate 
results in the universal region.

\subsection{Radiatively induced first-order phase transitions}

We now turn to a more complicated system,
a theory of two scalar fields. 
It provides a framework within which we can test
the reliability of our approach
in the case of two fluctuating fields. The 
evolution equation for the potential resembles very closely 
the ones appearing in gauged Higgs theories, with the additional
advantage that the approximations needed in the derivation
of this equation are more transparent. We expect
the qualitative conclusions for the region of validity of
Langer's picture of homogeneous nucleation to be valid 
for gauged Higgs theories as well. 
The most interesting feature of the two-scalar models is 
the presence of radiatively induced first-order phase transitions.
Such transitions usually take place when the mass of a certain field 
is generated through the expectation value of another. The fluctuations
of the first field can induce the appearance of new minima in the 
potential of the second, resulting in first-order phase transitions
~\cite{Col73}.
As we have already discussed, the problem of double-counting 
the effect of fluctuations is particularly acute 
in such situations. The introduction of a coarse-graining
scale $k$ resolves this problem, by separating the high-frequency 
fluctuations of the system which may be responsible for the 
presence of the second minimum through the Coleman-Weinberg mechanism,
from the low-frequency ones which are relevant for tunneling. 

Similarly to the one-field case, we approximate
the effective average action as
\beq
\Gammak = 
\int d^3x \left\{ 
\frac{1}{2} \left( \partial^{\mu} \pha~\partial_{\mu} \pha~  
+ \partial^{\mu} \phb~\partial_{\mu} \phb  \right) 
+ U_k(\pha,\phb) \right\}.
\label{twoeleven} \eeq
The evolution equation for the potential
can be written in the form~\cite{second,twoscalar}
\bea \nonumber
\frac{\partial}{\partial k^2} \left[ U_k(\pha,\phb) - U_k(0,0)\right] &= &
-\frac{1}{8 \pi} \left[
\sqrt{k^2 + M^2_1(\pha,\phb)} 
-\sqrt{k^2 + M^2_1(0,0)}~+ \right.\\ 
&&\phantom{-\frac{1}{8 \pi} }\hspace{-0.3em}\left.+
\sqrt{k^2 + M^2_2(\pha,\phb)}
-\sqrt{k^2 + M^2_2(0,0)} \right],
\label{evpot} \eea
where $M^2_{1,2}(\pha,\phb) $ are the two eigenvalues of the field-dependent mass matrix,
given by
\beq
M^2_{1,2}(\pha,\phb) 
=\frac{1}{2} \left[
U_{11}+U_{22}
\pm  \sqrt{
\left( U_{11}- U_{22} \right)^2
+ 4 U_{12}^2 }
\right],
\label{massm} \eeq
with
$U_{i j}\equiv \partial^2U_k/\partial\phi_i\partial\phi_j$.
The only neglected corrections to eq.~(\ref{evpot})
are related to the 
wave-function renormalization of the fields. 
We expect these corrections to be small, as the
anomalous dimension is
$\eta \approx 0.035-0.04$.
We consider models with 
the symmetry $\phb \leftrightarrow - \phb$ throughout
this paper. This means that the expressions for the mass eigenvalues
simplify along the $\pha$-axis:
$M^2_1=\partial^2 U_k/\partial \phi_1^2$, 
$M^2_2=\partial^2 U_k/\partial \phi_2^2$. 

We always choose parameters such that minima of the potential  
are located along the $\pha$-axis.
The saddle-point configuration satisfies eq.~(\ref{eom}) 
along the $\pha$-axis and has $\phb=0$. 
The bubble-nucleation rate is derived in complete analogy to
the one-field case and is given by 
\bea
I&=&A_{1k} A_{2k} \exp({-S_k})
\nonumber \\
A_{1k}&=& \frac{E_0}{2\pi}\left(\frac{S_k}{2\pi}\right)^{3/2}
\left|
\frac{\det'\left[-\partial^2+U_{11}(\phibounce(r)) \right]}
{\det \left[ -\partial^2+k^2 +U_{11}(\phibounce(r))\right]}
~\frac{\det\left[-\partial^2+k^2+U_{11}(0) \right]}
{\det\left[-\partial^2+U_{11}(0)\right]}
\right|^{-1/2},
\nonumber \\
A_{2k}&=& \left|
\frac{\det\left[-\partial^2+U_{22}(\phibounce(r)) \right]}
{\det \left[ -\partial^2+k^2 +U_{22}(\phibounce(r))\right]}
~\frac{\det\left[-\partial^2+k^2+U_{22}(0) \right]}
{\det\left[-\partial^2+U_{22}(0)\right]}
\right|^{-1/2}.
\label{rrate2} \eea
The calculation of the various determinants 
proceeds very similarly to the previous subsection.
The details are given in ref.~\cite{third}.

\unitlength1cm

\begin{figure}[t]
\begin{center}\hspace{-5mm}
\begin{picture}(16,11)
\putps(0,-0.4)(0,-0.3){fcubic}{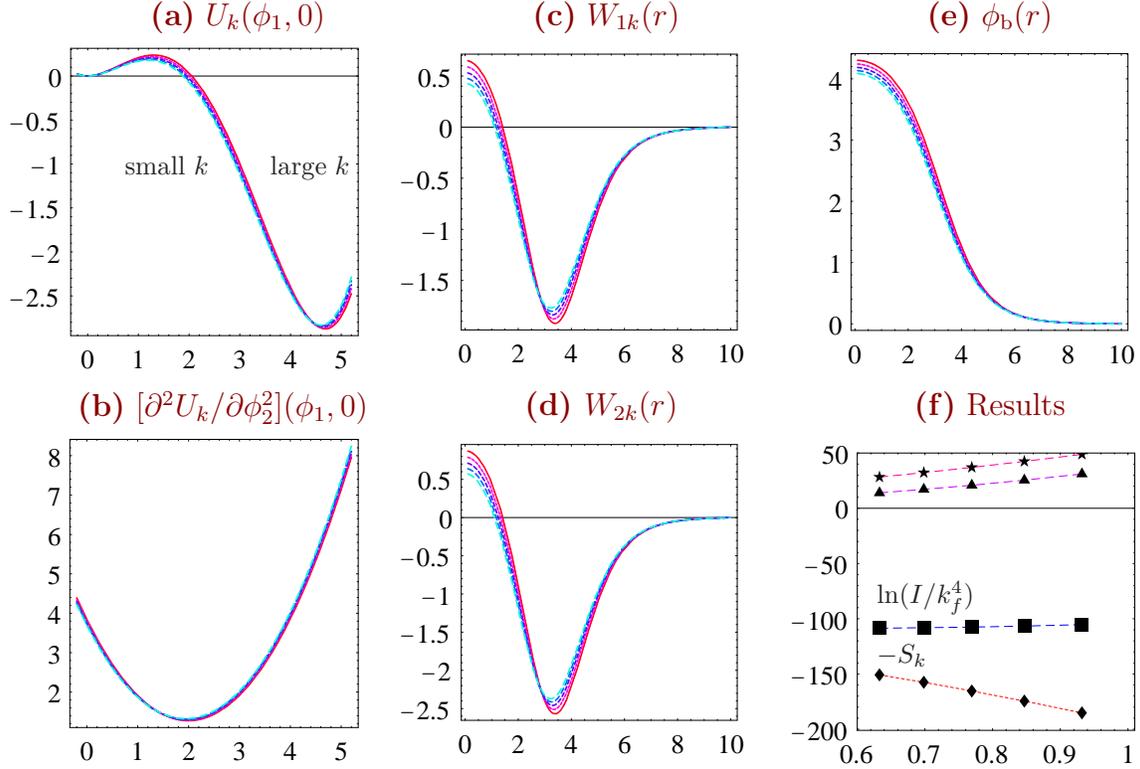}\Red
\put(3.5,10){\makebox(0,0){{\bf (a)} $U_k(\phi_1,0)$}}
\put(8.4,10){\makebox(0,0){{\bf (c)} $W_{1k}(r)$}}
\put(13.5,10){\makebox(0,0){{\bf (e)} $\phi_{\rm b}(r)$}}
\put(3.3,4.8){\makebox(0,0){{\bf (b)} $[\partial^2 U_k/\partial\phi_2^2](\phi_1,0)$}}
\put(8.4,4.8){\makebox(0,0){{\bf (d) $W_{2k}(r)$} }}
\put(13.5,4.8){\makebox(0,0){{\bf (f)} Results}}
\Black
\put(2,7.9){\footnotesize small $k\qquad$ large $k$}
\put(12,1.4){\footnotesize $-S_k$}
\put(12,2.2){\footnotesize $\ln(I/k_f^4)$}
\end{picture}
\caption[SP]{\em 
Nucleation rate for multicomponent models. We show the scale dependence
of various quantities necessary for 
 the computation of the nucleation rate. 
The microscopic
 potential is given by eq.~(\ref{pot1}) with $m_{2k_0}^2=-m_{1k_0}^2
=0.1~k_0^2$, 
$\lambda_{k_0}=g_{k_0}=0.1~k_0$, $J_{k_0}=0.6~k_0^{5/2}$.
The coarse graining scale $k$ varies between
 $k_i=e^{-0.8} k_0$ and $k_f=e^{-1.2} k_0$.
All dimensionful quantities are given in units of $k_f$.
In fig.~\ref{fig2.1}f we plot the saddle-point action (diamonds),
the two prefactors $\ln ( A_{1k}/k^4_f )$ (stars) and
 $\ln ( A_{2k} )$ (triangles),
and the nucleation rate $\ln(I/k_f^4)$ (squares)
as a function of $k/\sqrt{U_{11}(\phi_t,0)}$.
\label{fig2.1}}
\end{center}\end{figure}

In fig.~\ref{fig2.1} we present results for a 
class of models defined through the
potential 
\beq
U_{k_0}(\pha,\phb)= -J_{k_0}\pha
+\frac{1}{2} m^2_{1k_0}\pha^2 + \frac{1}{2} m^2_{2k_0}\phb^2
+ \frac{1}{8}\lx_{k_0}\left( \pha^4 + \phb^4 \right) 
+ g_{k_0} \pha^2 \phb^2.
\label{pot1} \eeq
The term linear in $\pha$ can be removed through an appropriate shift
of $\pha$.
This would introduce additional terms $\sim \pha^3$ 
and $\sim \pha \phb^2$. 
In fig.~\ref{fig2.1}a 
we present the evolution of $U_k(\pha)\equiv U_k(\pha,0)$ 
for $m^2_{1k_0}=-0.1~k^2_0$, $m^2_{2k_0}=0.1~k^2_0$, 
$\lx_{k_0}=g_{k_0}=0.1~k_0$ and $J_{k_0}=0.6~k_0^{5/2}$. We always shift
the location of the false vacuum to zero. 
The evolution of $U_{22}(\pha)\equiv\partial^2U_k/\partial\phb^2(\pha,0)$
is displayed in fig.~\ref{fig2.1}b. 
The solid lines correspond to $k_i/k_0=e^{-0.8}$, while the
line with longest dashes (that has the smallest barrier height)
corresponds to $k_f/k_0=e^{-1.2}$. 
The potential and the field have been
normalized with respect to $k_f$, so that they are of order 1. 
The profile of the critical bubble $\phibounce(r)$
is plotted in fig.~\ref{fig2.1}e in units of $k_f$
for the same sequence of scales.  
The quantities $W_{1k}(r)=U_{11}(\phibounce(r))-U_{11}(0)$
and $W_{2k}(r)=U_{22}(\phibounce(r))-U_{22}(0)$ are plotted in
figs.~\ref{fig2.1}c and \ref{fig2.1}d respectively.

Our results for the nucleation rate are presented in fig.~\ref{fig2.1}f.
The horizontal axis corresponds to $k/\sqrt{U_{11}(\phi_t})$,
i.e. the ratio of the scale $k$
to the square root of the positive curvature of the potential along the 
$\pha$-axis at the true vacuum. 
The latter quantity gives the mass of the field $\pha$
at the absolute minimum. 
Typically, when $k$ crosses below this mass 
the massive fluctuations of the fields
start decoupling (in all the examples we present 
the mass of $\phb$ is of the same order or larger
than that of $\pha$ at the absolute minimum) 
and the evolution of the convex parts of
the  potential slows down and eventually stops.
The dark diamonds give the negative of the action $S_k$ 
of the saddle point at the scale $k$. We observe a strong 
$k$ dependence of this quantity.
The stars in fig.~\ref{fig2.1}d indicate the values of 
$\ln ( A_{1k}/k^4_f )$ and the triangles those of $\ln ( A_{2k} )$,
where the two prefactors $A_{1k}$, $A_{2k}$ are defined 
in eqs.~(\ref{rrate2}). 
Again a significant $k$ dependence is observed. 
The dark squares give our results for 
$\ln(I/k^4_f ) 
= -S_k+\ln  (A_{1k}A_{2k}/k^4_f )$. 
This quantity has a very small
$k$ dependence, which confirms our expectation that the 
nucleation rate should be independent of the scale $k$.
The small residual dependence on $k$ can be used to estimate the 
contribution of the next order in the expansion around the saddle point.
This contribution is expected to be smaller than the first-order correction
$\ln (A_{1k}A_{2k}/k^4_f )$.

\begin{figure}[t]
\begin{center}\hspace{-5mm}
\begin{picture}(16,11)
\putps(0,-0.4)(0,-0.4){fradmaxg}{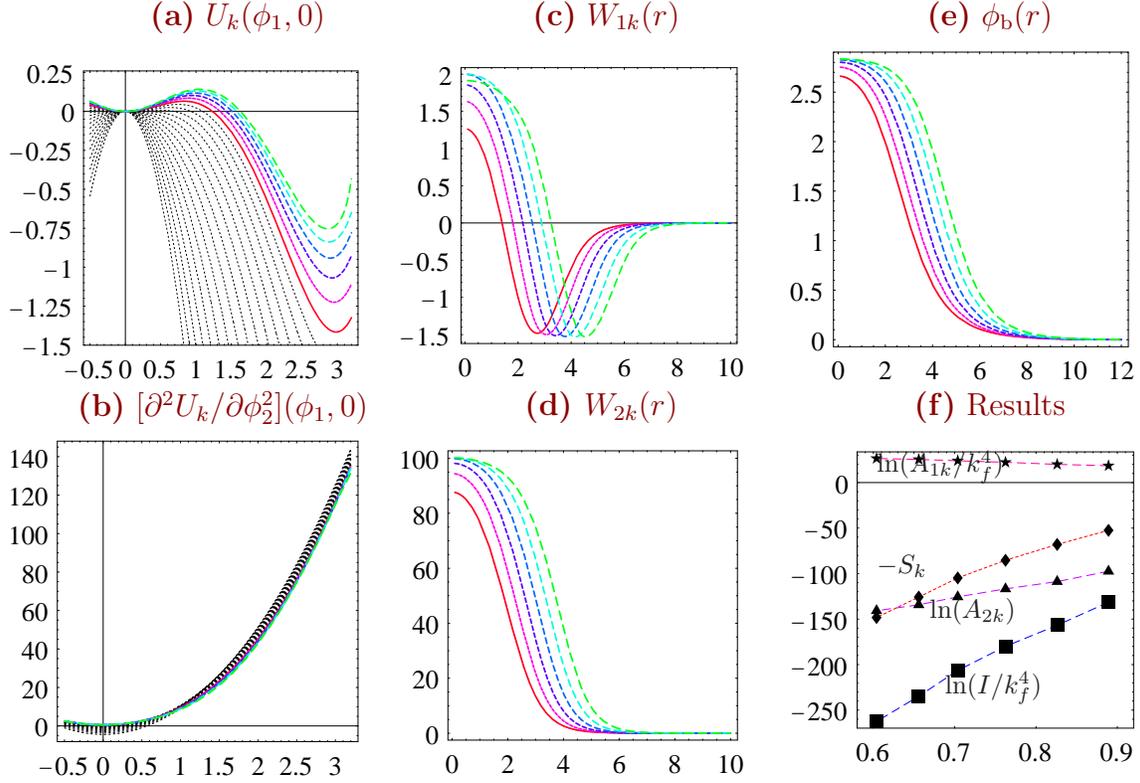}\Red
\put(3.5,10){\makebox(0,0){{\bf (a)} $U_k(\phi_1,0)$}}
\put(8.4,10){\makebox(0,0){{\bf (c)} $W_{1k}(r)$}}
\put(13.5,10){\makebox(0,0){{\bf (e)} $\phi_{\rm b}(r)$}}
\put(3.3,4.8){\makebox(0,0){{\bf (b)} $[\partial^2 U_k/\partial\phi_2^2](\phi_1,0)$}}
\put(8.4,4.8){\makebox(0,0){{\bf (d) $W_{2k}(r)$} }}
\put(13.5,4.8){\makebox(0,0){{\bf (f)} Results}}
\Black
\put(12,3.94){\footnotesize $\ln (A_{1k}/k_f^4)$}
\put(12,2.6){\footnotesize $-S_k$}
\put(12.7,2){\footnotesize $\ln (A_{2k})$}
\put(12.9,1){\footnotesize $\ln(I/k_f^4)$}
\end{picture}
\caption[SP]{\em 
Nucleation rate for a fluctuation induced first order transition. We
show the same quantities as  
in fig.~\ref{fig2.1} for 
 a model with initial potential given by eq.~(\ref{pot3})
with $\phi^2_{0k_0}=1.712~k_0$, $\lx_{k_0}=0.01~k_0$ and
$g_{k_0}=0.2~k_0$.
The calculation is performed between the scales $k_i=e^{-4.7} k_0$ and $k_f=e^{-5.2} k_0$.
All dimensionful quantities are given in units of $k_f$.
The strong scale dependence of the nucleation rate indicates the limits
of this calculation.
\label{fig2.4}}
\end{center}\end{figure}

We now turn to the discussion of radiatively induced first-order
phase transitions. 
An example can be observed in a model with
a potential
\beq
U_{k_0}(\pha,\phb)= 
 \frac{\lx_{k_0}}{8}
\left[ ( \pha^2-\phi_{0k_0}^2 )^2 
+      ( \phb^2-\phi_{0k_0}^2 )^2 \right] 
+ \frac{g_{k_0}}{4}  \pha^2 \phb^2,
\label{pot3} \eeq
with $\phi^2_{0k_0}=1.712~k_0$, 
$\lx_{k_0}=0.01~k_0$ and $g_{k_0}=0.2~k_0$. Since $J_{k_0}=0$,
in this case the ``classical potential'' $U_{k_0}$ only shows
second-order transitions independence on the classical parameters
$\phi^2_{0k_0}, \lambda_{k_0}, g_{k_0}$. 
Our results for this model  are presented in fig.~\ref{fig2.4}.
In fig.~\ref{fig2.4}a we plot a large part of the evolution of 
$U_k(\pha)$. The initial potential has only one minimum along the positive
$\pha$-axis (and the equivalent ones under the 
the symmetries
$\pha \leftrightarrow - \pha$, 
$\phb \leftrightarrow - \phb$, 
$\pha \leftrightarrow \phb$) and a maximum at the origin. 
In the sequence of potentials depicted by
dotted lines we observe the appearance of a new minimum
at the origin at some point in the evolution
(at $k/k_0 \approx e^{-4.4}$). This minimum is generated by
the integration of fluctuations of the $\phb$ field, whose mass depends on
$\pha$ through the last term in eq.~(\ref{pot3})
(the Coleman-Weinberg mechanism). 
In fig.~\ref{fig2.4}b it can be seen that the mass term of the 
$\phb$ field at the origin turns positive at the same value of $k$.
This is a consequence of the 
$\pha \leftrightarrow \phb$ symmetry of the potential.
We calculate the nucleation rate using the potentials of the
last stages of the evolution. 
The solid lines correspond to $k_i/k_0=e^{-4.7}$, while the
line with longest dashes 
corresponds to $k_f/k_0=e^{-5.2}$. 
In figs.~\ref{fig2.4}b--\ref{fig2.4}e 
we observe that the mass of the $\phb$ fluctuations in
the interior of the critical bubble is much larger than the other
mass scales of the problem, which are comparable to $k_f$. 
This is a consequence of our choice of couplings $g/\lx=20$. Such a large ratio
of $g/\lx$ is necessary for a strongly first-order
phase transition to be radiatively induced. Unfortunately, this range of
couplings also leads to large values for the $\phb$ mass and, as a result, to
values of 
$|\ln ( A_{2k})|$ that are comparable or larger than the
saddle-point action $S_k$, even though $\ln ( A_{1k}/k^4_f )$
remains small. As a result, the saddle-point approximation breaks down
and the predicted nucleation rate $I/k^4_f$ is strongly $k$ dependent.

One may wonder if it
is possible to obtain a convergent expansion around the saddle point
by considering models with smaller values of $g$. 
This question was addressed in ref.~\cite{third}. For smaller values of 
the ratio $g/\lx$, a weaker first-order phase transition is observed.
The expansion around the saddle point is
more problematic in this case. Not only
$|\ln ( A_{2k} )|$ is larger than the saddle-point 
action $S_k$, but the prefactor $\ln ( A_{1k}/k^4_f )$, associated with
the fluctuations of the $\pha$ field, becomes now comparable to $S_k$.
No region of the parameter space that leads
to a convergent saddle-point expansion for the nucleation rate was found.  

The above results are not surprising. The radiative corrections to the 
potential and the pre-exponential factor have a very similar form 
of fluctuation determinants. When the radiative corrections are large enough
to modify the bare potential and generate a new minimum, the 
pre-exponential factor should be expected to be important also. 
More precisely, the 
reason for this behaviour can be traced to the form of the differential
operators in the prefactor.
The prefactor associated with the field $\phb$
involves the ratio 
$\det(-\partial^2+m^2_{2}+W_{2k}(r))/\det(-\partial^2+m^2_{2})$, with
$m^2_{2}=U_{22}(0)$ and $W_{2k}(r)=U_{22}\left(\phi_b(r)\right)
-U_{22}(0)$. In units in which $\phi_b(r)$ is of order 1, 
the function $W_{2k}(r)$ takes very large positive values
near $r=0$ (see figs.~\ref{fig2.4}). This is a consequence of the
large values of $g$ that are required for
the appearance of a new minimum in the potential.
As a result, the lowest
eigenvalues of the operator $\det(-\partial^2+m^2_{2}+W_{2k}(r))$ are 
much larger 
than those of $\det(-\partial^2+m^2_{2})$. This induces a large suppression
of the nucleation rate. 
In physical terms, this implies that the deformations of the critical
bubble in the $\phb$ direction cost excessive amounts of free energy. 
As these fluctuations are inherent to the system, the total nucleation 
rate is suppressed when they are taken into account properly.
The implications for these finding for cosmological phase transitions, such
as the electroweak, were discussed in refs.~\cite{third,fifth}.

\subsection{Testing the approach through numerical simulations}

As a final application of our formalism we consider
(2+1)-dimensional theories at non-zero temperature. 
These theories provide
a test of several points of our approach that depend strongly on the
dimensionality, such as the form of the 
evolution equation of the potential, the nature of the
ultraviolet divergences of the fluctuation determinants, 
and the $k$ dependence of
the saddle-point action and prefactor. The
complementarity between the $k$ dependence of $S_k$ and $A_k$ is a 
crucial requirement for the nucleation rate $I$ to be $k$ independent.
Another strong motivation stems from the existence of 
lattice simulations of nucleation for (2+1)-dimensional systems
\cite{twopone}.

As before, we work within an effective
model after dimensional reduction. In two dimensions
the evolution equation for the potential
takes the form~\cite{fourth}
\beq 
\frac{\partial}{\partial k^2} \left[ U_k(\phi) - U_k(0)\right] = 
-\frac{1}{8 \pi} \left[
\ln \left( 1 + \frac{U''_k(\phi)}{k^2} \right) 
- \ln \left( 1 + \frac{U''_k(0)}{k^2} \right) \right].
\label{evpot2d} \eeq
An approximate solution of this equation is given by 
\beq
U_{k}(\phi) \approx V(\phi) + \frac{1}{8\pi} V''(\phi)
- \frac{1}{8\pi} \left(k^2+V''(\phi) \right)
\ln \left(\frac{k^2+V''(\phi)}{m^2}
\right).
\label{incond} \eeq
The potential $V(\phi)$ is taken
\beq
V(\phi) =
\frac{m^2}{2} \phi^2
+\frac{\gamma}{6} \phi^3
+\frac{\lambda}{8}  \phi^4,
\label{eq:two22} \eeq
with
\beq
\frac{\gamma}{m^2}=-\sqrt{\theta},\qquad
\frac{\lambda}{m^2}=\frac{1}{3}\, \theta\, \hat{\lambda},
\label{param} \eeq
in order to match the parameters $\theta$ and $\hat\lambda$ of
the renormalized theory simulated in 
ref.~\cite{twopone}. The potential of 
the simulated model is calculated through
lattice perturbation theory.
The dimensionless coupling that controls the
validity of the perturbative expansion is $\hat\theta \hat\lambda/3$. As a result, 
this expansion is expected to break down for $\hat\theta \gta 3 /\hat
\lambda$. 
Similarly, eq.~(\ref{incond}) is an approximate solution of eq.~(\ref{evpot2d})
only for $\hat\theta \gta 3 /\hat\lambda$.

\begin{figure}[t]
\begin{center}\hspace{-5mm}
\begin{picture}(17,11)
\putps(0,-0.4)(0,-0.5){figRes}{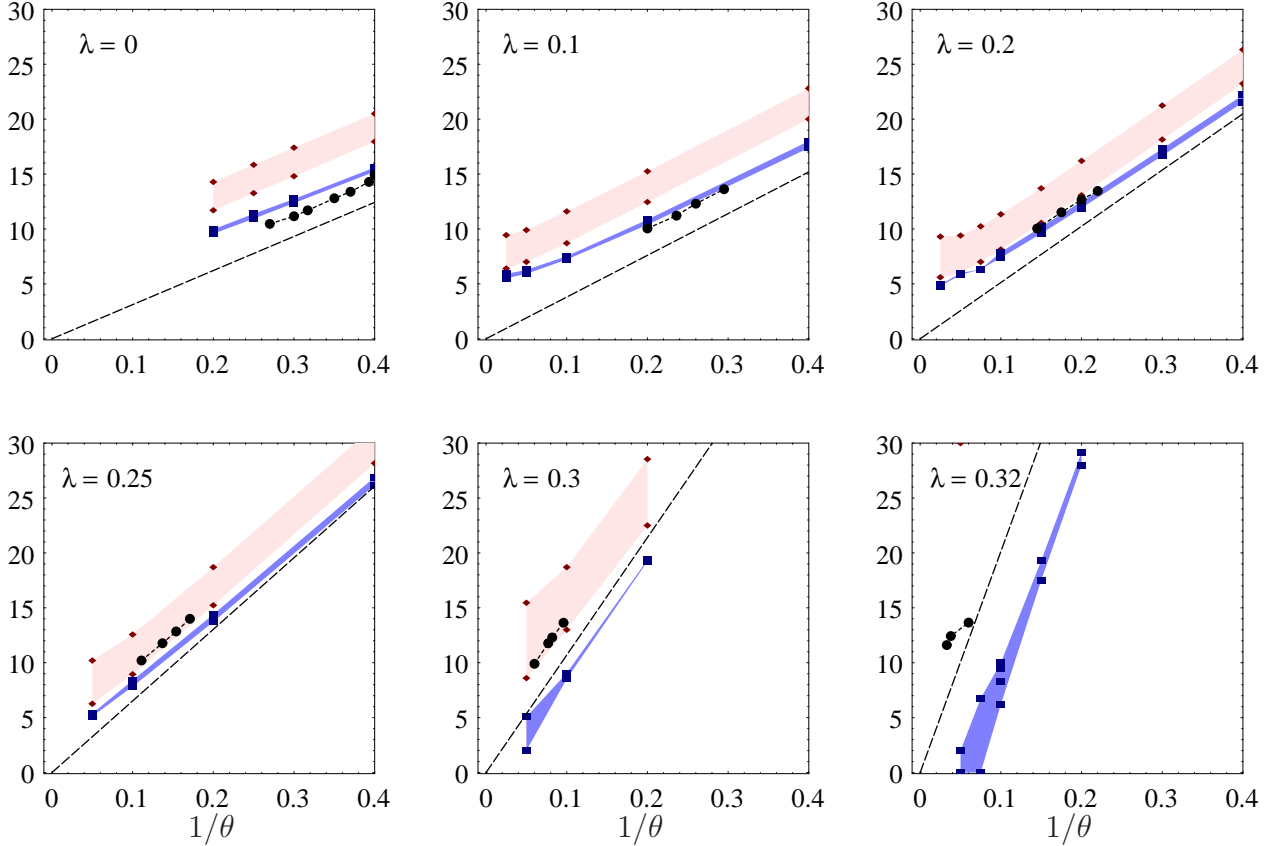}
\put(3.2,-0.5){\makebox(0,0){$1/\theta$}}
\put(8.9,-0.5){\makebox(0,0){$1/\theta$}}
\put(14.6,-0.5){\makebox(0,0){$1/\theta$}}
\end{picture}
\vspace{1cm}
\caption[SP]{\em Comparison of our method with lattice studies:
Diamonds denote the saddle-point action $S_k$ 
and squares the bubble-nucleation rate 
$-\ln\left(I/m^3\right)$ 
for $k=1.2\,k_f$ and $2\,k_f$ (shaded regions). 
Dark circles denote the results for the nucleation rate from
the lattice study of ref.~\cite{twopone}. 
Finally, the dashed straight lines
correspond to the action of 
the saddle point computed from the 
potential of eq.~(\ref{eq:two22}).
\label{fig4.2}}
\end{center}\end{figure}

The calculation of bubble-nucleation rates proceeds in complete analogy
to the (3+1)-dimensional case at non-zero temperature. The technical details
are given in ref.~\cite{fourth}.
In fig.~\ref{fig4.2} 
we present a comparison of results obtained through our method
with the lattice results of fig.~1 of ref.~\cite{twopone}. 
For each of several values of $\hat\lx$ we vary the parameter $\theta$
and determine the couplings $\gamma$, $\lambda$ according to eqs.~(\ref{param}).
The coarse-grained potential is then given by eq.~(\ref{incond}) for 
$k \geq k_f$.
The diamonds denote the saddle-point action $S_k$. 
For every choice of $\hat\lx$, $\theta$ we determine $S_k$
at two scales: $1.2\,k_f$ and $2\,k_f$. The light-grey region between
the corresponding points gives an indication of the
$k$ dependence $S_k$.
The bubble-nucleation rate $-\ln\left(I/m^3\right)$ is denoted 
by dark squares. 
The dark-grey region between the values obtained at 
$1.2\,k_f$ and $2\,k_f$ gives a good check of the convergence of the
expansion around the saddle point. If this region is thin, the
prefactor is in general small and
cancels the $k$ dependence of the action.
The dark circles denote the results for the nucleation rate from
the lattice study of ref.~\cite{twopone}. The dashed straight lines
correspond to the action of
the saddle point computed from the `tree-level'
potential of eq.~(\ref{eq:two22}).

For $\hat\lx=0$
the values of $-\ln(I/m^3)$ computed at $1.2\,k_f$ and $2\,k_f$
are equal to a very good approximation. This
confirms the convergence of the expansion around the saddle-point and 
the reliability of the calculation. The $k$ dependence of
the saddle-point action is canceled by the prefactor, so that the 
total nucleation rate is $k$ independent. Moreover, the
prefactor is always significantly smaller than the
saddle-point action.  
The circles indicate the results of
the lattice simulations of ref.~\cite{twopone}. 
The agreement with the lattice predictions is 
good. More specifically, it is clear that the contribution of
the prefactor is crucial for the correct determination of the
total bubble-nucleation rate. Similar conclusions can be drawn 
for $\hat\lx=0.1$ and $\hat\lx=0.2$.

For larger values of $\hat \lx$ the lattice simulations have been performed
only for $\theta$ significantly larger than 1. For smaller
$\theta$, nucleation events become too rare to be observable 
on the lattice. Also the matching between the lattice and the renormalized actions becomes imprecise for large $\theta$. 
This indicates that we should expect
deviations of our results from the lattice ones. 
These deviations start becoming apparent for the value $\hat\lx=0.25$, for
which the lattice simulations were performed with $\theta \sim 10$--20.
For $\hat\lx\geq 0.3$ the lattice results are in a region in which 
the expansion around the saddle point is not reliable any more.

For $\hat \lx=0.32$
the breakdown of the expansion around the saddle point is apparent
for $1/\theta \lta 0.12$.
The $k$ dependence of the predicted bubble-nucleation rate is 
strong\footnote{The additional squares for $1/\theta=0.1$ correspond
to results from the numerical integration of eq.~(\ref{evpot}).}.
The prefactor becomes comparable to the saddle-point
action and the higher-order corrections are expected to be large.
The $k$ dependence of $S_k$ is very large. For this reason we 
have not given values of $S_k$ in this case.

The comparison of our results with data from lattice simulations constitutes
a stringent quantitative test of our method. In the region where
the renormalized action for the 
lattice model is known, the data of fig.~\ref{fig4.2}, 
provide a strong confirmation of the reliability of our approach.
Another test has been carried out as well. In ref.~\cite{sixth}
a comparison has been made with the results of the thin-wall approximation
in three dimensions. Very good agreement has been found, which provides
additional support for the validity of the method.

\section{Quantum statistics for fermions and bosons}
\label{QuantStat}
\subsection{Quantum universality}

At low temperature the characteristic energies are near the ground
state energy. Only a few states are important and one expects
the effects of quantum mechanical coherence to become important.
The low temperature region is therefore the quantum domain. In particular,
the limit $T\to 0$ projects on the ground state and single excitations
of it. In particle physics this is the vacuum, and the single
excitations correspond to the particles. The opposite is classical
statistics. For large $T$ many excitations contribute in thermal
equilibrium. These thermal fluctuations destroy the quantum
mechanical coherence. The high temperature region is
the classical domain. This also holds for quantum field theories.
Their properties in thermal equilibrium are dominated by classical
aspects for high $T$.

For bosons (and vanishing chemical potential $\mu$) there is another
interesting limit. If the characteristic length scale of a physical
process becomes very large, many individual local modes must
be involved. Again, the quantum mechanical coherence becomes unimportant
and classical physics should prevail. Translated to characteristic
momenta $\vec q$ one concludes that the limit $|\vec q|\to 0$ belongs
to the classical domain, typically well described by classical fields.

In thermodynamic equilibrium a characteristic length scale is
given by the correlation length $\xi=m_R^{-1}$. This provides
us with a dimensionless combination $\xi T$ for an assessment
of the relevance of quantum statistics. For $\xi T\gg 1$ a classical
treatment should be appropriate, whereas for $\xi T\ll1$ quantum
statistics becomes important. For second-order phase transitions
the correlation length diverges. The universal critical phenomena
are therefore always described by classical statistics! This extends
to first-order transitions with small enough $\xi T$. On the other
hand the temperature may be much smaller than the microphysical
momentum scale $\Lambda$. The modes with momenta $(\pi T)^2
<q^2<\Lambda^2$ are governed by quantum statistics. Correspondingly,
the renormalization flow in the range $(\pi T)^2<k^2<\Lambda^2$ is
determined by the partial fixed points of the quantum system. This
introduces a new type of universality since much of the microscopic
information is lost in the running from $\Lambda$ to $\pi T$.
Subsequently, the difference in the flow for $k>\pi T$ and $k<\pi T$
leads to a crossover phenomenon\footnote{Such crossover phenomena
have also been discussed in the framework of ``environmentally friendly
renormalization'' \cite{EFRG}.} to classical statistics. 
For second-order phase transitions and $m_R\ll T\ll\Lambda$ not only
the standard universal critical exponents and amplitude ratios can
be computed. Also the (classically non-universal) critical amplitudes
can be predicted as a consequence of the new type of universality.
One may call this phenomenon ``quantum
universality''.

More formally, one can express the quantum trace in the partition
function (1.1) by a functional integral in D+1 dimensions, with D the
number of space dimensions of the classical theory. The basic
ingredient is Feynman's path integral in Euclidean space. Using
insertions of a complete set of eigenstates
\beq\label{7.1}
1=\int d\chi_n|\chi_n><\chi_n|\eeq
one has (for a single degree of freedom)
\beq\label{7.2}
Z=tr\ e^{-\beta H}=\prod^{N-1}_{n=0}\int d\chi_n\quad  
<\chi_n|e^{-\frac{\beta}{N}H}|\chi_{n+1}>
\eeq
with $\chi_N\equiv\chi_0$. Introducing the Euclidean time $\tau=n\beta/N,\   
\chi_n\equiv \chi(\tau),\ \chi(\beta)=\chi(0)$ one obtains for $N\to\infty, \  
d\tau=\beta/N$ a functional integral representation of $Z$
\bea\label{7.3}
Z&=&\int D\chi\ e^{-S[\chi]}\nonumber\\
e^{-S[\chi]}&=&\prod_n<\chi(\tau)|e^{-Hd\tau}|\chi(\tau+d\tau)>\eea
This formulation is easily generalized to the case where $\chi$ carries
additional indices or depends on coordinate or momentum variables.
If $\chi$ is a field in $D$ dimensions, we encounter a $D+1$
dimensional functional integral. The additional dimension corresponds
to the Euclidean time $\tau$ and is compactified on a torus with
circumference $\beta=1/T$.
The action $S[\chi]$ can be evaluated by
standard methods for the limit $d\tau\to 0$. For the example of a Hamiltonian
depending on coordinate and momentum-type operators $Q, P$ with
$[Q(\vec x), P(\vec y)]=i\delta(\vec x-\vec y)$
\beq\label{7.5}
H=\int d^3 x\{\frac{1}{2}P^2(x)+V(Q(x))+\frac{1}{2}
\vec\nabla Q(x)\vec \nabla Q(x)\}\eeq
one finds
\beq\label{7.6}
S=\int d^4x\{\frac{1}{2} \partial_\mu\chi(x)\partial_\mu\chi(x)+V
(\chi(x))\}\eeq
with $x=(\tau,\vec x),\ \partial_\mu=(\partial_\tau,\vec\nabla)$.
Up to the dimensionality this is identical to
eq. (1.4) if $V$ is a quartic polynomial. In particular,
eqs. (\ref{7.5}), (\ref{7.6}) describe the Hamiltonian
and the action for a scalar quantum field theory.

Bosonic fields obey the periodicity condition
$\chi(\vec x,\beta)=\chi(\vec x,0)$ and can therefore be expanded
in ``Matsubara modes'' with $j\in Z\!Z$
\beq\label{7.4}
\chi(\vec x,\tau)=\sum_j\chi_j(\vec x)e^{\frac{2\pi i}{\beta}
j\tau}\eeq
Correspondingly, the zero components of the momenta $q_\mu=(q_0,\vec q)$ are
given by the discrete Matsubara frequencies
\beq\label{7.5a}
q_0=2\pi jT\eeq
This discreteness
is the only difference between quantum field theory in thermal
equilibrium and in the vacuum. (For the vacuum $T\to 0$ and $q_0$ becomes
a continuous variable.) Eq. (\ref{7.5a}) is the only point where the
temperature enters into the quantum field theoretical formalism.
Furthermore, the discreteness of $q_0$ constitutes the only difference
between quantum statistics for a system in $D$ space dimensions and
classical statistics for a corresponding system in $D+1$ dimensions.

In thermal equilibrium a characteristic value of $q_0$ may be
associated with $m_R=\xi^{-1}$, where $m_R$ is the smallest
renormalized mass of the system. For $T/m_R\ll 1$ the discreteness
of $q_0$ may be neglected and the quantum system is essentially
determined by the ground state. According to our general discussion
the opposite limit $T/m_R\gg 1$ should be dominated by classical
statistics, at least for the ``infrared sensitive'' quantities
(i.e. except for the relevant parameters in critical phenomena).
For the action (\ref{7.6}) this is easily seen in perturbation
theory by realizing that the propagator
\beq\label{7.6a}
(q_\mu q_\mu+m_R^2)^{-1}=(\vec q^2+(2\pi jT)^2 + m_R^2)^{-1}\eeq
has a masslike term $\sim T^2$ for all $j\not=0$. For large $T$ the
contributions from the $j\not=0$ Matsubara modes are therefore
strongly suppressed in the momentum integrals for the fluctuation
effects and may be neglected. In the limit where only the
$j=0$ mode contributes in the functional integral (\ref{7.3}),
we are back to a classical functional integral in $D$ dimensions.
(The $\tau$-integration in eq. (\ref{7.6a}) results in a simple
factor $\beta$ which can be reabsorbed by a rescaling of $\chi$.)
This ``dimensional reduction'' \cite{DR} from $D+1$ to $D$ dimensions
is characteristic for the crossover from the quantum domain
to the classical domain as the characteristic momenta given by
$m_R$ fall below $T$. We will see in section 7.3 how the flow
equations realize this transition in a simple and elegant way.

We also want to discuss  the quantum statistics of fermionic
systems. They can be treated in parallel to the bosonic system,
with one major modification: The classical fields $\chi(\vec x)$
are replaced by anticommuting Grassmann variables $\psi(\vec x)$
\beq\label{7.7}
\{\psi_a(\vec x),\ \psi_b(\vec y)\}=0\eeq
Correspondingly, they are antiperiodic in $\beta$, i. e.
$\psi(\vec x, \beta)=-\psi(\vec x, 0)$. In the next subsection
7.2 we  generalize the exact renormalization group equation
to Grassmann variables. Applications of 

\subsection{Exact flow equation for fermions}
\label{FlowFerm}

The exact flow equation
(\ref{2.17}) can be extended to fermions in a straightforward way except 
for one important minus sign \cite{Wet90-1,CKM97-1}. 
We write the fermionic infrared cutoff term as
\beq\label{femionicIR}
 \Delta S_k^{(F)} = \int \frac{d^dq}{(2\pi)^d} \ol\zeta(q) R_{kF}(q)
 \zeta(q) 
\eeq
where $\ol{\zeta}$, $\zeta$ are euclidean Dirac or Weyl spinors in even
dimensions and all spinor indices have been suppressed. 
(There appears an additional factor of $1/2$ in the rhs of 
(\ref{femionicIR}) if $\ol{\zeta}$ and $\zeta$ are related or obey 
constraints, as for the case 
of Majorana and Majorana-Weyl spinors in those dimensions where this
is possible. For Dirac spinors or the standard
$2^{d/2-1}$ component Weyl spinors this factor is absent.)

In complete analogy to the case of bosonic fields $\chi$ (cf.\ sections
\ref{AverAct}, \ref{ExactFlow}) we add the infrared cutoff term 
(\ref{femionicIR}) to the 
classical action $S$
\beq
S_k[\chi,\ol{\zeta},\zeta]=S[\chi,\ol{\zeta},\zeta]
+ \Delta S_k [\chi] + \Delta S_k^{(F)}[\ol{\zeta}, \zeta] \, .
\eeq 
The generating functional of connected Green functions $W_k[J,\ol{\eta},
\eta]$ is now a generating functional of the Grassmann valued sources
$\ol{\eta}$ and $\eta$ in addition to the scalar sources $J$
already contained in (\ref{2.5}).
The effective average action is then defined as
\bea
\label{GaFB}
\Ga_k[\phi,\ol{\psi},\psi,]&=&-W_k[J,\ol{\eta},\eta]
+\int d^dx \left[J(x)\phi(x)+\ol{\eta}(x) \psi(x)- \ol{\psi}(x)
\eta(x)\right]
\nonumber\\
&&-\Delta S_k [\phi]-\Delta S_k^{(F)}  [\ol{\psi},\psi,]
\eea
with 
\beq
\psi(x)=\frac{\delta W_k[J,\ol{\eta},\eta]}{\delta \ol{\eta}(x)}
\qquad , \qquad 
\ol{\psi}(x)=\frac{\delta W_k[J,\ol{\eta},\eta]}{\delta \eta(x)} \, .
\eeq
The matrix $\Gamma_k^{(2)}[\phi,\ol{\psi},\psi]$ of second functional
derivatives of $\Gamma_k$ has a block substructure in 
boson-antifermion-fermion space which we abbreviate as
$B-\ol{F}-F$:
\begin{displaymath}
\Gamma_k^{(2)}= \left( \begin{array}{ccc}
\Gamma_{k,BB}^{(2)} & \Gamma_{k,B\ol{F}}^{(2)} & \Gamma_{k,BF}^{(2)} \\
\Gamma_{k,\ol{F}B}^{(2)} & \Gamma_{k,\ol{F}\ol{F}}^{(2)} 
& \Gamma_{k,\ol{F}F}^{(2)} \\
\Gamma_{k,FB}^{(2)} & \Gamma_{k,F\ol{F}}^{(2)} & \Gamma_{k,FF}^{(2)} 
\end{array} \right)
\end{displaymath}
where
\beq
\Gamma_{k,BB}^{(2)}=\frac{\delta^2 \Gamma_k}{\delta \phi \delta \phi}
\quad , \quad
\Gamma_{k,B\ol{F}}^{(2)}=\frac{\delta^2 \Gamma_k}{\delta \phi \delta \ol{\psi}}
\quad , \quad
\Gamma_{k,\ol{F}F}^{(2)}=-\frac{\delta^2 \Gamma_k}{\delta \ol{\psi} 
\delta \psi}
\quad , \quad {\rm etc.}
\eeq 
It is now straightforward to generalize the exact flow equation
(\ref{2.17}) to include fermions and one finds
\bea
  \frac{\partial}{\partial k}\Ga_k[\phi,\ol{\psi},\psi] 
  &=&  \half {\rm STr}\left\{\left[
  \Ga_k^{(2)}[\phi,\ol{\psi},\psi]+{\rm \bf R}_k\right]^{-1}
  \frac{\partial}{\partial k} {\rm \bf R}_k\right\} 
\nonumber\\
  &=& \half\Tr\left\{\left[
  \Ga_k^{(2)}[\phi,\ol{\psi},\psi]+{\rm \bf R}_k\right]^{-1}_{BB}
  \frac{\partial}{\partial k} {R}_k\right\}
\nonumber\\
  &-& \Tr\left\{\left[
  \Ga_k^{(2)}[\phi,\ol{\psi},\psi]+{\rm \bf R}_k\right]^{-1}_{F\ol{F}}
  \frac{\partial}{\partial k} {R}_{kF}\right\} 
  \label{ExactFB}
\eea
where the block structure of ${\rm \bf R}_k$ in $B-\ol{F}-F$ space is
given by
\begin{equation}
{\rm \bf R}_k= \left( \begin{array}{ccc}
R_k & 0 & 0 \\
0 & 0 & R_{kF} \\
0 & -R_{kF} & 0
\end{array} \right)\; 
\end{equation}
and the minus sign in the fermionic trace arises from the
anticommutation of fermions in the step corresponding  
to eq.\ (\ref{2.25}).
We observe that the infrared cutoff does not mix bosons and fermions. 
For purely bosonic background fields the inverse propagator
$\Gamma_k^{(2)}$ is also block diagonal, and one obtains
separate contributions from fermions and bosons. This is not
true any more for fermionic background fields since the
inverse propagator has now off-diagonal pieces. 
Again, the right-hand side of the flow equation can be expressed
as the formal derivative $\tilde\partial_t$ acting on the IR-cutoff
in a one-loop diagram.

The infrared cutoff for fermions has to meet certain requirements
for various reasons. First of all, chiral fermions do not
allow a mass term. In order to remain consistent with chiral
symmetries (a necessity for neutrinos, for example), the infrared
cutoff must have the same Lorentz structure as the kinetic term,
i.e. $R_{kF}\sim\gamma^\mu q_\mu$ \cite{Wet90-1} or, at least,
not mix left-handed and right-handed fermions, e.g. $R_k\sim
\gamma^\mu\gamma^5 q_\mu$.\footnote{We use a Euclidean convention with  
$\{\gamma^\mu,\gamma^\nu \}=2 \delta^{\mu\nu}$ and 
$\gamma^5=-\gamma^0\gamma^1\gamma^2\gamma^3$ in four dimensions.}
On the other hand, for
$q^2\to 0$ the infrared cutoff should behave as
$R_{kF} \sim k$, e.g. $R_{kF}\sim k \gamma^\mu q_\mu/\sqrt{q^2}$. The
nonanalyticity of $\sqrt{q^2}$ may then be a cause of problems. We
will discuss here a few criteria for a chirally invariant infrared
cutoff and present an explicit example which is suitable for practical
calculations. 

First of all, the fermionic infrared cutoff term $\Delta S_k ^{(F)}$ should be
quadratic in the fermion fields as in (\ref{femionicIR}).
We next require that $\Delta S_k^{(F)}$
should respect all symmetries of the kinetic term for free fermions.
We will include here chiral symmetries and Lorentz invariance. (Gauge
symmetries may be implemented by covariant derivatives in a
background gauge field \cite{RW93-1}.) 
The symmetry requirement implies in a momentum space
representation \mbox{($\psi(x)=\int \frac{d^dq}{(2\pi)^d} e^{iqx} \psi(q)$)}
\beq
 \Delta S_k^{(F)} = -\int \frac{d^dq}{(2\pi)^d} \ol{\psi}(q) Z_{\psi,k}
 q\sslash r_F\left(\frac{q^2}{k^2}\right)\psi(q) \; , 
\eeq
i.e.\ $R_{kF} \equiv - Z_{\psi,k} q\sslash r_F$,
where $\ol{\psi}$, $\psi$ are Dirac spinors and 
$q\sslash=q_\mu\gamma^\mu$.
The wave-function renormalization
$Z_\psi$ is chosen for convenience such that it matches with a
fermion kinetic term of the form 
\beq
\Gamma^{\rm kin}_k[\ol{\psi},\psi]=-\int \frac{d^dq}{(2\pi)^d}
\ol{\psi}(q) Z_{\psi,k}q\sslash \psi(q) \; .
\eeq
The third
condition requires that $\Delta S_k^{(F)}$ acts effectively as an infrared
cutoff. This means that for $k\to\infty$ the combination $Z_\psi
r_F(\frac{q^2}{k^2})$ should diverge for all values of $q^2$. This
divergence should also occur for finite $k$ and $q^2/k^2 \to 0$ and
be at least as strong as $\left( k^2/q^2\right)^{1/2}$. As a fourth
point we remark that $\Gamma_k$ becomes the effective action in the limit
$k\to 0$ only if $\dsp{\lim_{k\to 0}\Delta S_k^{(F)}=0}$. This should hold for all
Fourier modes separately, i.e.\ for $\dsp{\lim_{k\to 0} Z_{\psi,k}={\rm
const.}}$ one requires
\beq
 \lim_{k\to 0}r_F \left(\frac{q^2}{k^2}\right)q\sslash=0 \; .
 \label{Cond3}
\eeq
We furthermore request that (\ref{Cond3}) also holds in
the limit $q^2\to 0$. Together with the third condition this implies
exactly
\beq
 \lim_{q^2/k^2\to 0} r_F\left(\frac{q^2}{k^2}\right) \sim
 \left(\frac{q^2}{k^2}\right)^{-\half} \; .
 \label{Cond4}
\eeq
The requirement (\ref{Cond4}) implies a smooth behavior of $R_{kF}$
for $q^2/k^2\to 0$. However, the nonanalyticity of $r_F$
at $q^2=0$ requires a careful choice of $r_F$ in order to circumvent 
problems. We note that
$r_F$ appears in connection with the fermion propagator from
$\Gamma_k$. The combination which will appear in calculations is
\beq
 P_F = q^2 \left( 1+r_F\right)^2 \; .
\eeq
Up to the wave-function renormalization, $P_F$ corresponds to the
squared inverse propagator of a free massless fermion in the presence
of the infrared cutoff. We will require that $P_F$ and therefore
$(1+r_F)^2$
is analytic in $q^2$ for all $q^2\geq 0$. A reasonable choice is
\beq
 P_F =\frac{q^2}{1-\exp\left\{
 -\frac{q^2}{k^2}\right\}} \; .
 \label{PF}
\eeq
We will employ this choice in the treatment of 
Nambu--Jona-Lasinio type models in section \ref{Fermions}, where
we also discuss another choice for $P_F$.

\subsection{Thermal equilibrium and dimensional reduction}

The extension of the flow equations to
non--vanishing temperature $T$ is straightforward~\cite{TW93-1}. The 
(anti--)periodic boundary conditions for (fermionic) bosonic fields in
the Euclidean time direction \cite{Kap} leads to the replacement
\begin{equation}
  \label{AAA120}
  \int\frac{d^d q}{(2\pi)^d}f(q^2)\to
  T\sum_{j\in\ZZZ}\int\frac{d^{d-1}\vec{q}}{(2\pi)^{d-1}}
  f(q_0^2(j)+\vec{q}^{\,2})
\end{equation}
in the trace of the flow equation (\ref{2.32}) 
when represented as a momentum
integration. One encounters a discrete spectrum of Matsubara frequencies
for the zero component
$q_0(j)=2j \pi T$ for bosons and $q_0(j)=(2j+1) \pi T$ for fermions.
Hence, for $T>0$ a four--dimensional QFT can be interpreted as a
three--dimensional model with each bosonic or fermionic degree of
freedom now coming in an infinite number of copies labeled by
$j\in\ZZZ$ (Matsubara modes). Each mode acquires an additional
temperature dependent effective mass term $q_0^2(j)$ except
for the bosonic zero mode for which $q^2_0(0)$ vanishes. At high
temperature all massive Matsubara modes decouple from
the dynamics of the system. In this case, one therefore expects to observe an
effective three--dimensional theory with the bosonic zero mode as the
only relevant degree of freedom. 
One may visualize this behavior by noting that for a given characteristic
length scale $l$ much larger than the inverse temperature $\beta$ 
the compact 
Euclidean ``time'' dimension cannot be resolved anymore. 

This phenomenon of dimensional reduction can be observed
directly from the non-perturbative flow equations. 
The replacement
(\ref{AAA120}) in (\ref{2.32}) manifests itself in the flow equations
only through a change to
$T$--dependent threshold functions.  For instance, the dimensionless
threshold functions $l_n^d(w;\eta_\Phi)$ defined in eq.~(\ref{2.45}) are
replaced by
\begin{equation}
  \label{AAA200}
  l_n^d \left( w,\frac{T}{k};\eta_\Phi \right) \equiv
  \frac{n+\delta_{n,0}}{4}v_d^{-1}k^{2n-d}
  T\sum_{j\in\ZZZ}\int
  \frac{d^{d-1}\vec{q}}{(2\pi)^{d-1}}
  \left(\frac{1}{Z_{\Phi,k}}\frac{\partial R_k(q^2)}{\partial t}\right)
  \frac{1}{\left[P(q^2)+k^2 w\right]^{n+1}}
\end{equation}
where $q^2=q_0^2+\vec{q}^{\,2}$ and $q_0=2\pi j T$. In the
limit $k\gg T$ the sum over Matsubara modes approaches the integration
over a continuous range of $q_0$ and we recover the zero temperature
threshold function $l_n^d(w;\eta_\Phi)$.  In the opposite limit $k\ll
T$ the massive Matsubara modes ($j\neq0$) are suppressed and we expect
to find a $d-1$ dimensional behavior of $l_n^d$. In fact, one obtains
from~(\ref{AAA200})
\begin{equation}
  \label{AAA201}
  \begin{array}{rclcrcl}
    \dsp{l_n^d(w,T/k;\eta_\Phi)} &\simeq& \dsp{
      l_n^{d}(w;\eta_\Phi)}
    &{\rm for}& \dsp{T\ll k}\; ,\nnn
    \dsp{l_n^d(w,T/k;\eta_\Phi)} &\simeq& \dsp{
      \frac{T}{k}\frac{v_{d-1}}{v_d}
      l_n^{d-1}(w;\eta_\Phi)}
    &{\rm for}& \dsp{T\gg k}\; .
  \end{array}
\end{equation}

For the choice of the infrared cutoff function $R_k$
eq.~(\ref{2.15}) for bosons and eq.~(\ref{PF}) for fermions 
the contribution of the temperature-dependent massive
Matsubara modes to
$l_n^d(w,T/k;\eta_\Phi)$ is exponentially suppressed for $T\ll k$.  
Nevertheless, all bosonic threshold
functions are proportional to $T/k$ for $T\gg k$ whereas those with
fermionic contributions vanish in this limit. 
This behavior is demonstrated \cite{BJW97-1} in figure
\ref{Thresh} where we have plotted the quotients
$l_1^4(w,T/k)/l_1^4(w)$ and $l_1^{(F)4}(w,T/k)/l_1^{(F)4}(w)$ of
bosonic and fermionic threshold functions, respectively.
\begin{figure}
\unitlength1.0cm
\begin{center}
\begin{picture}(13.,18.0)

\put(0.0,9.5){
\epsfysize=11.cm
\rotate[r]{\epsffile{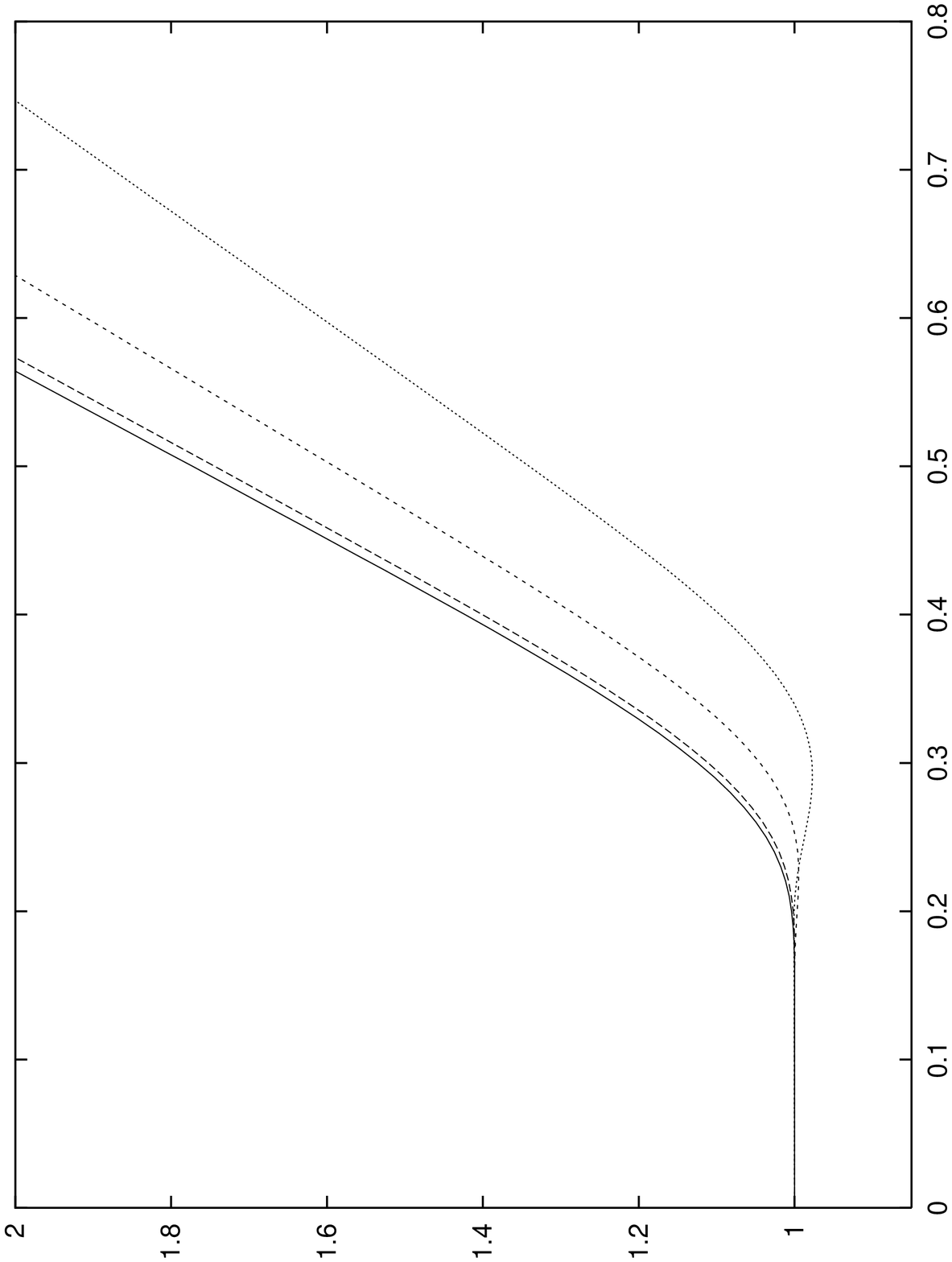}}
}
\put(-1.0,16.5){\bf $\dsp{\frac{l_1^4\left(w,\dsp{\frac{T}{k}}\right)}
    {l_1^4(w)}}$}
\put(1.5,16.5){\bf $\dsp{(a)}$}

\put(0.0,0.5){
\epsfysize=11.cm
\rotate[r]{\epsffile{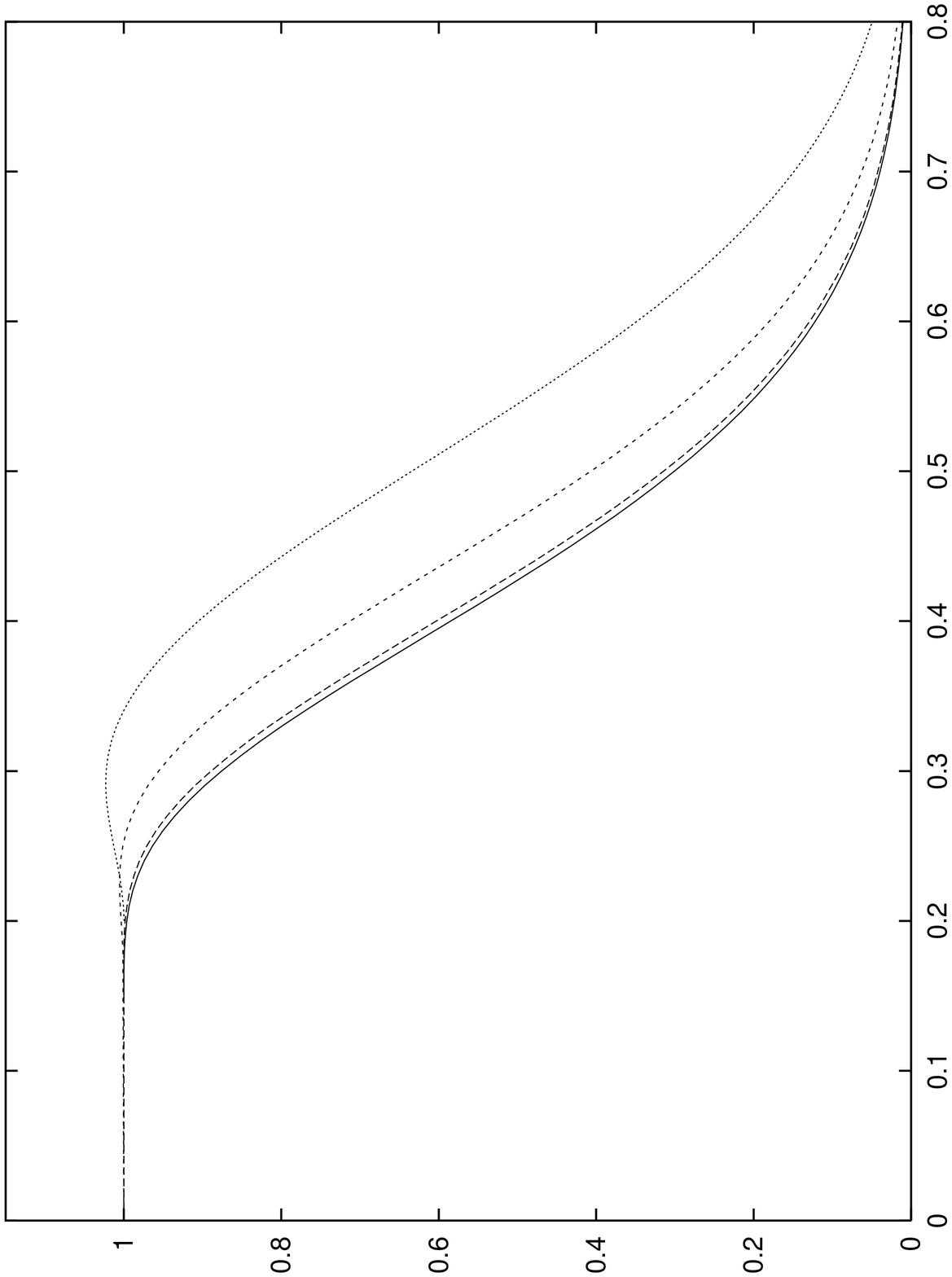}}
}
\put(-1.2,7.5){\bf
  $\dsp{\frac{l_1^{(F)4}\left(w,\dsp{\frac{T}{k}}\right)}
    {l_1^{(F)4}(w)}}$}
\put(6.1,-0.2){\bf $\dsp{T/k}$}
\put(1.5,7.5){\bf $\dsp{(b)}$}
\end{picture}
\end{center}
\caption{\em Effective dimensional reduction:
The plot shows the temperature 
  dependence of the bosonic (a) and the fermionic (b) threshold
  functions $l_1^4(w,T/k)$ and $l_1^{(F)4}(w,T/k)$, respectively, for
  different values of the dimensionless mass term $w$. 
  We have normalized them to the $T=0$ threshold functions. The solid line
  corresponds to $w=0$ whereas the dotted ones correspond to $w=0.1$,
  $w=1$ and $w=10$ with decreasing size of the dots.  For $T \gg k$
  the bosonic threshold function becomes proportional to $T/k$ whereas
  the fermionic one tends to zero.  In this range the theory with
  properly rescaled variables behaves as a classical
  three--dimensional theory.}
\label{Thresh}
\end{figure}
One observes that for $k\gg T$ both threshold functions essentially
behave as for zero temperature. For growing $T$ or decreasing $k$ this
changes as more and more Matsubara modes decouple until finally all
massive modes are suppressed. The bosonic threshold function $l^4_1$
shows for $k \ll T$ the linear dependence on $T/k$ derived in
eq.~(\ref{AAA201}).  In particular, for the bosonic excitations the
threshold function for $w\ll1$ can be approximated with reasonable
accuracy by $l_n^4(w;\eta_\Phi)$ for $T/k<0.25$ and by
$(4T/k)l_n^3(w;\eta_\Phi)$ for $T/k>0.25$. The fermionic threshold
function $l_1^{(F)4}$ tends to zero for $k\ll T$ since there is no
massless fermionic zero mode, i.e.~in this limit all fermionic
contributions to the flow equations are suppressed.  On the other
hand, the fermions remain quantitatively relevant up to $T/k\simeq0.6$
because of the relatively long tail in figure~\ref{Thresh}b.  
The formalism of the average action automatically provides
the tools for a smooth decoupling of the massive Matsubara modes as
the momentum scale $k$ is lowered from $k\gg T$ to $k\ll T$.  It therefore
allows one to directly link the four--dimensional quantum field theory
at low $T$ to the effective three--dimensional high--$T$-theory.

Whereas for $k\gg T$ the theory is most efficiently
described in terms of standard four--dimensional fields $\Phi$ a
choice of rescaled three--dimensional variables
$\Phi_{3}=\Phi/\sqrt{T}$ becomes better adapted for $k\ll T$.
Accordingly, for high temperatures one will use the rescaled 
dimensionless potential
\begin{equation}
  \label{CCC01}
  u_{3}(t,\tilde{\rho}_{3})=\frac{k}{T}
  u(t,\tilde{\rho})\; ;\;\;\;
  \tilde{\rho}_{3}=\frac{k}{T}\tilde{\rho}\; .
\end{equation}
For numerical calculations at non--vanishing temperature one can
exploit the discussed behavior of the threshold functions by using the
zero temperature flow equations in the range, say, $k\ge10T$. For smaller
values of $k$ one can approximate the infinite Matsubara sums
(cf.~eq.~(\ref{AAA200})) by a finite series so that the numerical
uncertainty at $k=10T$ is better than a given value. This approximation
becomes exact in the limit $k\ll10T$.

\subsection{The high-temperature phase transition for the $\phi^4$ 
quantum field theory}
 
The formalism of the previous sections can be applied to the
phase transition of the four-dimensional $O(N)$-symmetric $\phi^4$ 
theory at non-vanishing temperature \cite{TW93-1,RTW} (see also
\cite{EFRG,LS,ABP,UNO,BR99} for studies 
using similar techniques).
We consider models with spontaneous symmetry breaking 
at zero temperature, and investigate the
restoration of symmetry as the temperature is raised.
We specify the action together with some high momentum 
cutoff $\Lambda \gg T$ so that the theory is properly 
regulated. We then solve the evolution equation for the average 
potential for different values of the temperature. 
For $k \rightarrow 0$ 
this solution provides all the relevant features of the temperature-dependent 
effective potential. In order not to complicate our discussion,
we neglect the wave-function renormalization and
consider a very simple ansatz for the potential, in which
we keep only the quadratic and quartic terms. Improved accuracy can
be obtained by using more extended truncations of the average action.
The flow equations for the rescaled minimum of the potential
$\kappa(k,T)=\rho_0(k,T)/k^2$ and the quartic coupling $\lambda(k,T)$ 
are
\begin{eqnarray}
  \label{eq:7.1}
  \dsp{\partial_t\kappa} &=& \dsp{
    \beta_\kappa=-2 \kappa+
    \frac{1}{16\pi^2}\Bigg\{3l_1^4\left( 2\lambda\kappa,\frac{T}{k} \right)+
    (N-1)l_1^4\left( \frac{T}{k} \right)\Bigg\}
    }\\[2mm]
  \label{eq:7.2}
  \dsp{\partial_t\lambda} &=& \dsp{
    \beta_\lambda=
    \frac{1}{16\pi^2}\lambda^2
    \Bigg\{9l_2^4\left( 2\lambda\kappa,\frac{T}{k} \right)+
    (N-1)l_2^4\left( \frac{T}{k} \right)\Bigg\}
    }\; .
\end{eqnarray}

The initial conditions are 
determined by the 
``short distance values" 
$\rrz(k=\Lambda)$ and $\lb(k=\Lambda)$
that
correspond to the minimum and the quartic coupling of the bare
potential.
We then have to compute the evolution, starting at $k=\Lambda$ and
following the renormalization flow towards $k=0$.
This procedure has to be 
followed for $T=0$ and then to be repeated for $T\not=0$ in order to 
relate the zero and non-zero temperature 
effective potential of the same theory.
Since the running of the parameters is the same in the zero and non-zero 
temperature case for $k \gg T$
we actually do not need to compute the evolution in this range of $k$. Our 
strategy is equivalent to the following procedure: 
We start with the zero-temperature theory at $k=0$ taking  
the renormalized parameters as input. We 
subsequently integrate the zero-temperature flow equations 
``up" to $k=T/\theta_1$, where  $\theta_1$ is chosen such
that $l_n^4\left( w,{T}/{k} \right) = l_n^4(w)$ to a good approximation
for $k>T/\theta_1$. For bosons $\theta_1 \lta 0.1$ is sufficient. 
We can now use the values of the running 
parameters at $k=T/\theta_1$ as initial conditions for the non-zero temperature
flow equations and integrate them ``down" 
to $k=0$. In this way we obtain the 
renormalized 
parameters at non-zero temperature in terms of the renormalized parameters
at zero temperature.

As we discussed in the previous subsection, 
the threshold functions simplify considerably for $k<T/\theta_2$,
with $\theta_2 \simeq 0.4$ for bosons. 
In this range $l_n^4\left( w,{T}/{k} \right) = 4 l_n^3(w) T/k$
to a good approximation and we expect
an effective three-dimensional
evolution.
We can define the dimensionless quantities
\bea 
\tilde{\kappa}(k,T) & = &~\frac{\rho'(k,T)}{k} = \frac{\rrzkt}{k T} \nonumber \\
\tilde{\lambda}(k,T) & = &~\frac{\lambda'(k,T)}{k} = \lx(k,T) \frac{T}{k}.
\label{finin} \eea
where $\rho'$ and $\lambda'$ have the canonical three-dimensional
normalization.
In terms of these quantities the flow equations read for $k<T/\theta_2$:
\bea
\frac{d \kt}{dt} & = & - \kt +
\frac{1}{4 \pi^2} \bigl\{ (N-1) l^3_1 + 3 l^3_1(2 \lt \kt) \bigr\}
\label{fitena} \\
\frac{d \lt}{dt} & = &~
- \lt
+ \frac{1}{4 \pi^2} \lt^2 \bigl\{ (N-1)l^3_2  +
9 l^3_2(2 \lt \kt) \bigr\}.
\label{fitenb} 
\eea
The main qualitative difference of the last equations from those of
the zero-temperature theory arises from
the term $-\lt$ in the rhs of eq.~(\ref{fitenb}), which is
due to the dimensions of $\lambda'$. In consequence, the 
dimensionless quartic coupling $\lt$ is not infrared free. Its 
behaviour with $k \rightarrow 0$ is characterized by an 
approximate fixed point for the region where $\kt$ varies only 
slowly. Taken together, the 
pair of differential equations 
for $(\lt, \kt)$ 
has an exact fixed point 
($\tilde{\kappa}_{*}$, $\lt_{*}$) corresponding to 
the phase transition.

By using $l^4_n(w,T/k)=l^4_n(w)$ 
 for $k > T/\bar \theta$ and $l_n^4(w,T/k)=4l^3_n(w)T/k$ for $k<T/\bar
\theta$, with $\bar \theta$ between $\theta_1$ and $\theta_2$,
our procedure simplifies even further. It may be summarized 
as ``run up in four dimensions, run down in three dimensions", with a 
matching of the $k$-dependent couplings at the scale $T /\bar\theta,
~\bar\theta
\approx0.25$.
In practice we take the ``threshold correction" from the 
different running for $ T / \theta_2 < k < T/ \theta_1$
into account numerically. 
In the case that $\rrzkt$ becomes zero at some non-zero 
$\ks$ we continue with the equations for the symmetric regime 
with boundary conditions $m^2 (\ks,T)=0$ and  
$\lb (\ks,T)$ given by its value at the end of the running in the 
spontaneously broken regime.\\

\begin{figure}[h]
\unitlength1.0cm
\begin{center}
\begin{picture}(13.,9.)
\put(0,0.){
\epsfysize=10.cm
\epsfxsize=12.cm
\epsffile{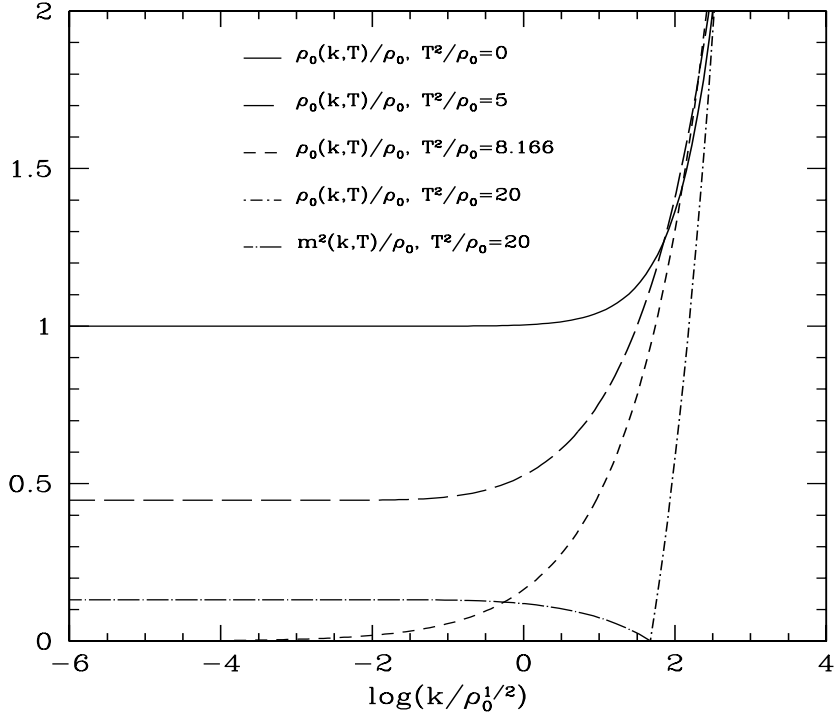}
}
\end{picture}
\end{center}
\vspace*{-0.5cm}
\caption[]{\em 
The evolution of the minimum of the potential 
$\rho_0(k,T)$ at various temperatures. For $T > \tcr$
the evolution of the mass term $m^2(k,T)$ in the symmetric regime
is also displayed. $N=1$ and $\lx_R=0.1$.
}
\label{fig7.1}
\end{figure}

\begin{figure}[h]
\unitlength1.0cm
\begin{center}
\begin{picture}(13.,9.)
\put(0,0.){
\epsfysize=10.cm
\epsfxsize=12.cm
\epsffile{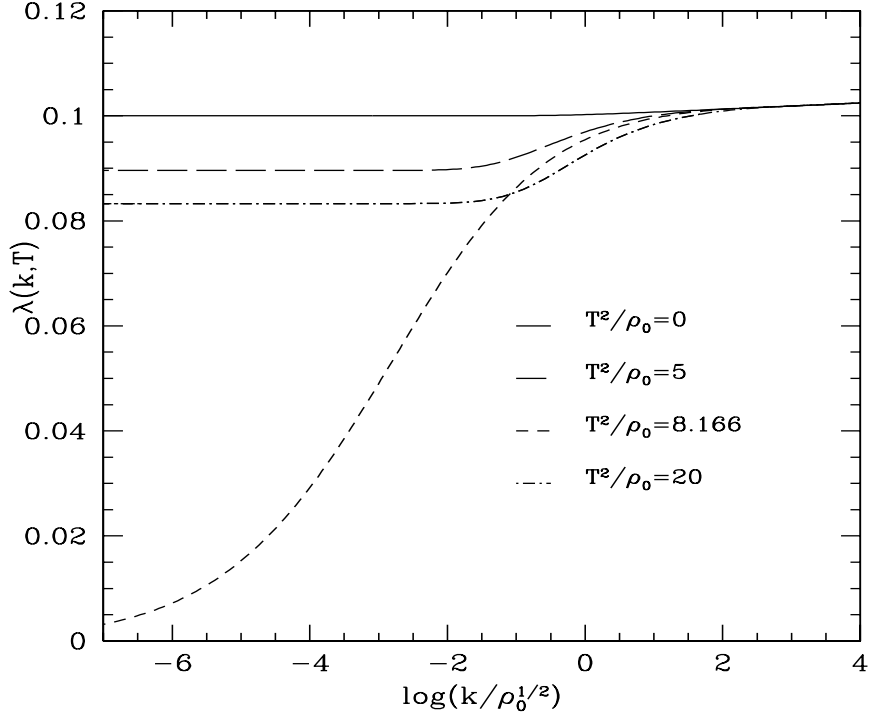}
}
\end{picture}
\end{center}
\caption[]{\em Scale dependence of the quartic coupling 
$\lx(k,T)$ at various temperatures. The strong drop near the critical
temperature is characteristic for the critical behavior.
}
\label{fig7.2}
\end{figure}

\begin{figure}[h]
\unitlength1.0cm
\begin{center}
\begin{picture}(13.,9.)
\put(0,0.){
\epsfysize=10.cm
\epsfxsize=12.cm
\epsffile{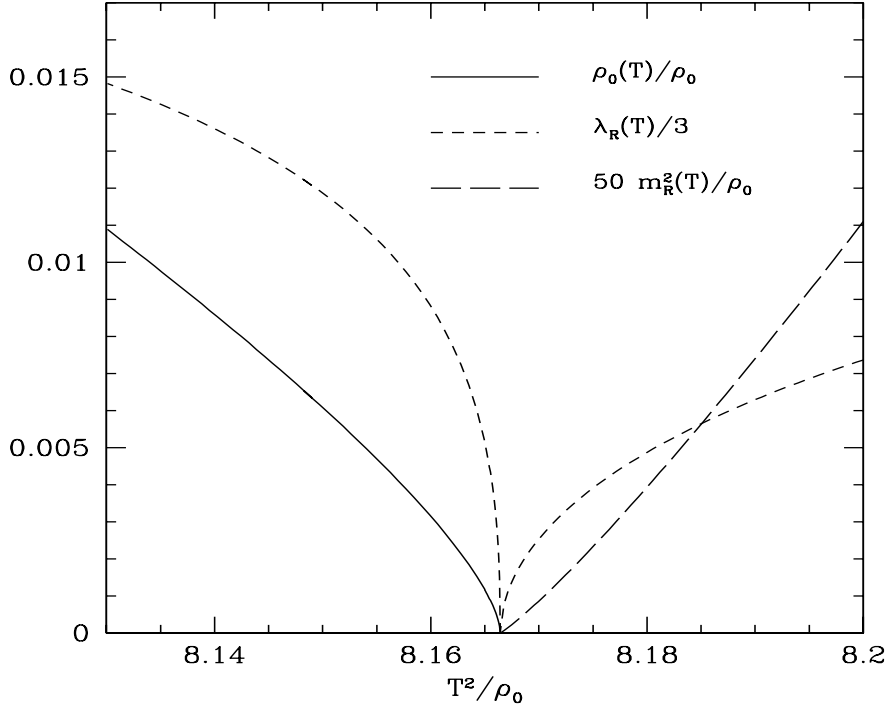}
}
\end{picture}
\end{center}
\caption[]{\em High temperature symmetry restoration: We show the
temperature dependence of the  order parameter
 $\rho_0(T)$, the renormalized mass $m^2_R(T)$ and the 
quartic coupling  $\lx_R(T)$
near $\tcr$. The critical behavior of a second order phase transition
 is apparent.
}
\label{fig7.3}
\end{figure}

\begin{figure}[h]
\unitlength1.0cm
\begin{center}
\begin{picture}(13.,9.)
\put(0,0.){
\epsfysize=10.cm
\epsfxsize=12.cm
\epsffile{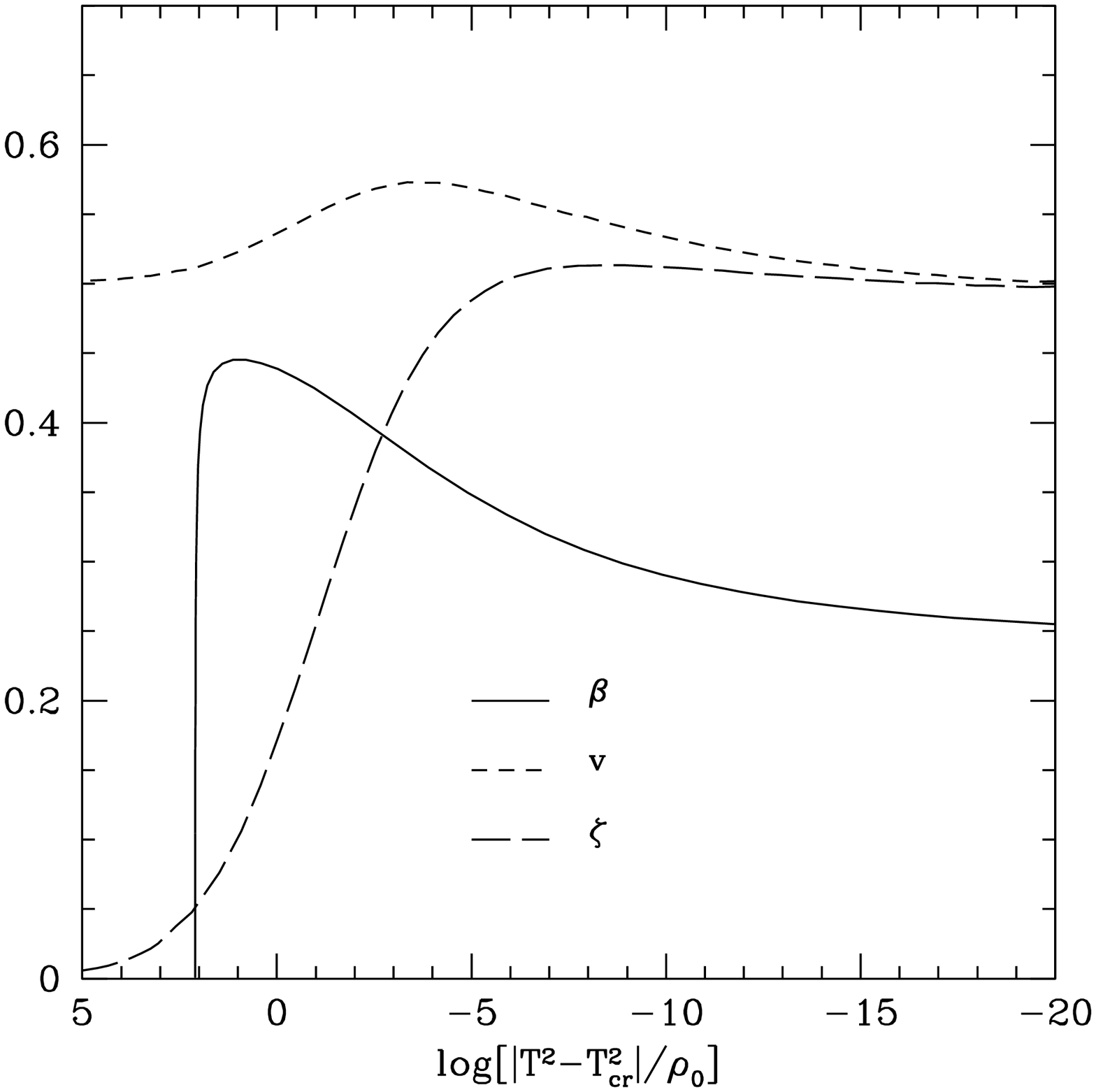}
}
\end{picture}
\end{center}
\caption[]{\em Temperature dependent effective 
 critical exponents $\beta, \nu$ and $\zeta$. As the critical temperature
 is approached
$(T\to \tcr)$ they become equal to the critical exponents of
the zero-temperature three-dimensional theory.
}
\label{fig7.4}
\end{figure}

The results of the numerical integration of the flow equations for $N=1$
are presented in figs.~\ref{fig7.1}, \ref{fig7.2}
for a zero-temperature theory with renormalized quartic coupling
$\lx_R=0.1$.
The solid line in fig.~\ref{fig7.1} displays the ``quadratic 
renormalization'' of 
the minimum of the 
zero-temperature average potential.
At non-zero temperature (dashed lines) we notice 
the deviation from the zero temperature behaviour. 
For low temperatures, in the limit $k \rightarrow 0$, $\rrzkt$ reaches
an asymptotic value $\rho_0(0,T) < \rho_0(0) \equiv \rrz$.
This value corresponds to the vacuum expectation value 
of the non-zero temperature 
theory and we denote it by
$\rrzt=\rrz(0,T)$.
At a specific temperature $T_{cr}$, $\rrzt$ becomes zero and this 
signals the restoration of symmetry for $T \geq T_{cr}$.
The running of $\lx(k)$, $\lx(k,T)$ 
is shown in fig.~\ref{fig7.2}. 
We observe the logarithmic running of $\lx(k)$ (solid line) 
which is stopped by the mass 
term. For non-zero temperatures $\lx(k,T)$ deviates from the 
zero temperature running and
reaches a non-zero value in the limit $k \rightarrow 0$.
We observe that $\lx(k,T)$ runs to zero for $T \rightarrow T_{cr}$
\cite{parisi,TW93-1}.
For $T > T_{cr}$ the running in the spontaneously broken regime ends 
at a non-zero $\ks$, at which $\rrz(\ks,T)$ equals zero. 
From this point on we continue the evolution in the symmetric regime.
The running of the mass term at the origin $\mbkt$ is depicted in 
fig.~\ref{fig7.1}, while the evolution of 
$\lx(k,T)$ proceeds continuously in the new 
regime as shown in fig.~\ref{fig7.2}.

The procedure of ``running up in four dimensions" and ``running down in
three dimensions" 
provides the connection between the renormalized quantities 
at zero and non-zero temperature. We define the zero temperature theory
in terms of the location of the minimum $\rrz$ and the renormalized 
quartic coupling $\lx_R=\lx(0)$. 
Through the solution of the evolution equations we obtain 
$\rrzt$ and $\lx_R(T)=\lx(0,T)$ 
for non-zero temperatures $T < T_{cr}$. For $T \geq T_{cr}$ the symmetry is
restored ($\rrzt=0$) and the non-zero temperature theory is described 
in terms of $\mbzt=m^2(0,T)$ and $\lbzt$.
In fig.~\ref{fig7.3} we plot ${\rrzt}/{\rrz}$, $\lbzt$ and 
${\mbzt}/{T^2}$ as a function of temperature for $N=1$
and $\lx_R=0.1$.
As the temperature increases towards 
$T_{cr}$ we observe a continuous transition from the spontaneously broken 
to the symmetric phase. This clearly indicates a 
second order phase transition.
The renormalized quartic coupling $\lbzt$ remains close to its
zero temperature value $\lx_R$ for a large range of temperatures and 
drops quickly to zero at $T=T_{cr}$.
Recalling our parametrization of the average potential in terms of its
successive $\rho$ derivatives at the minimum, we conclude that, at $T_{cr}$,
the first
non-zero term in the expression for the effective potential  
is the $\phi^6$ term (which we have neglected in our truncated solution).
For $T \gg T_{cr}$ the coupling 
$\lbzt$ quickly grows to approximately its zero temperature 
value $\lx_R$, while $\mbzt$ asymptotically becomes 
proportional to $T^2$ as $T \rightarrow \infty$. In the temperature
range where $\lambda_R(T)\lta 0.5 \lambda_R(0)$
the fluctuations are important and the universal critical behavior 
becomes dominant (cf. sect. \ref{intro2}).

The value of the critical temperature $T_{cr}$ in terms of the 
zero temperature quantities has been calculated in the context of
``naive" perturbation theory \cite{lindept, doljak, weinberg}. 
It was found that it is given by 
$T^2_{cr}= {24\rrz}/{(N+2)} $, independently of the quartic coupling
in lowest order.
This prediction was confirmed in ref. \cite{TW93-1}
for various values  of $N$. 
Another parameter which can be compared with the perturbative 
predictions is $\mbzt$ in the limit $T \rightarrow \infty$.
In ref. \cite{TW93-1} it was shown that the quantity 
$\left[ \frac{\mbzt }{\lbzt (N+2) T^2} \right]^{-1}$ 
becomes equal to 24 for ${T^2}/{\rrz} \to \infty$ and small 
$\lx_R$, again in agreement
with the perturbative result.

The most important aspect of our approach is related to the infrared 
behaviour of the theory for $T \rightarrow \tcr$. 
The temperature dependence of $\rrzt, \lbzt, \mbzt$ near $\tcr$ is presented in
fig.~\ref{fig7.3}
($\lx_R=0.1$). 
We have already mentioned 
that all the above quantities become zero at $T=\tcr$. What becomes
apparent in this figure is a critical behaviour which can be
characterized by critical exponents. Following 
the notation of statistical mechanics, we parametrize the critical behaviour
of $\rrzt$ and $ \mbzt$ as
\bea
\rrzt & \propto &(\tcr^2-T^2)^{2 \beta} \nonumber \\
\mbzt & \propto &(T^2-\tcr^2)^{2 \nu}.
\label{fifiv} \eea
We also define a critical exponent $\zeta$ for $\lbzt$ in the symmetric 
regime:
\beq
\lbzt \propto (T^2- \tcr^2)^{\zeta}.
\label{fisix} \eeq
These exponents are plotted as function of the logarithm of 
$|T^2-\tcr^2|$ in fig.~\ref{fig7.4}. 
We notice that in the limit $T \rightarrow  \tcr$ 
the critical exponents approach asymptotic values. These are 
independent of $\lx_R$ and therefore fall into universality classes
determined only by $N$. They are equal to the critical exponents of the 
zero temperature three-dimensional theory.
This fact can be understood by recalling that the evolution in the
high-temperature region is determined by an effective three-dimensional theory 
whose phase diagram 
has a fixed point corresponding to the phase transition. 
For $T \rightarrow \tcr$ the evolution of $\rrzkt, \lx(k,T)$ in the 
high temperature region is given by a line in the phase diagram 
very close to the critical line. In this case $\rrzkt, \lx(k,T)$
spend an arbitrarily long ``time" $t$ 
close to the fixed point and, as a result,  
lose memory of their ``initial values" $\rrz(T/\theta_2, T),
\lx(T/\theta_2,T)$. 
The critical behaviour is determined solely by the fixed point. 
Our crude truncation for the potential results in the 
values $\beta=0.25$, $\nu=\zeta=0.5$, which can be compared
with the more accurate ones presented in section \ref{secsec}
(cf. tables \ref{critexp1} and \ref{IsingExponenten}).
Notice that these values satisfy the correct scaling law
$\nu = 2 \beta$
in the limit of zero wave-function renormalization.

The critical behaviour of $\lbzt$ is related to the resolution of the
problem of the infrared divergences which cause the breakdown of the
``naive" perturbative expansion in the limit $T \rightarrow \tcr$
\cite{lindept,weinberg}.
The infrared problem is manifest in the presence of higher order 
contributions to the effective potential which contain increasing
powers of ${\lx_R(T) T}/{k}$, where $k$ is the effective 
infrared cutoff of the theory. If the evolution of $\lx(k,T)$ 
is omitted and $\lx_R(T)$ is approximated by its zero-temperature value
$\lx_R$, these contributions diverge and the perturbative expansion
breaks down. 
A similar situation appears for the zero-temperature three-dimensional
theory in the critical region \cite{parisi}.
In this case the problem results from an effective expansion
in terms of the quantity 
${u}/{[M^2-M_{cr}^2]^{1/2}}$, where 
$u$ is the bare three-dimensional
quartic coupling and $[M^2-M_{cr}^2]^{1/2}$
is a measure of the distance from the point where the phase 
transition occurs as it is approached from the symmetric phase. 
The two situations can be seen to be of
identical nature by simply remembering that the high-temperature
four-dimensional 
coupling $\lx$ corresponds to an effective three-dimensional
coupling $\lx T$ and 
that the effective infrared cutoff in the symmetric phase is 
equal to $\smbzt$.
In the three-dimensional
case the problem has been resolved \cite{parisi} by a reformulation of the 
calculation in terms of an effective parameter ${\lb_3}/{m}$,
where $\lb_3$ is the renormalized 1-PI four point function 
in three dimensions (the renormalized quartic coupling) and 
$m$ the renormalized mass (equal to the inverse correlation length).
It has been found 
\cite{parisi,Zin93-1} that the above quantity has an infrared stable fixed
point in the critical region $m \rightarrow 0.$ No infrared divergences
arise within this approach. Their only residual effect is detected in the 
strong renormalization of $\lb_3$.
In our scheme the problem is formulated in terms of the effective 
dimensionless parameters
$ 
\tilde{\kappa}(k,T) = {\rrzkt}/{k T},
\tilde{\lambda}(k,T) = \lx(k,T) {T}/{k}
$
(see equations (\ref{finin})),
for which a fixed point corresponding to the phase transition is found.
The critical behaviour is determined by this fixed point in the limit
$k \rightarrow 0$. 
Everything remains finite in the vicinity of the critical temperature,
and the only memory of the infrared divergences is reflected in the 
strong renormalization of $\lbzt$ near $\tcr$. 
We conclude that the infrared problem 
disappears if formulated in terms of the appropriate 
renormalized quantities. 
When expressed in the correct language, it 
becomes simply a manifestation of the strong renormalization 
effects in the critical region.
In ref.~\cite{TW93-1}
the quantity ${\lbzt T}/{\smbzt}$
was calculated in the limit $T \rightarrow \tcr$. It reaches
a universal asymptotic value depending only on $N$.
For $N=1$ we find ${\lbzt T}/{\smbzt}=6.8$ within our crude truncation,
to be compared with the more accurate result $\lambda_R(T)T/m_R(T)
\approx 8$ shown in table \ref{IsingUnivAmpTab}.
Moreover, the existence of this asymptotic value 
explains the equality of the critical exponents $\nu$ and $\zeta$.

Finally, we point out that the ``non-universal quantities'' as the critical
temperature or the non-universal amplitudes are completely determined
by the renormalized zero-temperature couplings $\lambda_R$,
$\rho_{0}$. No additional free parameters (amplitudes) appear
in fig. \ref{7.1}. The microphysics at the scale $\Lambda$ may not be known
precisely, similarly to the situation often encountered in
statistical mechanics. Nevertheless, most of the memory of
the microphysics is already lost at the scale $k\approx T$, except
from the relevant parameter $\rho_{0}$ and the marginal
coupling $\lambda_R$. This predictive power for the amplitudes
is an example of quantum universality.

\section{Fermionic models}
\label{Fermions}

\subsection{Introduction \label{introferm}}

Shortly after the discovery of asymptotic freedom \cite{GWP}
it was realized \cite{CP75} that 
at sufficiently high temperature or density the theory
of strongly interacting elementary particles, quantum
chromodynamics (QCD), differs in
important aspects from the corresponding zero temperature or
vacuum properties. A phase transition at some critical
temperature $T_c$ or a relatively sharp crossover may separate the
high and low temperature physics \cite{MO96-1}. At nonzero 
baryon density QCD is expected to have a rich phase structure
with different possible phase transitions as the density varies
\cite{KrishRev,FrankLec,PhaseDia,HJSSV}. 
 
We will concentrate in the following on properties of the
chiral phase transition in QCD and consider
an application of the average action method to
an effective fermionic model.  
The vacuum of QCD contains a condensate of 
quark--antiquark pairs, $\langle \bar{\psi} \psi\rangle \not =0$, which 
spontaneously breaks the (approximate) chiral symmetry of QCD and has 
profound implications for the hadron 
spectrum. At high temperature this condensate is expected to
melt, i.e.\ $\langle \bar{\psi} \psi \rangle \simeq 0$, which signals
the chiral phase transition. We will investigate the chiral phase
transition within 
the linear quark meson model \cite{Ju95-7} for two quark flavors. 
Truncated nonperturbative flow equations are derived at nonzero  
temperature \cite{BJW97-1} and chemical potential \cite{BJWchemRG}. 
Whereas the renormalization group flow leads to 
spontaneous chiral symmetry breaking in vacuum, the symmetry gets restored 
in a second order phase transition at high temperature for vanishing 
quark mass. The description \cite{BJW97-1} 
covers both the low temperature chiral 
perturbation theory domain of validity as well as the high temperature 
domain of critical phenomena. In particular, we obtain 
a precise estimate of the universal equation of state in the vicinity 
of critical points. We explicitly 
connect the physics at zero temperature and realistic quark mass with 
the universal behavior near the critical temperature $T_c$ and the chiral 
limit. An important property will be the observation that certain
low energy properties are effectively independent of the details
of the model even away from the phase transition. This behavior 
is caused by a strong attraction of the renormalization group
flow to approximate partial infrared fixed points 
\cite{Ju95-7,BJW97-1}. Within this approach
at high density we find \cite{BJWchemRG} a chiral symmetry 
restoring first order transition. The 
results imply the presence of a critical endpoint in the phase diagram 
in the universality class of the three dimensional Ising model 
\cite{BerRaj99,HJSSV}. The universal properties of this endpoint
have been discussed in section \ref{IsingModel}. For details of
the QCD aspects of this approach see the reviews 
\cite{Faro} and \cite{PhaseDia} for a discussion of the phase diagram. 
Similar nonperturbative renormalization group studies of  
QCD motivated models can be found in \cite{HeatKern,BRscfer,AMSTT}.
Field theories with scalars and fermions have been investigated
using similar techniques in \cite{Magg89,CHL,LP94,KS}.   

\subsection{Linear quark meson model} 

We consider a
Nambu--Jona-Lasinio type model for QCD in which quarks interact via 
effective fermionic interactions,  
with $a,b,c,d=1,2$
flavors and $i,j=1,2,3$ colors. The model is defined at some
``high'' momentum scale $k_{\Phi}\approx 600-700$ MeV
\beq
\label{FermModel}
 \Gamma_{k_{\Phi}}[\bar{\psi},\psi] = \int d^4x \,\,
 \{\bar{\psi}_a^i(x) Z_{\psi,k_{\Phi}}\left[
 i \gamma^{\mu}\partial_{\mu}+ m(x) \gamma_5\right]\psi^a_i(x)\}
 + \Gamma_{k_{\Phi}}^{({\rm int})}[\bar{\psi},\psi]
\eeq
with given fermion
wave function renormalization constant $Z_{\psi,k_{\Phi}}$.
The curled brackets around the fermion bilinears in
(\ref{FermModel}) indicate contractions over the Dirac spinor indices 
which are suppressed. For later purposes
we allow for a non-constant mass term $m(x)\gamma_5$ and
we concentrate at the end on equal constant current quark masses
$m=\frac{1}{2}(m_u+m_d)$.\footnote{We use
chiral conventions in which the (real) mass term is multiplied
by $\gamma_5$. The more
common version of the fermion mass term can be 
obtained by a chiral rotation.}  
The fermions interact via a four-fermion interaction which in
momentum space is given by
\bea
 \Gamma_{k_{\Phi}}^{({\rm int})}[\bar{\psi},\psi] &=&
 -\frac{1}{8}\int\left(\prod_{l=1}^4
 \frac{d^4 p_l}{(2\pi)^4}\right)
 \left(2\pi\right)^4\delta(p_1+p_2-p_3-p_4) 
 \bar{h}_{k_\Phi}^2 G(p_l) \nonumber\\
&&\Big[
 \left\{\bar\psi^i_a(-p_1) i(\tau^z)^a_b \psi_i^b(p_2)\right\}
 \left\{\bar\psi^j_c(p_4) i(\tau_z)^c_d \psi_j^d(-p_3)\right\} \nonumber\\
&& + \left\{\bar\psi^i_a(-p_1) \gamma^5 \psi_i^a(p_2)\right\}
 \left\{\bar\psi^j_b(p_4) \gamma^5 \psi_j^b(-p_3)\right\}
 \Big] \; .
\label{momint}
\eea
We consider a momentum dependent four-fermion interaction
\beq
G^{-1}=\ol{m}^2_{k_{\Phi}}+Z_{\Phi,k_{\Phi}} (p_1+p_2)^2
\eeq
The action is invariant under the chiral flavor group $SU(2)_L\times
SU(2)_R$ except for the mass term.

Let us define composite fields
\bea
O^z[\bar{\psi},\psi;q] &=& - \frac{1}{2}\int \frac{d^4 p}{(2\pi)^4}
\bar{h} \bar\psi^i_a(p) i (\tau^z)^a_b \psi_i^b(q+p) \, , \nonumber\\
O_{(5)}[\bar{\psi},\psi;q] &=& - \frac{1}{2} \int \frac{d^4 p}{(2\pi)^4}
\bar{h} \bar\psi^i_a(p) \gamma^5 \psi_i^a(q+p) \, 
\eea 
and rewrite
\bea
\label{ComFields}
 \Gamma_{k_{\Phi}}[\bar{\psi},\psi] &=& -\int \frac{d^4 q}{(2\pi)^4} \Big\{
 Z_{\psi,k_{\Phi}} \bar{\psi}_a^i(q)  
 \gamma^{\mu}q_{\mu} \psi^a_i(q) 
 + 2 \bar{h}^{-1} Z_{\psi,k_{\Phi}} m(q) O_{(5)}[\bar{\psi},\psi;-q]  
\nonumber \\
&& + \frac{1}{2} G(q^2) \Big( O^z[\bar{\psi},\psi;-q]
 O_z[\bar{\psi},\psi;q] +  O_{(5)}[\bar{\psi},\psi;-q]
 O_{(5)}[\bar{\psi},\psi;q] \Big) \Big\} \; .
\eea
It is advantageous to consider an equivalent formulation
which introduces bosonic collective fields with 
the quantum numbers of the fermion bilinears 
appearing in (\ref{ComFields}). 
This amounts to replace the effective action (\ref{FermModel}) by
\bea
\label{HubStr}
\hat\Gamma_{k_{\Phi}}[\bar{\psi},\psi; s,\pi]&\equiv&
\Gamma_{k_{\Phi}}[\bar{\psi},\psi]
+\int\frac{d^4{q}}{(2\pi)^4} 
\Big(s(-q) - 
2 Z_{\psi,k_{\Phi}} \bar{h}^{-1}  m(-q) - O_{(5)}(-q) G(q^2)\Big)
\nonumber\\
&& \Big(s(q) - 2 Z_{\psi,k_{\Phi}}\bar{h}^{-1} m(q) 
- O_{(5)}(q) G(q^2)\Big) \frac{1}{2 G(q^2)}\\
&&+\int\frac{d^4{q}}{(2\pi)^4} 
\Big(\pi^z(-q) - O^z(-q) G(q^2)\Big)
\Big(\pi_z(q) - O_z(q) G(q^2)\Big) \frac{1}{2 G(q^2)} \nonumber \, .
\eea
The scalar fields $\pi^z$ and $s$ have the quantum numbers
of the pions and the $\sigma$-field.
The terms added to $\Gamma_{k_{\Phi}}[\bar{\psi},\psi]$ are quadratic 
in the fermion bilinears. They cancel the original four--fermion 
interaction (\ref{momint}) and introduce 
a Yukawa interaction between fermions and collective fields, as well 
as a propagator term for the collective fields:
\bea
\label{TheModel}
\hat\Gamma_{k_{\Phi}}[\bar{\psi},\psi; s,\pi]
&=& \int \frac{d^4 q}{(2\pi)^4} \Big\{
 -Z_{\psi,k_{\Phi}} \bar{\psi}_a^i(q) 
 \gamma^{\mu}q_{\mu} \psi^a_i(q) \\ 
&& + \frac{1}{2}\left[Z_{\Phi,k_{\Phi}} q^2 + \ol{m}^2_{k_{\Phi}}\right] 
\Big(s(-q)s(q)+\pi^z(-q)\pi_z(q)\Big)  
- \jmath (-q) s(q) \nonumber\\
&&+ \int \frac{d^4 p}{(2\pi)^4} \bar{\psi}^i_a(p)
\frac{\bar{h} \gamma_5}{2} 
\Big(s(q)\delta_b^a+i\pi^z(q)(\tau_z)_b^a \gamma^5\Big) 
\psi^b_i(p-q)
\Big\} \nonumber
\eea
where we dropped field independent terms in (\ref{HubStr}). 
The above replacement of $\Gamma_{k_{\Phi}}[\bar{\psi},\psi]$
by $\hat\Gamma_{k_{\Phi}}[\bar{\psi},\psi; s,\pi]$ corresponds 
to a Hubbard-Stratonovich 
transformation in the defining functional integral
for the effective action, in which
the collective fields are introduced 
by inserting identities into the functional integral.
The introduction of collective fields in the context
of flow equations is discussed in refs.\ \cite{EW94-1,BJW97-1}. 
The equivalence of $\hat\Gamma_{k_{\Phi}}[\bar{\psi},\psi; 
s,\pi]$ with the
original formulation in terms of fundamental fields only is
readily established by solving the field equations for 
$s$ and $\pi$ 
\bea\label{8.9}
\frac{\delta \hat\Gamma_{k_{\Phi}}[\bar{\psi},\psi; s,\pi]}
{\delta s}&=0&
\quad, \quad s_0=G O_{(5)}[\bar{\psi},\psi] 
+2 Z_{\psi,k_{\Phi}} \bar{h}^{-1}  m \nonumber \\
\frac{\delta \hat\Gamma_{k_{\Phi}}[\bar{\psi},\psi; s,\pi]}
{\delta \pi^z}&=0&
\quad, \quad \pi^z_0=G O^z[\bar{\psi},\psi]
\eea
and inserting the solution in the effective action $\hat\Gamma_{k_{\Phi}}[\bar{\psi},\psi; s_0,\pi_0]
=\Gamma_{k_{\Phi}}[\bar{\psi},\psi]\,$. The $SU(2)_L\times SU(2)_R$ 
symmetry is most manifest in a $2\times 2$ matrix notation
\beq\label{8.11}
\Phi\equiv \frac{1}{2}\left(s+i\pi^z\tau_z \right)
\eeq
 
The quark mass term in the original fermionic description appears now as a
source term which is proportional to $m$ 
\beq\label{8.10}
\jmath (q)\equiv 2 \bar{h}^{-1} Z_{\psi,k_{\Phi}} m(q) 
(\ol{m}^2_{k_{\Phi}}+Z_{\Phi,k_{\Phi}} q^2) \; .
\eeq
We will be mainly interested in momentum independent sources
$\jmath(q\equiv 0)$ or constant fermion masses. 
Since the chiral symmetry breaking is linear in $\Phi$ we
can define a chirally symmetric effective action
\beq\label{8.10a}
\dsp{{\Gamma}_k[\psi,\Phi]} = \dsp{
 \hat{\Gamma}_k[\bar\psi,\psi;s,\pi] + \int d^4 x
 \, \jmath\, \tr \Phi}\eeq
The flow of $\Gamma_k$ will conserve the chiral symmetries. The
explicit chiral symmetry is now reflected by the field equation
\beq\label{8.10b}
\frac{\delta\Gamma_k}{\delta s}=j\eeq
Knowledge of $\Gamma_0$ for an arbitrary constant field
$\Phi={\rm diag}(\overline{\sigma}_0,\overline{\sigma}_0)$ 
contains the information
on the model for arbitrary quark masses. Spontaneous chiral 
symmetry breaking manifests itself by 
$\overline{\sigma}_0\not=0$ for $j\to 0$.
Our approach allows therefore for a simple
unified treatment of spontaneous
and explicit chiral symmetry breaking.

The effective action at the scale
$k_{\Phi}$ specifies the ``initial condition'' for the 
renormalization flow of the average action $\Gamma_k$. 
For scales $k < k_{\Phi}$
we allow for a more general
form and consider a truncation $(\rho=(s^2+\pi^z\pi_z)/2=
\tr \Phi^\dagger\Phi)$
\begin{eqnarray}
\label{QMAnsatz}
  \dsp{{\Gamma}_{k}[\psi,\Phi]} &=& \dsp{
    \int d^4x \Bigg\{
    i Z_{\psi,k} \bar{\psi}_a^i \gamma^{\mu}\partial_{\mu} \psi^a_i
    +\bar{h}_k {\bar{\psi}}_a^i \left[ \frac{1+\gamma^5}{2} {\Phi^a}_b
    - \frac{1-\gamma^5}{2} {(\Phi^{\dagger})^a}_b\right] \psi^b_i
    }\nnn 
  && \dsp{
    \quad \qquad +Z_{\Phi,k} 
    \partial_{\mu}(\Phi^\dagger)_{ab}\partial^{\mu}\Phi^{ba}
    +U_k({\rho})\Bigg\} 
    }\, .
\end{eqnarray}  
which takes into account the
most general field dependence of the $O(4)$-symmetric average 
potential $U_k$. Here $Z_{\Phi,k}$ and $Z_{\psi,k}$ denote scale 
dependent wave function renormalizations for the bosonic fields and
the fermionic fields, respectively. We note that in our
conventions the scale dependent Yukawa coupling $\bar{h}_k$ is real.

In terms of the renormalized expectation value
\begin{equation}
  \label{AAA100}
  \sigma_{0}=Z_{\Phi}^{1/2}\overline{\sigma}_{0}\; 
\end{equation}
we obtain the following expressions for quantities
as the pion decay constant $f_{\pi}$, chiral condensate
$\VEV{\ol{\psi}\psi}$, constituent quark mass $M_{q}$ and
pion and sigma mass, $m_{\pi}$ and $m_{\sigma}$, respectively 
($d=4$) \cite{BJW97-1}
\begin{equation}
  \label{AAA65}
  \begin{array}{rcl}
    \dsp{f_{\pi,k}} &=& \dsp{2\sigma_{0,k}}\; ,\nnn
    \dsp{\VEV{\ol{\psi}\psi}_k} &=& \dsp{
      -2\ol{m}^2_{k_\Phi}\left[Z_{\Phi,k}^{-1/2}
      \sigma_{0,k}-{m}\right]}\; ,\nnn
      \dsp{M_{q,k}} &=& \dsp{
        h_k\sigma_{0,k}}\; ,\nnn
      \dsp{m^2_{\pi,k}} &=& \dsp{
        Z_{\Phi,k}^{-1/2}
        \frac{\ol{m}^2_{k_\Phi}{m}}{\sigma_{0,k}}=
        Z_{\Phi,k}^{-1/2}\frac{\jmath}{2\sigma_{0,k}}}\; ,\nnn
      \dsp{m_{\sigma,k}^2} &=& \dsp{
        Z_{\Phi,k}^{-1/2}
        \frac{\ol{m}^2_{k_\Phi}{m}}{\sigma_{0,k}}+
        4\lambda_k\sigma_{0,k}^2}\; .
  \end{array}
\end{equation}
Here we have defined the dimensionless, renormalized couplings
\begin{equation}
  \label{AAA102}
  \begin{array}{rcl}
    \dsp{\lambda}_k &=& \dsp{Z_{\Phi,k}^{-2}
      \frac{\partial^2U_k}{\partial\rho^2}(\rho=2\ol{\sigma}_{0,k}^2)}\; ,\nnn
    \dsp{h_k} &=& \dsp{
      Z_{\Phi,k}^{-1/2}Z_{\psi,k}^{-1}\ol{h}_k}\; .
  \end{array}
\end{equation}
We are interested in the ``physical values'' of the
quantities (\ref{AAA65}) in the limit $k\to 0$ where the infrared
cutoff is removed, i.e.\ $f_{\pi}=f_{\pi,k=0}$,
$m_{\pi}^2=m_{\pi,k=0}^2$, etc.\\

\subsection{Flow equations and infrared stability}
\label{FlowEquationsAndInfraredStability}


The dependence of the effective action
$\Gamma_k$ on the infrared cutoff scale $k$ 
is given by the exact flow equation (\ref{ExactFB}) or (\ref{2.17}) 
for fermionic fields $\psi$ (quarks) and bosonic fields $\Phi$ (mesons)
\cite{Wet93-2,Wet90-1}, $(t=\ln (k/k_{\Phi}))$
\begin{equation}
  \label{frame}
  \frac{\partial}{\partial t}\Gamma_k[\psi,\Phi] = \frac{1}{2}{\rm Tr} 
  \left\{ \frac{\partial R_{kB}}
    {\partial t} \left(\Gamma^{(2)}_k[\psi,\Phi]+R_k\right)^{-1}  \right\} 
    -{\rm Tr} \left\{ \frac{\partial R_{kF}}
    {\partial t} \left(\Gamma^{(2)}_k[\psi,\Phi]+R_k\right)^{-1} 
  \right\} \, . 
\end{equation}
Here $\Gamma_k^{(2)}$ is the matrix of second functional derivatives of
$\Gamma_k$ with respect to both fermionic and bosonic field components. The
first trace in the rhs of~(\ref{frame}) effectively runs only
over the bosonic degrees of freedom. It implies a momentum integration and
a summation over flavor indices. The second trace runs over the
fermionic degrees of freedom and contains in addition a summation over
Dirac and color indices. The infrared cutoff function $R_k$ has a 
block substructure
with entries $R_{kB}$ and $R_{kF}$ for the bosonic and the fermionic fields,
respectively (cf.\ section \ref{FlowFerm}).
We compute the flow equation for the effective potential $U_k$ from
equation (\ref{frame}) using the ansatz (\ref{QMAnsatz}) for $\Gamma_k$.
The bosonic contribution to the running effective potential corresponds 
exactly to eq.\ (\ref{2.32}) for the scalar $O(4)$ model in lowest order
of the derivative expansion.
The fermionic contribution to the evolution equation for the
effective potential can be computed without much additional effort 
from (\ref{QMAnsatz}) since the fermionic fields appear only
quadratically. The respective flow equation is obtained by taking
the second functional derivative evaluated at $\psi=\bar{\psi}=0$.

For the study of phase transitions it
is convenient to work with rescaled, dimensionless and renormalized variables. 
We introduce (with a generalization to arbitrary dimension $d$)
\begin{equation}
  \label{AAA190}
  u(t,\tilde{\rho})\equiv k^{-d}U_k(\rho)\; ,\;\;\;
  \tilde{\rho}\equiv Z_{\Phi,k} k^{2-d}\rho\; ,\;\;\;
  h_k=Z_{\Phi,k}^{-1/2}Z_{\psi,k}^{-1} k^{d-4} \bar{h}_k\; .
\end{equation}
Combining the bosonic and the fermionic contributions
one obtains the flow equation \cite{BJW97-1} 
\begin{equation}
  \begin{array}{rcl}
    \dsp{\frac{\partial}{\partial t}u} &=& \dsp{
      -d u+\left(d-2+\eta_\Phi\right)
      \tilde{\rho}u^\prime}\\[2mm]
    &+& \dsp{
      2v_d\left\{
      3l_0^d(u^\prime;\eta_\Phi)+
      l_0^d(u^\prime+2\tilde{\rho}u^{\prime\prime};\eta_\Phi)-
      2^{\frac{d}{2}+1} 
      3 l_0^{(F)d}(\frac{1}{2}\tilde{\rho}h^2;\eta_\psi)
      \right\} }\; . \label{udl}
  \end{array}
\end{equation}
Here $v_d^{-1}\equiv2^{d+1}\pi^{d/2}\Gamma(d/2)$ and primes denote
derivatives with respect to $\tilde{\rho}$. 
The bosonic, $l_0^d$, and the fermionic, $l_0^{(F)d}$,
threshold functions are defined
in section \ref{SecThresh} and appendix \ref{ThreshApp}.   
The first two terms of
the second line in (\ref{udl}) denote the contributions from
the pions and the $\sigma$ field, and the last 
term corresponds to the fermionic contribution from the $u,d$ quarks.

Eq.~(\ref{udl}) is a partial differential equation for the effective
potential $u(t,\tilde{\rho})$ which has to be supplemented by the flow
equation for the Yukawa coupling $h_k$ and expressions for the anomalous
dimensions, where
\begin{equation}  
\eta_\Phi=\frac{d}{d t}(\ln Z_{\Phi,k}) \quad,\quad 
\eta_\psi =\frac{d}{d t}(\ln Z_{\psi,k}) \; .
\end{equation}
Here the wave function renormalizations are evaluated for a $k$-dependent
background field $\rho_{0,k}$ or
$\kappa\equiv k^{2-d}Z_{\Phi,k}\rho_{0,k}^2$  
determined by the condition
\begin{equation}
  \label{AAA90}
  u^\prime(t,\kappa)=\frac{\jmath_0}{\sqrt{2\kappa}}
  k^{-\frac{d+2}{2}}Z_{\Phi,k}^{-1/2}\equiv
  \epsilon_g \; ,
\end{equation}
with $\jmath_0$ some fixed source. For a study of realistic
quark masses the optimal choice is given by (\ref{8.10}),
whereas an investigation of the universal critical behavior for
$m_q\to 0$ should employ $j_0=0$.
Equation (\ref{AAA90}) allows us to follow the flow of $\kappa$
 according to
\begin{eqnarray}
  \label{AAA91}
  \dsp{\frac{d}{d t}\kappa} &=& \dsp{
    \frac{\kappa}{\epsilon_g+2\kappa\lambda}
    \Bigg\{\left[\eta_\Phi-d-2\right]\epsilon_g-
    2\frac{\partial}{\partial t}u^\prime(t,\kappa)\Bigg\} }
\end{eqnarray}
with $\lambda\equiv u^{\prime\prime}(t,\kappa)$.  We also define the Yukawa
coupling at $\tilde{\rho}=\kappa$ and its flow equation
reads~\cite{Ju95-7,LW}
\begin{eqnarray}
  \dsp{\frac{d}{dt} h^2} & = & \dsp{ (d-4+2\eta_\psi+\eta_\phi)\, h^2 } \nnn
  & - & \dsp{ 2 v_d h^4 \, \left\{3 \, { l^{(FB)d}_{1,1}}
(\hal h^2\kappa,\epsilon_g;
      \eta_\psi,\eta_\Phi) - { l^{(FB)d}_{1,1}}
(\hal h^2\kappa,\epsilon_g+2\lambda\kappa;
      \eta_\psi,\eta_\Phi) \right\} } \nnn
  & + & \dsp{ 4 v_d h^4 \kappa \left\{ 3 \lambda \, { l^{(FB)d}_{1,2}}(\hal
      h^2\kappa,\epsilon_g;
      \eta_\psi,\eta_\Phi) \right. } \nnn
  & - & \dsp{ \left. \left( 3\lambda+2\kappa u'''(\kappa)\right) \, 
      { l^{(FB)d}_{1,2}}
(\hal h^2\kappa,\epsilon_g+2\lambda\kappa;\eta_\psi,\eta_\Phi)
    \right\} } \nnn
  & + & \dsp{ 2 v_d h^6 \kappa \, \left\{ 3 { l^{(FB)d}_{2,1}}(\hal
      h^2\kappa,\epsilon_g;\eta_\psi,\eta_\Phi) - \, { l^{(FB)d}_{2,1}}(\hal
      h^2\kappa,\epsilon_g+2\lambda\kappa;\eta_\psi,\eta_\Phi) \right\}} \, .
 \label{AAA70}
\end{eqnarray}
Similarly, the scalar anomalous dimension is infered from
\begin{eqnarray}
    \dsp{\eta_\Phi} &\equiv& \dsp{
      -\frac{d}{d t}\ln Z_{\Phi,k}}=
    \dsp{ 4 \frac{v_d}{d} \,
    \left\{ 4\kappa \lambda^2 \, 
      { m^d_{2,2}}(\epsilon_g,\epsilon_g+2\lambda\kappa;\eta_\Phi) 
    \right. }\nnn 
  & + & \dsp{\left. 2^{\frac{d}{2}} N_c h^2 { m^{(F)d}_4}(\hal
      h^2\kappa;\eta_\psi) + 2^{\frac{d}{2}-1} N_c h^4 \kappa \, 
      { m^{(F)d}_2}(\hal h^2\kappa;\eta_\psi)\right\}}  
\end{eqnarray}
and the quark anomalous dimension reads
\begin{equation}
  \label{AAA69}
  \begin{array}{rcl}
    \dsp{\eta_\psi} &\equiv& \dsp{
      -\frac{d}{d t}\ln Z_{\psi,k}=
      2\frac{v_d}{d}h^2\Bigg\{
      3m_{1,2}^{(F B)d}(\frac{1}{2}h^2\kappa,
      \epsilon_g;\eta_\psi,\eta_\Phi) }\\[2mm]
    &+& \dsp{
      m_{1,2}^{(F B)d}(\frac{1}{2}h^2\kappa,
      \epsilon_g+2\lambda\kappa;
      \eta_\psi,\eta_\Phi)
      \Bigg\}\; , }
  \end{array}
\end{equation}
which constitutes a linear set of equations for the anomalous dimensions.  
The threshold functions $l_{n_1,n_2}^{(FB)d}$, $m_{n_1,n_2}^d$, $m_2^{(F)d}$,
$m_4^{(F)d}$ and $m_{n_1,n_2}^{(FB)d}$ are specified in appendix
\ref{ThreshApp}.


Most importantly for practical applications to QCD, the system of flow equations 
for the effective potential $U_k(\rho)$, the Yukawa
coupling $h_k$ and the wave function renormalizations $Z_{\Phi,k}$,
$Z_{\psi,k}$ exhibits
an approximate partial fixed point~\cite{Ju95-7,BJW97-1}. 
For a small initial
value of the scalar wave function renormalization,
$Z_{\Phi,k_\Phi}\ll 1$ at the scale $k_\Phi$, one observes a large renormalized
meson mass term $Z_{\Phi,k_\Phi}^{-2} U_{k_\Phi}'$ and a large 
renormalized Yukawa coupling 
$h_{k_\Phi}=Z_{\Phi,k_\Phi}^{-1/2}\bar{h}_{k_\Phi}$ (for $Z_{\psi,k_\Phi}=1$). 
In this case, for the initial running one can neglect in the flow
equations all scalar contributions with threshold functions involving
the large meson masses.  This yields the simplified equations 
\cite{BJW97-1,Ju95-7}
for the rescaled quantities ($d=4,v_4^{-1}=32\pi^2$)
\begin{equation}
  \label{AAA110}
  \begin{array}{rcl}
    \dsp{\frac{\partial}{\partial t}u} &=& \displaystyle{
      -4u+\left(2+\eta_\Phi\right)
      \tilde{\rho}u^\prime
      -\frac{N_c}{2\pi^2}
      l_0^{(F)4}(\frac{1}{2}\tilde{\rho}h^2)\; ,
      }\nnn
    \displaystyle{\frac{d}{d t}h^2} &=& \displaystyle{
      \eta_\Phi h^2 \; ,
      }\nnn
      \displaystyle{\eta_\Phi} &=& \displaystyle{
        \frac{N_c}{8\pi^2} h^2\; ,
        }\nnn
      \dsp{\eta_\psi} &=& \dsp{0}\; .
    \end{array}
\end{equation}
Of course, this approximation is only valid
for the initial range of running below $k_\Phi$ before the
(dimensionless) renormalized scalar mass squared
$u^\prime(t,\tilde{\rho}=0)$ approaches zero near the chiral symmetry
breaking scale.  The system (\ref{AAA110}) is exactly soluble 
and we find \cite{BJW97-1}
\begin{equation}
  \label{AAA113}
  \begin{array}{rcl}
    \dsp{h^2(t)} &=& \dsp{
      Z_\Phi^{-1}(t)=
      \frac{h_I^2}{1-\frac{N_c}{8\pi^2}h_I^2 t}\; ,
      }\nnn
    \dsp{u(t,\tilde{\rho})} &=& \dsp{
      e^{-4t}u_I(e^{2t}\tilde{\rho}\frac{h^2(t)}{h_I^2})-
      \frac{N_c}{2\pi^2}\int_0^t d r e^{-4r}
      l_0^{(F)4}(\frac{1}{2}h^2(t)\tilde{\rho}e^{2r}) }\; .
  \end{array}
\end{equation}
Here $u_I(\tilde{\rho})\equiv u(0,\tilde{\rho})$ denotes the effective
average potential at the scale $k_\Phi$ and $h_I^2$ is the
initial value of $h^2$ at $k_\Phi$, i.e. for $t=0$. To make the behavior
more transparent we
consider an expansion of the initial value effective potential
$u_I(\tilde{\rho})$ in powers of $\tilde{\rho}$ around
$\tilde{\rho}=0$
\begin{equation}
  \label{AAA140}
  u_I(\tilde{\rho})=
  \sum_{n=0}^\infty
  \frac{u_I^{(n)}(0)}{n!}\tilde{\rho}^n \; .
\end{equation}
Expanding also $l_0^{(F)4}$ in eq.~(\ref{AAA113}) in powers of its
argument one finds for $n>2$
\begin{equation}
  \label{LLL00}
  \dsp{\frac{u^{(n)}(t,0)}{h^{2n}(t)}} = \dsp{
    e^{2(n-2)t}\frac{u_I^{(n)}(0)}{h_I^{2n}}+
    \frac{N_c}{\pi^2}
    \frac{(-1)^n (n-1)!}{2^{n+2}(n-2)}
    l_n^{(F)4}(0)
    \left[1-e^{2(n-2)t}\right]}\; .
\end{equation}
For decreasing $t\to-\infty$ the initial values $u_I^{(n)}$ become
rapidly unimportant and $u^{(n)}/h^{2n}$ approaches a fixed point.
For $n=2$, i.e., for the quartic coupling, one finds
\begin{equation}
  \label{LLL01}
  \frac{u^{(2)}(t,0)}{h^2(t)}=
  1-\frac{1-\frac{u_I^{(2)}(0)}{h_I^2}}
  {1-\frac{N_c}{8\pi^2}h_I^2 t}
\end{equation}
leading to a fixed point value $(u^{(2)}/h^2)_*=1$. As a consequence
of this fixed point behavior the system looses all its ``memory'' on
the initial values $u_I^{(n\ge2)}$ at the compositeness scale
$k_\Phi$! 
Furthermore, the attraction to partial infrared fixed points continues
also for the range of $k$ where the scalar fluctuations cannot be
neglected anymore.

On the other hand, the initial value
for the bare dimensionless mass parameter
\begin{equation}
  \label{AAA142}
  \frac{u_I^\prime(0)}{h_I^2}=
  \frac{\ol{m}^2_{k_\Phi}}{k_\Phi^2}
\end{equation}
is never negligible.   
In other words, for $h_I\to\infty$ the infrared behavior of the linear quark
meson model will depend (in addition to the value of the compositeness
scale $k_\Phi$ and the quark mass ${m}$) only on one parameter,
$\ol{m}^2_{k_\Phi}$. One can therefore add higher scalar self-interactions
to $\Gamma_{k_{\Phi}}$ in eq.\ (\ref{TheModel}) without changing
the result much. We have numerically verified this feature by
starting with different values for the quartic scalar self-interaction 
$u_I^{(2)}(0)$. Indeed, the differences in the physical observables were 
found to be small. For definiteness, the 
numerical analysis of the full system of flow equations \cite{BJW97-1} 
is performed with the idealized initial value
$u_I(\tilde{\rho})=u_I^\prime(0)\tilde{\rho}$ in the limit
$h_I^2\to\infty$. Deviations from
this idealization lead only to small numerical deviations in the
infrared behavior of the linear quark meson model as long as say 
$h_I\gtrsim 15$.

\subsection{High temperature chiral phase transition}
\label{TheQuarkMesonModelAtTNeq0}

Strong interactions in thermal equilibrium at high temperature $T$ 
differ in important aspects from the well tested vacuum or zero
temperature properties. A phase transition at some critical
temperature $T_c$ or a relatively sharp crossover may separate the
high and low temperature physics~\cite{MO96-1}. 
It was realized early that the transition should be
closely related to a qualitative change in the chiral condensate
according to the general observation that spontaneous symmetry
breaking tends to be absent in a high temperature situation. A series
of stimulating contributions~\cite{PW84-1,RaWi93-1,Raj95-1} pointed
out that for sufficiently small up and down quark masses, $m_u$ and
$m_d$, and for a sufficiently large mass of the strange quark, $m_s$,
the chiral transition is expected to belong to the universality class
of the $O(4)$ Heisenberg model. It
was suggested~\cite{RaWi93-1,Raj95-1} that a large correlation length
may be responsible for important fluctuations or lead to a disoriented
chiral condensate. One main question we are going
to answer using non-perturbative flow equations for the linear
quark meson model is: How
small $m_u$ and $m_d$ would have to be in order to see a large
correlation length near $T_c$ and if this scenario could be realized
for realistic values of the current quark masses.
  
In order to solve our model we need to specify the
``initial condition'' $\Gamma_{k_{\Phi}}$ for the renormalization
flow of $\Gamma_{k}$. We will choose in the following a 
normalization of $\psi,\Phi$ such that
$Z_{\psi,k_\Phi}=\bar{h}_{k_\Phi}=1$. We therefore need as initial
values at the scale $k_\Phi$ the scalar wave function renormalization
$Z_{\Phi,k_\Phi}$ and the shape of the potential $U_{k_\Phi}$. We will
make here the important assumption that $Z_{\Phi,k}$ is small at the
compositeness scale $k_\Phi$ (similarly to what is usually assumed in
Nambu--Jona-Lasinio--like models) 
\begin{equation}
Z_{\Phi,k_\Phi} \ll 1 \; . \label{compcon}
\end{equation}
This results in a large value of
the renormalized Yukawa coupling
$h_k=Z_{\Phi,k}^{-1/2}Z_{\psi,k}^{-1}\bar{h}_k$. A large value of
$h_{k_\Phi}$ is phenomenologically suggested by the comparably large
value of the constituent quark mass $M_q$. The latter is related to
the value of the Yukawa coupling for $k\to0$ and the pion decay
constant $f_\pi=92.4\MeV$ by $M_q=h f_\pi/2$ (with $h=h_{k=0}$), and
$M_q\simeq300\MeV$ implies $h^2/4\pi\simeq3.4$. For increasing $k$ the
value of the Yukawa coupling grows rapidly for $k\gtrsim M_q$.  Our
assumption of a large initial value for $h_{k_\Phi}$ is therefore
equivalent to the assumption that the truncation (\ref{QMAnsatz}) can be
used up to the vicinity of the Landau pole of $h_k$. The existence of
a strong Yukawa coupling enhances the predictive power of our approach
considerably.  It implies a fast approach of the running couplings to
partial infrared fixed points as shown in section 
\ref{FlowEquationsAndInfraredStability} \cite{Ju95-7,BJW97-1}. 
In consequence, the
detailed form of $U_{k_\Phi}$ becomes unimportant, except for the
value of one relevant parameter corresponding to the scalar mass term
$\ol{m}^2_{k_\Phi}$. In this work we fix $\ol{m}^2_{k_\Phi}$ such
that $f_\pi=92.4\MeV$ for $m_\pi=135\MeV$. The value$f_\pi=92.4\MeV$ 
(for $m_\pi=135\MeV$) sets our unit
of mass for two flavor QCD which is, of course, not directly
accessible by observation. Besides $\ol{m}^2_{k_\Phi}$ (or $f_\pi$)
the other input parameter used in this work is the constituent quark
mass $M_q$ which determines the scale $k_\Phi$ at which $h_{k_\Phi}$
becomes very large. We consider a range $300\MeV\lesssim M_q\lesssim 350\MeV$
and find a rather weak dependence of our results on the precise value
of $M_q$. The results presented in the following are for $M_q=303\MeV$.

We first consider the model at nonzero temperature $T$.
The case for nonvanishing baryon number density will be presented
in section \ref{ChiralDensity}.  
Figure \ref{ccc_T} shows our
results \cite{BJW97-1} for the chiral condensate
$\VEV{\bar{\psi}\psi}$ as a function of the temperature
$T$ for various values of the average quark mass 
${m}=(m_u+m_d)/2$.
\begin{figure}
\unitlength1.0cm

\begin{center}
\begin{picture}(13.,7.0)

\put(0.0,0.0){
\epsfysize=11.cm
\rotate[r]{\epsffile{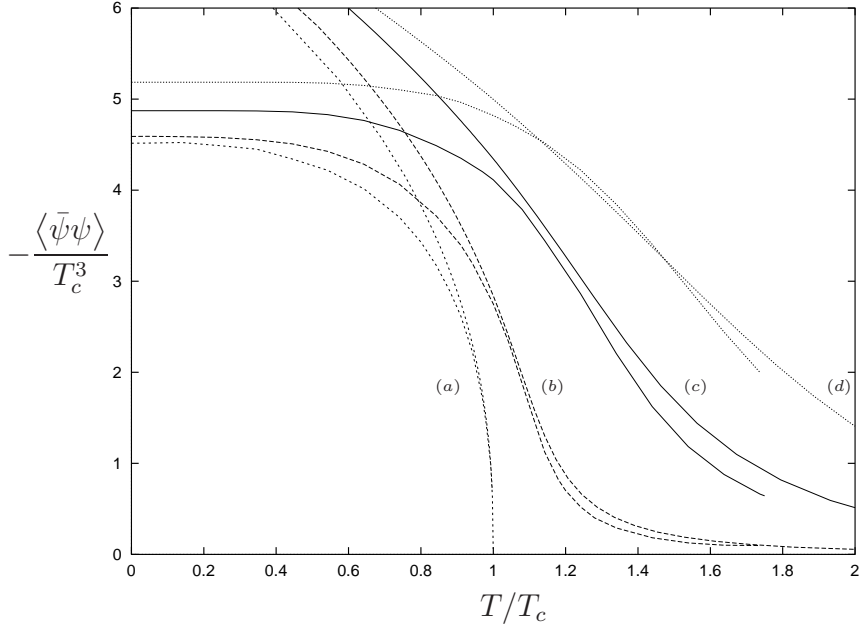}}}
\put(-0.5,4.2){\bf $\dsp{-\frac{\VEV{\bar{\psi}\psi}}{T_{c}^3}}$}
\put(5.8,-0.5){\bf $\dsp{T/T_{c}}$}
\put(5.2,2.5){\tiny $(a)$}
\put(6.6,2.5){\tiny $(b)$}
\put(8.5,2.5){\tiny $(c)$}
\put(10.4,2.5){\tiny $(d)$}

\end{picture}
\end{center}
\caption{\em Chiral equation of state for a phase transition or crossover in
two-flavor QCD.
The plot shows the chiral condensate
  $\VEV{\bar{\psi}\psi}$ as a function of temperature $T$.  Lines
  $(a)$, $(b)$, $(c)$, $(d)$ correspond at zero temperature to
  $m_\pi=0,45\MeV,135\MeV,230\MeV$, respectively. For each pair of
  curves the lower one represents the full $T$--dependence of
  $\VEV{\bar{\psi}\psi}$ whereas the upper one shows for comparison the
  universal scaling form of the equation of state for the $O(4)$
  Heisenberg model (cf.\ fig.\ \ref{scalfunc}).
 The critical temperature for zero quark mass is
  $T_c=100.7\MeV$. The chiral condensate is normalized at a scale
  $k_{\Phi}\simeq 620\MeV$.}
\label{ccc_T}
\end{figure}
Curve $(a)$ gives the temperature dependence of $\VEV{\bar{\psi}\psi}$
in the chiral limit ${m}=0$. We first consider only the lower
curve which corresponds to the full result.
One observes that the order parameter $\VEV{\bar{\psi}\psi}$
goes continuously (but non--analytically)
to zero as $T$ approaches the critical temperature in the massless
limit $T_c=100.7 \MeV$. The transition from the phase with
spontaneous chiral symmetry breaking to the symmetric phase
is second order. The curves $(b)$, $(c)$
and $(d)$ are for non--vanishing values of the average current quark
mass ${m}$. The transition turns into a smooth crossover. 
Curve $(c)$ corresponds to ${m}_{\rm phys}$ or,
equivalently, $m_\pi(T=0)=135\MeV$. The
transition turns out to be much less dramatic than for ${m}=0$. We
have also plotted in curve $(b)$ the results for comparably small
quark masses $\simeq1\MeV$, i.e.~${m}={m}_{\rm phys}/10$, for
which the $T=0$ value of $m_\pi$ equals $45\MeV$. The crossover is
considerably sharper but a substantial deviation from the chiral limit
remains even for such small values of ${m}$.

For comparison, the upper curves in figure
\ref{ccc_T} use the universal scaling form of the equation of state
of the {\em three dimensional} $O(4)$--symmetric Heisenberg model which
has been computed explicitly in section \ref{uceos}. 
The scaling equation of state in
terms of the chiral condensate for the general case of a temperature
and quark mass dependence is  
\begin{equation}
  \label{XXX30}
  \VEV{\bar{\psi}\psi}=
  -\ol{m}^2_{k_\Phi}T_c
  \left(\frac{\jmath/T_c^3}{f(x)}\right)^{1/\delta}+
    \jmath
\end{equation}
as a function of $T/T_c=1+x(\jmath/T_c^3 f(x))^{1/\beta\delta}$. 
The curves shown in figure \ref{ccc_T} correspond to quark masses
${m}=0$, ${m}={m}_{\rm phys}/10$, ${m}={m}_{\rm
  phys}$ and ${m}=3.5{m}_{\rm phys}$ or, equivalently, to zero
temperature pion masses $m_\pi=0$, $m_\pi=45\MeV$, $m_\pi=135\MeV$ and
$m_\pi=230\MeV$, respectively. We see perfect agreement 
of both curves in the chiral limit for $T$
sufficiently close to $T_c$ which is a manifestation
of universality and the phenomenon of dimensional reduction. 
In particular, we reproduce the critical exponents of the $O(4)$--model
given in table \ref{critex} of section \ref{uceos}.
Away from the chiral limit we find for a realistic pion mass
that the $O(4)$ universal equation 
of state provides a reasonable approximation for $\VEV{\bar{\psi}\psi}$
in the crossover region $T=(1.2-1.5)T_c$.  

In order to facilitate comparison with lattice simulations which are typically
performed for larger values of $m_\pi$ we also present results for
$m_\pi(T=0)=230\MeV$ in curve $(d)$. One may define a ``pseudocritical
temperature'' $T_{pc}$ associated to the smooth crossover as the
inflection point of $\VEV{\bar{\psi}\psi}(T)$. 
Our results for $T_{pc}$ are presented in 
table \ref{tab11} for the four different values of ${m}$ or, 
equivalently, $m_\pi(T=0)$.
\begin{table}
\begin{center}
\begin{tabular}{|c||c|c|c|c|} \hline
  $\qquad{{m_\pi}/{\MeV}}\quad$ &
  $0\qquad$ &
  $45\qquad$ &
  $135\qquad$ &
  $230\qquad$
  \\[1.0mm] \hline
  $\qquad{{T_{pc}}/{\MeV}}\quad$ &
  $100.7\qquad$ &
  $\simeq110\qquad$ &
  $\simeq130\qquad$ &
  $\simeq150\qquad$
  \\[1mm] \hline
\end{tabular}
\caption{\em Critical and
  ``pseudocritical'' temperature for various values of the zero
  temperature pion mass. Here $T_{pc}$ is defined as the
  inflection point of $\VEV{\bar{\psi}\psi}(T)$.}
\label{tab11}
\end{center}
\end{table}
The value for the pseudocritical temperature for $m_{\pi}=230 \MeV$
compares well with the lattice results for two flavor QCD. 
This may be taken as an indication that the linear quark meson
model gives a reasonable picture of the chiral properties in two-flavor
QCD. An extension of the truncation for the linear quark meson model
may lead to corrections in the value of $T_c$, but we do not
expect qualitative changes of the overall picture.
One should mention, though, that
a determination of $T_{pc}$ according to this definition is subject to
sizeable numerical uncertainties for large pion masses as the curve in
figure \ref{ccc_T} is almost linear around the inflection point for
quite a large temperature range.  A problematic point in lattice
simulations is the extrapolation to realistic values of $m_\pi$ or
even to the chiral limit. Our results may serve here as an analytic
guide. The overall picture shows the approximate validity of the
$O(4)$ scaling behavior over a large temperature interval in the
vicinity of and above $T_c$ once the (non--universal) amplitudes are
properly computed. We point out that the link between the universal
behavior near $T_c$ and zero current quark mass on the one hand and the
known physical properties at $T=0$ for realistic quark masses on the
other hand is crucial to obtain all non--universal information
near $T_c$. 

A second important result is the temperature
dependence of the space--like pion correlation length
$m_\pi^{-1}(T)$. (We will often call $m_\pi(T)$ the temperature
dependent pion mass since it coincides with the physical pion mass for
$T=0$.) Figure \ref{mpi_T} shows $m_\pi(T)$ and one again observes the
second order phase transition in the chiral limit ${m}=0$. 
\begin{figure}
\unitlength1.0cm
\begin{center}
\begin{picture}(13.,7.0)

\put(0.0,0.0){
\epsfysize=11.cm
\rotate[r]{\epsffile{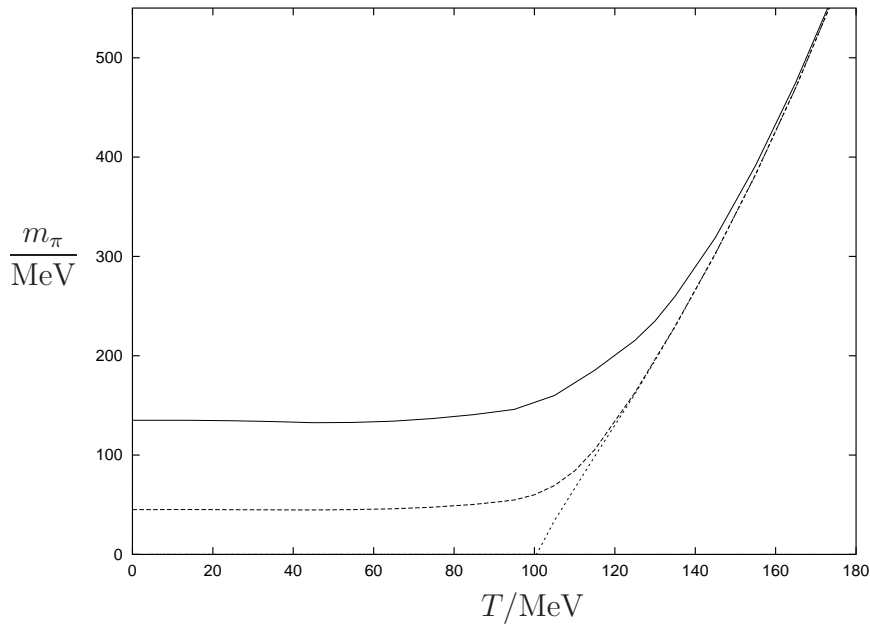}}
}
\put(-0.5,4.2){\bf $\dsp{\frac{m_\pi}{\MeV}}$}
\put(5.8,-0.5){\bf $\dsp{T/\MeV}$}

\end{picture}
\end{center}
\caption{\em Temperature dependence of the pion mass:
The plot shows $m_\pi$ as a function of
  temperature $T$ for three different values of the average light
  current quark mass ${m}$. The solid line corresponds to the
  realistic value ${m}={m}_{\rm phys}$ whereas the dotted line
  represents the situation without explicit chiral symmetry breaking,
  i.e., ${m}=0$. The intermediate, dashed line assumes 
  ${m}={m}_{\rm phys}/10$.}
\label{mpi_T}
\end{figure}
For $T<T_c$ the pions are massless Goldstone bosons whereas for
$T>T_c$ they form with the sigma a degenerate vector of
$O(4)$ with mass increasing as a function of temperature.  For
${m}=0$ the behavior for small positive $T-T_c$ is characterized
by the critical exponent $\nu$, i.e.
$m_\pi(T)=\left(\xi^+\right)^{-1}T_c \left( (T-T_c)/T_c\right)^\nu$
and we obtain $\nu=0.787$, $\xi^+=0.270$. For ${m}>0$ we find that
$m_\pi(T)$ remains almost constant for $T\lesssim T_c$ with only a very
slight dip for $T$ near $T_c/2$. For $T>T_c$ the correlation length
decreases rapidly and for $T\gg T_c$ the precise value of ${m}$
becomes irrelevant. We see that the universal critical behavior near
$T_c$ is quite smoothly connected to $T=0$.  The full functional
dependence of $m_\pi(T,{m})$ allows us to compute the overall size
of the pion correlation length near the critical temperature and we
find $ m_\pi(T_{pc})\simeq 1.7 m_\pi(0)$ for the realistic value
${m}_{\rm phys}$. This correlation length is even smaller than the
vacuum ($T=0$) one and gives no indication for strong fluctuations of
pions with long wavelength.\footnote{For a QCD phase transition
far from equilibrium long wavelength modes of the pion field 
can be amplified \cite{RaWi93-1,Raj95-1}.}  We will discuss the possibility of
a tricritical point with a massless excitation in the 
two--flavor case at non--zero
baryon number density 
in section \ref{highdensitypt}.   
We also point out that the present
investigation for the two flavor case does not take into account a
possible ``effective restoration'' of the axial $U_A(1)$ symmetry
at high temperature \cite{PW84-1,Shu94-1}.

\subsection{Renormalization flow at nonzero chemical potential}
\label{ChiralDensity}

At nonzero temperature and chemical potential $\mu$ associated to the 
conserved quark number we consider the following ansatz for $\Gamma_k$ 
\begin{eqnarray}
  \label{truncation} 
  \Gamma_k &=& \dsp{
    \int^{1/T}_0 dx^0\int d^3x \Bigg\{
    i \bar{\psi}^a_i (\gamma^{\mu}\partial_{\mu} + \mu \gamma^0) \psi_a^i
    +\bar{h}_k {\bar{\psi}}^a_i \left[ \frac{1+\gamma^5}{2} {\Phi_a}^b
    - \frac{1-\gamma^5}{2} {(\Phi^{\dagger})_a}^b\right] \psi_b^i
    }\nnn 
  && \dsp{
    \qquad\qquad\qquad \quad +Z_{\Phi,k} 
    \partial_{\mu}\Phi^*_{ab}\partial^{\mu}\Phi^{ab}
    +U_k({\rho};\mu,T)\Bigg\} 
    }\, .
\end{eqnarray}
A nonzero $\mu$ to lowest order results in the term $\sim i\mu
\bar{\psi}^a\gamma^0\psi_a$ appearing in the rhs
of~(\ref{truncation}). We neglect here the running
of the fermionic wave function renormalization constant and the dependence of
$Z_{\Phi,k}$ and $\bar{h}_k$ on $\mu$ and $T$. The temperature dependence 
of $Z_{\Phi,k}$ and $\bar{h}_k$, which has been taken into account for the 
results presented in 
section \ref{TheQuarkMesonModelAtTNeq0}, is indeed small \cite{BJW97-1}. 
We also neglect a possible difference in normalization of the quark 
kinetic term and the baryon number current.  

There is a substantial caveat concerning the approximation (\ref{truncation}) 
at nonzero density. At sufficiently high density diquark condensates 
form, opening up a gap at the quark Fermi surfaces 
\cite{CSC,ARW1}. In 
order to describe this phenomenon in the present framework, the above
ansatz for $\Gamma_k$ has to be extended to include diquark
degrees of freedom. However, a nonzero diquark condensate only
marginally affects the equation of state 
for the chiral condensate \cite{BerRaj99}. In particular, 
in the two flavor case the inclusion of diquark 
degrees of freedom hardly changes the behavior of the chiral 
condensate at nonzero 
density and the order of the transition to the chirally symmetric 
phase. 
For NJL-type models diquark condensation is 
suppressed at low density by the presence of the chiral condensate
\cite{BerRaj99,CarterDiakonov}. 
We therefore expect the ansatz (\ref{truncation}) to give
a good description of the restoration of chiral symmetry
within the present model. It is straightforward to generalize
our method to include diquark degrees of freedom. 
For simplicity we concentrate 
here on the chiral properties and neglect the superconducting
properties at high densities.
 
We employ the same exponential infrared
cutoff function for the bosonic fields $R_{kB}$ (\ref{2.15}) 
as in the previous section at nonzero temperature.
At nonzero density a mass--like fermionic infrared cutoff simplifies
the computations considerably compared to an exponential
cutoff like (\ref{PF}) because of the trivial momentum
dependence. In presence of a chemical
potential $\mu$ we use 
\begin{equation}
  \label{fermir}
  R_{kF}=-(\gamma^\mu q_\mu - i \mu \gamma^0) r_{kF} \, . 
\end{equation}
The effective squared inverse fermionic propagator is then of
the form
\begin{eqnarray}
  \label{fermprop}
  P_{kF} &=& \dsp{
    [(q_0-i\mu)^2+\vec{q}^{\,2}](1+r_{kF})^2
    } \nnn 
  &=& \dsp{
    (q_0-i \mu)^2+\vec{q}^{\,2}+k^2 \Theta 
    (k_{\Phi}^2-(q_0-i\mu)^2-\vec{q}^{\,2})
    }\; ,
\end{eqnarray}
where the second line defines $r_{kF}$ and one observes that
the fermionic infrared cutoff acts as an additional mass--like
term $\sim k^2$. 

We compute the flow equation for the effective potential $U_k$ from
equation (\ref{frame}) using the ansatz (\ref{truncation}) for $\Gamma_k$.
The only explicit
dependence on the chemical potential $\mu$ appears in the fermionic
contribution to the flow equation for $U_k$, whereas the derivation of the
bosonic part strictly follows section \ref{FlowEquationsAndInfraredStability}.  
It is instructive to consider the fermionic part of the flow equation
in more detail and 
to perform the summation of the Matsubara modes explicitly for the
fermionic part. Since the flow equations only involve one momentum
integration, standard techniques for one loop expressions apply \cite{Kap}
and we find
\begin{eqnarray}
  \lefteqn{
    \frac{\partial }{\partial k}U_{kF}(\rho;T,\mu) = -8 N_c 
    \int\limits_{-\infty}^{\infty}\,\,\,
    \frac{d^4q}{(2 \pi)^4} \frac{k \, \Theta (k_{\Phi}^2-q^2)}
    {q^2+k^2  + h_k^2 \rho/2}  
    +4 N_c \int\limits_{-\infty}^{\infty}
    \frac{d^3\vec{q}}{(2 \pi)^3} 
    \frac{k}{\sqrt{\vec{q}^{\,2}+k^2+h_k^2 \rho/2}} 
    }\nnn && \dsp{
    \hspace*{-12truemm}\times
    \left\{ \frac{1}
      {\dsp \exp\left[(\sqrt{\vec{q}^{\,2}+k^2+h_k^2 \rho/2}-\mu)/T\right]+1}
      +\frac{1}
      {\dsp \exp\left[(\sqrt{\vec{q}^{\,2}+k^2+h_k^2 \rho/2}+\mu)/T\right]+1}
      \label{uapprox}
    \right\} }
\end{eqnarray}
For simplicity, we sent here $k_{\Phi}\to\infty$ in the $\mu,T$ dependent
second integral. This is justified by the fact that in the
$\mu,T$ dependent part the high momentum modes are exponentially
suppressed.
 
For comparison, we note that within the present approach one obtains
standard mean field theory results for the free energy if the
meson fluctuations are neglected, $\partial U_{kB}/\partial k \equiv 0$,
and the Yukawa coupling is kept constant, $h_k=h$ in (\ref{uapprox}). The
remaining flow equation for the fermionic contribution could then easily be
integrated with the (mean field) initial condition 
$U_{k_{\Phi}}(\rho)=\ol{m}_{k_\Phi}^2\rho$. In the following we will
concentrate on the case of vanishing temperature.  
We find (see below) that a mean field
treatment yields relatively good estimates only for the
$\mu$--dependent part of the free energy $U(\rho;\mu)-U(\rho;0)$. On
the other hand, mean field theory does not give a very reliable
description of the vacuum properties
encoded in $U(\rho;0)$. The latter are important for a
determination of the order of the phase transition at $\mu \not = 0$.

In the limit of vanishing temperature one expects and observes a
non--analytic behavior of the $\mu$--dependent integrand of the fermionic
contribution (\ref{uapprox}) to the flow equation for $U_k$ because of the
formation of Fermi surfaces. Indeed, the explicit $\mu$--dependence of the
flow equation reduces to a step function
\begin{eqnarray}
  \label{dtuf0} 
  \dsp{\frac{\partial }{\partial k}U_{kF}(\rho;\mu) =} &-& \dsp{
    8 N_c 
    \int\limits_{-\infty}^{\infty}\,\,\,
    \frac{d^4q}{(2 \pi)^4} \frac{k \Theta (k_{\Phi}^2-q^2)}
    {q^2+k^2  + h_k^2 \rho/2}  
    }\nnn 
  &+& \dsp{
    4 N_c \int\limits_{-\infty}^{\infty}
    \frac{d^3\vec{q}}{(2 \pi)^3} 
    \frac{k}{\sqrt{\vec{q}^{\,2}+k^2+h_k^2 \rho/2}} \,\,\,\,
    \Theta\! \left( \mu-\sqrt{\vec{q}^{\,2}+k^2+h_k^2 \rho/2} \right) 
    }\, .
\end{eqnarray}
The quark chemical potential $\mu$ enters the bosonic part of
the flow equation only implicitly through the meson mass terms
$U_k'(\rho;\mu)$ and $U_k'(\rho;\mu) + 2 \rho U_k''(\rho;\mu)$ for the
pions and the $\si$--meson, respectively.  For scales $k > \mu$ the
$\Theta$--function in (\ref{dtuf0}) vanishes identically and there is no
distinction between the vacuum evolution and the $\mu\not = 0$ evolution.
This is due to the fact that our infrared cutoff adds to the effective
quark mass $(k^2+h_k^2 \rho/2)^{1/2}$. For a chemical potential smaller
than this effective mass the ``density'' $-\partial U_k/\partial \mu$
vanishes whereas for larger $\mu$ one can view
$\mu=[\vec{q}_F^{\,2}(\mu,k,\rho)+k^2+h_k^2 \rho/2]^{1/2}$ as an effective
Fermi energy for given $k$ and $\rho$. A small infrared cutoff $k$ removes
the fluctuations with momenta in a shell close to the physical Fermi
surface\footnote{If one neglects the mesonic fluctuations one can
  perform the $k$--integration of the flow equation~(\ref{dtuf0}) in the
  limit of a $k$--independent Yukawa coupling. One recovers (for
  $k_\Phi^2\gg k^2+h^2\rho/2,\mu^2$) mean field theory results except for a
  shift in the mass, $h^2\rho/2\to h^2\rho/2+k^2$, and the fact that modes
  within a shell of three--momenta
  $\mu^2-h^2\rho/2-k^2\le\vec{q}^{\,2}\le\mu^2-h^2\rho/2$ are not yet
  included. Because of the mass shift the cutoff $k$ also suppresses the
  modes with $q^2<k^2$. For $k>0$ no infrared singularities appear
in the computation of $U_k$ and its $\rho$-derivatives.}
$\mu^2-h_k^2\rho/2-k^2<q^2<\mu^2-h_{k=0}^2\rho/2$. Our flow equation
realizes the general idea~\cite{Pol92-1} that for $\mu\neq0$ the lowering
of the infrared cutoff $k\to 0$ should correspond to an approach to the
physical Fermi surface. For a computation of the meson effective potential
the approach to the Fermi surface in (\ref{dtuf0}) proceeds from below and
for large $k$ the effects of the Fermi surface are absent.  By lowering $k$
one ``fills the Fermi sea''.

As discussed in section \ref{FlowEquationsAndInfraredStability}
the observed fixed point behavior in the symmetric regime
allows us to fix the model by
only two phenomenological input parameters and we use 
$f_{\pi}=92.4\MeV$ and $300\MeV\lesssim M_q\lesssim 350\MeV$.
The results for the evolution in vacuum 
\cite{Ju95-7,BJW97-1} show that for
scales not much smaller than $k_{\Phi}\simeq 600 \MeV$ chiral symmetry
remains unbroken. This holds down to a scale of about
$k_{\chi SB} \simeq 400\MeV$
at which the meson potential $U_k(\rho)$ develops a minimum at
$\rho_{0,k}>0$ even for a vanishing source,
thus breaking chiral symmetry spontaneously.  Below the
chiral symmetry breaking scale the running couplings are
no longer governed by
the partial fixed point. In particular, for $k \lesssim k_{\chi SB}$ the
Yukawa coupling $h_k$ and the meson wave function renormalization
$Z_{\Phi,k}$ depend only weakly on $k$ and approach their infrared values.
At $\mu \not = 0$ we will follow the evolution from $k=k_{\chi SB}$ to
$k=0$ and neglect the $k$--dependence of $h_k$ and $Z_{\Phi,k}$ in this
range.  According to the above discussion the initial value 
$U_{k_{\chi SB}}$ is $\mu$--independent for $\mu < k_{\chi SB}$. 
We solve the flow equation for $U_k$ numerically as a 
partial differential equation for the potential
depending on the two variables $\rho$ and $k$ for given $\mu$ 
\cite{BJWchemRG}. Nonzero current quark masses result in
a pion mass threshold and effectively stop the renormalization
group flow of renormalized couplings at a scale around $m_{\pi}$. 

In the fermionic part (\ref{dtuf0}) of the flow equation the
vacuum and the $\mu$--dependent term contribute with opposite signs. This
cancellation of quark fluctuations with momenta below the Fermi
surface is crucial for the restoration of chiral symmetry at high
density\footnote{The renormalization group 
investigation of a linear sigma model in $4-\epsilon$
dimensions misses this property \cite{HS}.}.  
In vacuum, spontaneous chiral symmetry breaking is induced in
our model by quark fluctuations which drive the scalar mass
term $U_k'(\rho=0)$ from positive to negative values at the scale $k =
k_{\chi SB}$.  (Meson fluctuations have the tendency to restore chiral
symmetry because of the opposite relative sign,
cf.~(\ref{frame}).) As the chemical potential becomes larger than the
effective mass $(k^2+h_k^2 \rho/2)^{1/2}$ quark fluctuations with momenta
smaller than $\vec{q}_F^{\,2}(\mu,k,\rho)=\mu^2-k^2-h_k^2 \rho/2$
are suppressed.  Since $\vec{q}_F^{\,2}$ is monotonically
decreasing with $\rho$ for given $\mu$ and $k$ the origin of the effective
potential is particularly affected.  We will see in the next section that
for large enough $\mu$ this leads to a second minimum of
$U_{k=0}(\rho;\mu)$ at $\rho=0$ and a chiral symmetry restoring first order
transition.

\subsection{High density chiral phase transition}
\label{highdensitypt}

In vacuum or at zero density the effective potential $U$
as a function of $\sigma=\sqrt{Z_{\Phi,{k=0}}\, \rho/2}$ has its 
minimum at a nonvanishing value
$f_{\pi}/2$ corresponding to spontaneously broken chiral symmetry.
As the quark chemical potential $\mu$ increases, $U$ can develop different
local minima. The lowest minimum corresponds to the state of lowest free
energy and is favored.  In Figure~\ref{potps}
\begin{figure}[t]
  \unitlength1.0cm
  \begin{center}
  \begin{picture}(13.0,8.0)
  \put(0.0,0.5){
  \epsfysize=8.cm
  \epsfbox[140 525 525 760]{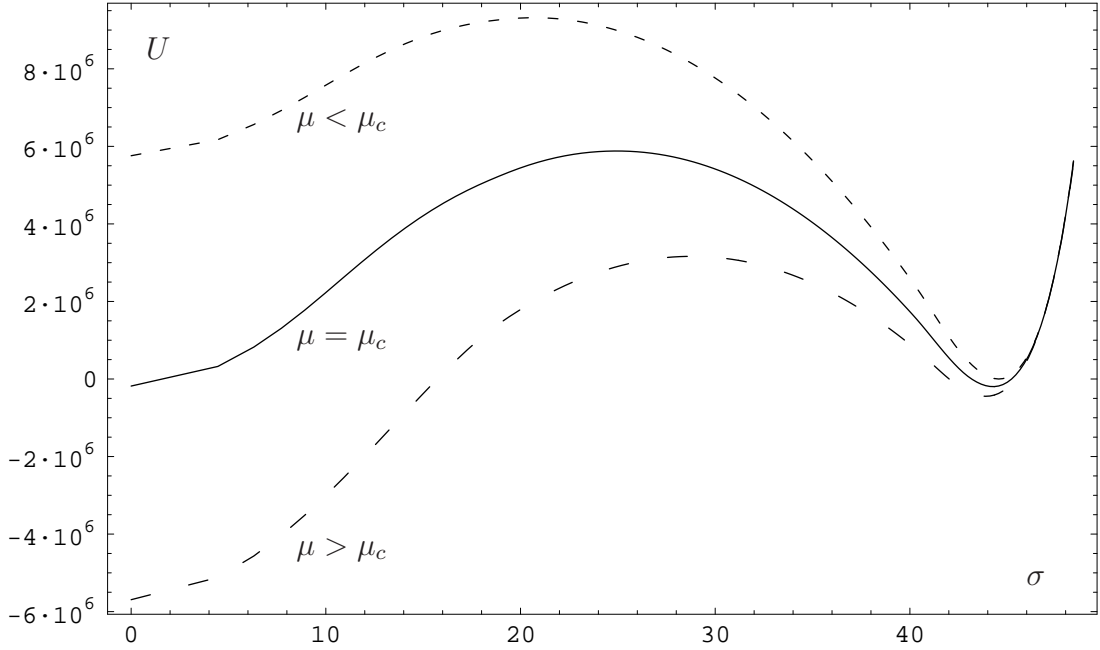}}
  \put(0.8,7.8){$\dsp{U}$}
  \put(12.5,0.8){$\dsp{\sigma}$}
  \put(2.8,6.9){$\dsp{\mu<\mu_c}$}
  \put(2.8,4){$\dsp{\mu=\mu_c}$}
  \put(2.8,1.2){$\dsp{\mu>\mu_c}$}
  \end{picture}
  \end{center}
\caption{\em High density chiral phase transition in the NJL model.
The zero temperature effective potential $U$ ( in ${\rm MeV}^4$)
  as a function of $\sigma=(Z_{\Phi,{k=0}}\, \rho/2)^{1/2}$ ( in ${\rm MeV}$)
  is shown for different chemical
  potentials. One observes two degenerate minima for a critical chemical
  potential $\mu_c/M_q=1.025$ corresponding to a first order phase
  transition at which two phases have equal pressure and can coexist
  ($M_q=316.2 \MeV$). \label{potps}}
\end{figure}
we plot the free energy as a function of $\sigma$ for different values
of the chemical potential $\mu=322.6, 324.0,325.2$ MeV.  Here the effective
constituent quark mass is $M_q=316.2 \MeV$.
We observe that for $\mu < M_q$ the potential at its minimum does not
change with $\mu$.  Since
\begin{equation}
  n_q=-\frac{\partial U}{\partial \mu}_{|{\rm min}}
\end{equation}
we conclude that the corresponding phase has zero density. In contrast,
for a chemical potential larger than $M_q$ we find a low density 
phase where chiral symmetry is still 
broken. The quark number density as a function of $\mu$ is shown in
Figure~\ref{densityps}. One clearly observes the non--analytic behavior
at $\mu=M_q$ which denotes the ``onset'' value for nonzero density. 
\begin{figure}[t]
  \unitlength1.0cm
  \begin{center}
  \begin{picture}(13.0,8.0)
  \put(0.3,0.){
  \epsfysize=8.cm
  \epsfbox[125 415 560 700]{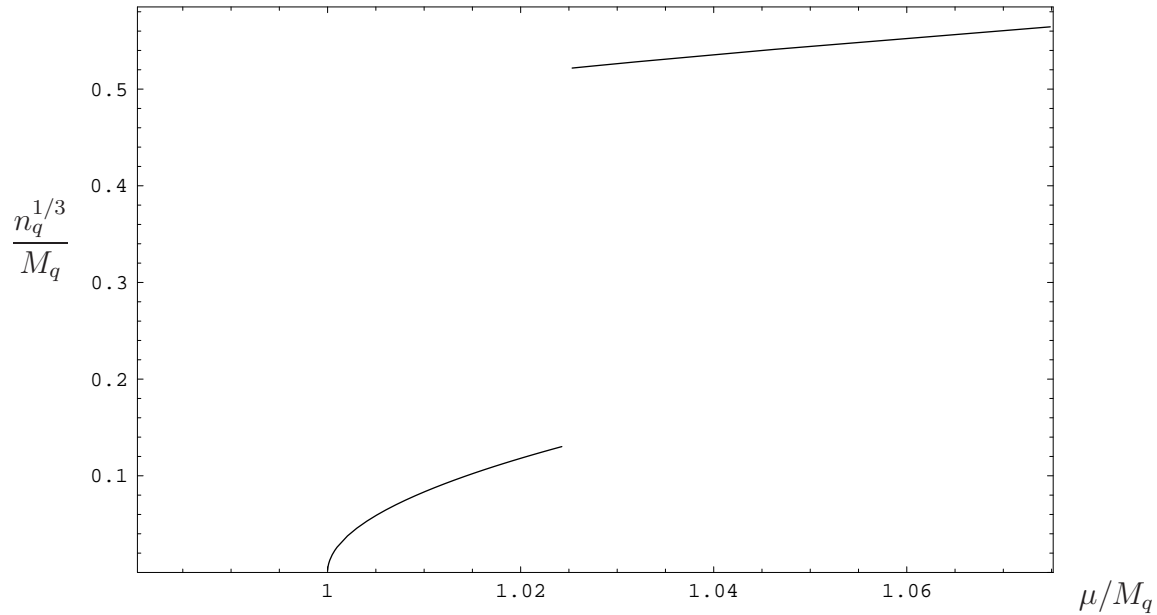}}
  \put(-1.2,5.2){$\dsp{\frac{n_q^{1/3}}{M_q}}$}
  \put(13.0,0.5){$\dsp{\mu/M_q}$}
  \end{picture}
  \end{center}
\caption{\em Density discontinuity in the NJL model at $T=0$.
The plot shows $n_q^{1/3}$, where $n_q$ denotes
  the quark number density as a function of $\mu$ in units of the effective
  constituent quark mass ($M_q=316.2\MeV$).} 
  \label{densityps}
\end{figure}
From Figure \ref{potps} one also notices the appearance of an additional
local minimum at the origin of $U$.  As the pressure $p=-U$ increases in
the low density phase with increasing $\mu$, a critical value $\mu_c$ is
reached at which there are two degenerate potential minima. Before $\mu$
can increase any further the system undergoes a first order phase
transition at which two phases have equal pressure and can coexist.  In the
high density phase chiral symmetry is restored as can be seen from the
vanishing order parameter for $\mu>\mu_c$. 
We note that the
relevant scale for the first order transition is $M_q$. 
For this reason we
have scaled our results for dimensionful quantities in units of $M_q$. 

For the class of quark meson models considered here (with $M_q/f_{\pi}$ in
a realistic range around $3$ -- $4$) the first order nature of the high
density transition has been clearly established. In particular, these
models comprise the corresponding Nambu--Jona-Lasinio models where the
effective fermion interaction has been eliminated by the introduction of
auxiliary bosonic fields. In summary, we find that the linear quark meson model
exhibits in the chiral
limit a high temperature second order chiral transition at zero chemical 
potential (cf.\ section \ref{TheQuarkMesonModelAtTNeq0}) and a first order 
high density chiral transition at zero temperature. By
continuity these transitions meet at a tricritical point in the
($\mu,T$)--plane.  Away from the chiral limit, the second
order chiral transition turns into a smooth crossover. The first order line
of transitions at low temperatures now terminates in a critical endpoint
in the Ising universality class with long--range 
correlations~\cite{BerRaj99,HJSSV}. 

It is an interesting question to what extent the phase
diagram for the NJL-type models reflects features of
two-flavor QCD. The prominent property of QCD which is missing
in the NJL-models is confinement which binds the quarks to color-neutral
baryons. As a consequence one should use for QCD baryons instead
of quarks as the relevant fermionic degrees of freedom for low 
momenta or low $k$. For the high temperature behavior at zero 
density the effects of confinement are presumably not too important.
The reason is the effective decoupling of the quarks for $k
\stackrel{\scriptstyle<}{\sim}$ 300 MeV due to their constituent
mass. This decoupling would only be enhanced by the binding
to nucleons and we do not expect qualitative changes\footnote{There
may be some quantitative influence of this effect on the value
of $T_c$, however.}. The situation is different for the high
density behavior at zero temperature \cite{BJW99chem}. The mass of the 
fermions has an important influence on their Fermi surface. The high
density transition in the NJL-type models describes in the chiral
limit a transition from a ``constituent quark liquid'' at low
density to a ``massless quark liquid'' at high density. Especially
the low density constituent quark liquid has no direct correspondence
in QCD where one rather encounters a gas of nucleons and a liquid 
of nuclear matter.

\paragraph{Acknowledgements:}
The work of J.B.\ is supported in part by funds provided by the 
U.S.~Department of Energy (D.O.E.) under cooperative research 
agreement \# DE-FC02-94ER40818. The work of N.T.\ is supported by 
the E.C.\ under contract Nos.\ ERBFMRXCT960090 and ERBFMBICT983132.
The work of C.W.\ is supported in part by the Deutsche 
Forschungsgemeinschaft and by the E.C.\ contract ERBFMRX-CT97-0122.

\appendix

\section[\hspace*{2.1cm} Threshold functions]{Threshold functions}
\applabel{ThreshApp}

In this appendix we list the various definitions of 
threshold functions appearing in the flow equations and the
expressions for the anomalous dimensions. They involve the
inverse scalar average propagator  for the $IR$ cutoff (\ref{2.15})
\begin{equation}
  \label{BBB01}
  P(q)=q^2+Z_{\Phi,k}^{-1}R_k(q)=
  \frac{q^2}{1-\exp\left\{-\frac{q^2}{k^2}\right\}}
\end{equation}
and the corresponding fermionic function $P_F$ which can be chosen
as (cf.\ section \ref{FlowFerm})
\begin{equation}
  \label{JJJ000}
  P_F(q)=P(q)\equiv q^2\left(1+r_F(q)\right)^2\; .
\end{equation}
We abbreviate
\begin{equation}
  \label{LLL23}
  x= q^2\; ,\;\; P(x)\equiv P(q)\; ,\;\;
  \dot{P}(x)\equiv\frac{\partial}{\partial x}P(x)\; ,\;\;
  \tilde{\partial}_t \dot{P}\equiv
  \frac{\partial}{\partial x}\tilde{\partial}_t P\; ,
\end{equation}
etc., and define
\begin{equation}
  \label{LLL24}
  \dsp{\tilde{\partial}_t}
  \equiv \dsp{
    \frac{1}{Z_{\Phi,k}}\frac{\partial R_k}{\partial t}
    \frac{\partial}{\partial P} }
  + \dsp{
    \frac{2}{Z_{\psi,k}} \frac{P_F}{1+r_F}
    \frac{\partial \left[Z_{\psi,k} r_F\right]}{\partial t}
    \frac{\partial}{\partial P_F} }\; .
\end{equation}
The bosonic threshold functions read
\begin{equation}
  \label{LLL20}
  \begin{array}{rcl}
    \dsp{l_n^d (w;\eta_\Phi)} &=& \dsp{
      l_n^d (w) - \eta_\Phi\hat{l}_n^d (w)
      }\nnn
    &=& \dsp{
      \frac{n+\delta_{n,0}}{2} 
      k^{2n-d} \int_0^\infty d x\, x^{\frac{d}{2}-1}
      \left(\frac{1}{Z_{\Phi,k}} \frac{\partial R_k}{\partial t}\right)
        \left( P+w k^2\right)^{-(n+1)} }\nnn
    \dsp{ l_{n_1,n_2}^d(w_1,w_2;\eta_\Phi)} &=& \dsp{
      l_{n_1,n_2}^d(w_1,w_2)
      -\eta_\Phi \hat{l}_{n_1,n_2}^d(w_1,w_2)}\nnn
    &=& \dsp{
      -\hal k^{2(n_1+n_2)-d}
      \int_0^\infty dx\, x^{\frac{d}{2}-1}
      \tilde{\partial}_t \left\{\left(
      P+w_1k^2\right)^{-n_1}
      \left( P+w_2k^2\right)^{-n_2} \right\}}
  \end{array}
\end{equation}
where $n,n_1,n_2\ge0$ is assumed. For $n\neq0$ the
functions $l_n^d$ may also be written as
\begin{equation}
  \label{LLL21}
  l_n^d (w;\eta_\Phi) =
  -\hal k^{2n-d} \int_0^{\infty} dx x^{\frac{d}{2}-1}
  \tilde{\partial}_t \left( P+w k^2\right)^{-n}\; .
\end{equation}
The fermionic integrals $l_n^{(F)d} (w;\eta_\psi)=l_n^{(F)d} (w)-
\eta_\psi\check{l}_n^{(F)d} (w)$ are defined analogously as
\begin{equation}
  \label{JJJ001}
  \begin{array}{rcl}
    \dsp{l_n^{(F)d} (w;\eta_\psi)} &=& \dsp{
      \left(n+\delta_{n,0}\right)
      k^{2n-d} \int_0^\infty d x\, x^{\frac{d}{2}-1}
      \frac{1}{Z_{\psi,k}}\frac{P_F}{1+r_F}
      \frac{\partial\left[ Z_{\psi,k}r_F\right]}{\partial t}
        \left(P+w k^2\right)^{-(n+1)} }\; .
  \end{array}
\end{equation}
Furthermore one has
\begin{equation}
  \label{LLL22}
  \begin{array}{rcl}
    \dsp{l_{n_1,n_2}^{(FB)d}(w_1,w_2;\eta_\psi,\eta_\Phi)} &=& \dsp{
      l_{n_1,n_2}^{(FB)d}(w_1,w_2)
      -\eta_\psi \check{l}_{n_1,n_2}^{(FB)d}(w_1,w_2)
      -\eta_\Phi \hat{l}_{n_1,n_2}^{(FB)d}(w_1,w_2) 
      }\nnn
    && \dsp{ \hspace{-2cm}
      = -\hal k^{2(n_1+n_2)-d}
      \int_0^\infty dx\, x^{\frac{d}{2}-1}
      \tilde{\partial}_t\left\{
      \frac{1}{[P_F(x)+k^2w_1]^{n_1} [P(x)+k^2w_2]^{n_2} } \right\}
      }\nnn
    \dsp{m_{n_1,n_2}^d (w_1,w_2;\eta_\Phi)} &\equiv& \dsp{
      m_{n_1,n_2}^d (w_1,w_2) - \eta_\Phi 
      {m}_{n_1,n_2}^d (w_1,w_2) 
      }\nnn
    && \dsp{\hspace{-2cm} 
      = -\hal k^{2(n_1+n_2-1)-d}
      \int_0^\infty dx\, x^{\frac{d}{2}}
      \tilde{\partial}_t \left\{
      \frac{\dot{P} (x)}
      {[P(x)+k^2 w_1]^{n_1} }
      \frac{\dot{P} (x)}
      {[P(x)+k^2 w_2]^{n_2}} \right\} 
      }\nnn
  \dsp{m_2^{(F)d} (w;\eta_\psi)} &=& \dsp{
    m_2^{(F)d} (w)-\eta_\psi \check{m}_2^{(F)d} (w)
    }\nnn
  &=& \dsp{
    -\hal k^{6-d}
    \int_0^\infty dx\, x^{\frac{d}{2}}
    \tilde{\partial}_t 
    \left( \frac{\dot{P}_F(x)}{[P_F(x)+k^2w]^2}\right)^2
    }\nnn
    \dsp{m_4^{(F)d} (w;\eta_\psi)} &=& \dsp{
      m_4^{(F)d} (w)-\eta_\psi \check{m}_4^{(F)d} (w) 
      }\nnn
    &=& \dsp{ 
      -\hal k^{4-d}
      \int_0^\infty dx\, x^{\frac{d}{2}+1}
      \tilde{\partial}_t \left(
      \frac{\partial}{\partial x}
      \frac{1+r_F(x)}{P_F(x)+k^2w}\right)^2
      \label{m4Fd} 
      }\nnn
    \dsp{m_{n_1,n_2}^{(FB)d}(w_1,w_2;\eta_\psi,\eta_\Phi)} &=& \dsp{
      m_{n_1,n_2}^{(FB)d}(w_1,w_2)
      -\eta_\psi \check{m}_{n_1,n_2}^{(FB)d}(w_1,w_2)
      -\eta_\Phi {m}_{n_1,n_2}^{(FB)d}(w_1,w_2) 
      }\nnn
    && \dsp{ \hspace{-2cm}
      = -\hal k^{2(n_1+n_2-1)-d}
      \int_0^\infty dx\, x^{\frac{d}{2}}
      \tilde{\partial}_t \left\{
      \frac{1+r_F(x)}{[P_F(x)+k^2w_1]^{n_1}}
      \frac{\dot{P}(x)}{[P(x)+k^2w_2]^{n_2}} \right\} \; . 
      }
  \end{array}
\end{equation}
The dependence of the threshold functions on the anomalous dimensions
arises from the $t$--derivative acting on $Z_{\Phi,k}$ and
$Z_{\psi,k}$ within $R_k$ and $Z_{\psi,k}r_{F}$, respectively.  We
furthermore use the abbreviations
\begin{equation}
  \label{LLL25}
  \begin{array}{rcl}
  \dsp{l_n^d(\eta_\Phi)\equiv l_n^d(0;\eta_\Phi)} &,& \dsp{
    l_{n}^{(F)d}(\eta_\psi)\equiv l_{n}^{(F)d}(0;\eta_\psi)}\nnn
  \dsp{l_n^d(w)\equiv l_n^d(w;0)} &,& \dsp{
  l_n^d\equiv l_n^d(0;0)}
  \end{array}
\end{equation}
etc.~and note that in four dimensions the integrals
\begin{equation}
  \label{LLL26}
  l_2^4(0,0)=l_2^{(F)4}(0,0)=
  l_{1,1}^{(FB)4}(0,0)=m_4^{(F)4}(0)=m_{1,2}^{(FB)4}(0,0)=1
\end{equation}
are independent of the particular choice of the infrared cutoff.

\section[\hspace*{2.1cm} Anomalous dimension in the sharp cutoff limit]
{Anomalous dimension in the sharp cutoff limit}
\applabel{adsc}

It is instructive to evaluate $\xi_k$ as defined by eq. (\ref{102})
in the sharp cutoff limit.
In this limit one has
\beq\label{106}
M^{-1}_0(\rho,q^2)=(Z_k(\rho,q^2)q^2+U_k'(\rho))^{-1}\Theta(q^2-k^2)\eeq
and
\beq\label{107}
\frac{\partial_tR_k(p)}{M^2_0(\rho,p^2)}=\frac{2}{Z_k}(z(\rho)
+u'(\rho))^{-1}\delta(p^2-k^2)\eeq
with similar expressions for $M^{-1}_1$ and $\partial_t R_k/M^2_1$.
The momentum integration in $\partial_tG^{-1}$ (\ref{101})
therefore reduces to an angular integration for the angle between
$p$ and $q$, $(pq)=|p|\ |q|\cos\theta$.
For an evaluation of $\partial_tG^{-1}(\rho,k^2)$
one has $q^2=p^2=k^2$ and
\beq\label{108}
M^{-1}_0(\rho,(p+q)^2)=[2k^2Z_k(\rho,2k^2(1+\cos\theta))\ (1
+\cos\theta)+U_k']^{-1}\Theta(1+2\cos\theta)\eeq
Let us define for $p^2=q^2=k^2,(pq)=k^2
\cos\theta$
\bea\label{110}
&&\tilde\lambda_k^{(1)}(\rho,s)=Z^{-2}_kk^{d-4}\lambda_k^{(1)}(
\rho;p,q)\ ,\
\tilde\lambda_k^{(2)}(\rho,s)=Z^{-2}_kk^{d-4}\lambda_k^{(2)}(
\rho;q,-q,p),\\
&&\tilde\lambda_k^{(3)}(\rho,s)=Z^{-2}_kk^{d-4}\lambda_k^{(2)}(
\rho;q,p,-q)\ ,\
\tilde\gamma_k^{(2)}(\rho,s)=Z^{-3}_kk^{2d-6}\gamma_k^{(2)}(
\rho;p,-p,q))\nonumber\eea
where we have introduced the variable
\beq\label{1100}
s=2(1+\cos\theta)\eeq
With
\beq\label{111}
\lambda^{(1)}(\rho;-q-p,q)=Z_k^2k^{4-d}\tilde\lambda^{(1)}(\rho,
2-\sqrt s)\eeq
one finds\footnote{For $d\not= 3$ the integration measure contains an
additional Jacobian $J^{(d)}(s)$.}
 for $d=3$
\bea\label{112}
&&\partial_tG^{-1}(\rho,k^2)=\frac{1}{2}v_dk^2Z_k
\Bigl[4\tilde\rho\int^4_1ds\nonumber\\
&&\{(\tilde\lambda^{(1)}(\tilde\rho,s))^2(z+\tilde\rho\tilde y+
u'+2\tilde\rho u'')^{-1}[s(z+\ \Delta z(\tilde\rho,s))+u']^{-1}\nonumber\\
&&+(\tilde\lambda^{(1)}(\tilde\rho,2-\sqrt s))^2(z+u')^{-1}[s(
z+\tilde\rho\tilde y+\Delta z(\tilde\rho,s)+\tilde\rho\Delta\tilde y
(\tilde\rho,s))+u'+2\tilde\rho u'']^{-1}\}\nonumber\\
&&-\int^4_0ds\{(z+u')^{-1}((N-1)\tilde\lambda^{(2)}(\tilde\rho,s)
+2\tilde\lambda^{(3)}(\tilde\rho,s))\nonumber\\
&&+(z+\tilde\rho\tilde y+u'+2\tilde\rho u'')^{-1}
(\tilde\lambda^{(2)}(\tilde\rho,s)+2\tilde\rho\tilde\gamma
^{(2)}(\tilde\rho,s))\}\Bigr]\eea
Next one uses the relations (\ref{G9})
\bea\label{113}
\tilde\lambda^{(1)}(\tilde\rho,s)&=&u''(\tilde\rho)+
\frac{s}{2}(z'(\tilde\rho)+\Delta z'(\tilde\rho,\frac{s}{2}))
+\frac{1}{2}\tilde y(\tilde\rho)+\Delta\tilde\lambda^{(1)}
(\tilde\rho,s)\nonumber\\
\tilde\lambda^{(2)}(\tilde\rho,s)&=&u''(\tilde\rho)+2z'(\tilde\rho)
+\Delta\tilde\lambda^{(2)}(\tilde\rho,s)\nonumber\\
\tilde\lambda^{(3)}(\tilde\rho,s)&=&u''(\tilde\rho)+(2-s)(z'(\tilde
\rho)+\Delta z'(\tilde\rho,2-s))+\frac{1}{2}s(\tilde y(\tilde\rho)
+\Delta\tilde
y(\tilde\rho,s))+\Delta\tilde\lambda^{(3)}(\tilde\rho,s)\nonumber\\
\tilde\gamma^{(2)}(\tilde\rho,s)&=&u'''(\tilde\rho)+z''
(\tilde\rho)+\frac{1}{2}
\tilde y'(\tilde\rho)+\Delta\tilde\gamma^{(2)}(\tilde\rho,s)\eea
where
\bea\label{114}
\Delta\tilde\lambda^{(1)}(\tilde\rho,0)&=&0\ ,\ \Delta\tilde\lambda
^{(2,3)}(\tilde\rho)=\frac{1}{4}\int^4_0ds\Delta\tilde\lambda
^{(2,3)}(\tilde\rho,s),\nonumber\\
\Delta\tilde\gamma^{(2)}(\tilde\rho)&=&\frac{1}{4}\int^4_0ds
\Delta\tilde\gamma^{(2)}(\tilde\rho,s)\eea
and finally finds the exact expression
\bea\label{115}
&&\xi_k(\tilde\rho,1)=-2v_d\tilde\rho\int^4_1 ds\{(u''+\frac{s}{2}(z'+
\Delta z'(\frac{s}{2}))+\frac{1}{2}\tilde  
y+\Delta\tilde\lambda^{(1)}(s))^2\nonumber\\
&&[s(z+\Delta z(s))+u']^{-1}[z+\tilde\rho\tilde y+u'+2\tilde\rho u'']^{-1}
\nonumber\\
&&+(u''+(1-\frac{\sqrt{s}}{2})(z'+\Delta z'(2-\sqrt s))+\frac{1}{2}\tilde  
y+\Delta\tilde\lambda^{(1)}(2-\sqrt s))^2\nonumber\\
&&[s(z+\tilde\rho\tilde y
+\Delta z(s)+\tilde\rho\Delta\tilde y(s))+u'+2\tilde\rho u'']^{-1}[z+u']^{-1}
+2v_d\{(z+u')^{-1}\nonumber\\
&&[2u''+(N+3-2s)z'+
2(2-s)\Delta z'(2-s)+s\tilde y+s\Delta \tilde y(s)+(N-1)
\Delta\tilde\lambda
^{(2)}+2\Delta\tilde\lambda^{(3)}]\nonumber\\
&&-(z+\tilde\rho\tilde y+u'+2\tilde\rho u'')^{-1}[2u''-z'-
2\tilde\rho z''+\tilde y-\Delta\tilde\lambda^{(2)}-2\tilde\rho
\Delta\tilde \gamma^{(2)}]\}\eea
The first order in the hybrid derivative expansion neglects
the momentum-dependent corrections $\Delta\tilde\lambda^{(i)}, \Delta
\tilde\gamma^{(2)},\Delta z$ and $\Delta\tilde y$. This yields the
evolution equation for $z(\tilde\rho)$
\bea\label{116}
\partial_tz&=&\eta z+(d-2+\eta)\tilde\rho z'+2v_d\tilde\rho\
(u''+\sigma_zz'+\frac{1}{2}\tilde y)^2\nonumber\\
&&\{\frac{1}{z}\ln\left(\frac{4z+u'}{z+u'}\right)
(z+\tilde\rho\tilde y+u'+2\tilde\rho u'')^{-1}\nonumber\\
&&+
\frac{1}{z+\tilde\rho\tilde y}\ln\left(\frac{4(z+\tilde\rho\tilde y)
+u'+2\tilde\rho u''}{z+\tilde\rho\tilde y+u'+2
\tilde\rho u''}\right)(z+u')^{-1}\}
\nonumber\\
&&-2v_d\{(z+u')^{-1}[2u''+(N+2)z'+\frac{1}{2}\tilde y]\nonumber\\
&&-(z+\tilde\rho\tilde y+u'+2\tilde\rho u'')^{-1}
[2u''-z'-2\tilde\rho z''+\tilde y]\}\eea
where we have replaced for simplicity in the third term the
correct $s$-integration of the terms $\sim z'(z')^2$ by an approximate
expression with $\sigma_z\approx0.5-1$. The anomalous dimension
reads for $\kappa>0$

\bea\label{117}
\eta&=&(1+\kappa z_0')^{-1}\Bigl\{[(2-d)\kappa-\partial_t\kappa]
z_0'\nonumber\\
&&-2v_d\kappa(\lambda+\sigma_zz_0'+\frac{1}{2}\tilde y_0)^2
\Bigl((1+\kappa \tilde y_0+2\lambda\kappa)^{-1}\ln 4\nonumber\\
&&+(1+\kappa\tilde y_0)^{-1}
\ln\left(\frac{4(1+\kappa\tilde y_0)+2\lambda\kappa}{1+\kappa\tilde  
y_0+2\lambda\kappa}\right)\Bigr)\nonumber\\
&&+2v_d\Bigl[\frac{2\lambda(2\lambda\kappa+\kappa\tilde y_0)}
{1+2\lambda\kappa+\kappa\tilde y_0}+(N+2)z_0'+\frac{1}{2}
\tilde y_0\nonumber\\
&&+(1+2\lambda\kappa+\kappa\tilde y_0)^{-1}(z_0'
+2\kappa z_0''-\tilde y_0)\Bigr]\Bigr\}\eea
where we have defined
\beq\label{118}
\lambda=u''(\kappa)\ ,\ z_0'=z'(\kappa)\ ,\ z_0''=z''(\kappa)\ ,\
\tilde y_0=\tilde y(\kappa)\eeq
We observe that eq. (\ref{116}) is well defined as long
as $z,z+u',z+\tilde\rho\tilde y$ and $z+\tilde\rho\tilde y+u'+
2\tilde\rho u''$ remain all positive. These are the same
conditions as those required for a consistent flow of the potential.
The problems that a sharp cutoff engenders for a definition
of the anomalous dimension at $q^2=0$ are avoided by the use
of the hybrid derivative expansion with the definition (\ref{103}).

For large $N$ the characteristic scaling $\tilde\rho\sim N,\ u'\sim 1,\
u''\sim 1/N$ implies for the solution of eq. (\ref{116})
$z'\sim 1/N^2, z''\sim a/N^3$. On the other hand the evolution equation
for the inverse radial propagator $\tilde G^{-1}
(\rho,q^2)=M_1(\rho,q^2)-R_k(q)$ leads to $\tilde y\sim 1/N$
and $\tilde y$ contributes in order $\eta\sim 1/N$. Taking
nevertheless $\tilde y=0$ for a first discussion, one finds
the anomalous dimension
\beq\label{119}
\eta=2v_d\lambda^2\kappa\left(\frac{4}{1+2\lambda\kappa}
-\frac{\ln 4}{1+2\lambda\kappa}-\ln\left(\frac{4+2
\lambda\kappa}{1+2\lambda\kappa}\right)\right)+2v_dNz_0'-\kappa z_0'\eeq
For $d=3$ one may insert for the scaling solution the leading
expression $\lambda\kappa=1$ (cf. (\ref{89d}))
so that
\beq\label{120}
\eta=\frac{1}{N}\{\frac{4}{3}-\frac{1}{3}\ln 4-\ln 2\}=
\frac{0.178}{N}\eeq
For a computation of the exact expression for $\eta$ in
order $1/N$ one needs to include effects from $\tilde y_0\not=0$. Also
the contribution $-2(\partial\Delta z(\kappa,y)/\partial y)(y=1)$
has to be added in order $1/N$. In the same approximation as above
the scaling solution for $z(\tilde\rho)$ obeys
the differential equation
\bea\label{121}
&&\tilde\rho z'+2v_d\tilde\rho(u'')^2\Bigl\{(1+u'+2\tilde\rho u'')^{-1}
\ln\left(\frac{4+u'}{1+u'}\right)\nonumber\\
&&+(1+u')^{-1}\ln
\left(\frac{4+u'+2\tilde\rho u''}{1+u'+2\tilde\rho u''}\right)\Bigr\}
\nonumber\\
&&-2v_d
\left\{\frac{2u''+Nz'}{1+u'}-\frac{2u''}{1+u'+2\tilde\rho u''}\right\}=
-\eta z\eea
By differentiation with respect to $\tilde\rho$ eq. (\ref{121})
implies for the scaling regime
\bea\label{122}
&&z_0'(1+2v_dN\lambda)+z_0''(\kappa-2v_dN)=\nonumber\\
&&-2v_d\Bigl\{\frac{\lambda}{(1+2\lambda\kappa)^2}(\lambda
+2\gamma\kappa-\lambda\kappa(\lambda-2\gamma\kappa))\ln 4\nonumber\\
&&+\lambda(\lambda+2\gamma\kappa-\lambda^2\kappa)\ln\left(
\frac{4+2\lambda\kappa}{1+2\lambda\kappa}\right)\nonumber\\
&&-\lambda^2\kappa\left[\frac{1}{1+2\lambda\kappa}(\frac{15}{4}\lambda
+2\gamma\kappa)-\frac{3\lambda+2\gamma\kappa}{4+2\lambda\kappa}\right]
\nonumber\\
&&+2\lambda^2-2\gamma+\frac{2\gamma}{1+2\lambda\kappa}-
\frac{2\lambda(3\lambda+2\gamma\kappa)}{(1+2\lambda\kappa)^2}\Bigr\}
\eea
For $d=3, \ 2v_dN=\kappa,\ \lambda\kappa=1,\ \gamma\kappa=
2\lambda/3$, this yields
\beq\label{123}
N\kappa z_0'=\frac{89}{216}-\frac{4}{27}\ln 4-\frac{2}{3}
\ln2\eeq
and we see indeed $z_0'\sim 1/N^2$.


\end{document}